\documentclass[preprint, times]{elsarticle}

\usepackage{amssymb}
\usepackage{subfigure} 
\usepackage{graphicx}
\usepackage{amsmath}
\usepackage{algorithmic}
\usepackage{algorithm}
\usepackage{bm}
\usepackage{geometry}
\usepackage{ntheorem}
\usepackage{graphicx}
\usepackage{array}
\usepackage{makecell}
\usepackage{hyperref}
\usepackage{xcolor}
\usepackage{soul}

\RequirePackage{CJKnumb}

\journal{elsevier}

\begin{document}

\begin{frontmatter}



\title{Hybrid deep learning and iterative methods for accelerated solutions of viscous incompressible flow}


\author[inst1]{Heming Bai}
\author[inst1]{Xin Bian\corref{cor1}}
\ead{bianx@zju.edu.cn}

\affiliation[inst1]{organization={State Key Laboratory of Fluid Power and Mechatronic Systems, Department of Engineering Mechanics, Zhejiang University},
            city={Hangzhou},
            postcode={310027}, 
            country={China}}

\cortext[cor1]{Corresponding author}

\begin{abstract}
The pressure Poisson equation, central to the fractional step method in incompressible flow simulations, incurs high computational costs due to the iterative solution of large-scale linear systems. To address this challenge, we introduce HyDEA~(Hybrid Deep lEarning line-search directions and iterative methods for Accelerated solutions), a novel framework that synergizes deep learning with classical iterative solvers. It leverages the complementary strengths of a deep operator network~(DeepONet) -- capable of capturing large-scale features of the solution -- and the conjugate gradient~(CG) or a preconditioned conjugate gradient (PCG)~(with Incomplete Cholesky, Jacobi, or Multigrid preconditioner) method, which efficiently resolves fine-scale errors. 
Specifically,
within the framework of line-search methods, the DeepONet predicts search directions to accelerate convergence in solving sparse, symmetric-positive-definite linear systems,
while the CG/PCG method ensures robustness through iterative refinement.
The framework seamlessly extends to flows over solid structures via the decoupled immersed boundary projection method, preserving the linear system's structure.
Crucially, the DeepONet is trained on {\it fabricated} linear systems rather than flow-specific data,
endowing it with inherent generalization across geometric complexities and Reynolds numbers without retraining.
Benchmarks demonstrate HyDEA's superior efficiency and accuracy over the CG/PCG methods for flows with no obstacles, single/multiple stationary obstacles, and one moving obstacle -- using {\it fixed network weights}. 
Remarkably, HyDEA also exhibits super-resolution capability: although the DeepONet is trained on a $128 \times 128$ grid for Reynolds number $Re=1000$, the hybrid solver delivers accurate solutions on a $512 \times 512$ grid for $Re=10,000$ via interpolation, despite discretizations mismatch. In contrast, a purely data-driven DeepONet fails for complex flows, underscoring the necessity of hybridizing deep learning with iterative methods.
HyDEA's robustness, efficiency, and generalization across geometries, resolutions, and Reynolds numbers highlight its potential as a transformative solver for real-world fluid dynamics problems.

\end{abstract}




\end{frontmatter}


\section{Introduction}
\label{introduction}

Computational fluid dynamics~(CFD) has revolutionized our understanding of fluid mechanics and enabled innovative real-world applications through advanced numerical methods~\cite{wendt2009}. 
However, solving the Navier-Stokes~(NS) equations remains computationally expensive, especially for unsteady or geometrically complex flows that require substantial resources.
This challenge is compounded by the necessity of performing separate simulations for each flow condition, rendering conventional CFD approaches inefficient for engineering applications like active flow control~\cite{rabault2019artificial} and structural shape optimization~\cite{viquerat2021direct}. These limitations underscore the pressing need for novel techniques to accelerate simulations and minimize redundant computations across similar flow scenarios.

Deep learning has demonstrated remarkable success in fields such as computer vision~\cite{CV_He} and natural language processing~\cite{zhang2016NLP}, and its strong predictive capabilities have recently extended to solving partial differential equations~(PDEs). Physics-informed neural networks~(PINNs)~\cite{2019PINNs,Cai2021}, for instance, embed PDE constraints directly into training via automatic differentiation, enabling solutions without relying solely on large datasets. Despite their promise, PINNs still face challenges in computational efficiency and accuracy in fluid dynamics~\cite{chuang2022experiencePINNs,karnakov2024solving, zhu2024physics}, calling for further methodological advances~\cite{Wang2022110768, Yu2022114823}.

Alternatively, data-driven surrogate models trained on high-fidelity data offer fast solutions to NS equations~\cite{lee2019data, nakamura2021r26}, bypassing costly simulations during deployment. However, these models often generalize poorly beyond their training distribution. 
Recent efforts to address this fall into two categories:\\
{\it 1. Novel architectures:} Operator learning frameworks~\cite{chenandchen1995Universal}, such as Deep operator network~(DeepONet)~\cite{lu2021DeepONet} and Fourier neural operator~\cite{li2020FNO}, approximate PDE solutions by encoding spatio-temporal dynamics.
They have been applied to Burgers' equation~\cite{lu2021DeepONet},  multiphase flows~\cite{Lin_Maxey_Li_Karniadakis_2021, wen2022uFNOmultiphase}, steady compressible flows~\cite{mao2021DeepMMNet}, and unsteady incompressible flows~\cite{bai2024DON, han2025unsteadyflowFNO}.
Yet, their performance remains tied to the distribution of the operator parameters sampled in the training dataset.\\
{\it 2. Training strategies:} Via pre-training and fine-tuning operations, meta-learning~\cite{zhang2023Metalearning}, self-supervised learning~\cite{xu2024pof},
and transfer learning~\cite{goswami2022} improve adaptability to new flow conditions.
However, they usually require supplementary data, which may be unavailable in practical applications. 

The critical limitation of data-driven models is their inability to generalize to out-of-training distribution, such as varying Reynolds numbers or modified boundary conditions that alter the underlying dynamics.
This recognition has spurred the development of {\it hybrid methods},
which integrate deep learning with classical numerical methods rather than outright eliminating the latter. 
For example, neural networks~\cite{yang2016data_poisson, tompson2017accelerating, ajuria2020, 2022chenJinLi} have been used to accelerate solution to the pressure Poisson equation~(PPE) in the fractional step methods\cite{chorin1967, perot1993_fractional_step_method}, a major computational bottleneck in incompressible NS simulations. 
However, these models are employed as end-to-end mapping for one shot at each instance and do not improve its own prediction with iterative efforts.
More importantly, they often rely on flow-specific data, leading to performance degradation for out-of-distribution scenarios, a limitation shared by purely data-driven models.
Kaneda et al.~\cite{kaneda2023DCDM} proposed the deep conjugate direction method~(DCDM), which trains on the PPE coefficient matrix rather than flow data, leading to superior generalization. 
While promising, DCDM has only been tested on Euler equations, leaving its efficacy for NS equations unexplored. Additionally, it suffers from boundary inaccuracies due to its voxel-based fluid-solid distinction.

Another fundamental challenge is {\it spectral bias}~\cite{rahaman2019spectral}: neural networks excel at capturing low-frequency features but struggle with high-frequency details. Whereas classical iterative methods, including standard relaxation methods and Krylov subspace techniques, efficiently attenuate high-frequency error components but converge slowly for low-frequency modes without heavily employing advanced techniques such as multigrid or proper preconditioning~\cite{Briggs2015, Golub2013}. These complementary behaviors have inspired a new paradigm of hybrid methods that combine data-driven models with numerical methods to exploit their respective strengths.
Notable advances include HINTs~\cite{zhang2024blending, kahana2023geometryHINTs}, Multi-scale neural computing~\cite{suo2025novel}, 
DeepONet-preconditioned Krylov methods~\cite{kopanivcakova2025deeponetHybrid}, and
PINN-MG~\cite{dong2024pinnMG},
but their direct application to nonlinear PDEs such as NS equations remains underexplored. Chen et al.~\cite{chen2025WTCNNMG} introduced a wavelet-based neural multigrid method for periodic flows. While the construction of training dataset is innovative, the mapping from source terms to pressure fields based on neural network remains susceptible to distribution shifts, such as unseen obstacles. Notably, the DCDM although effective for generalization due to matrix-derived training, depends exclusively on neural networks~\cite{kaneda2023DCDM}, leaving it vulnerable to spectral bias, which may reduce robustness and accuracy in complex flows.

Despite progress, no hybrid framework has successfully unified data-driven efficiency with numerical robustness while remaining independent of flow-specific training data. 
Addressing this gap could unlock scalable, high-fidelity solvers for fluid dynamics and beyond. To tackle this challenge, we propose HyDEA~(Hybrid Deep lEarning line-search directions and iterative methods for Accelerated solutions) -- a novel framework for solving the PPE in incompressible flow simulations.
Based on a finite difference discretization, the PPE amounts to a large, sparse, symmetric, positive-definite linear system of equations. 
The solution to these equations is formulated as an extremum-seeking problem of a quadratic function for the solution vector.
Within the framework of line-search methods~\cite{nocedal2006numerical}, a DeepONet is employed
to predict new search directions at each iteration, which is in sharp contrast to 
other classical line-search directions such as those derived from steepest descent~(SD) and conjugate gradient~(CG) methods~\cite{nocedal2006numerical, shewchuk1994introductionCG}.
It is termed as Deep learning Line-Search Method or DLSM in brevity, for later reference.
Furthermore, HyDEA integrates DLSM with the CG method or one of its preconditioned variants~(collectively referred to as CG-type methods), creating a synergistic approach that:
\begin{itemize}
\item Preserves generalization: training on fabricated linear systems biased toward the lower spectrum of the PPE matrix, avoiding flow-specific data;
\item Integrates DLSM with a CG-type method to balance neural acceleration and numerical stability;
\item Incorporates fluid-structure interaction with decoupled immersed boundary projection method (DIBPM)~\cite{li2016DIBPM},  which preserves the linear system's structure for extensive generalization.
\end{itemize}
Subsequently, HyDEA adopts {\it identical weights} for neural network and
demonstrates through a series of benchmarks that:
\begin{itemize}
 \item Achieves superior convergence rate compared to the standalone CG-type methods,
 \item Maintains generalization capability across diverse flow regimes with/without solid structures, and also with super-resolutions.
 \item Outperforms DCDM and DLSM in both stability and accuracy for complex flows.
\end{itemize}
By synergistically combing the strengths of neural networks and iterative methods, HyDEA represents a significant step toward scalable, high-fidelity PDE solver capable of handling complex multi-physics simulations. 

The remainder of this work is structured as follows: Section~\ref{method} presents the numerical foundation, beginning with the fractional step method for incompressible NS equations and DIBPM for solid structures, followed by a detailed exposition of the proposed HyDEA methodology. This section also covers the construction of training dataset, neural network architecture, and model training protocols. Section~\ref{result} demonstrates HyDEA’s performance through multiple benchmark cases with comprehensive analysis of convergence acceleration and generalization capability. Finally, Section~\ref{conclusion} concludes with key findings and outlines promising directions for future research.

\section{Methodology}
\label{method}

\subsection{Fractional step method for incompressible Navier-Stokes equations}
\label{Fractional step method}

The governing equations are the incompressible NS equations
expressed in non-dimensional form as
\begin{eqnarray}
\label{momentum equation}
 \frac{\partial{\mathbf{u}}}{\partial{t}} + \mathbf{u}\cdot\nabla_{}\mathbf{u} &=& -\nabla_{}p + \frac{1}{Re}\nabla_{}^{2}\mathbf{u},
\\
\label{con equation}
  \nabla_{}\cdot\mathbf{u} &=& 0,
\end{eqnarray}
where $\mathbf{u}$, $p$ and $Re$ represent the velocity vector, pressure and Reynolds number, respectively. 
The numerical solution of the NS equations is challenging,
as the pressure term is coupled to the velocity and needs to be updated to implicitly satisfy the incompressibility constraint. The fractional step method is a classical algorithm, which effectively decouples the velocity and pressure. Following Ref.~\cite{perot1993_fractional_step_method}, the NS equations can be discretized and derived into a system of algebraic equations as
\begin{eqnarray}
\label{Discretized equation}
 \begin{bmatrix}
  A & G \\
  D & 0 
\end{bmatrix}
\begin{bmatrix}
  \mathbf{u}^{n+1} \\
  \delta p 
\end{bmatrix} = 
\begin{bmatrix}
 \mathbf{r}^{n} + bc_{1} \\
 bc_{2} \\
\end{bmatrix},
\end{eqnarray}
where $A$ represents the implicit operator matrix for the advection-diffusion component, $D$ is the divergence operator matrix, $G$ is the gradient operator matrix, and $\mathbf{r}^{n}$ represents the explicit terms of the discrete momentum equation. Additionally, $bc_1$ and $bc_2$ are boundary condition vectors for the momentum equation and the incompressibility constraint, respectively. $\mathbf{u}^{n+1}$ is the unknown velocity vector. $\delta p=p^{n+1}-p^{n}$,  and this form can reduce the splitting error of the fractional step method~\cite{dukowicz1992_lowspliterror}.
The structure of $A$ and $\mathbf{r}^{n}$ critically depends on the specific time advancement scheme and mesh configuration. In this work, the NS equations are discretized by the finite difference method on a staggered-mesh, utilizing the explicit second-order Adams-Bashforth scheme for the convective terms and the implicit Crank-Nicolson method for the diffusion terms.

In view of Ref.~\cite{perot1993_fractional_step_method}, the fractional step method can be regarded as a block LU decomposition of Eq.~(\ref{Discretized equation}),
\begin{eqnarray}
\label{Fractional_blockLU}
 \begin{bmatrix}
  A & 0 \\
  D & -DA^{-1}G 
 \end{bmatrix}
 \begin{bmatrix}
  I & A^{-1}G \\
  0 & I 
 \end{bmatrix}
 \begin{bmatrix}
  \mathbf{u}^{n+1} \\
  \delta p 
 \end{bmatrix} = 
 \begin{bmatrix}
 \mathbf{r}^{n} + bc_{1} \\
 bc_{2} \\
 \end{bmatrix}.
\end{eqnarray}
The direct calculation of $A^{-1}$ in Eq.~(\ref{Fractional_blockLU}) is prohibitively expensive, thus an approximate solution for $A^{-1}$ is employed. The Taylor series expansion is applied to $A^{-1}$ and only its first term is kept, $A^{-1} \approx \Delta t$. Subsequently, Eq.~(\ref{Fractional_blockLU}) transforms to be
\begin{eqnarray}
\label{Fractional_blockLU_appro}
 \begin{bmatrix}
  A & 0 \\
  D & -\Delta tDG 
 \end{bmatrix}
 \begin{bmatrix}
  I & \Delta tG \\
  0 & I 
 \end{bmatrix}
 \begin{bmatrix}
  \mathbf{u}^{n+1} \\
  \delta p 
 \end{bmatrix} = 
 \begin{bmatrix}
  A & 0 \\
  D & -\Delta tDG 
 \end{bmatrix}
 \begin{bmatrix}
  \mathbf{u}^{\ast} \\
  \delta p 
 \end{bmatrix} =
 \begin{bmatrix}
 \mathbf{r}^{n} + bc_{1} \\
 bc_{2} \\
 \end{bmatrix}.
\end{eqnarray}
Eq.~(\ref{Fractional_blockLU_appro}) is typically rewritten in the form of three sequential steps:
\begin{eqnarray}
\label{firststep_fr}
  A\mathbf{u}^{\ast} &=& \mathbf{r}^{n} + bc_{1},
\\
\label{secondstep_fr}
  \Delta tDG\delta p &=& D\mathbf{u}^{\ast}-bc_{2},
\\
\label{thirdstep_fr}
  \mathbf{u}^{n+1} &=& \mathbf{u}^{\ast} - \Delta tG\delta p,
\end{eqnarray}
where $\mathbf{u}^{\ast}$ is the intermediate velocity vector. 
Eq.~(\ref{secondstep_fr}) represents the discrete PPE, where $DG$ is a symmetric-positive-definite matrix. 
Classical iterative methods, such as CG-type methods~\cite{nocedal2006numerical, shewchuk1994introductionCG}, can be employed to iteratively solve Eq.~(\ref{secondstep_fr}) for $\delta p$, which is the most computationally intensive part of the whole solution process. 

Finally, the pressure vector is updated as follow:
\begin{eqnarray}
\label{pressure update}
 p^{n+1} &=& p^{n} + \delta p.
\end{eqnarray}

\subsection{Decoupled immersed boundary projection method}
\label{DIBPM}

The immersed boundary method~(IBM) is a numerical technique for handling fluid-structure interactions, originally developed by Peskin et al.~\cite{peskin1972heartIBM} to simulate blood flow in cardiovascular systems. The method incorporates a forcing term $\mathbf{f}$ into NS equations as
\begin{eqnarray}
\label{momentum equation IBM}
 \frac{\partial{\mathbf{u}}}{\partial{t}} + \mathbf{u}\cdot\nabla_{}\mathbf{u} &=& -\nabla_{}p + \frac{1}{Re}\nabla_{}^{2}\mathbf{u}+\mathbf{f},
\\
\label{con equation IBM}
  \nabla_{}\cdot\mathbf{u} &=& 0.
\end{eqnarray}
Taira et al.~\cite{Taira_2007_IBPM} proposed the immersed boundary projection method~(IBPM) based on the fractional step method of Perot et al.~\cite{perot1993_fractional_step_method}. However, this proposition leads the pressure variable to couple with the forcing term, resulting in a substantially high computational cost. To address this issue, Li et al.~\cite{li2016DIBPM} developed the decoupled IBPM or DIBPM expressed in a two-step block LU decomposition, which effectively decouples the velocity, pressure and forcing term.

In the framework of DIBPM, fluid dynamics is expressed in Eulerian framework with Cartesian coordinates, while the immersed boundary is defined at Lagrangian points $\mathbf{X}(\mathbf{s},t)$, with $\mathbf{s}$ and $t$ denoting the curvilinear coordinates and time, respectively. The force and velocity at $\mathbf{X}(\mathbf{s},t)$ are denoted by $\mathbf{F}(\mathbf{s},t)$ and $\mathbf{U}(\mathbf{s},t)$. The variables defined at two coordinates impose each other with a delta function as \begin{eqnarray}
\label{H operator}
  \mathbf{f}(\mathbf{s},t) &=& \int_{\Gamma}\mathbf{F}(\mathbf{s},t)\delta(\mathbf{x}-\mathbf{X}(\mathbf{s},t))d\mathbf{s},
\\
\label{E operator}
  \mathbf{U}(\mathbf{s},t) &=& \int_{\Omega}\mathbf{u}(\mathbf{x},t)\delta(\mathbf{X}(\mathbf{s},t)-\mathbf{x})d\mathbf{x},
\end{eqnarray}
where $\Gamma$ is fluid-solid boundary in Lagrangian curvilinear coordinate, $\Omega$ is fluid domain in Eulerian coordinate. The discretization of Eq.~(\ref{H operator}) and Eq.~(\ref{E operator}) are represented by the operator matrices $H$ and $E$, which correspond to the discrete regularization and interpolation operators, respectively. In this work, the discrete delta function of Roma et al.~\cite{roma1999adaptive_delta} is utilized. Ultimately, the governing equations are discretized and expressed as a system of algebraic equations as
\begin{eqnarray}
\label{Discretized equation_IBM}
 \begin{bmatrix}
  A & -H & G \\
  E & 0  & 0 \\
  D & 0  & 0
\end{bmatrix}
\begin{bmatrix}
  \mathbf{u}^{n+1} \\
  \delta \mathbf{F}  \\
  \delta p 
\end{bmatrix} = 
\begin{bmatrix}
 \mathbf{r}^{n} + bc_{1} \\
 \mathbf{U_{B}}  \\
 bc_{2} 
\end{bmatrix},
\end{eqnarray}
where $\delta \mathbf{F}=\mathbf{F}^{n+1}-\mathbf{F}^{n}$, and $\mathbf{U_B}$ is the Lagrangian velocity on the immersed boundary. 

Expressed in a two-step block LU decomposition by DIBPM, the pressure decoupling process is initially conducted with Eq.~(\ref{Discretized equation_IBM}) being rewritten in the following form, 
\begin{eqnarray}
\label{Discretized equation_IBM_rewritten}
 \begin{bmatrix}
  \bar{A} & \bar{G} \\
  \bar{D} & 0 
\end{bmatrix}
\begin{bmatrix}
  \mathbf{z}^{n+1} \\
  \delta p 
\end{bmatrix} = 
\begin{bmatrix}
 \mathbf{\bar{r}}^{n} \\
 bc_{2} \\
\end{bmatrix}.
\end{eqnarray}
Here $\bar{A}=\begin{bmatrix}
  A & -H \\
  E & 0 
\end{bmatrix}$, $\bar{G}=\begin{bmatrix}
  G \quad
  0  
\end{bmatrix}^T$, $\bar{D}=\begin{bmatrix}
  D \quad
  0  
\end{bmatrix}$, $\mathbf{z}^{n+1}=\begin{bmatrix}
  \mathbf{u}^{n+1} \quad
  \delta \mathbf{F}  
\end{bmatrix}^T$, and $\mathbf{\bar{r}}^{n}=\begin{bmatrix}
  \mathbf{r}^{n}+bc_1 \quad
  \mathbf{U_{B}}  
\end{bmatrix}^T$.
Similar to the fractional step method without solid structure~\cite{perot1993_fractional_step_method}, the first block LU decomposition of Eq.~(\ref{Discretized equation_IBM_rewritten}) is performed,
\begin{eqnarray}
\label{Fractional_blockLU_appro_firstLU}
 \begin{bmatrix}
  \bar{A} & 0 \\
  \bar{D} & -\Delta t\bar{D}\bar{G} 
 \end{bmatrix}
 \begin{bmatrix}
  I & \Delta t\bar{G} \\
  0 & I 
 \end{bmatrix}
 \begin{bmatrix}
  \mathbf{z}^{n+1} \\
  \delta p 
 \end{bmatrix} = 
 \begin{bmatrix}
  \bar{A} & 0 \\
  \bar{D} & -\Delta t\bar{D}\bar{G} 
 \end{bmatrix}
 \begin{bmatrix}
  \mathbf{z}^{\ast} \\
  \delta p 
 \end{bmatrix} =
 \begin{bmatrix}
 \mathbf{\bar{r}}^{n} \\
 bc_{2} \\
 \end{bmatrix}.
\end{eqnarray}
Eq.~(\ref{Fractional_blockLU_appro_firstLU}) can be rewritten in the form of  three sequential steps as
\begin{eqnarray}
\label{firststep_fr_IBM1}
  \bar{A}\mathbf{z}^{\ast} &=& \bar{\mathbf{r}}^{n},
\\
\label{secondstep_fr_IBM1}
  \Delta t\bar{D}\bar{G}\delta p &=& \bar{D}\mathbf{z}^{\ast}-bc_{2},
\\
\label{thirdstep_fr_IBM1}
  \mathbf{z}^{n+1} &=& \mathbf{z}^{\ast} - \Delta t\bar{G}\delta p.
\end{eqnarray}
Subsequently, according to the specific definitions of $\bar{A}$, $\bar{G}$, $\bar{D}$,  $\mathbf{z}^{n+1}$ and $\bar{\mathbf{r}}^{n}$, Eq.~(\ref{firststep_fr_IBM1})-(\ref{thirdstep_fr_IBM1}) can be further rewritten as
\begin{eqnarray}
\label{firststep_fr_IBM2}
  \begin{bmatrix}
  A & -H \\
  E & 0 
  \end{bmatrix}
  \begin{bmatrix}
  \mathbf{u}^{\ast} \\
  \delta \mathbf{F} 
  \end{bmatrix} &=& 
  \begin{bmatrix}
 \mathbf{r}^n + bc_{1} \\
 \mathbf{U_{B}} \\
  \end{bmatrix}, \\
\label{secondstep_fr_IBM2}
  \Delta tDG\delta p &=& D\mathbf{u}^{\ast}-bc_{2},
\\
\label{thirdstep_fr_IBM2}
  \mathbf{u}^{n+1} &=& \mathbf{u}^{\ast} - \Delta tG\delta p.
\end{eqnarray}
As the intermediate velocity $\mathbf{u}^{\ast}$ remains coupled with $\delta \mathbf{F}$, the second LU block decomposition is employed for Eq.~(\ref{firststep_fr_IBM2}):
\begin{eqnarray}
\label{Fractional_blockLU_second}
 \begin{bmatrix}
  A & 0 \\
  E & \Delta tEH 
 \end{bmatrix}
 \begin{bmatrix}
  I & -\Delta tH \\
  0 & I 
 \end{bmatrix}
 \begin{bmatrix}
  \mathbf{u}^{\ast} \\
  \delta \mathbf{F} 
 \end{bmatrix} &=& 
 \begin{bmatrix}
 \mathbf{r}^{n} + bc_{1} \\
 \mathbf{U_{B}} \\
 \end{bmatrix}.
\end{eqnarray}
Eq.~(\ref{Fractional_blockLU_second}) can be rewritten in a similar form as Eq.~(\ref{firststep_fr_IBM1})-(\ref{thirdstep_fr_IBM1}) as
\begin{eqnarray}
\label{firststep_fr_secondLU_IBM}
  A\mathbf{u}^{\ast\ast} &=& \mathbf{r}^{n} + bc_{1},
\\
\label{secondstep_fr_secondLU_IBM}
  \Delta tEH\delta \mathbf{F} &=& \mathbf{U_{B}}-E\mathbf{u}^{\ast\ast},
\\
\label{thirdstep_fr_secondLU_IBM}
  \mathbf{u}^{\ast} &=& \mathbf{u}^{\ast\ast} + \Delta tH\delta \mathbf{F},
\end{eqnarray}
where $\mathbf{u}^{\ast\ast}$ is the intermediate velocity vector of $\mathbf{u}^{\ast}$.

In summary, the complete DIBPM computation process is outlined successively as follows:
\begin{eqnarray}
\label{1step_fr_secondLU_IBM_summary}
  A\mathbf{u}^{\ast\ast} &=& \mathbf{r}^{n} + bc_{1},
\\
\label{2step_fr_secondLU_IBM_summary}
  \Delta tEH\delta \mathbf{F} &=& \mathbf{U_{B}}-E\mathbf{u}^{\ast\ast},
\\
\label{3step_fr_secondLU_IBM_summary}
  \mathbf{u}^{\ast} &=& \mathbf{u}^{\ast\ast} + \Delta tH\delta \mathbf{F},
\\
\label{4step_fr_secondLU_IBM_summary}
 \Delta tDG\delta p &=& D\mathbf{u}^{\ast}-bc_{2},
\\
\label{5step_fr_secondLU_IBM_summary}
 \mathbf{u}^{n+1} &=& \mathbf{u}^{\ast} - \Delta tG\delta p,
\\
\label{6step_fr_secondLU_IBM_summary}
 p^{n+1} &=& p^{n} + \delta p, 
\\
\label{7step_fr_secondLU_IBM_summary}
 \mathbf{F}^{n+1} &=& \mathbf{F}^{n} + \delta \mathbf{F}.
\end{eqnarray}
The fractional step and DIBPM methods employed in this work are based on the open-source CFD solver $\mathtt{PetIBM}$~\cite{chuang2018petibm}.

\subsection{HyDEA: Hybrid Deep lEarning line-search directions and iterative methods for Accelerated solutions}
\label{HyDEA}

The PPE in Eq.~(\ref{secondstep_fr}) or (\ref{4step_fr_secondLU_IBM_summary}) can be expressed in a compact form:
\begin{eqnarray}
\label{discretLinearEquation}
 M\delta p = S,
\end{eqnarray}
where matrix $M \in \mathbf{R}^{a \times a}$ represents the discrete approximation of the Laplace operator, which is sparse, symmetric and positive-definite. $\delta p$ is unknown and $S$ is the source term, 
where $\delta p \in \mathbf{R}^a$ and $S \in \mathbf{R}^a$. The solution to Eq.~(\ref{discretLinearEquation}) is equivalent to solving a quadratic vector optimization problem~\cite{Golub2013, shewchuk1994introductionCG},
\begin{eqnarray}
\label{min_Vector_quadratic_F}
  \min_{\delta p \in \mathbf{R}^a} g(\delta p) \stackrel{\text{def}}{=} \frac{1}{2}\delta p^{T}M\delta p-S^{T}\delta p+c.
\end{eqnarray}
Line search methods~\cite{nocedal2006numerical} provide a framework for solving Eq.~(\ref{min_Vector_quadratic_F}), where solution $\delta p_{k}$ at $k$th iteration is progressively updated through a line-search direction $d_k$ with step length $\alpha_{k}$ to minimize the objective function $g(\delta p)$:
\begin{eqnarray}
\label{alpha_lineResearch}
 \alpha_{k} &=& \operatorname*{\arg\min}_{\alpha}g(\delta p_{k}+\alpha d_{k}) = \frac{r_{k}^{T}d_{k}}{d_{k}^T M d_{k}},
 \\
\label{lineResea_update}
 \delta p_{k+1} &=& \delta p_{k} + \alpha_{k}d_{k},
\end{eqnarray}
where $r_{k}=S-M\delta p_{k}$ denotes the iterative residual vector. 

A different choice of $d_{k}$ corresponds to a different variant in the family of line search methods. When $d_{k}$ is taken as $r_{k}$, it is known as steepest descent~(SD) method. However, the SD method exhibits suboptimal convergence performance. Furthermore, the conjugate direction~(CD) methods can be viewed as an extension of the SD method and they construct a set of $M$-orthogonal vectors \{$d_{0}, d_{1},\cdots, d_{k}$\}~($d_{i}^{T}Md_{j}=0$ for $i \neq j$) as line-search directions at each iteration through Gram-Schmidt~(GS) orthogonalization. The conjugate gradient~(CG) method represents a specific implementation of the CD methods, where each residual vector $r_{k}$ is employed in the GS process. 

The convergence performance of a line-search method is fundamentally determined by the quality of the selected line-search directions. Theoretically, if the line-search direction aligned exactly with the current iteration error $e_{k}=\delta p^{true}- \delta p_k$, the line-search method would converge to the exact solution in a single iteration. However, this theoretical scenario is practically unattainable since $e_{k}$ is unknown. Recognizing this fundamental limitation, we propose to utilize a deep neural network, namely DeepONet, to predict an error $e_{k}^{NN}$ as a line-search direction at each iteration. By combining $e_{k}^{NN}$ with the optimal step length given in Eq.~(\ref{alpha_lineResearch}), this strategy will prove to achieve rapid convergence within a few number of iterations. We denote this novel methodology as Deep learning Line-Search Method~(DLSM) for later reference.
More specifically, Eq.~(\ref{discretLinearEquation}) is transformed into the residual form at each iteration as
\begin{eqnarray}
\label{discretLinearEquation_delta}
 Me_{k} = M\Delta(\delta p) = r_{k}.
\end{eqnarray}
DLSM employs a deep neural network to provide the mapping between the $r_{k}$ and $e_{k}$, with the algorithmic implementation formally described in Algorithm 1. The DCDM of Kaneda et al.~\cite{kaneda2023DCDM} represents another example of a line-search method, where the line-search directions are made additionally $M$-orthogonal by a GS process. In the present work, both DLSM and DCDM employ identical neural network architecture, training dataset and training methodology. We shall demonstrate in Section~\ref{OsCylinder} that the GS orthogonalization is unnecessary, as a properly trained neural network already suggests effective line-search directions.

\begin{algorithm}[!h]
    \caption{DLSM: Deep learning Line-Search Method}
    \label{alg1}
    
    \begin{algorithmic}[1]
        \STATE $r_{0}=S-M\delta p_{0}$
        \STATE $k=0$
        
        \WHILE{$||r_{k}||_2>\epsilon$}

            \STATE $e_{k}^{NN}=NN(\frac{r_{k}}{||r_{k}||_2})$
            \STATE $k=k+1$
            \STATE $\alpha_{k-1}=\frac{r_{k-1}^{T}e^{NN}_{k-1}}{(e^{NN}_{k-1})^{T}Me^{NN}_{k-1}}$
            \STATE $\delta p_{k} = \delta p_{k-1} + \alpha_{k-1}e^{NN}_{k-1}$
            \STATE $r_{k} = S - M\delta p_{k}$
        \ENDWHILE
    \end{algorithmic}
\end{algorithm}

Due to the inherent limitations of neural networks in capturing high-frequency features~\cite{rahaman2019spectral}, DLSM is potentially incompetent to eliminate high-frequency errors, thus adversely affect the convergence performance as will be demonstrated in Sections~\ref{128Inter} and~\ref{OsCylinder}. 
Inspired by the work of Zhang et al.~\cite{zhang2024blending}, we propose HyDEA, a hybrid strategy that combines DLSM with iterative methods: the DLSM providing a superior global large-scale line-search directions, while iterative methods resolving small-scale errors. Specifically, we focus on employing CG-type methods as representative iterative methods in the HyDEA framework and meanwhile the performance of standalone CG-type methods shall be considered as the baseline for comparisons.

\begin{figure}[htbp]
\centering
  \includegraphics[scale=0.14]{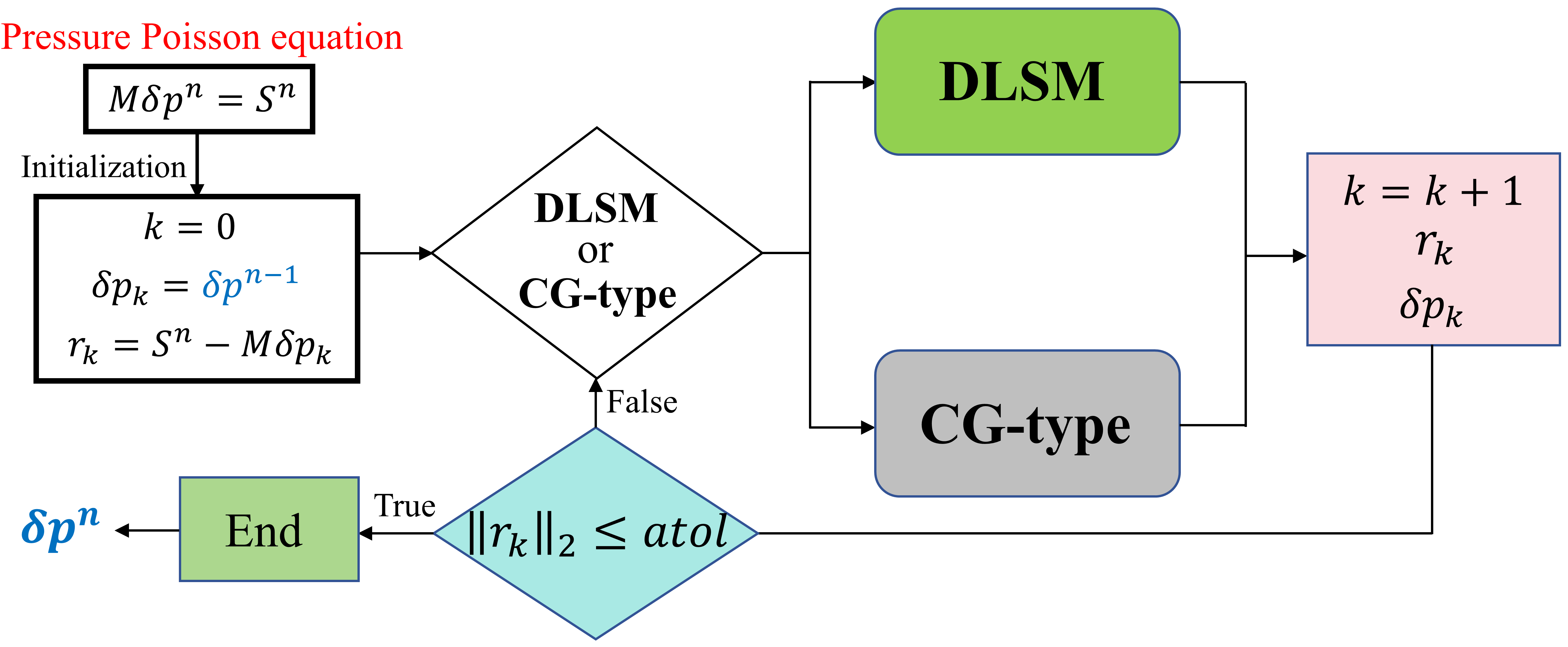}\\
  \caption{The workflow of HyDEA, $n$ denotes the time step for an unsteady CFD simulation, $k$ represents the iteration index for solving the discrete PPE at the current time step, and $atol$ is the predefined absolute tolerance. The initial value for iteration is taken from last time step: $\delta p_k=\delta p^{n-1}$. }\label{HyDEA_workflow}
\end{figure}

The workflow of HyDEA is presented in Fig.~\ref{HyDEA_workflow}, where it receives an initial solution and its corresponding residual for the current CFD time step. Subsequently, it selects to perform whether the DLSM or a CG-type method for the current iteration. After each iteration, the $L2$-norm of the updated residual is evaluated and if it falls below a predefined threshold, the iterative process terminates. Otherwise, the procedure repeats, progressively refining the solution until the convergence criterion is met. Notably, when the procedure switches from DLSM to a CG-type method, the CG-type method's iterative state is completely restarted from scratch, with the DLSM's final solution serving as the initial guess for the subsequent iterations.

\begin{figure}[htbp]
\centering
  \subfigure[]{
  \label{HyDEA_1}
  \includegraphics[scale=0.11]{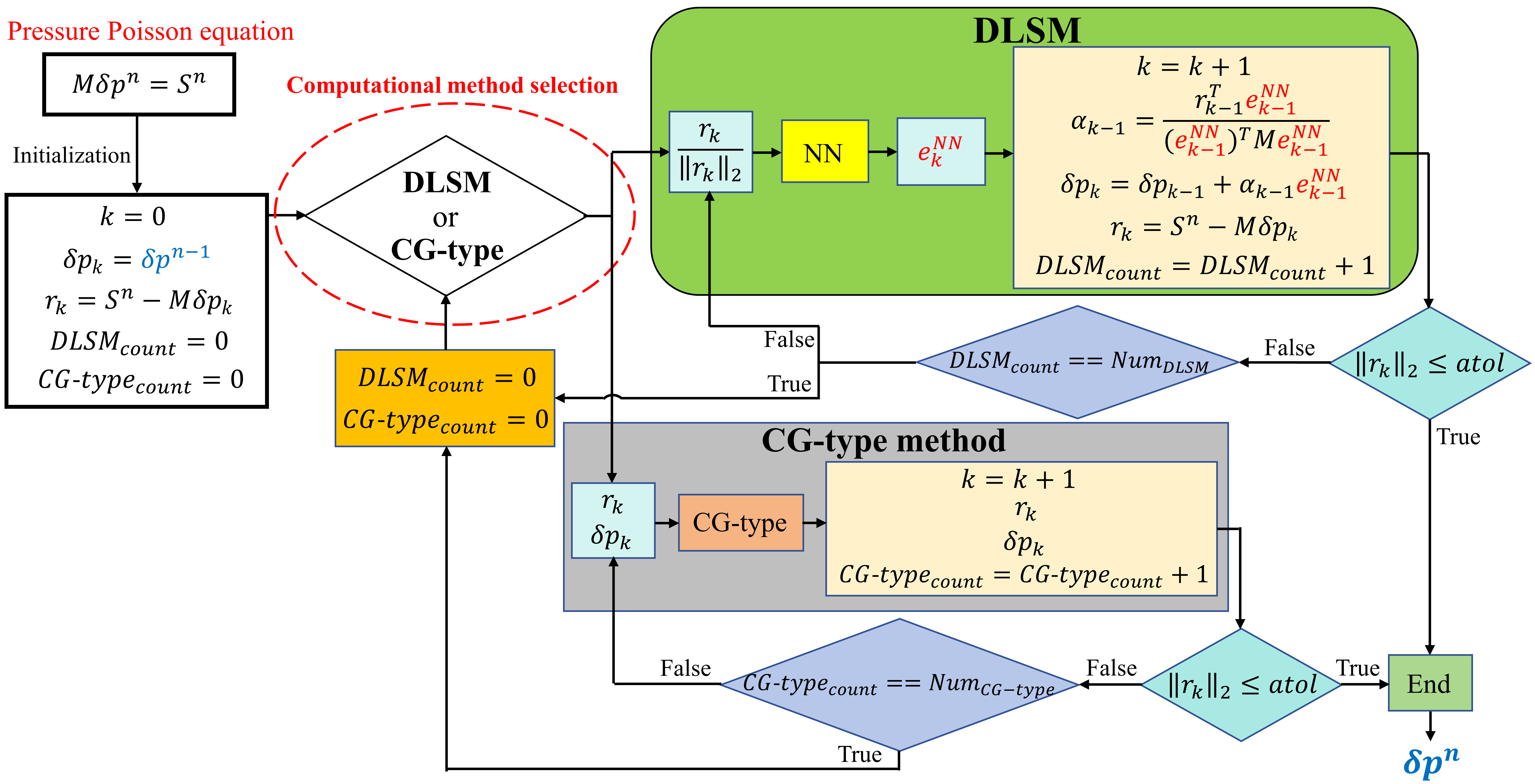}}
  \subfigure[]{
  \label{InitialPCG}
  \includegraphics[scale=0.12]{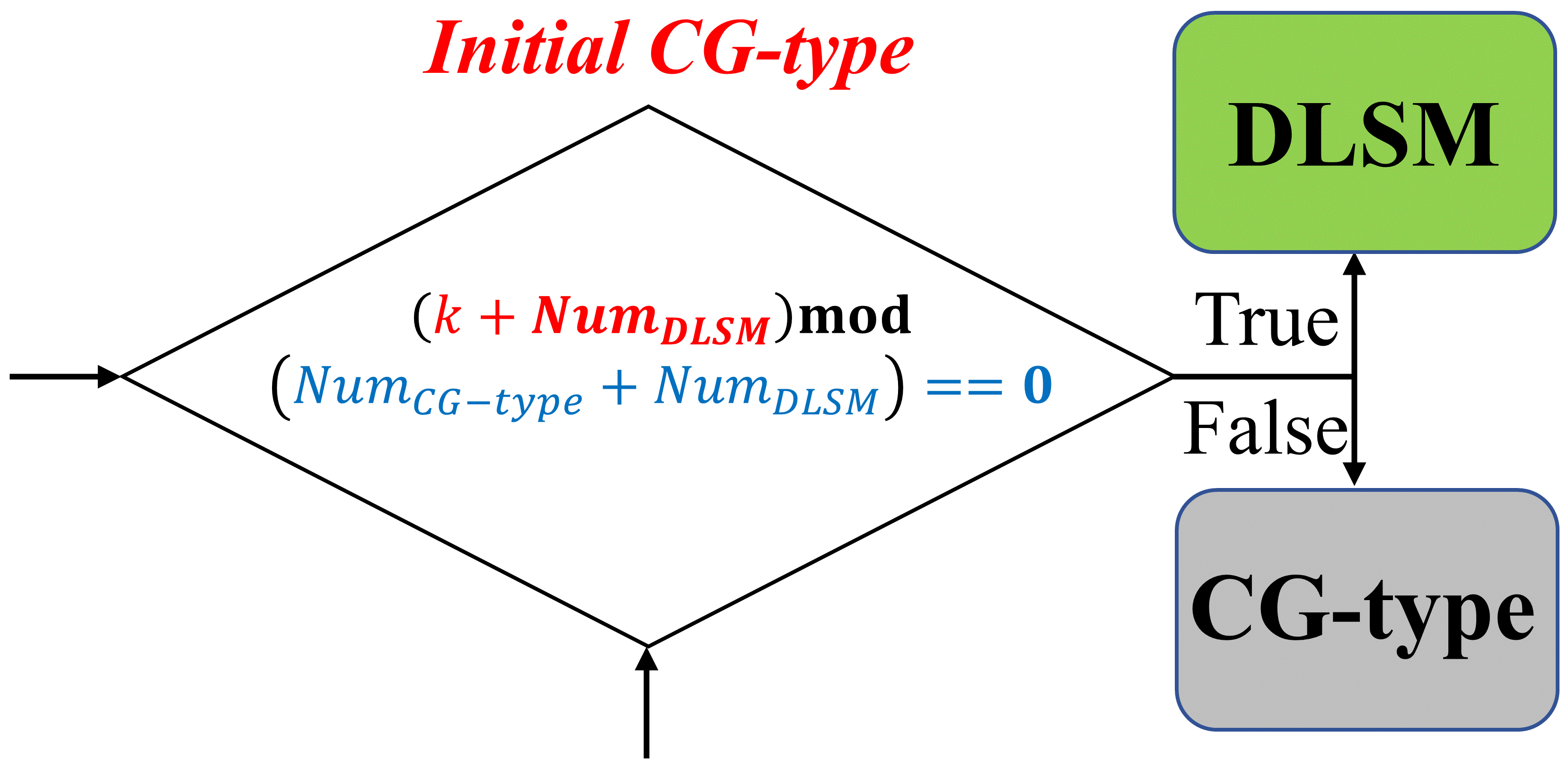}} 
  \subfigure[]{
  \label{InitialNN}
  \includegraphics[scale=0.12]{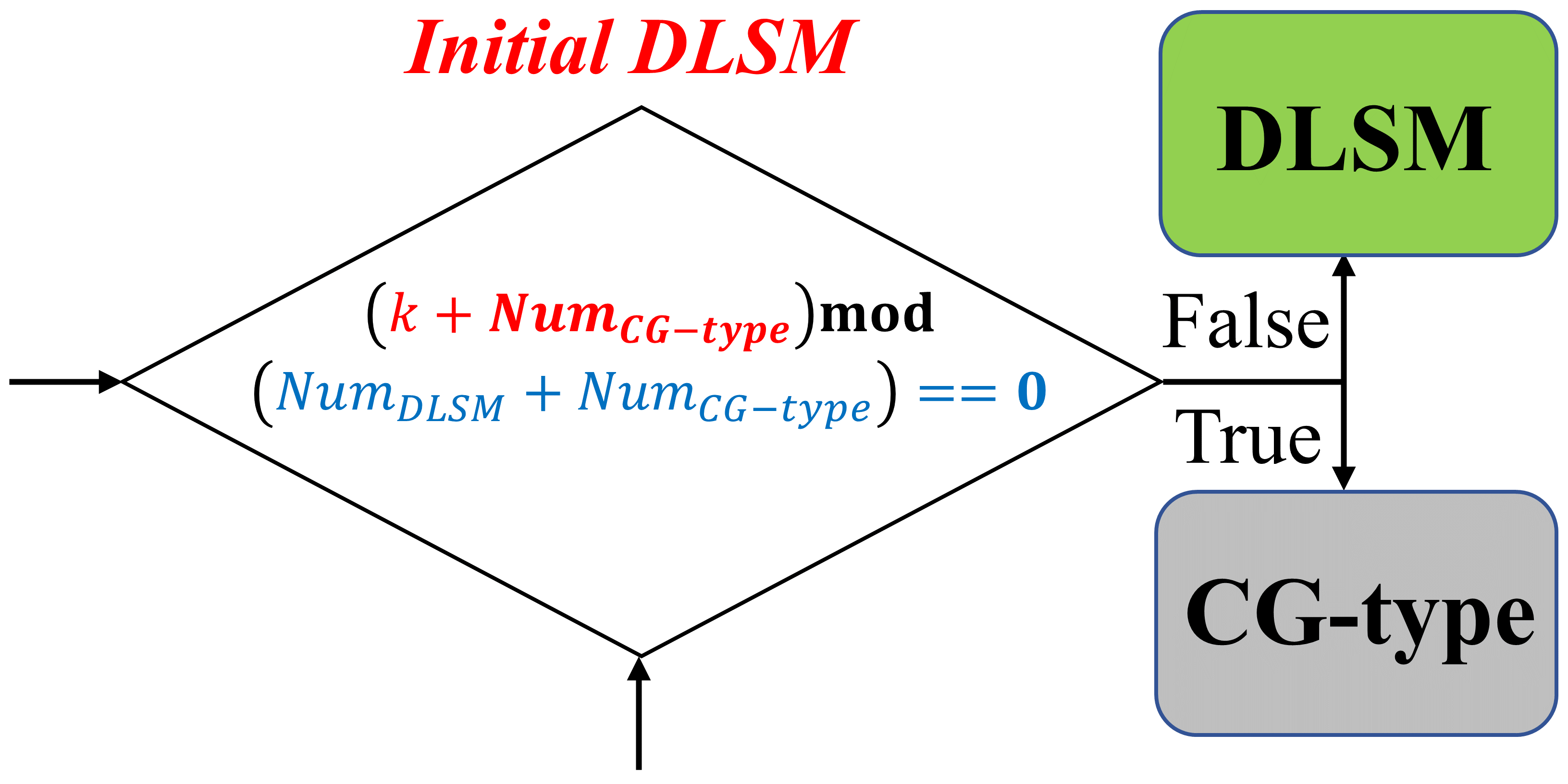}}
  \caption{The implementation details and two versions of HyDEA. (a) Detailed workflow. (b) For each round of the hybrid algorithm, a CG-type method is taken as the initial solver for maximum $Num_{CG-type}$ iterations followed by DLSM for maximum $Num_{DLSM}$ iterations. (c) For each round of the hybrid algorithm, DLSM is taken as the initial solver for maximum  $Num_{DLSM}$ iterations followed by a CG-type method for maximum $Num_{CG-type}$ iterations.}\label{HyDEA_process}
\end{figure}

Further details of HyDEA are illustrated in Fig.~\ref{HyDEA_process}. 
As which solver executed first is potentially essential,
HyDEA offers two distinct choices as indicated by the red-dashed circle in Fig.~\ref{HyDEA_1}.
The two implementations in Fig.~\ref{InitialPCG} and Fig.~\ref{InitialNN}
correspond to the two versions of HyDEA for {\it one round} of the hybrid algorithm: the former represents a CG-type method as the initial solver, while the latter employs DLSM as the initial solver.
Moreover, $k$ represents the current iteration number, and $mod$ is the modulo operation. 
For the two respective methods
${DLSM}_{count}$ and ${CG-type}_{count}$ record the numbers of consecutive iterations
while $Num_{DLSM}$ and $Num_{CG-type}$ denote the predefined maximum consecutive iterations.

\subsection{Dataset and model configuration}
\label{MODEL}

\subsubsection{Dataset construction}
\label{Dataset}

Following the method proposed by Kaneda et al.~\cite{kaneda2023DCDM}, random vectors biased toward the lower end of the spectrum of $M$ are constructed as the training dataset. This dataset construction method is fundamentally rooted in the convergence behavior of the CG method, for which we provide a comprehensive analysis in~\ref{appendixA}.

The critical step in dataset construction is to generate a set of approximate eigenvectors $Q_{m}=[\mathbf{q}_{0},\mathbf{q}_{1},\cdots,\mathbf{q}_{m-1}] \in \mathbf{R}^{a \times m}$~(where $m < a$), which can effectively represent the entire spectrum of $M$. Specifically, the Lanczos iteration first transforms $M$ to a low-dimensional tridiagonal matrix $T_{m}\in \mathbf{R}^{m \times m}$~\cite{lanczos1950iteration}, yielding the relation 
\begin{eqnarray}
\label{Loss}
 T_{m} = V^{T}_{m}MV_{m},
\end{eqnarray}
where $V_{m}\in \mathbf{R}^{a \times m}$ is the orthogonal matrix whose columns consist of the orthogonal vectors generated during the Lanczos iteration. 
Subsequently, the eigenvalues of $T_{m}$ form a diagonal matrix $\Lambda_{m}\in \mathbf{R}^{m \times m}$ with non-decreasing eigenvalues on the diagonal referred to as approximate eigenvalues or Ritz values of $M$.
Moreover, the eigenvectors of $T_{m}$ form a matrix $U_{m}\in \mathbf{R}^{m \times m}$.
Thereafter, $Q_{m}=V_{m} U_{m}$ forms the approximate eigenvectors of $M$, commonly known as Ritz vectors. 
Finally, the dataset is constructed as follows
\begin{eqnarray}
\label{b_construct}
 r^{i} &=& \frac{\sum_{j=0}^{m-1} c_{j}^{i} \mathbf{q}_{j}}{||\sum_{j=0}^{m-1} c_{j}^{i} \mathbf{q}_{j}||_{2}},
\\
\label{c_construct}
 c_{j}^i &=& \begin{cases}
 9 \cdot N(0,1), & \text{if } 0 \leq j \leq b \cdot m \\
  N(0,1), & otherwise
 \end{cases},
\end{eqnarray}
where $i$ denotes distinct indices for diverse data, $N(0,1)$ is a random number drawn from the normal distribution. The hyper-parameter $m$ denotes the number of Lanczos iterations, which directly determines the number of approximate eigenvalues and eigenvectors computed for $M$. The hyper-parameter $b$ controls the frequency characteristics of vectors $r$ in the dataset. Specifically, decreasing the value of $b$ enhances the relative dominance of low-frequency components in the dataset. 
Since values of $m$ and $b$ may affect the quality of the dataset and the overall performance of HyDEA,
we shall perform a sensitivity analysis of the two parameters.

\subsubsection{Deep operator network and model training}
\label{DON and training}
\begin{figure}[htbp]
\centering
  \includegraphics[scale=0.12]{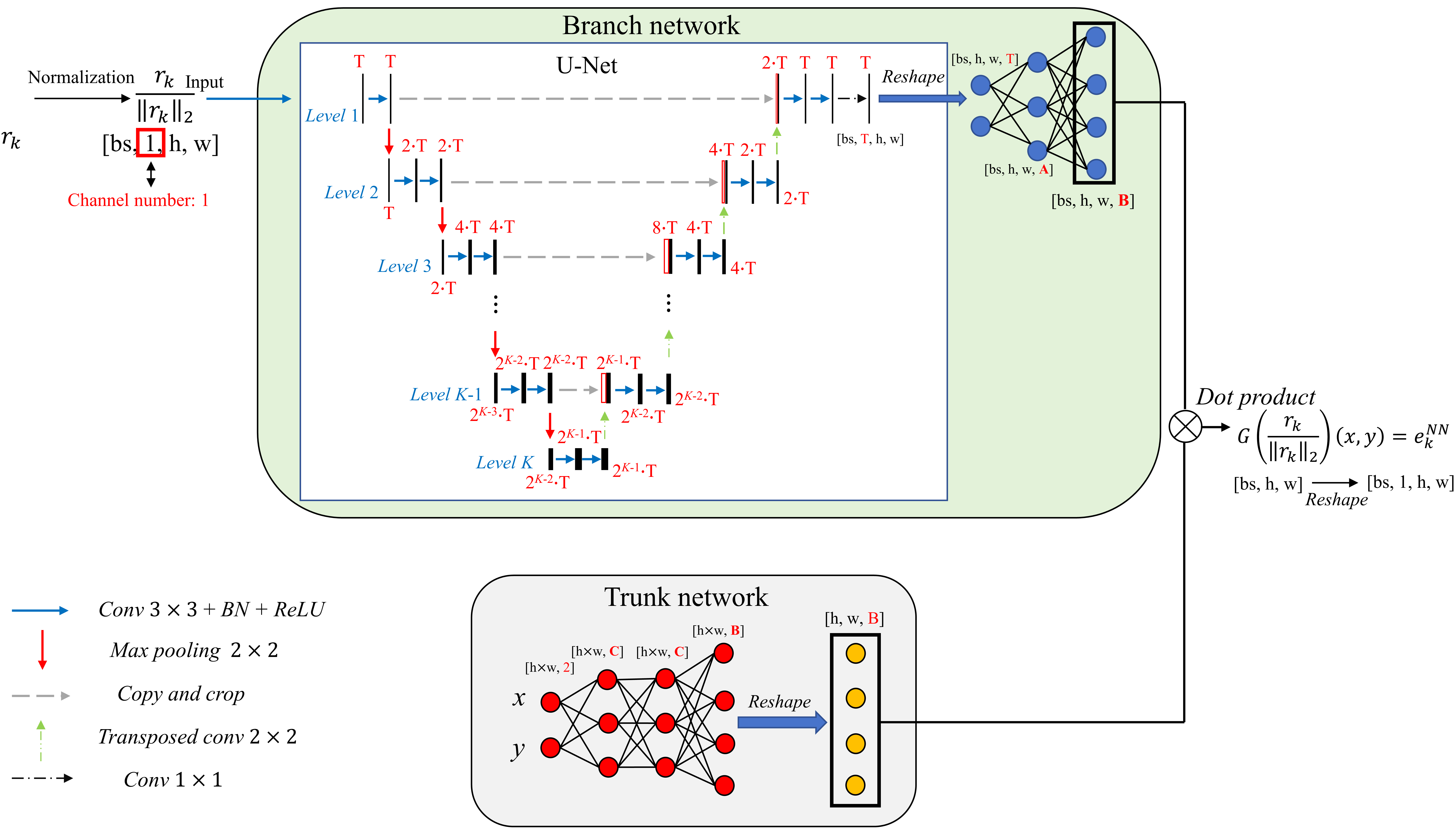}\\
  \caption{The DeepONet architecture.}\label{NN_structure}
\end{figure}

The mapping between $e_{k}$ and $r_k$ in Eq.~(\ref{discretLinearEquation_delta}) can be understood as an operator, which we employ the DeepONet to learn its biased version governed by the dataset.
The unstacked version of DeepONet employed here consists of a branch network and a trunk network~\cite{lu2021DeepONet} as illustrated in Fig.~\ref{NN_structure}. The computational domain is discretized using a uniform Cartesian grid. This representation allows the flow field to be naturally treated as an image, where each grid point corresponds to a pixel, with the values of $r_k$ at different grid points forming the pixel values. The U-Net architecture made of convolutional neural networks~(CNN), known for its exceptional performance in computer vision~\cite{2015U-Net}, is particularly suitable for processing such image-structured data. Therefore, the branch network employs a U-Net concatenated by a feedforward neural network~(FNN). The trunk network applies a simple FNN.
$[bs,c,h,w]$ is the feature map, where $bs$ is $batchsize$, $c$ represents the number of channels, $h$ and $w$ are the sizes of the input in two spatial directions. In U-Net, $T$, $2 \cdot T$, $4 \cdot T$, etc., represent the numbers of feature map channels at specific network layers and $T=40$ by default. $A$, $B$ and $C$ are the number of neurons in FNNs and are assigned as $100$, $200$ and $100$ by default, respectively. $K$ represents the depth of the U-Net architecture. A detailed analysis of these parameters are presented in~\ref{appendixB}. 

Any vector $r$ constructed by Eq.~(\ref{b_construct}) serves as input to the DeepONet, which predicts a vector $NN(r)$ governed by the loss function as
\begin{eqnarray}
\label{Loss}
Loss &=& \frac{1}{N} \sum_{i=1}^{N}[r_{i}-(M \cdot NN(r))_{i}]^{2},
\end{eqnarray}
where $N$ is the total number of the grid points. 
This loss function is more concise than the form given by Kaneda et al.~\cite{kaneda2023DCDM} and is similar to the form suggested in Lan et al.~\cite{lan2023PSDO}.
The random mini-batch~\cite{wandel2022spline} method is employed, with $bs=20$. 
The training process consists of $1000$ epochs and each epoch contains $1000$ iterations. The size of the training dataset is always $54,000$ irrespective of CFD resolutions. The Adam optimizer combined with sharpness aware minimization method~(SAM)~\cite{foret2020SAM} are utilized to update the neural network parameters. The cosine learning rate schedule is utilized and set to $0.0002$. The maximum parameter perturbation range in SAM is set to be $0.0002$~\cite{bai2024DON}.

\subsection{Technical details}

We summarize components, resolutions and applications of HyDEA in Table~\ref{tab:components}.
Each realization of HyDEA consists of a CG-type method and a DLSM or a DCDM.
CG-type methods include the vanilla CG and three preconditioned variants, namely,
incomplete Cholesky decomposition PCG~(ICPCG), Jacobi PCG~(JPCG), and Multigrid PCG~(MGPCG).
Moreover, both DLSM and DCDM have two resolutions: 
\begin{itemize}
    \item DLSM-1~(or DCDM-1) with DeepONet receiving input feature maps of a spatial resolution of $128 \times 128$;
    \item  DLSM-2~(or DCDM-2) with DeepONet receiving input feature maps of a spatial resolution of $192 \times 192$. 
\end{itemize}
For example, HyDEA~(CG+DLSM-1) means that HyDEA hybridizes the vanilla CG method with DLSM for the first resolution.  
The same acronym rule applies to HyDEA~(ICPCG+DLSM-1), HyDEA~(JPCG+DLSM-1), and HyDEA~(MGPCG+DLSM-1).
It also applies to four variants of HyDEA~(xxCG+DLSM-2) for the second resolution.

The same nomenclature also applies to the hybrid algorithm of a CG-type with DCDM, namely, HyDEA (ICPCG+DCDM-2) for the second resolution.

\begin{table}
    \renewcommand{\arraystretch}{1.5}
    \normalsize
    \centering
    \caption{Components and associated resolutions of HyDEA~(CG-type+DLSM) and (CG-type+DCDM) }
    \begin{tabular}{ccccc}
    
    \hline
      HyDEA~(CG-type+Deep learning)   &   DLSM-1 &  DLSM-2  & DCDM-1 & DCDM-2\\
    \hline
      Resolution &  $128\times 128$  & $192\times 192$ & $128\times 128$& $192\times 192$  \\
      CG     & \checkmark  &  \checkmark & - & -\\
      ICPCG  & \checkmark &  \checkmark  & - & \checkmark \\
      JPCG   &  \checkmark & \checkmark  & - & -\\
      MGPCG  &  \checkmark & \checkmark  & - & -\\
      Super-resolution & $640 \times 640 \downarrow$ & -  & $512 \times 512 \downarrow$ & -\\
      Solid structures & - & \checkmark & - & \checkmark  \\
    \hline
    \end{tabular}
    \label{tab:components}
\end{table}

DLSM, DCDM and associated DeepONet are implemented using $\mathtt{Python}$ and $\mathtt{PyTorch}$.
CG-type methods are adopted from the $\mathtt{PETSc}$ library in $C$~\cite{balay2019petsc} with its $\mathtt{Python}$ interface $\mathtt{petsc4py}$.

The interface between $\mathtt{PetIBM}$ and $\mathtt{Python}$ is implemented using the $\mathtt{Pybind11}$ library~\cite{Pybind11}. 

The DeepONet models are trained and deployed on a single NVIDIA GeForce RTX 4090
and all other computations are executed on a single CPU of Intel Xeon Silver 4210R. 

The code and data are available on GitHub at \url{https://github.com/HMB9666/HyDEA}.

\subsection{Error Analysis of HyDEA}
Prior to applying HyDEA to CFD, we perform an error analysis by solving a one-dimensional(1D) Poisson equation with a known, frequency-controllable solution:
\begin{eqnarray}
\label{1D_Poisson_analysis}
 \frac{d^{2}u}{dx^2} &=& f(x), \quad x \in \Omega=[-1,1]
\\
\label{boundary_condition_analysis}
  u(x) &=& g, \quad x \in \partial \Omega
\end{eqnarray}
where $u(x)$ denotes the field function to be computed. The source term $f(x)$ and the boundary values $g$ are prescribed based on the manufactured solution $u_{exact}$:
\begin{eqnarray}
\label{1D_Poisson_exact_analysis}
 u_{exact} =  \sin(x)-\frac{1}{4}\sin(4x)+\frac{1}{8}\sin(8x)-\frac{1}{16}\sin(16x)+ \frac{1}{24}\sin(24x)- \nonumber\\ 
 \frac{1}{48}\sin(48x)+ \frac{1}{128}\sin(128x)-\frac{1}{256}\sin(256x).
\end{eqnarray}

A finite difference discretization is applied to Eq.~(\ref{1D_Poisson_analysis}) on a uniform grid of 2000 nodes, utilizing a second-order central difference scheme for the Laplace operator. The resulting linear equation is solved using both the CG method and HyDEA (CG + DLSM) from a zero initial guess, with the termination residual expressed in $L2$-norm set to $\epsilon=10^{-7}$. To enable a clear observation of the distinct error evolution behaviors of both the CG and DLSM components within HyDEA, $Num_{CG-type}$ and $Num_{DLSM}$ are both set to 10. The neural network architecture employs the DeepONet as illustrated in Fig.~\ref{NN_structure}. The U-Net depth $K$ is set to 7, with the input feature map, coordinates, and convolutional kernels all adapted for 1D configuration. To prepare for the training dataset, we adopt the parameter values $m=800$ and $b=0.6$. All other hyperparameters and training settings remain consistent with those detailed in Section~\ref{DON and training}. 

\begin{figure}[htbp]
\centering
  \includegraphics[scale=0.24]{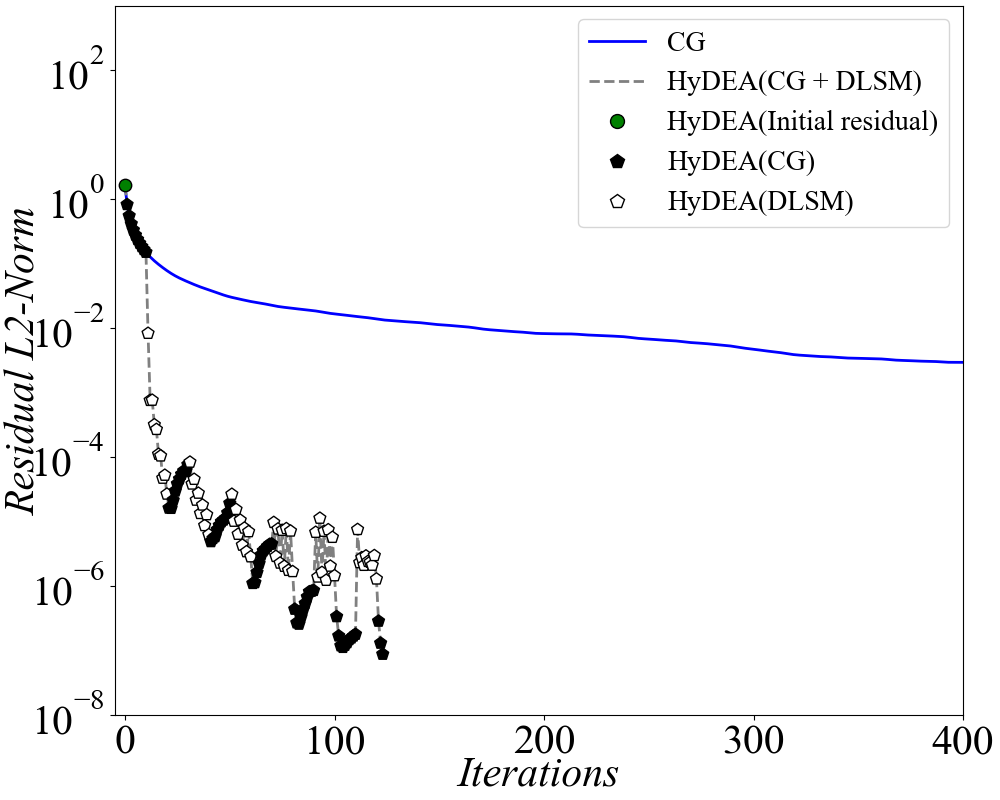}\\
  \caption{Iterative residuals of solving the 1D Poisson equation by CG and HyDEA (CG+DLSM).}\label{Rline_1D_Poisson}
\end{figure}

\begin{figure}[htbp]
\centering
  \includegraphics[scale=0.24]{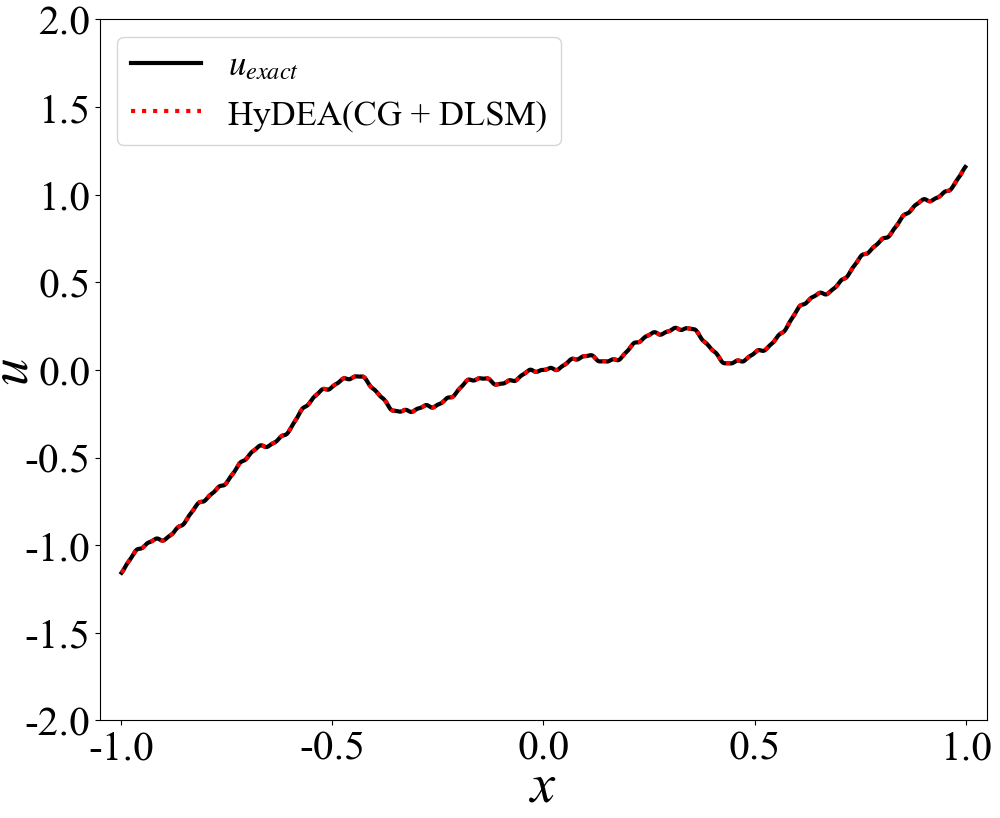}\\
  \caption{Comparison of the numerical solution obtained by HyDEA (CG+DLSM) and $u_{exact}$.}\label{result_compare}
\end{figure}

The iterative residuals of the 1D Poisson equation using both the CG method and HyDEA(CG+DLSM) are presented in Fig.~\ref{Rline_1D_Poisson}. HyDEA receives the same initial residual (shown in solid circle) as in CG, then executes $Num_{CG-type}=10$ CG iterations (solid pentagon), followed by $Num_{DLSM}=10$ DLSM iterations (empty pentagon), and continues this alternating pattern. In total, HyDEA reaches the predefined tolerance by 7 rounds of hybrid algorithm, outperforming the CG method alone. Furthermore, it can be observed that the iterative residual reduction of HyDEA in the initial-stage iteration is primarily driven by the DLSM component, while the CG component becomes the dominant contributor in the later-stage iteration. The accuracy of HyDEA is confirmed in Fig.~\ref{result_compare} by the close match between the numerical solution and $u_{exact}$.

\begin{figure}[htbp]
\centering
  \subfigure[]{
  \label{PSD_e_HyDEA}
  \includegraphics[scale=0.30]{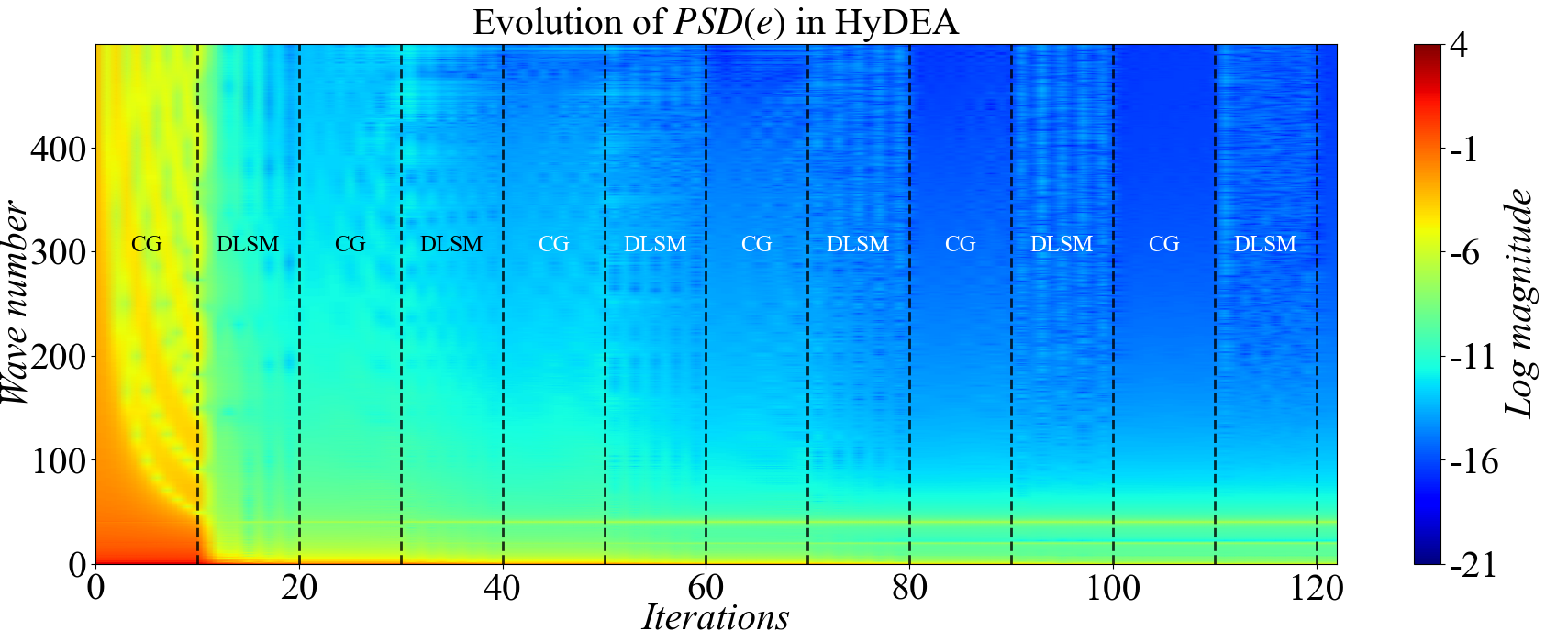}}
  \subfigure[]{
  \label{PSD_e_compare}
  \includegraphics[scale=0.28]{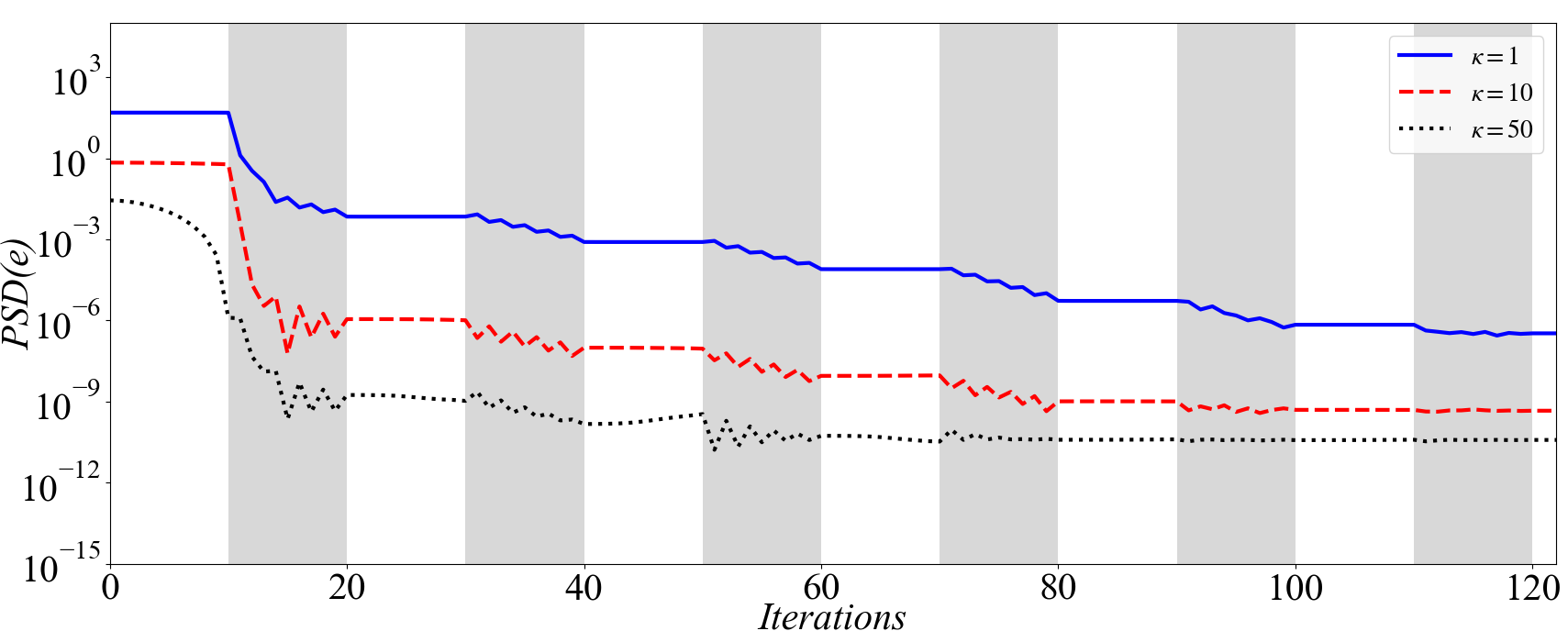}} 
  \caption{Evolution of the $PSD(e)$ for HyDEA (CG+DLSM). (a) Color map. (b) Evolution profiles at $\kappa=1$, $10$ and $50$, with gray areas marking the intervals of DLSM activity.}\label{PSD_e}
\end{figure}

Furthermore, the evolution of the iterative error's power spectral density ($PSD(e)$) for HyDEA (CG+DLSM) as a color map is shown in Fig.~\ref{PSD_e_HyDEA}, while the corresponding evolution profiles at wavenumbers $\kappa=1$, $10$ and $50$ are presented in Fig.~\ref{PSD_e_compare}. The results reveal that the DLSM component in HyDEA acts as the primary driver for eliminating low-wavenumber error energy during the initial-stage iteration. Nevertheless, in the later-stage iteration, DLSM excites growth and oscillations in high-wavenumber error energy, which is subsequently eliminated effectively after switching to the CG method. This observation is consistent with the evolution of iterative residuals shown in Fig.~\ref{Rline_1D_Poisson}, which implies that the complementary behavior of CG and DLSM in HyDEA leads to a rapid decay of the iterative residuals.

\section{Results and discussions}
\label{result}

We shall systematically evaluate the performance of HyDEA through four cases of benchmarks governed by the incompressible fluid dynamics. 

Case 1: Two-dimensional~(2D) lid-driven cavity flows are considered in Section~\ref{cavity} and more specifically 
\begin{itemize}
 \item in Section~\ref{cavityRe1000}  HyDEA~(CG-type+DLSM-1) is evaluated for $Re=1000$.
 \item in Section~\ref{cavityRe3200}  HyDEA~(CG-type+DLSM-2) is evaluated for $Re=3200$.
 \item in Section~\ref{128Inter}  DLSM-1 further applies in mismatched-discretization scenarios via bilinear interpolation upto resolution $640 \times 640$, and HyDEA~(CG-type+DLSM-1) is evaluated for $Re=10,000$ as a super-resolution application.
\end{itemize}

In Cases 2-4, the decoupled immersed boundary projection method~(DIBPM) is activated and HyDEA is evaluated for fluid-structure interactions 
with one stationary circular cylinder in Section~\ref{1Cylinder}, two stationary elliptical cylinders in Section~\ref{2Cylinder}, and one moving cylinder in Section~\ref{OsCylinder}, respectively.
Without re-training or fine-tuning of network parameters, DLSM-2 applies universally in all three cases and the generalization capacity of HyDEA~(CG-type + DLSM-2) is evaluated.

As presented in Section~\ref{HyDEA}, there are two sequential orders for one round of HyDEA algorithm and they are
\begin{enumerate}
 \item A CG-type method is taken as the initial solver and followed by DLSM shown Fig.~\ref{InitialPCG};
 \item DLSM is taken as the initial solver and followed by a CG-type method shown in Fig.~\ref{InitialNN}.
\end{enumerate}
We will employ the first version for most of the applications due to its better robustness.
The reasoning is as follows: the initial residual vector $r_0$ is arbitrary and does not necessarily tend to the lower end of the spectrum of $M$, compromising the efficacy of neural network-based computation, as in the second version of HyDEA. 
In contrast, the first version of HyDEA adopts a CG-type method during initial iterations,
which results in a "smoothed" residual to boost the strength of neural network for later iterations.
We defer a fair comparison of the two versions in the context of a complex flow scenario with one moving cylinder in Section~\ref{OsCylinder}.

\subsection{Case 1: 2D lid-driven cavity flow}
\label{cavity}

\subsubsection{$Re=1000$}
\label{cavityRe1000}

\begin{figure}[htbp]
\centering
  \includegraphics[scale=0.13]{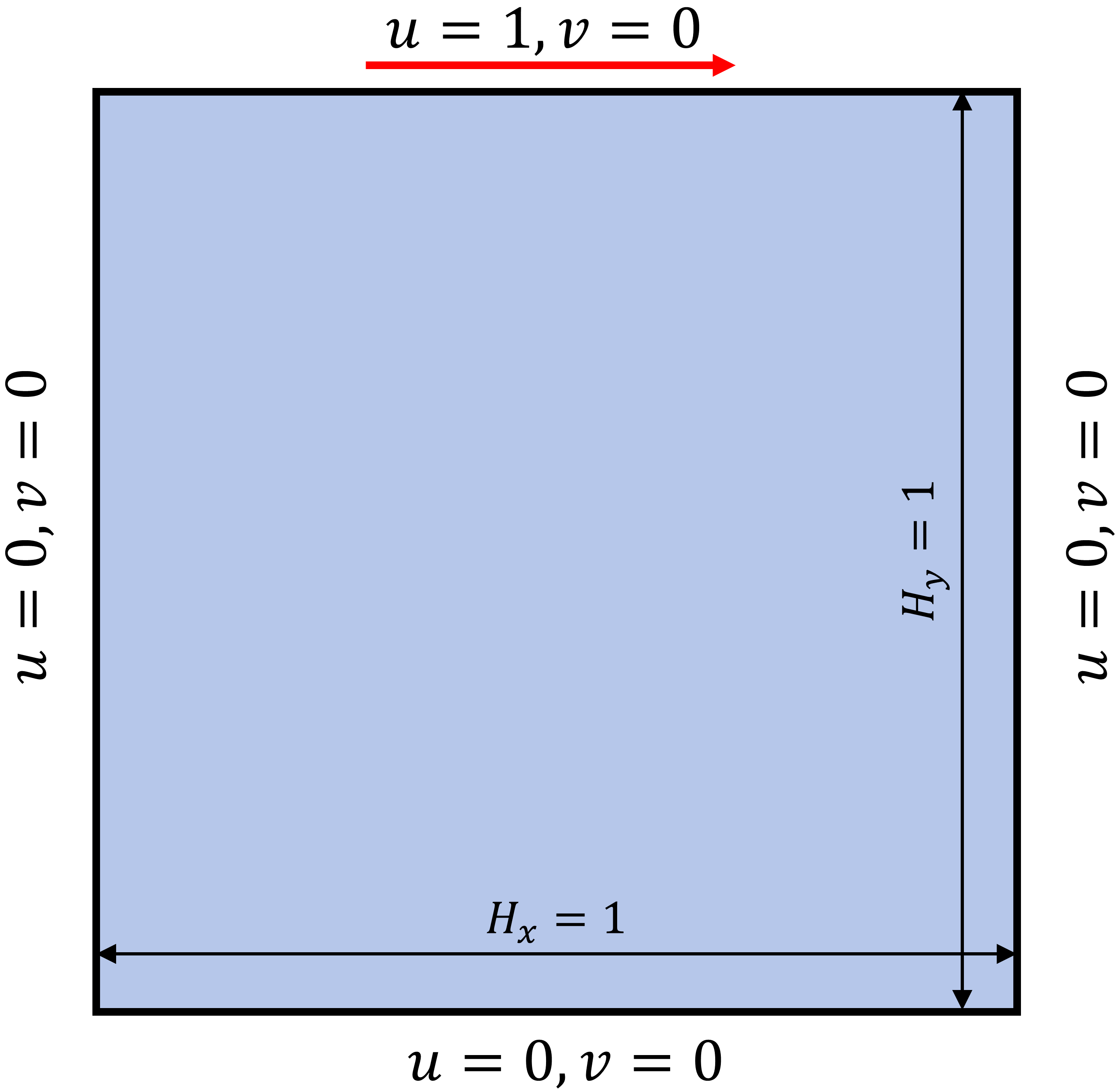}\\
  \caption{Schematic diagram of 2D lid-driven cavity flow.}\label{Cavity_domain}
\end{figure}

The geometric configuration and boundary conditions of the flow are illustrated in Fig.~\ref{Cavity_domain}. The computational domain is a square with side length $H_x=H_y=H=1$. 
The x-component of the velocity for the top lid is set to be $u=1$ and all other velocity components of the boundaries are $0$. The kinematic viscosity $\nu$ is $0.001$, and the CFD time step~ $\Delta t=0.004$. Therefore, Reynolds number $Re=uH/\nu=1000$.
The computational domain is discretized using a uniform Cartesian grid with resolution of $128 \times 128$. 
Consequently, the size of the input feature map for the branch network of DeepONet is set to $h=128$ and $w=128$. The U-Net depth $K$ is set to 5 for this case and all subsequent sections, as it provides satisfactory empirical performance, with no further parameter analysis presented.

The training dataset is prepared with the parameter values $m = 3000$ and $b = 0.6$, which required 268 seconds for construction and 18 hours for model training.  
A systematic sensitivity analysis of their influence on HyDEA's performance is presented in~\ref{appendixC}.
Furthermore, we set $Num_{CG-type}=3$ and $Num_{DLSM}=2$ as the maximum executive iterations
of the two respective solvers in each round of the hybrid algorithm and remove a detailed analysis on the two values into \ref{appendixD}.
The termination residual expressed in $L2$-norm for the iterative calculation is set to $\epsilon=10^{-6}$, 
being consistent with the official benchmark of $\mathtt{PetIBM}$. The influence of this parameter on the performance of HyDEA is discussed in~\ref{appendixE}.

\begin{figure}[htbp] 
 \centering  
  \subfigure[]{
  \label{128_Rline_CG_10steps}
  \includegraphics[scale=0.21]{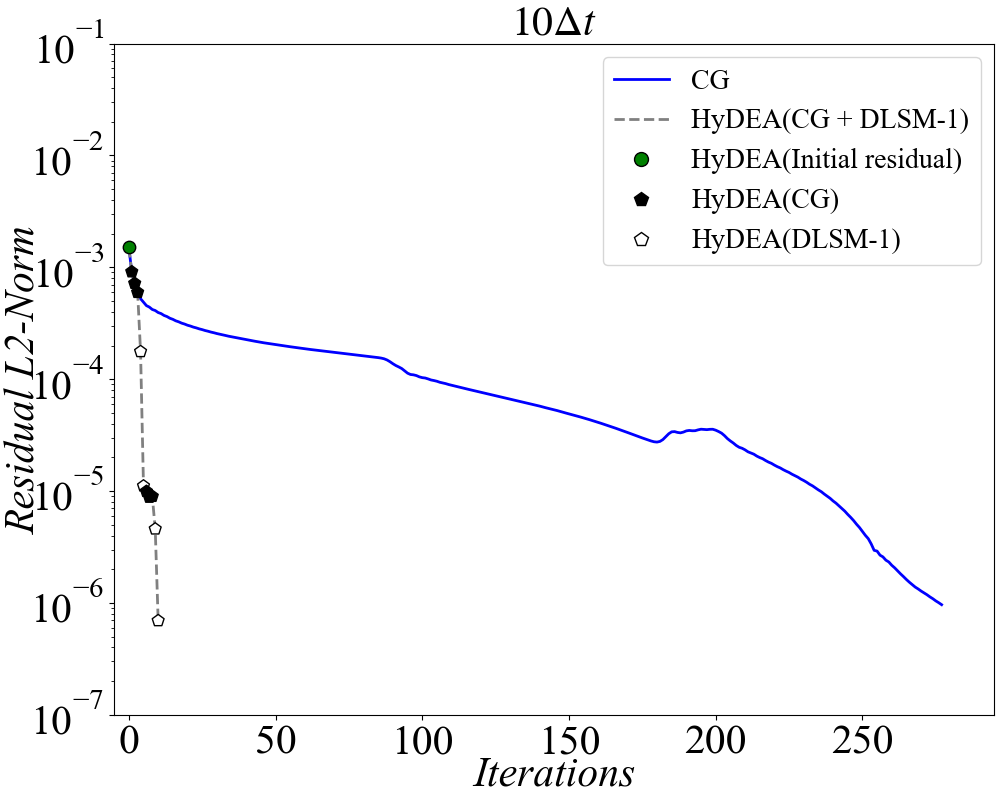}}
  \subfigure[]{
  \label{128_Rline_CG_100steps}
  \includegraphics[scale=0.21]{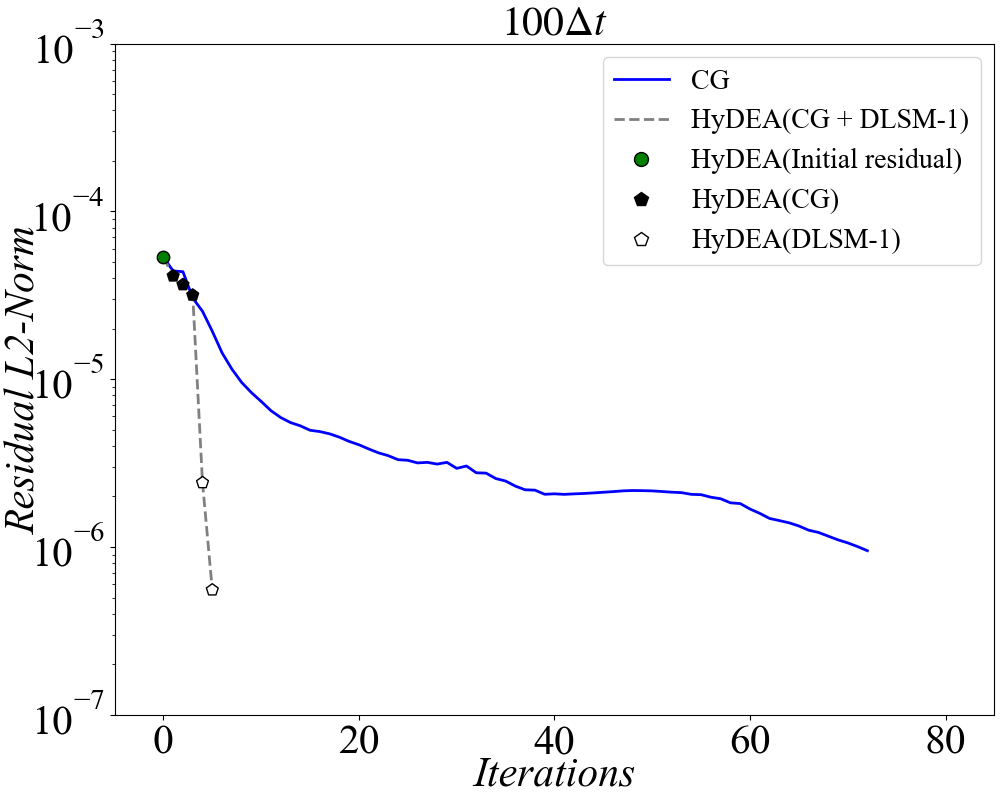}}
  \subfigure[]{
  \label{128_Rline_CG_1000steps}
  \includegraphics[scale=0.21]{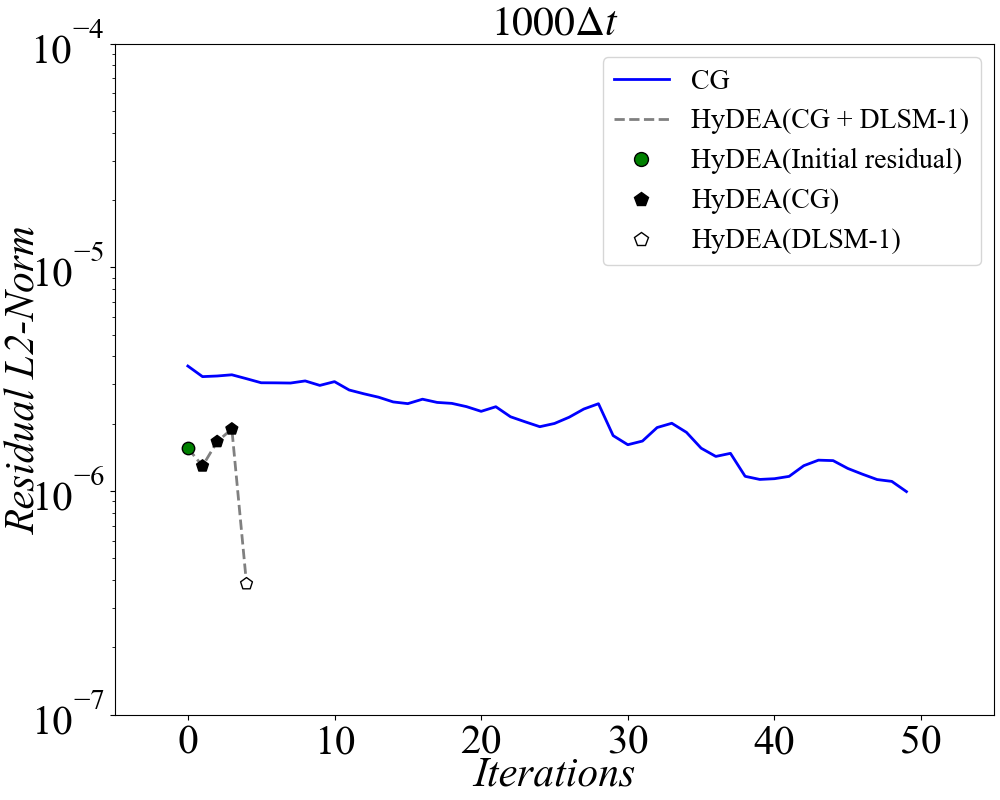}}
  \subfigure[]{
  \label{128_Rline_IC_10steps}
  \includegraphics[scale=0.21]{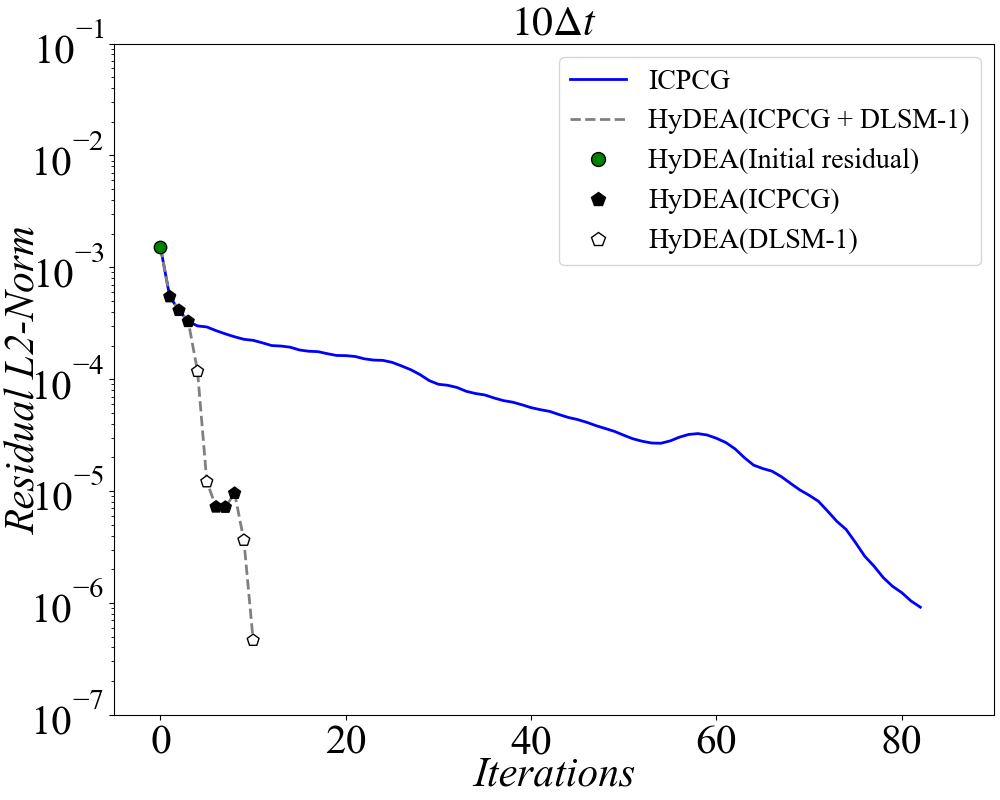}}
  \subfigure[]{
  \label{128_Rline_IC_100steps}
  \includegraphics[scale=0.21]{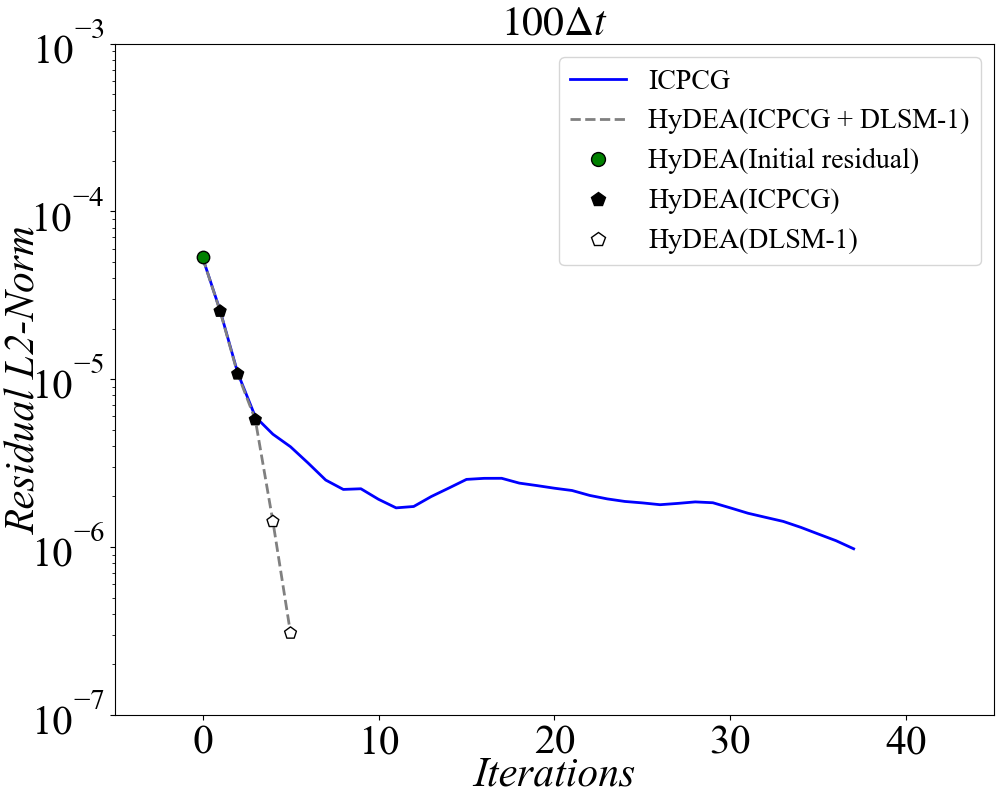}}
  \subfigure[]{
  \label{128_Rline_IC_1000steps}
  \includegraphics[scale=0.21]{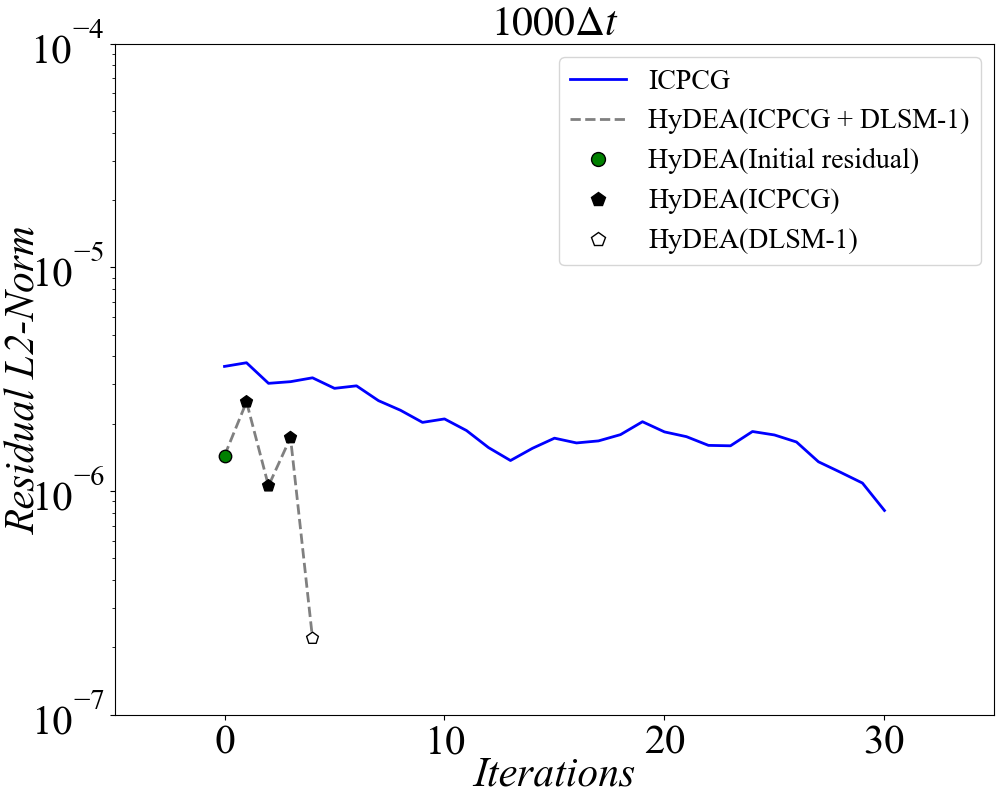}}
  \subfigure[]{
  \label{128_Rline_J_10steps}
  \includegraphics[scale=0.21]{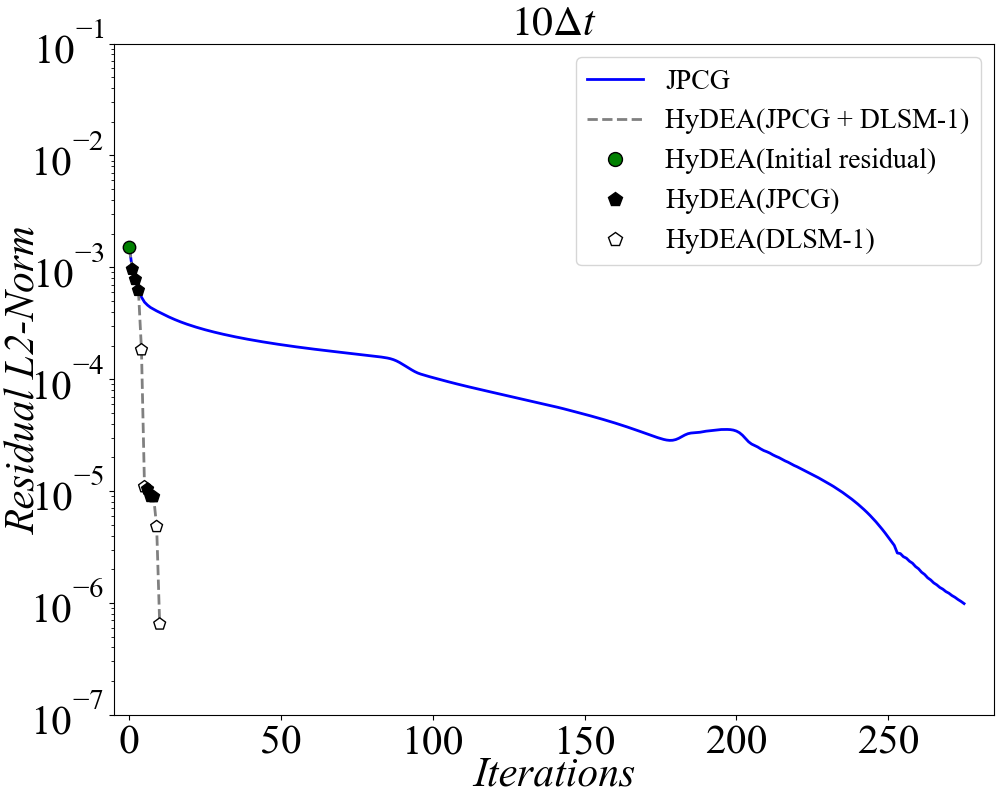}}
  \subfigure[]{
  \label{128_Rline_J_100steps}
  \includegraphics[scale=0.21]{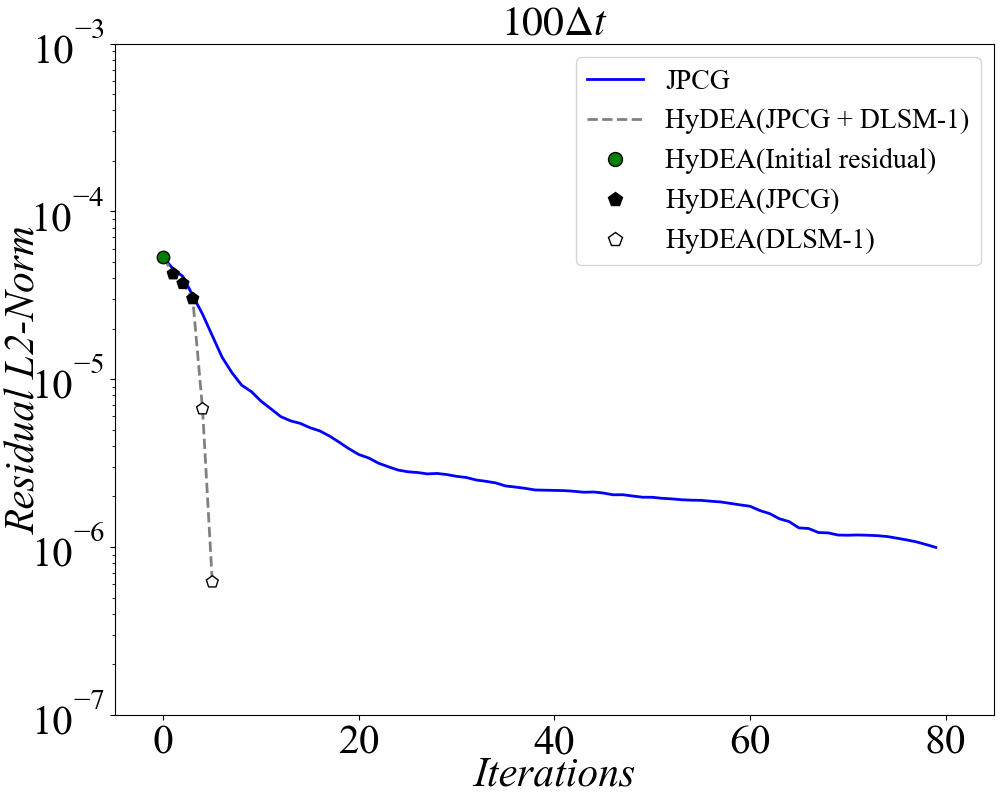}}
  \subfigure[]{
  \label{128_Rline_J_1000steps}
  \includegraphics[scale=0.21]{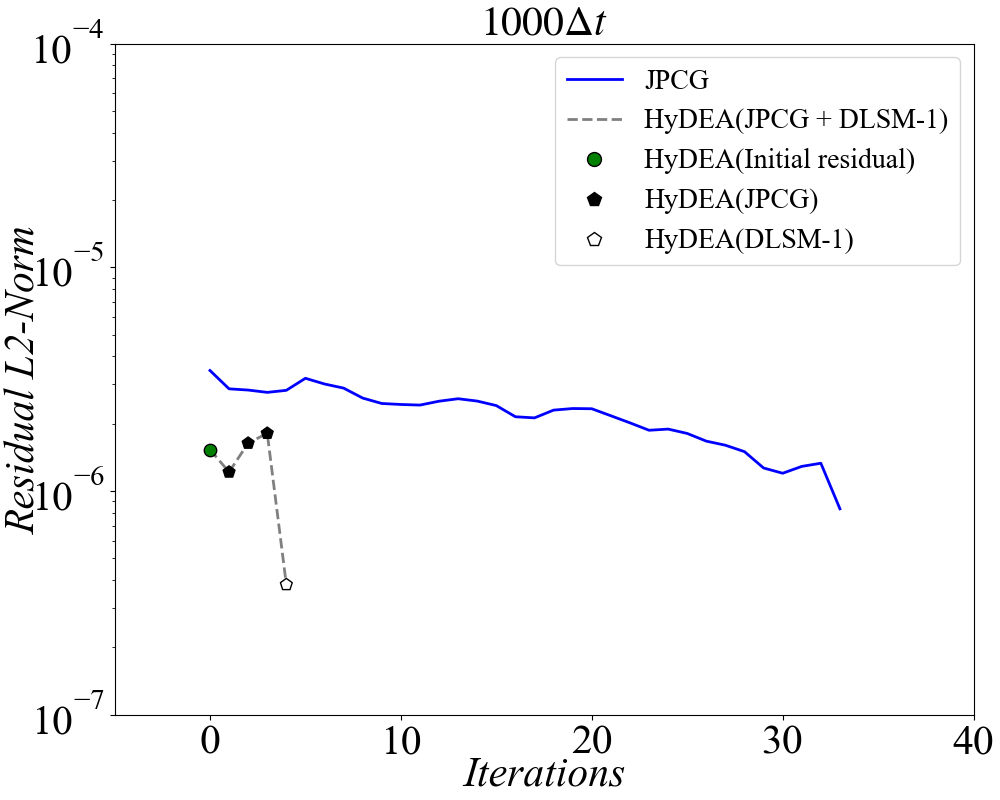}}
  \subfigure[]{
  \label{128_Rline_MG_10steps}
  \includegraphics[scale=0.21]{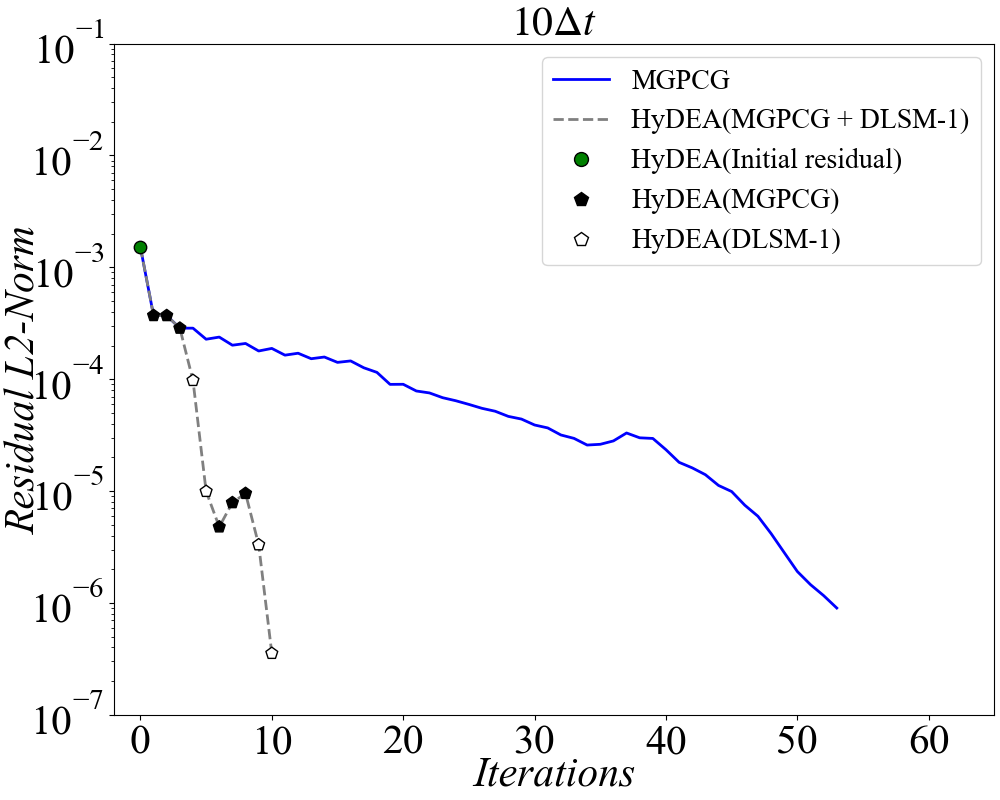}}
  \subfigure[]{
  \label{128_Rline_MG_100steps}
  \includegraphics[scale=0.21]{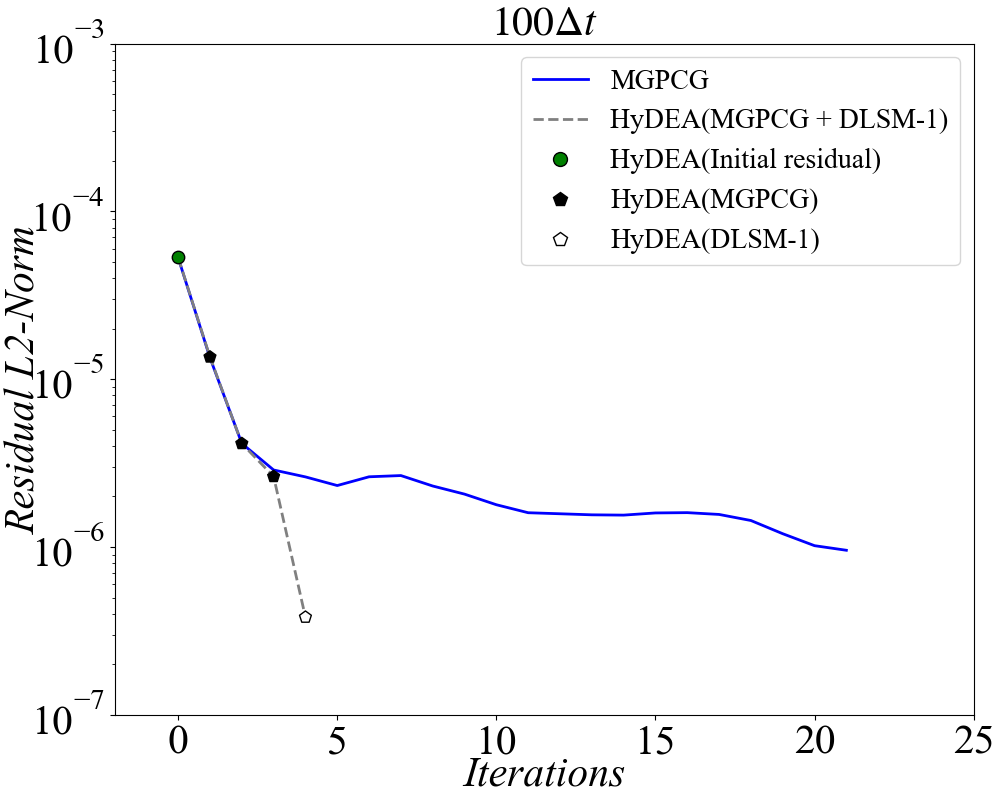}}
  \subfigure[]{
  \label{128_Rline_MG_1000steps}
  \includegraphics[scale=0.21]{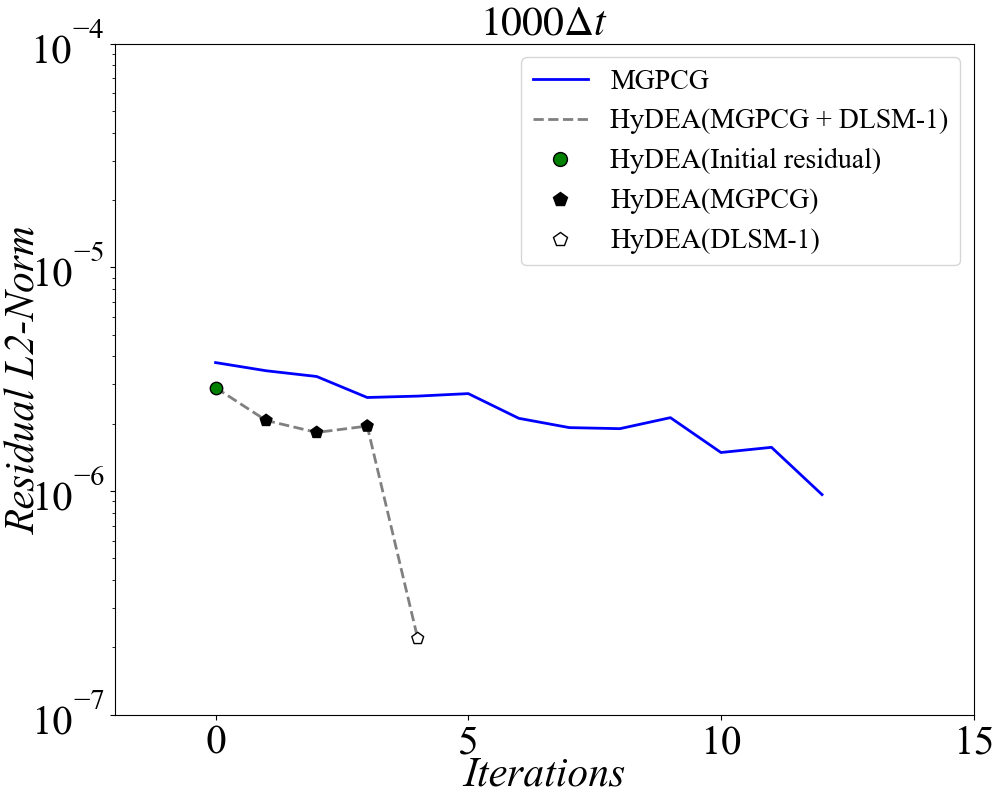}}
  \caption{Iterative residuals of solving the PPE at the $10th$, $100th$ and $1000th$ time steps for 2D lid-driven cavity flow at $Re=1000$. (a)-(c) HyDEA~(CG+DLSM-1). (d)-(f) HyDEA~(ICPCG+DLSM-1). (g)-(i) HyDEA~(JPCG+DLSM-1). (j)-(l) HyDEA~(MGPCG+DLSM-1).}\label{128_Rline_3+2}
\end{figure}

The iterative residuals of the PPE using both HyDEA~(CG-type method+DLSM-1) and various CG-type methods are compared at three representative time steps in Fig.~\ref{128_Rline_3+2}.
For example, in Fig.~\ref{128_Rline_3+2}(a), (b) and (c) the iterative residuals using CG method alone provide baseline performance in (blue) solid lines,
where it takes over $250$, $70$ and $45$ iterations to achieve the predefined tolerance
at $t=10 \Delta t$, $100\Delta t$ and $1000 \Delta t$, respectively.
The required number of iterations decreases as the flow develops towards the steady state.
At $t=10 \Delta t$, HyDEA receives a similar initial residual (shown in solid circle) as in CG,
then executes $Num_{CG-type}=3$ CG iterations (solid pentagon), followed by $Num_{DLSM}=2$ DLSM-1 iterations (empty pentagon);
and again $Num_{CG-type}=3$ CG iterations and then $Num_{DLSM}=2$ DLSM-1 iterations.
In total, HyDEA reaches the predefined tolerance by 2 rounds of the hybrid algorithm
with $10$ iterations in total, a significant reduction compared to $250$ iterations of the CG method alone.
Similarly, HyDEA finishes in 1 round of the hybrid algorithm for $t=100\Delta t$ with $5$ iterations in total and in less then 1 round for $t=1000\Delta t$ with $4$ iterations in total.
Note that at $t=1000\Delta t$, the initial residual of HyDEA appears apparently
different from that of the CG method, as they have a completely different history
of residuals from previous time steps.

Furthermore, we conduct similar comparisons between three standalone preconditioned CG methods: ICPCG, JPCG, and MGPCG with HyDEA integrating the corresponding PCG methods.
As shown in Fig.~(\ref{128_Rline_3+2})(d)-(l), HyDEA always reaches the predefined tolerance with maximum 2 rounds of the hybrid algorithm or maximum $10$ iterations,
an overall significant reduction of iteration numbers compared to their counterparts.

Table~\ref{acceleration ratio 128} summarizes the wall-time acceleration ratio and total computational time for solving the PPE by HyDEA relative to the corresponding CG-type methods over $10,000$ consecutive time steps. Overall HyDEA~(CG-type+DLSM-1) demonstrates a noticeable acceleration compared to its counterpart CG-type variant alone. 

\begin{table}[htbp]
\renewcommand{\arraystretch}{1.5}
\normalsize
\centering
\caption{The wall-time acceleration ratio and computational time for solving the PPE by HyDEA for a duration of $10,000\Delta t$ at $Re=1000$. }
\begin{tabular}{ccccc}
\hline
  HyDEA  &   (CG+DLSM-1) & (ICPCG+DLSM-1) & (JPCG+DLSM-1) & (MGPCG+DLSM-1)  \\
\hline
   Acceleration ratio  & $\times3.83$ & $\times1.88$ & $\times3.56$ & $\times1.64$  \\
   Computational time (s)  &  71 &  83 &  75  &  167 \\
   \makecell{Computational time of \\ DLSM-1 (s)} &   59  &   57  &  60  &  54   \\
  \makecell{Computational time of \\ DeepONet (s)} & 39 & 38 & 39 & 35 \\
   \makecell{GPU-CPU transfer \\ time (s)} &  4.62  &  4.48   &  4.67   &   3.93    \\
\hline
\end{tabular}
\label{acceleration ratio 128}
\end{table}

Some comments are in order;
the reduction of computational time in Table~\ref{acceleration ratio 128} is less impressive than the reduced number of iterations in Fig.~\ref{128_Rline_3+2}. This performance discrepancy stems from multiple factors. 
Foremost, the DLSM component in HyDEA is implemented in $\mathtt{Python}$  and $\mathtt{PyTorch}$. The interpreted nature of the language, combined with the complexity of the neural network architecture, thereby limiting the overall computational efficiency, as clearly demonstrated in Table~\ref{acceleration ratio 128}. Additionally, the bidirectional tensor transfers between the GPU (neural networks) and CPU (iterative methods) introduce minor communication overhead. 
While these technical factors limit the current computational efficiency of HyDEA,
they also represent a clear opportunity for future optimization. 
We emphasize that the present work primarily focuses on establishing the algorithmic framework of HyDEA and examine its generalization capability.

\begin{figure}[htbp] 
 \centering  
  \subfigure[]{
  \label{U_IC_compareGhia}
  \includegraphics[scale=0.21]{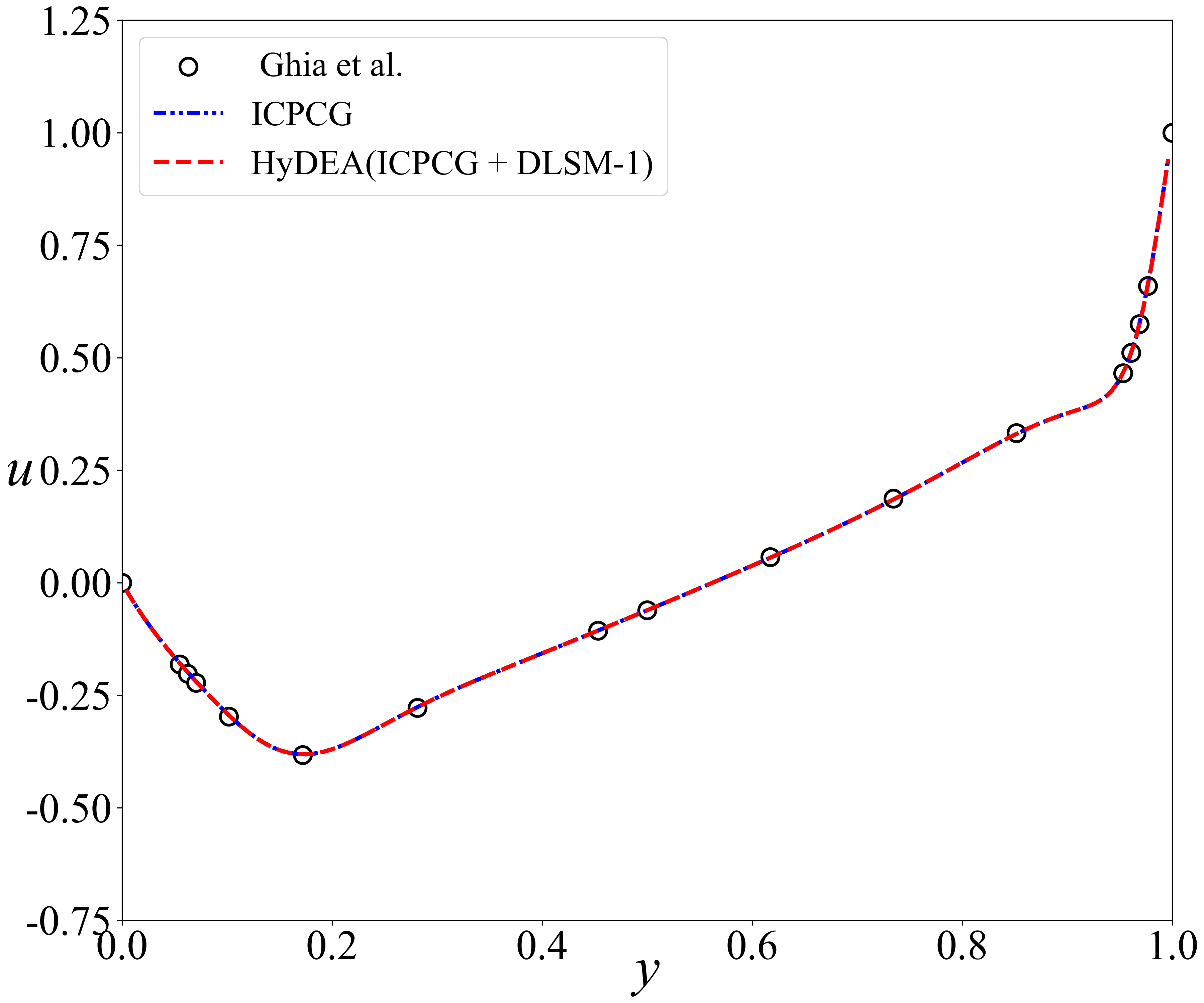}}
  \subfigure[]{
  \label{V_IC_compareGhia}
  \includegraphics[scale=0.21]{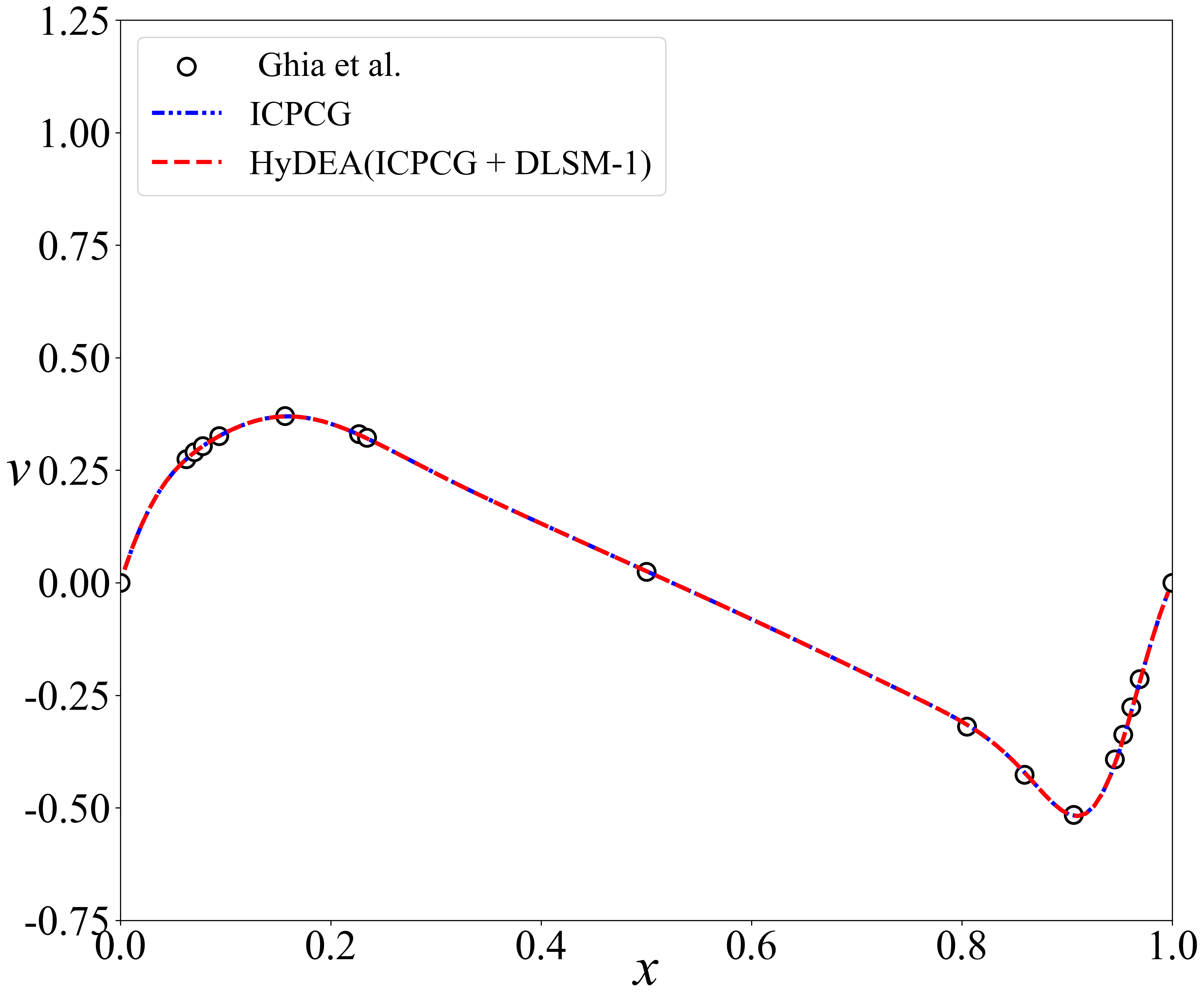}}
 \caption{Profiles of $u$ along $x=0.5$ and $v$ along at $y=0.5$ at steady state for 2D lid-driven cavity flow at $Re=1000$ by ICPCG and HyDEA~(ICPCG+DLSM-1). (a) $u$. (d) $v$.}
 \label{ComparewithGhia}
\end{figure}

\begin{figure}[htbp] 
 \centering  
  \subfigure[]{
  \label{U_IC_1500step_Re1000}
  \includegraphics[scale=0.129]{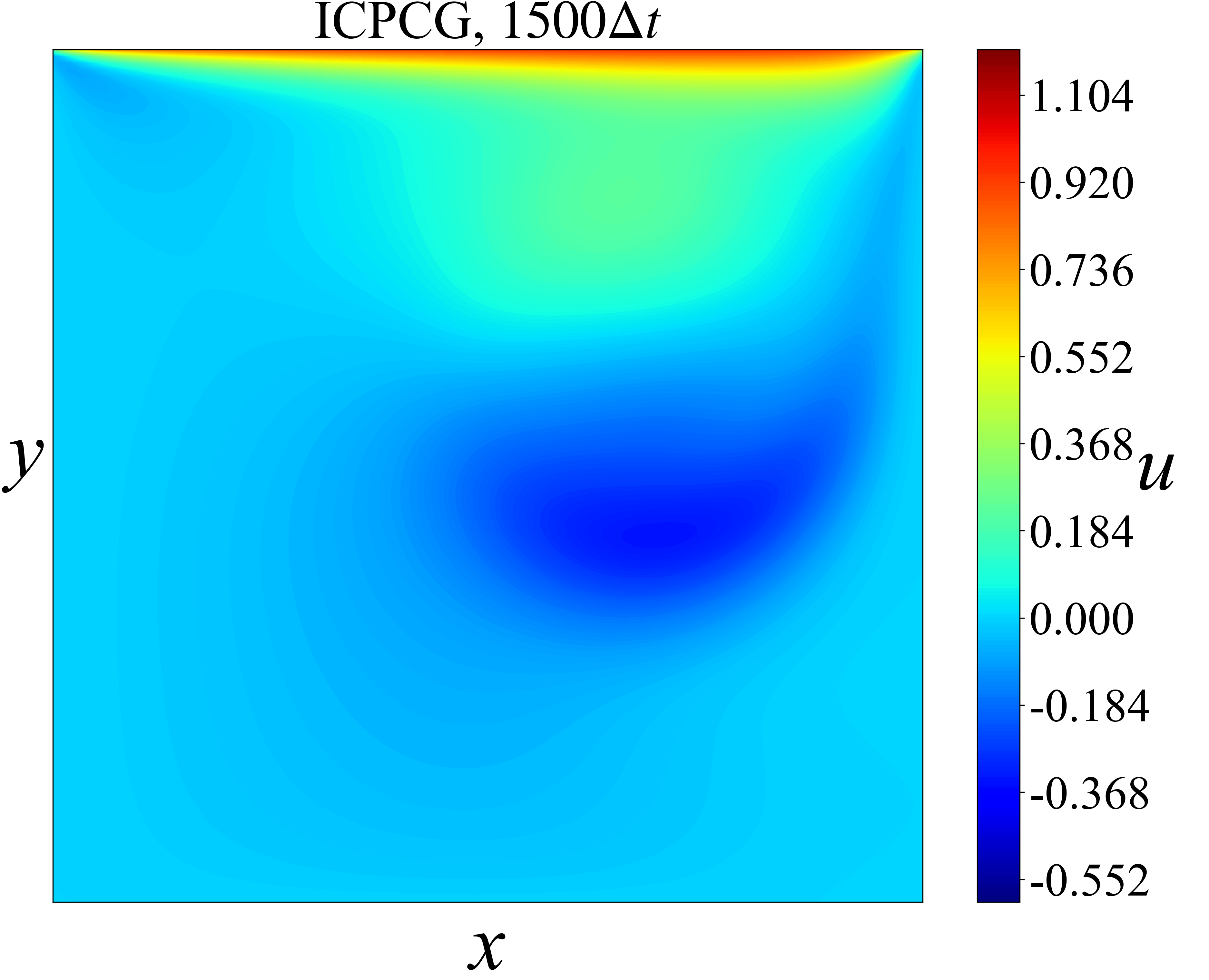}}
  \subfigure[]{
  \label{U_HIC_1500step_Re1000}
  \includegraphics[scale=0.129]{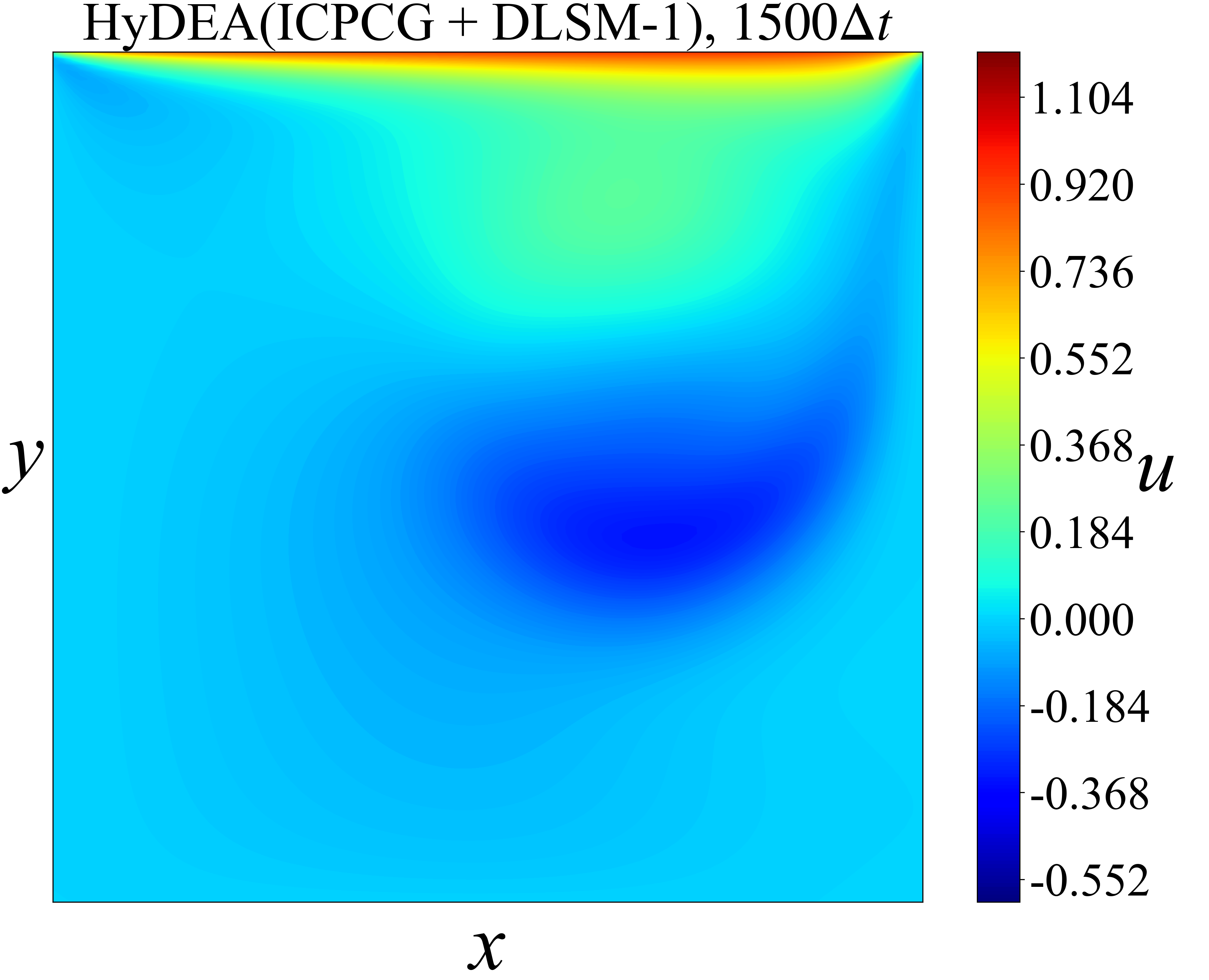}}
  \subfigure[]{
  \label{V_IC_1500step_Re1000}
  \includegraphics[scale=0.129]{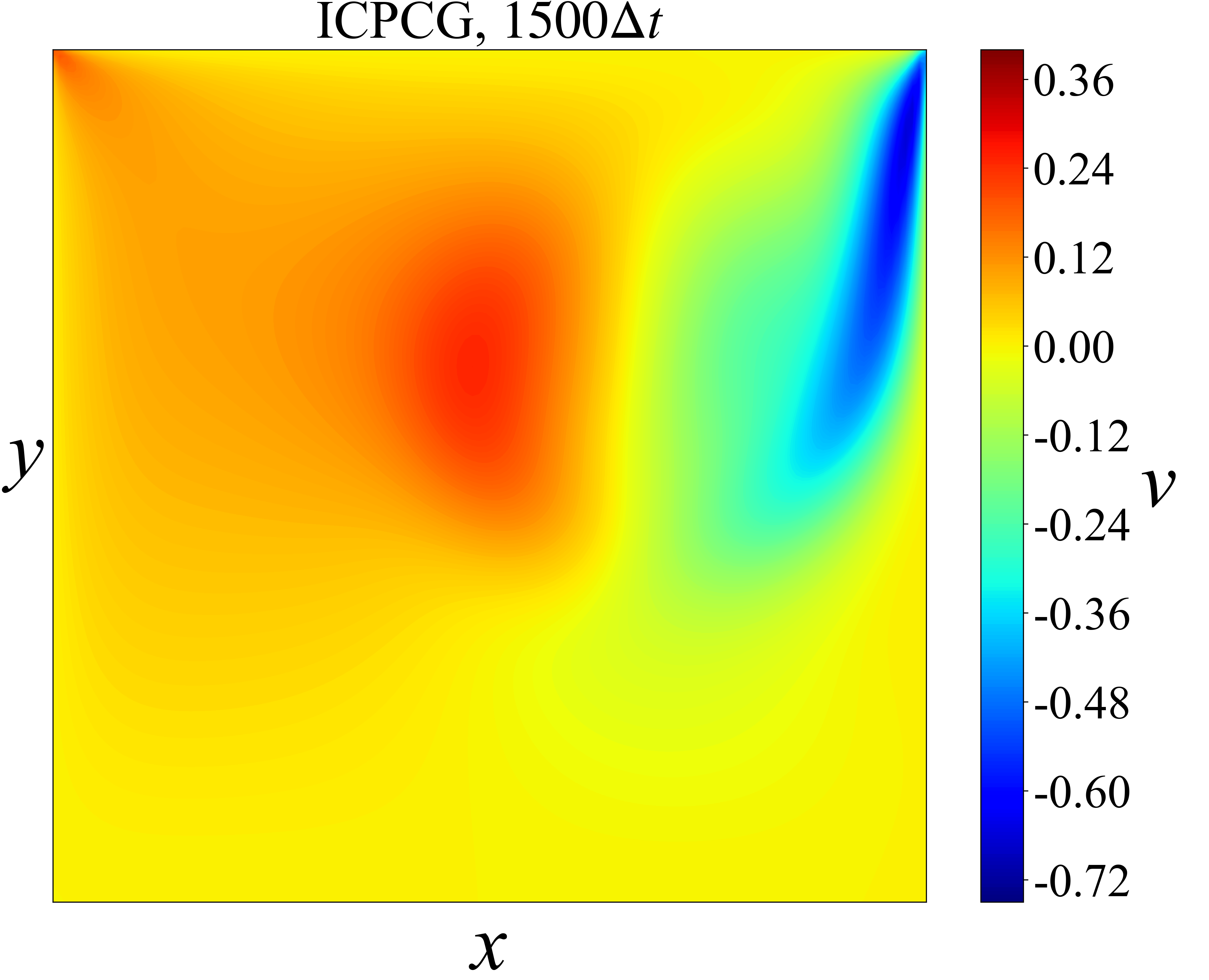}}
  \subfigure[]{
  \label{V_HIC_1500step_Re1000}
  \includegraphics[scale=0.129]{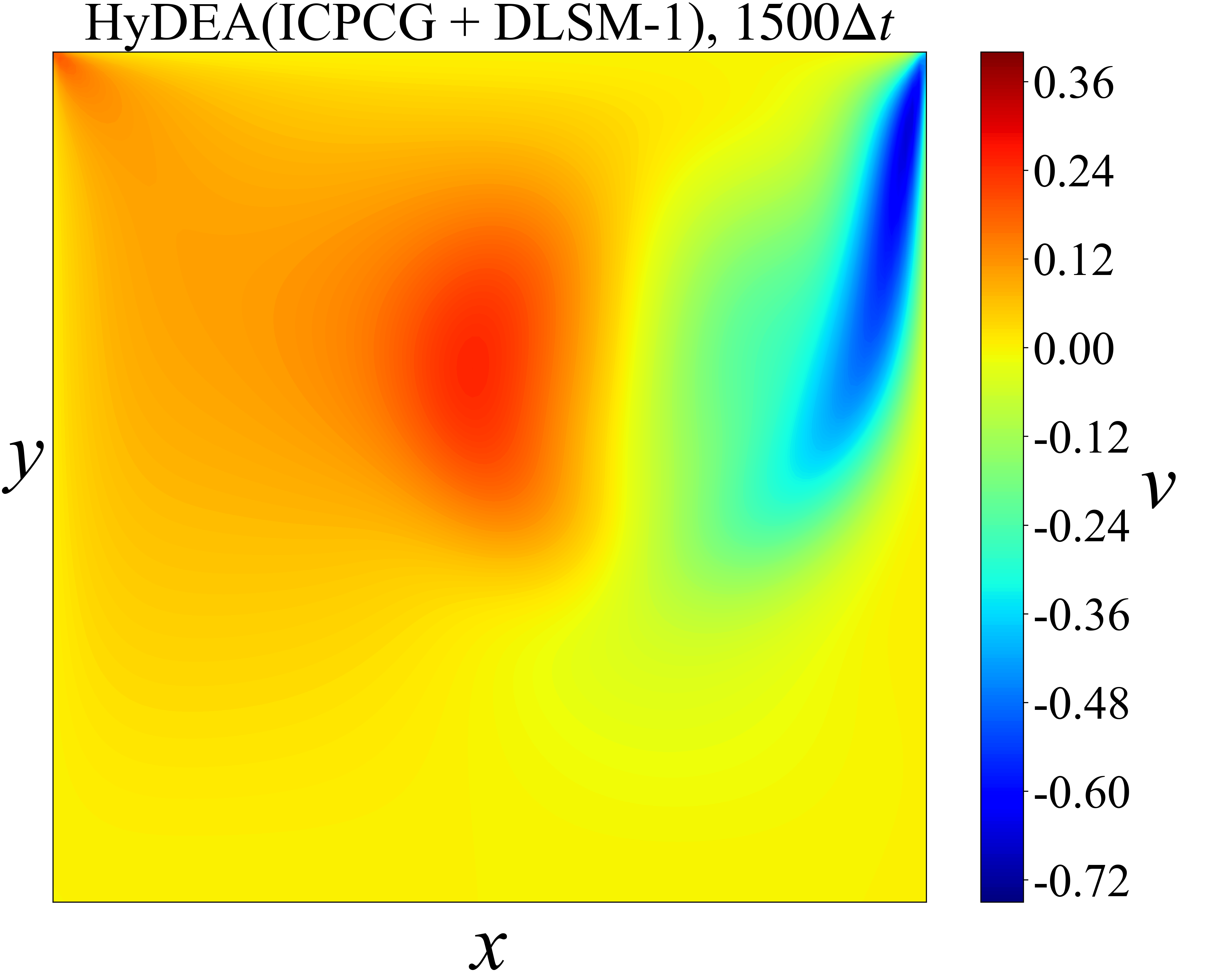}}
  \subfigure[]{
  \label{U_IC_10000step_Re1000}
  \includegraphics[scale=0.129]{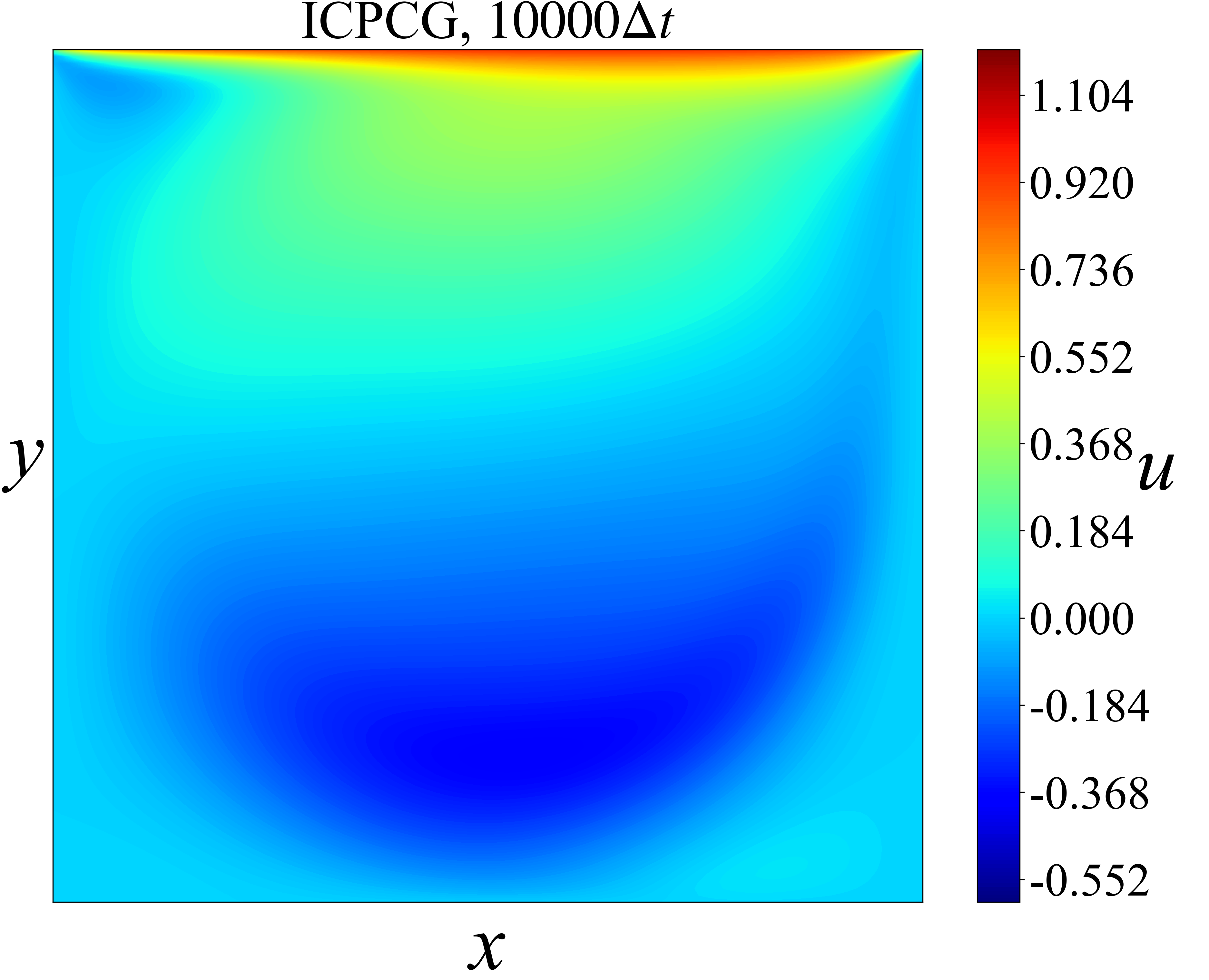}}
  \subfigure[]{
  \label{U_HIC_10000step_Re1000}
  \includegraphics[scale=0.129]{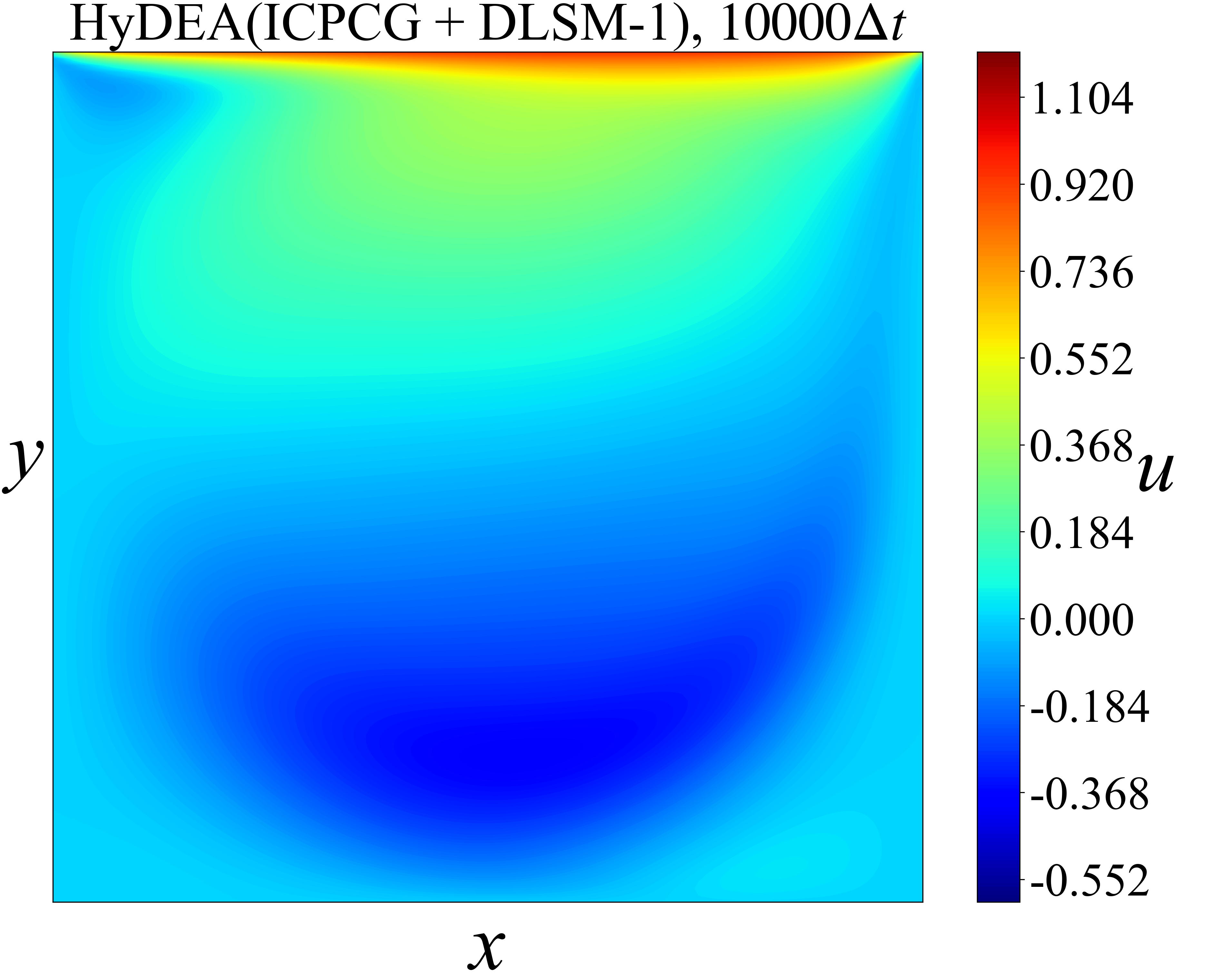}}
  \subfigure[]{
  \label{V_IC_10000step_Re1000}
  \includegraphics[scale=0.129]{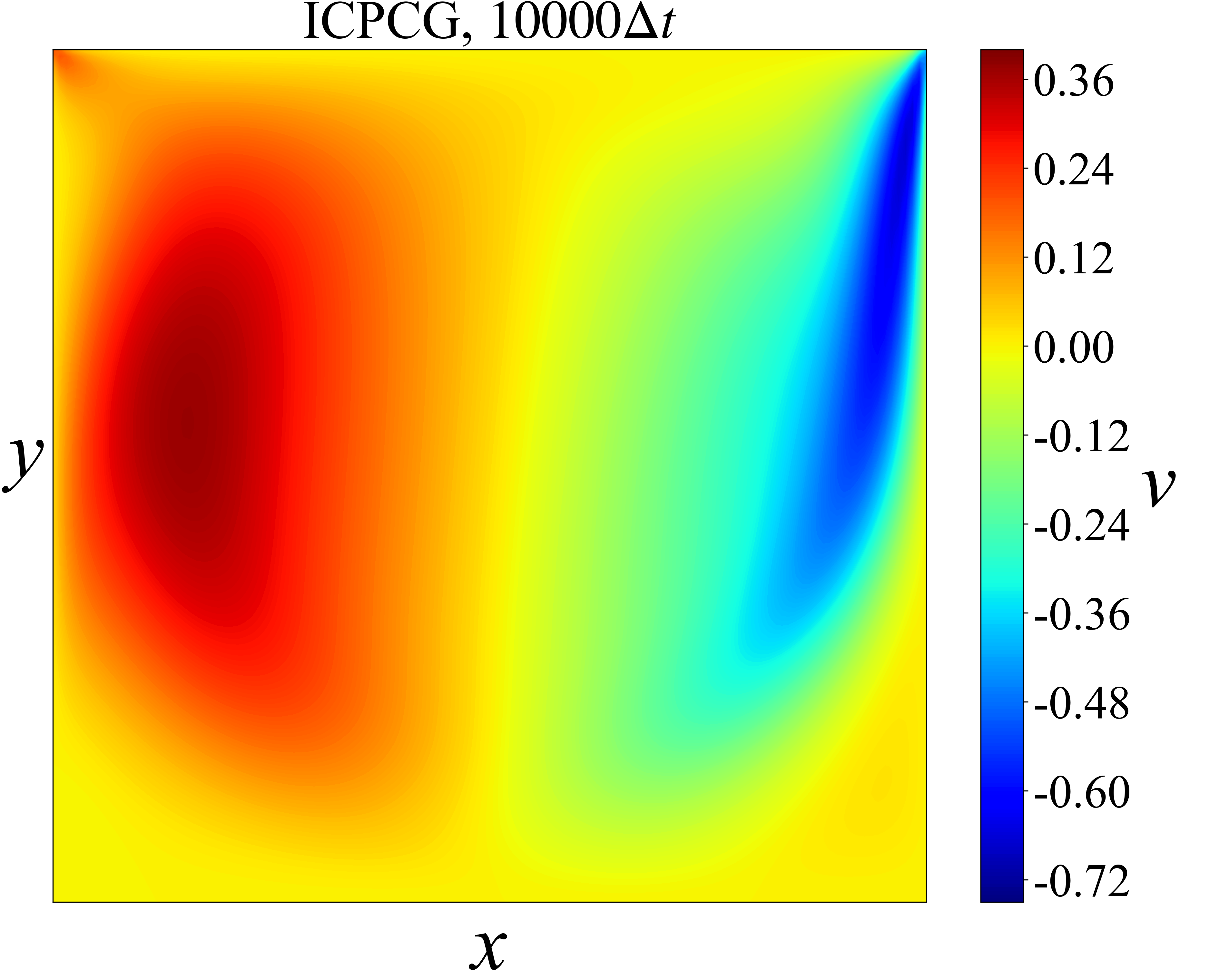}}
  \subfigure[]{
  \label{V_HIC_10000step_Re1000}
  \includegraphics[scale=0.129]{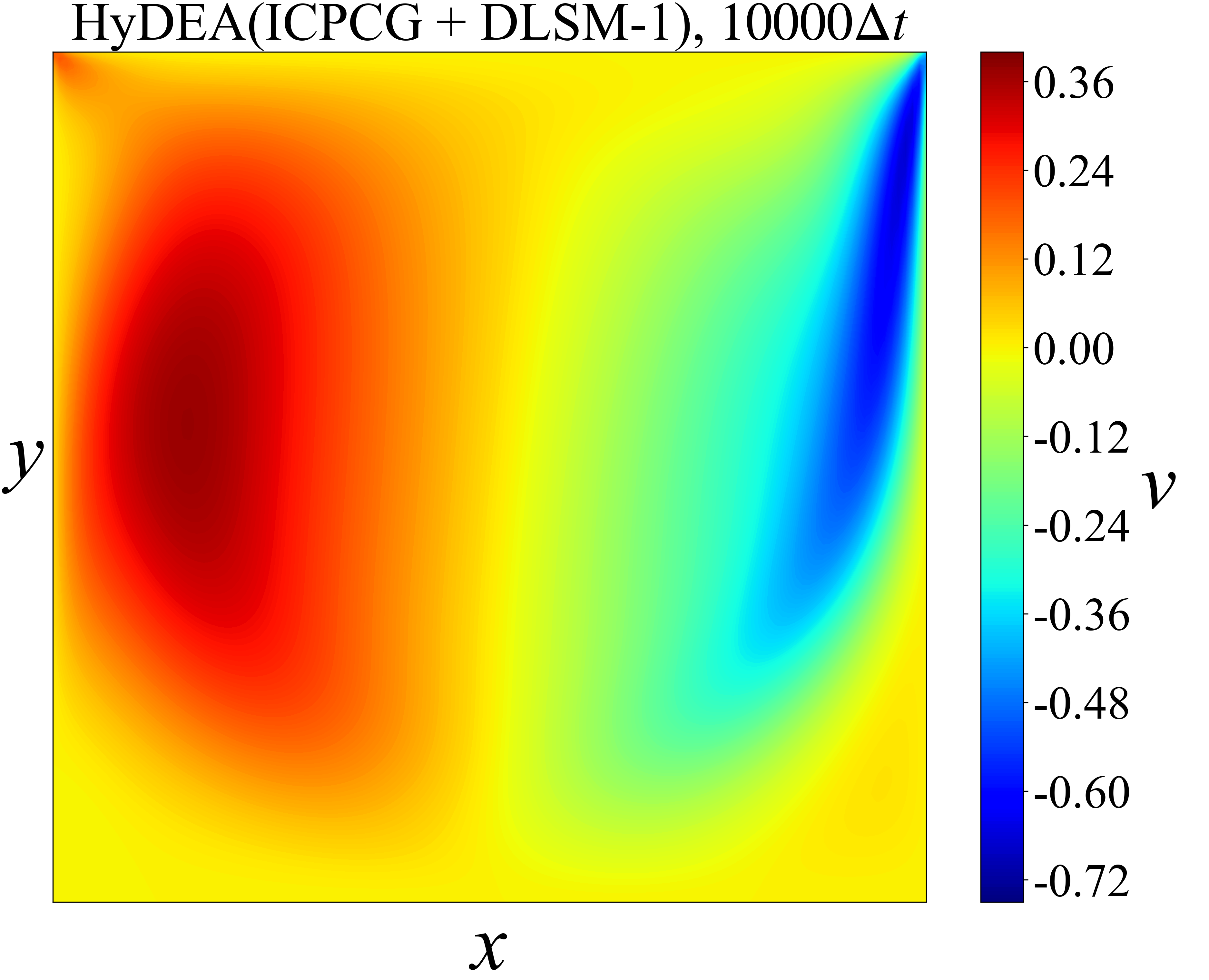}}
 \caption{Velocity fields for 2D lid-driven cavity flow at $Re=1000$ by ICPCG and HyDEA~(ICPCG+DLSM-1). (a)-(d) $u$ and $v$ at the $1500th$ time step. (e)-(h) $u$ and $v$ at the $10,000th$ time step.}
 \label{Flowfield128}
\end{figure}

Taking the ICPCG and HyDEA~(ICPCG+DLSM-1) as an example, velocity profiles at steady state are compared with the results of Ghia et al.~\cite{ghia1982}, as illustrated in Fig.~\ref{ComparewithGhia}. 
The excellent agreement demonstrates a high accuracy of present computational results. 
Furthermore, Fig.~\ref{Flowfield128} presents the velocity fields at the $1500th$ and $10,000th$ time steps, confirming that the temporal evolution of the flow field is accurately captured.

\subsubsection{$Re=3200$} 
\label{cavityRe3200}

Based on Section~\ref{cavityRe1000}, we adjust the configuration with $\nu=0.0003125$ and $\Delta t=0.002$ so that $Re=3200$.
To accommodate the change, the grid resolution becomes $192 \times 192$ and consequently, the size of the input feature map for the branch network of DeepONet is set to $h=192$ and $w=192$.

The training dataset is prepared with the parameter values $m = 7000$ and $b = 0.6$, which required 3138 seconds for construction and 41 hours for model training. 
A systematic sensitivity analysis of their influence on HyDEA's performance is also provided in~\ref{appendixC}.
Furthermore, we set $Num_{CG-type}=3$ and $Num_{DLSM}=2$ as the maximum executive iterations of the two respective solvers in each round of the hybrid algorithm. 
The termination criterion is again with $\epsilon=10^{-6}$.

\begin{figure}[htbp] 
 \centering  
  \subfigure[]{
  \label{192_Rline_3addM_CG_10steps}
  \includegraphics[scale=0.21]{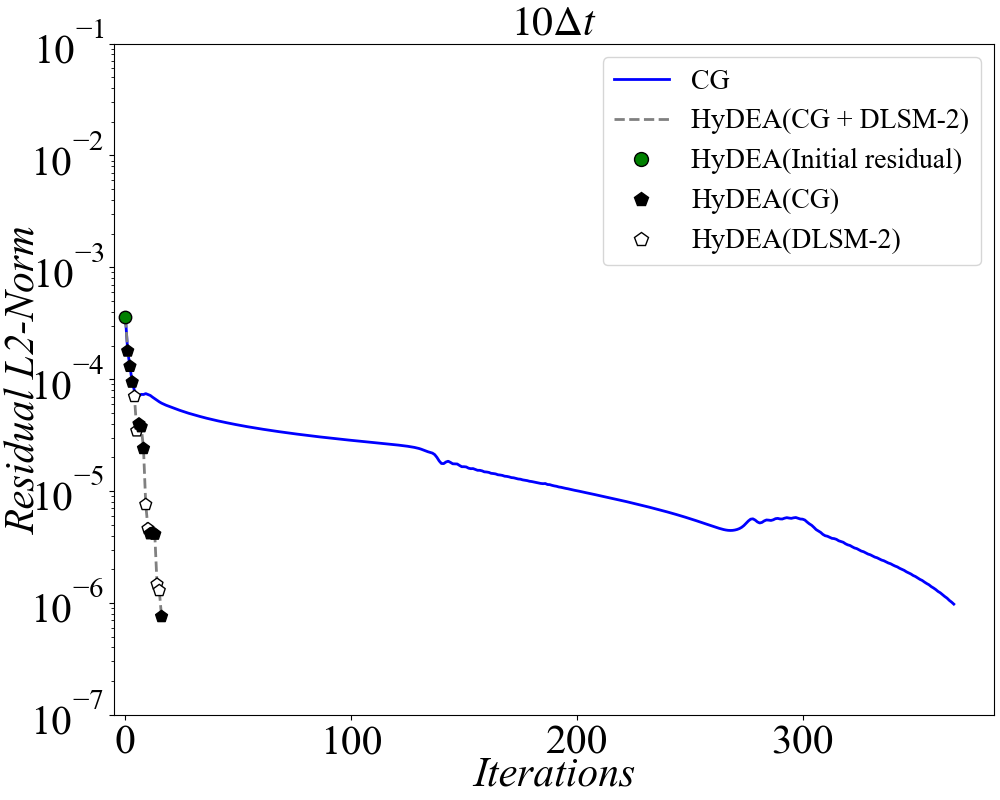}}
  \subfigure[]{
  \label{192_Rline_3addM_CG_100steps}
  \includegraphics[scale=0.21]{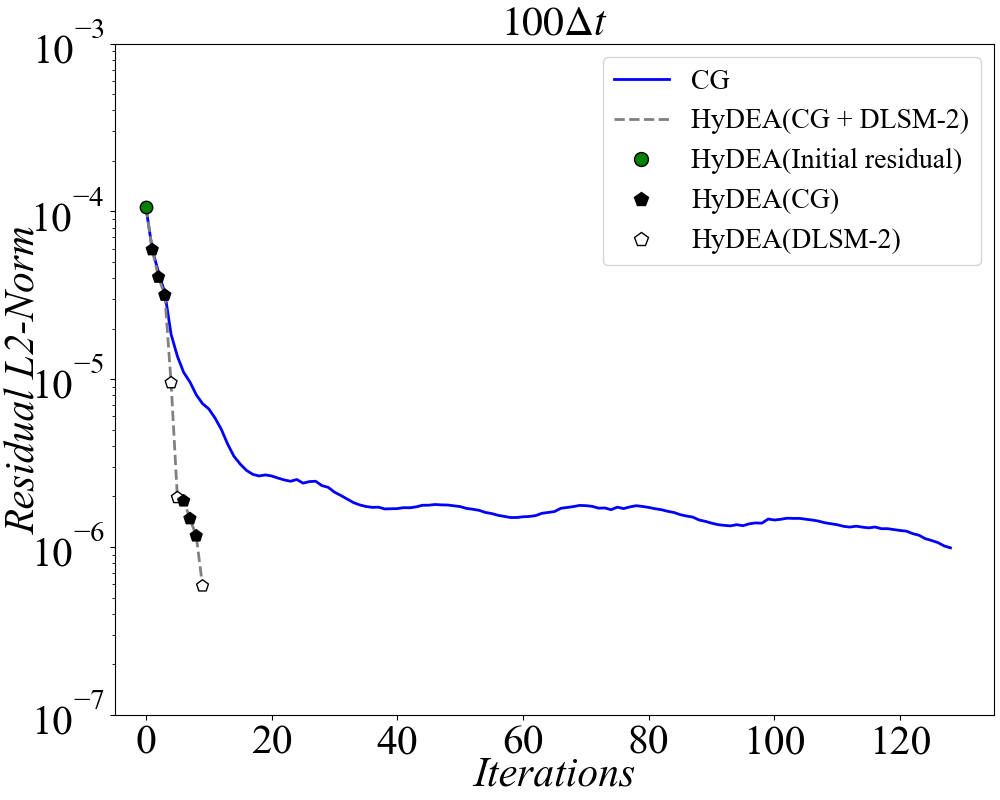}}
  \subfigure[]{
  \label{192_Rline_3addM_CG_1000steps}
  \includegraphics[scale=0.21]{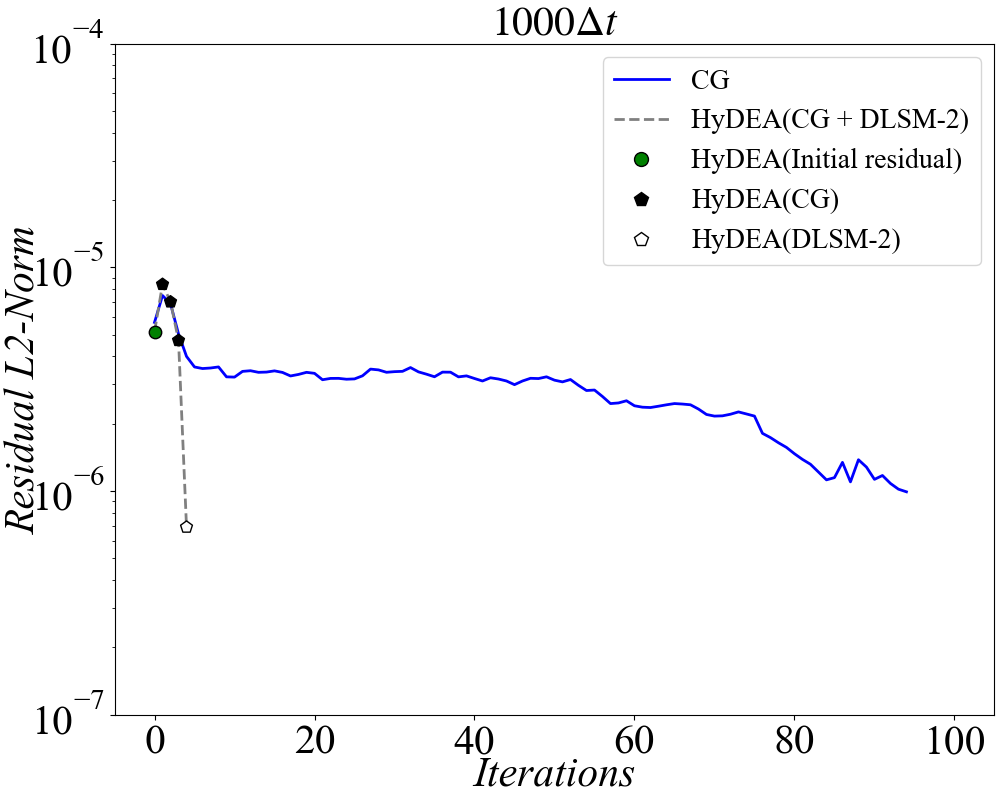}}
  \subfigure[]{
  \label{192_Rline_3addM_IC_10steps}
  \includegraphics[scale=0.21]{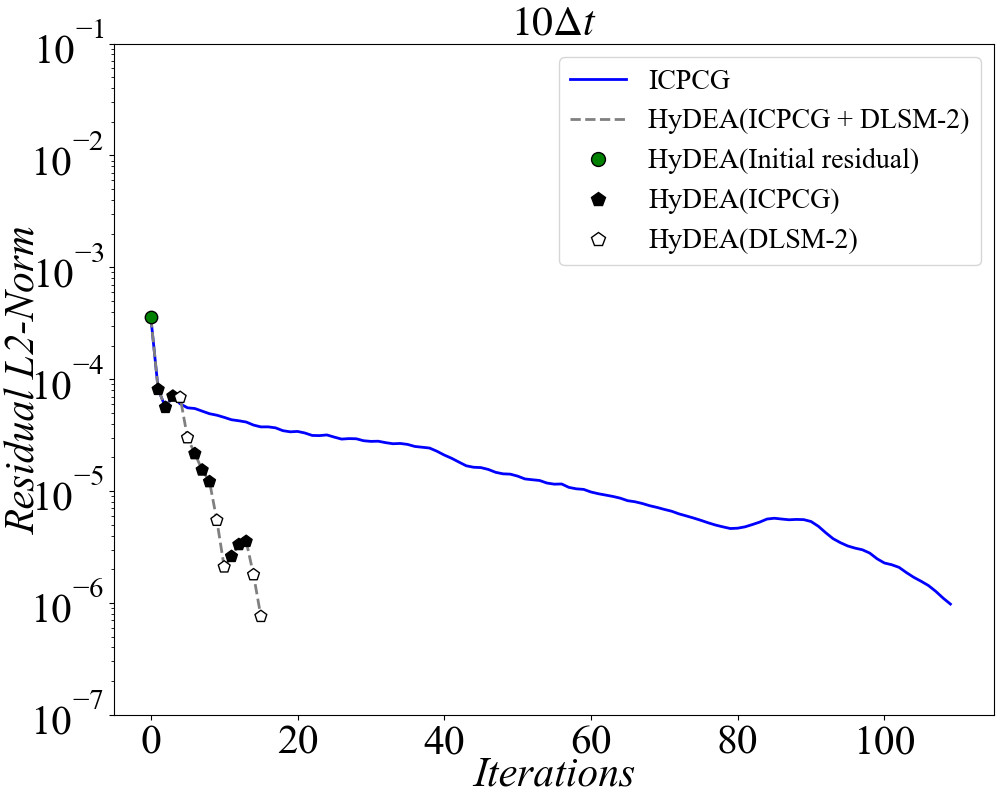}}
  \subfigure[]{
  \label{192_Rline_3addM_IC_100steps}
  \includegraphics[scale=0.21]{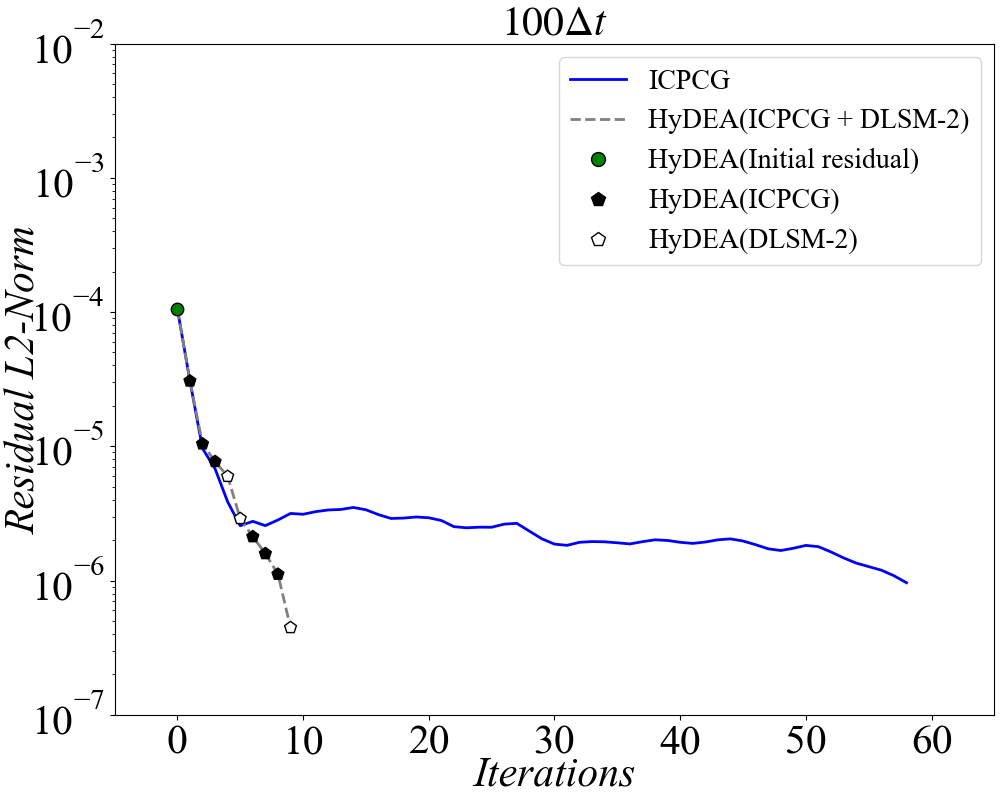}}
  \subfigure[]{
  \label{192_Rline_3addM_IC_1000steps}
  \includegraphics[scale=0.21]{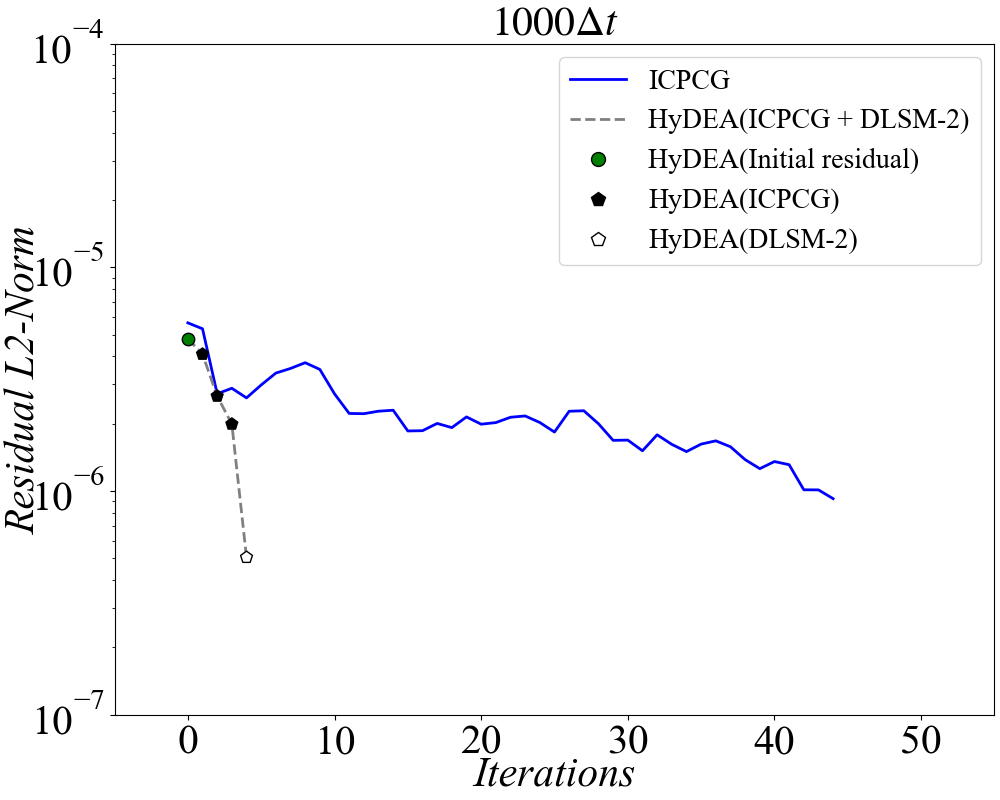}}
  \caption{Iterative residuals of solving the PPE at the $10th$, $100th$ and $1000th$ time steps for 2D lid-driven cavity flow at $Re=3200$. (a)-(c) HyDEA~(CG+DLSM-2). (d)-(f) HyDEA~(ICPCG+DLSM-2).}\label{192_Rline_3add2}
\end{figure}

To avoid over-extensive comparisons,
we only present results of HyDEA~(CG+DLSM-2), HyDEA~(ICPCG+DLSM-2)
and their counterparts.
The iterative residuals of the PPE solution at the $10th$, $100th$ and $1000th$ time steps are shown in Fig.~\ref{192_Rline_3add2}. 
Overall HyDEA reaches the predefined toleration within $4$ rounds of the hybrid algorithm
with maximum around $16$ iterations in total.
HyDEA demonstrates significantly reduced number of iterations compared to CG-type methods alone, being consistent with the results in Section~\ref{cavityRe1000}.

Table~\ref{acceleration ratio 192} summarizes the wall-time acceleration ratio and computational time of solving the PPE by HyDEA relative to the corresponding CG-type methods over $10,000$ consecutive time steps. Overall HyDEA~(CG-type+DLSM-2) demonstrates a noticeable acceleration compared to its counterpart CG-type variant alone.
Notably, the acceleration ratios become more significant than those for the low-resolution in Section~\ref{cavityRe1000}, highlighting HyDEA's enhanced potential for high-resolution flow simulations.

\begin{table}[htbp]
\renewcommand{\arraystretch}{1.5}
\normalsize
\centering
\caption{The wall-time acceleration ratio and computational time for solving the PPE by HyDEA for a duration of $10,000\Delta t$ at $Re=3200$. }
\begin{tabular}{ccccc}
\hline
  HyDEA  &   (CG+DLSM-2) & (ICPCG+DLSM-2) & (JPCG+DLSM-2) & (MGPCG+DLSM-2) \\
\hline
   Acceleration ratio  & $\times5.96$ & $\times2.82$ & $\times5.64$ & $\times2.08$ \\
   Computational time (s)  &  131 &  162 &  134  & 342 \\
    \makecell{Computational time of \\ DLSM-2 (s)}  &   100  &   103  &  102  &  97  \\
   \makecell{Computational time of \\ DeepONet (s)} & 62 & 63 & 63 & 60 \\
  \makecell{GPU-CPU transfer \\ time (s)} &   9.56  &  8.69  &  9.71   &  8.54  \\
\hline
\end{tabular}
\label{acceleration ratio 192}
\end{table}

\begin{figure}[htbp] 
 \centering  
  \subfigure[]{
  \label{Re3200_U_compareGhia}
  \includegraphics[scale=0.21]{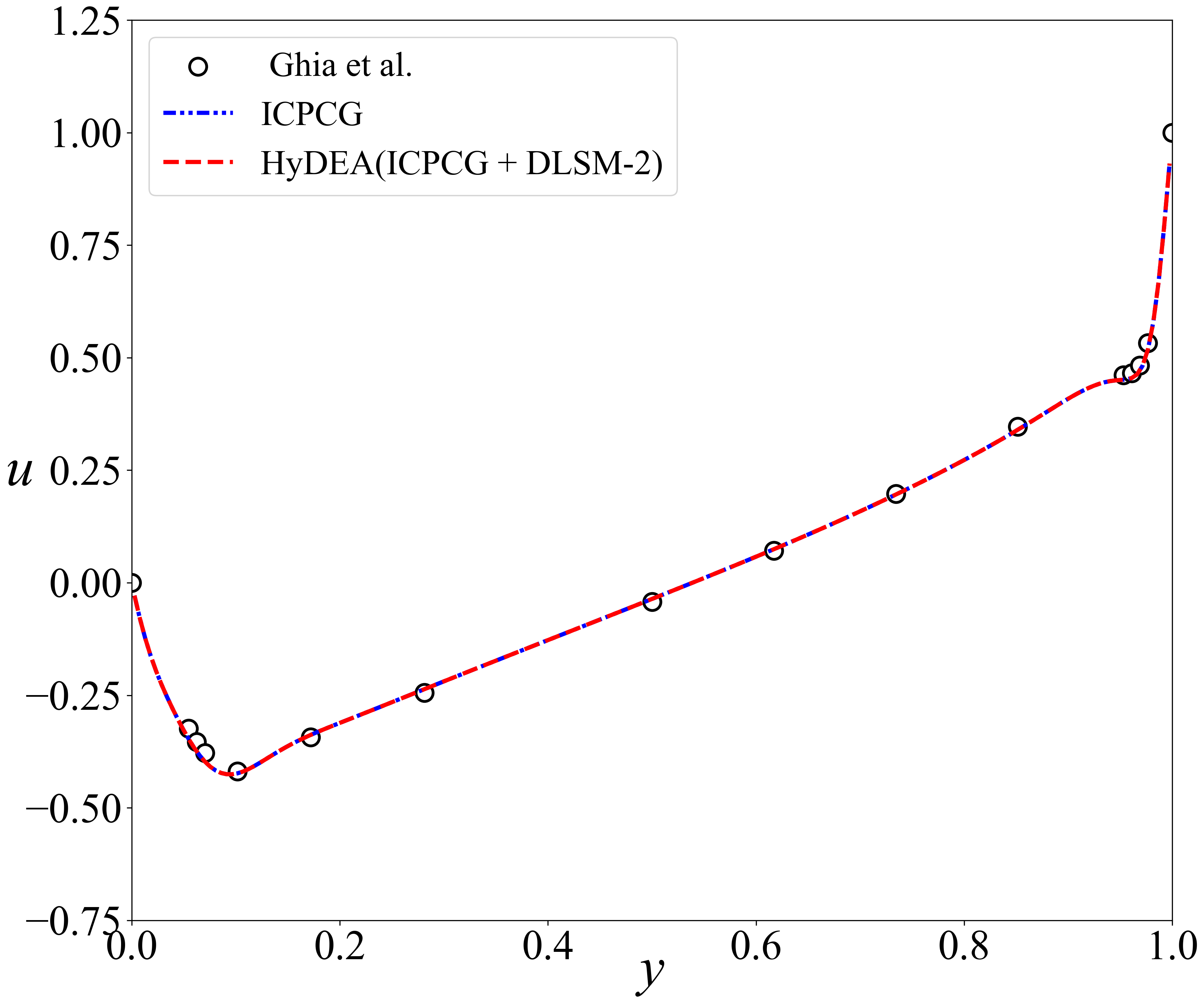}}
  \subfigure[]{
  \label{Re3200_V_compareGhia}
  \includegraphics[scale=0.21]{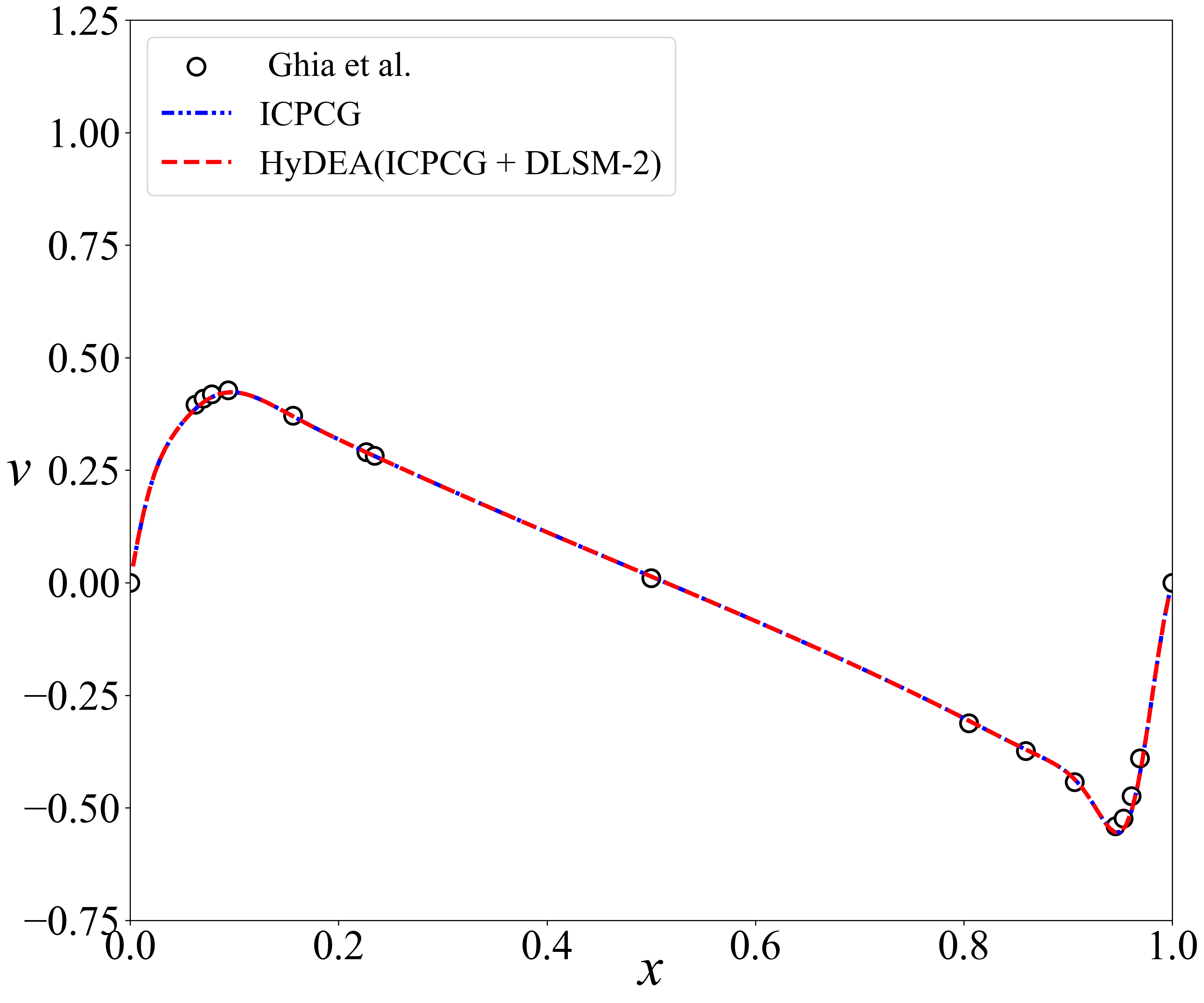}}
 \caption{Profiles of $u$ along $x=0.5$ and $v$ along $y=0.5$ at stedy state for 2D lid-driven cavity flow at $Re=3200$ by ICPCG and HyDEA~(ICPCG+DLSM-2). (a) $u$. (b) $v$.}
 \label{Re3200_Ghia_compare}
\end{figure}

\begin{figure}[htbp] 
 \centering  
  \subfigure[]{
  \label{U_IC_1500step_Re3200}
  \includegraphics[scale=0.129]{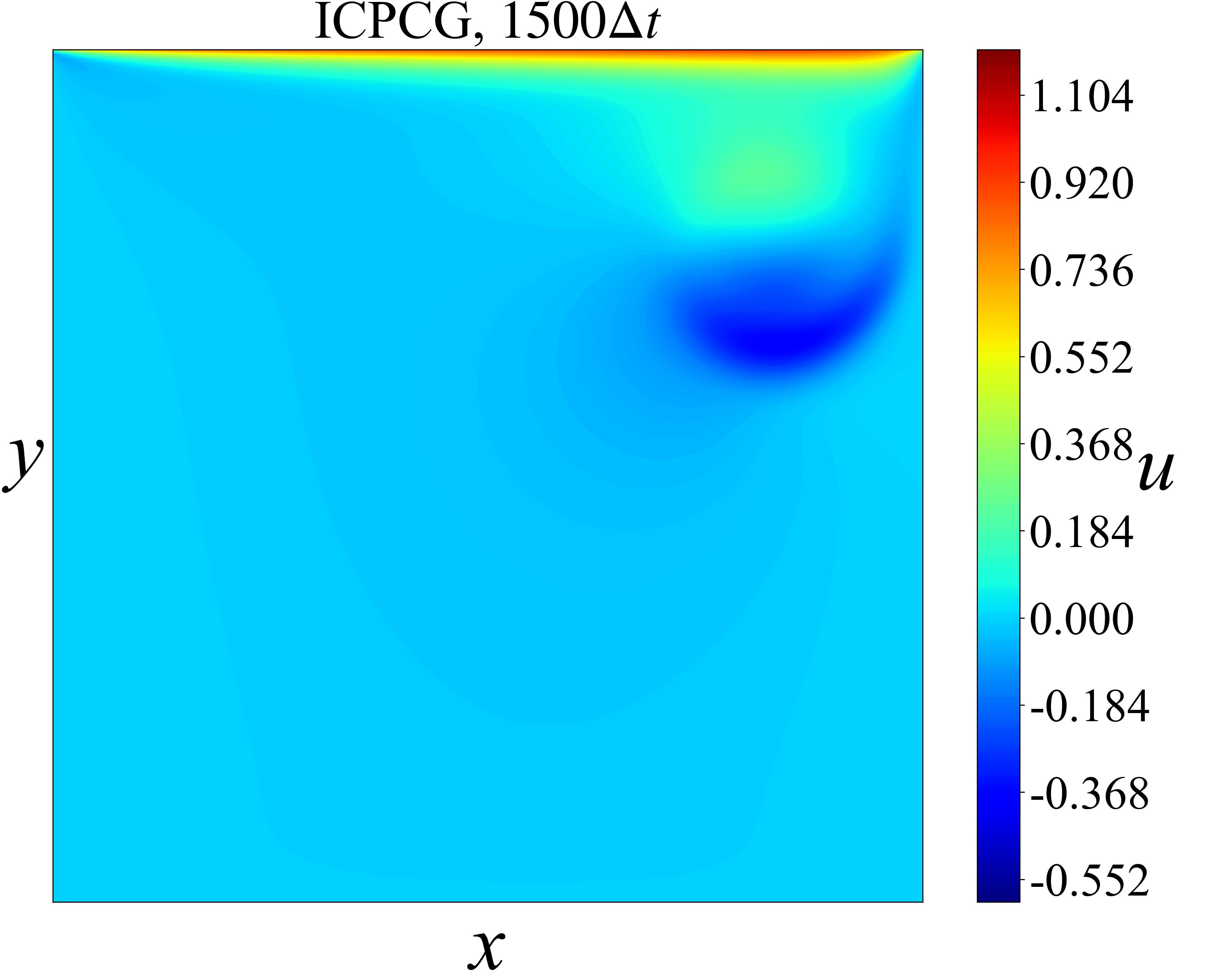}}
  \subfigure[]{
  \label{U_HIC_1500step_Re3200}
  \includegraphics[scale=0.129]{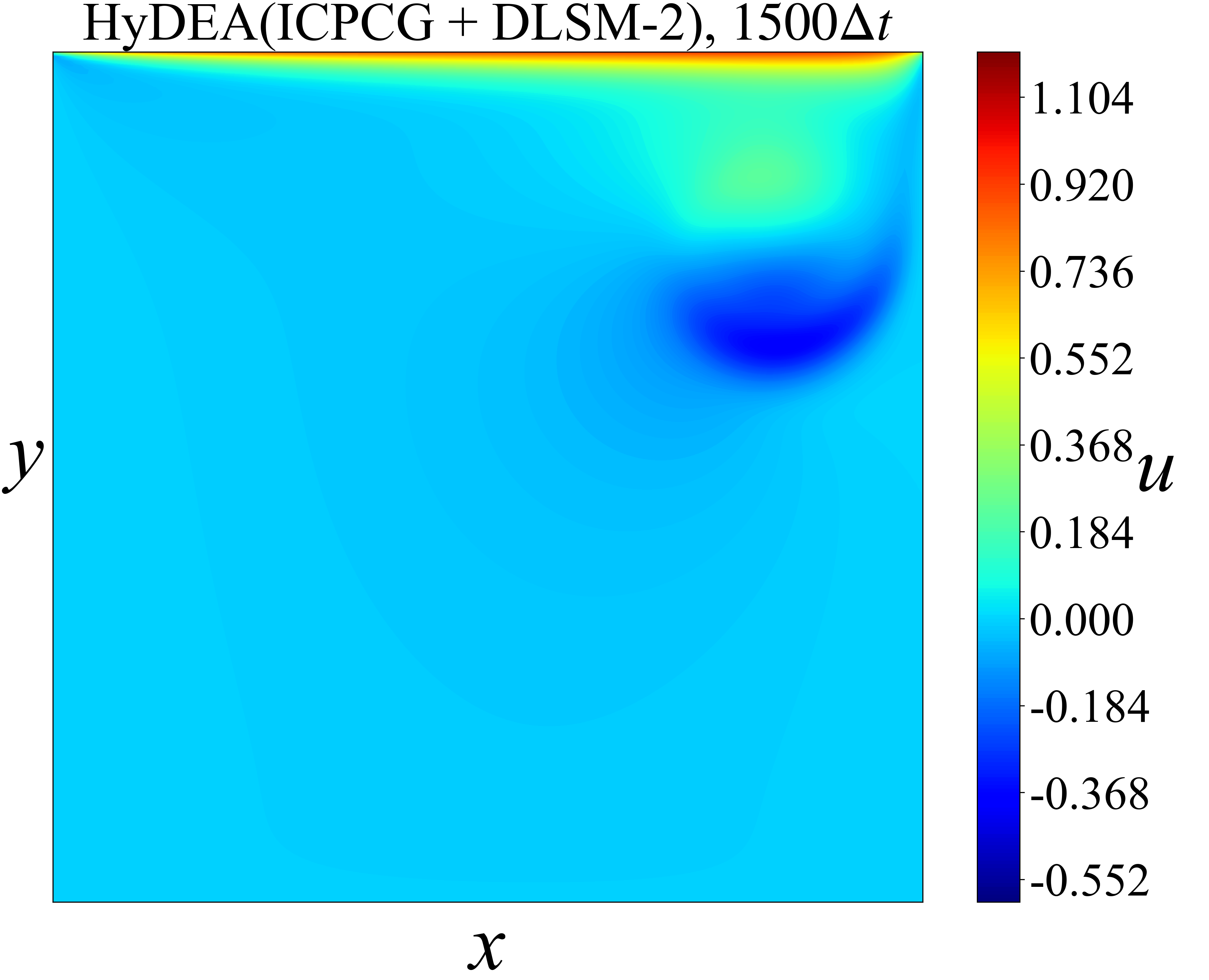}}
  \subfigure[]{
  \label{V_IC_1500step_Re3200}
  \includegraphics[scale=0.129]{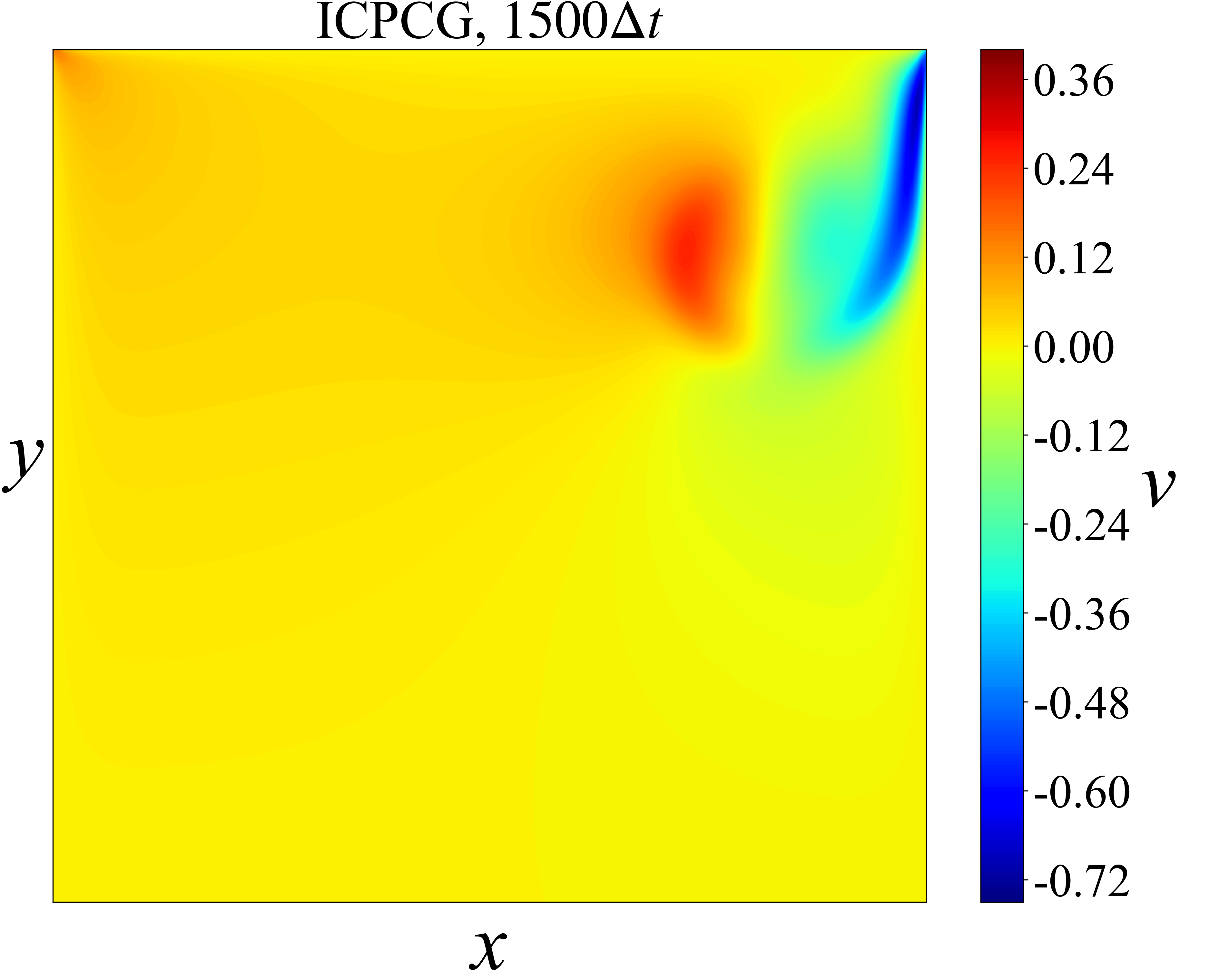}}
  \subfigure[]{
  \label{V_HIC_1500step_Re3200}
  \includegraphics[scale=0.129]{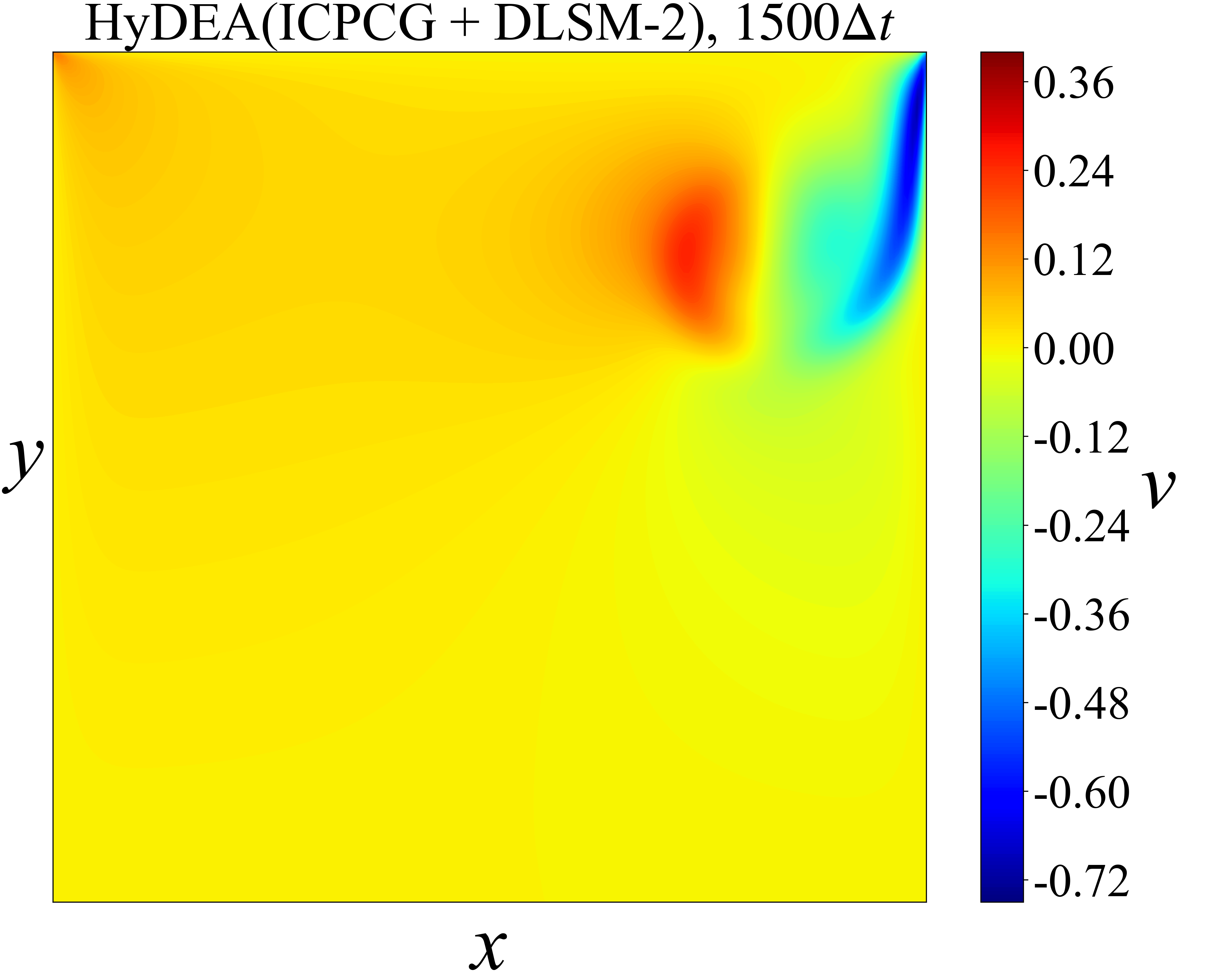}}
  \subfigure[]{
  \label{U_IC_10000step_Re3200}
  \includegraphics[scale=0.129]{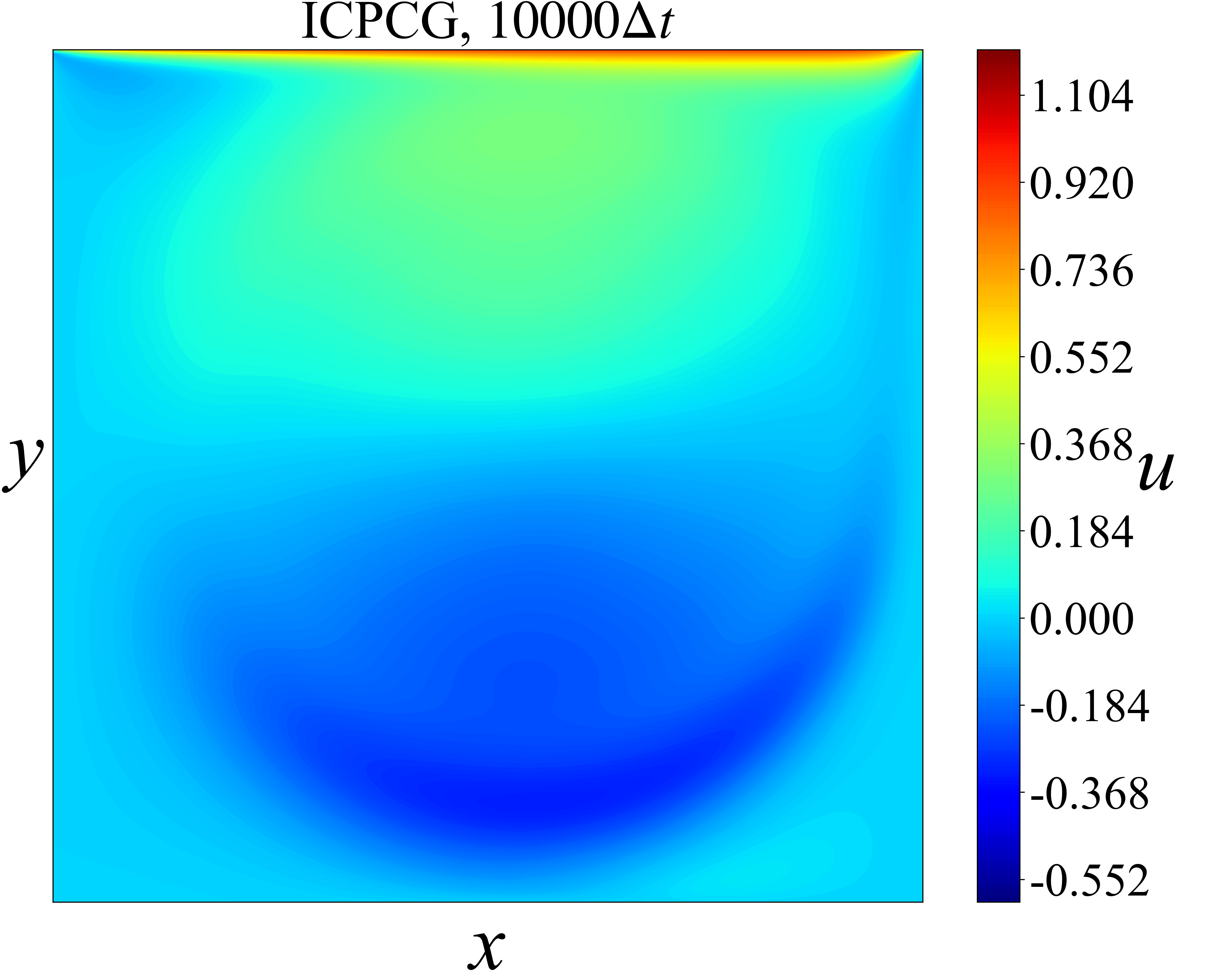}}
  \subfigure[]{
  \label{U_HIC_10000step_Re3200}
  \includegraphics[scale=0.129]{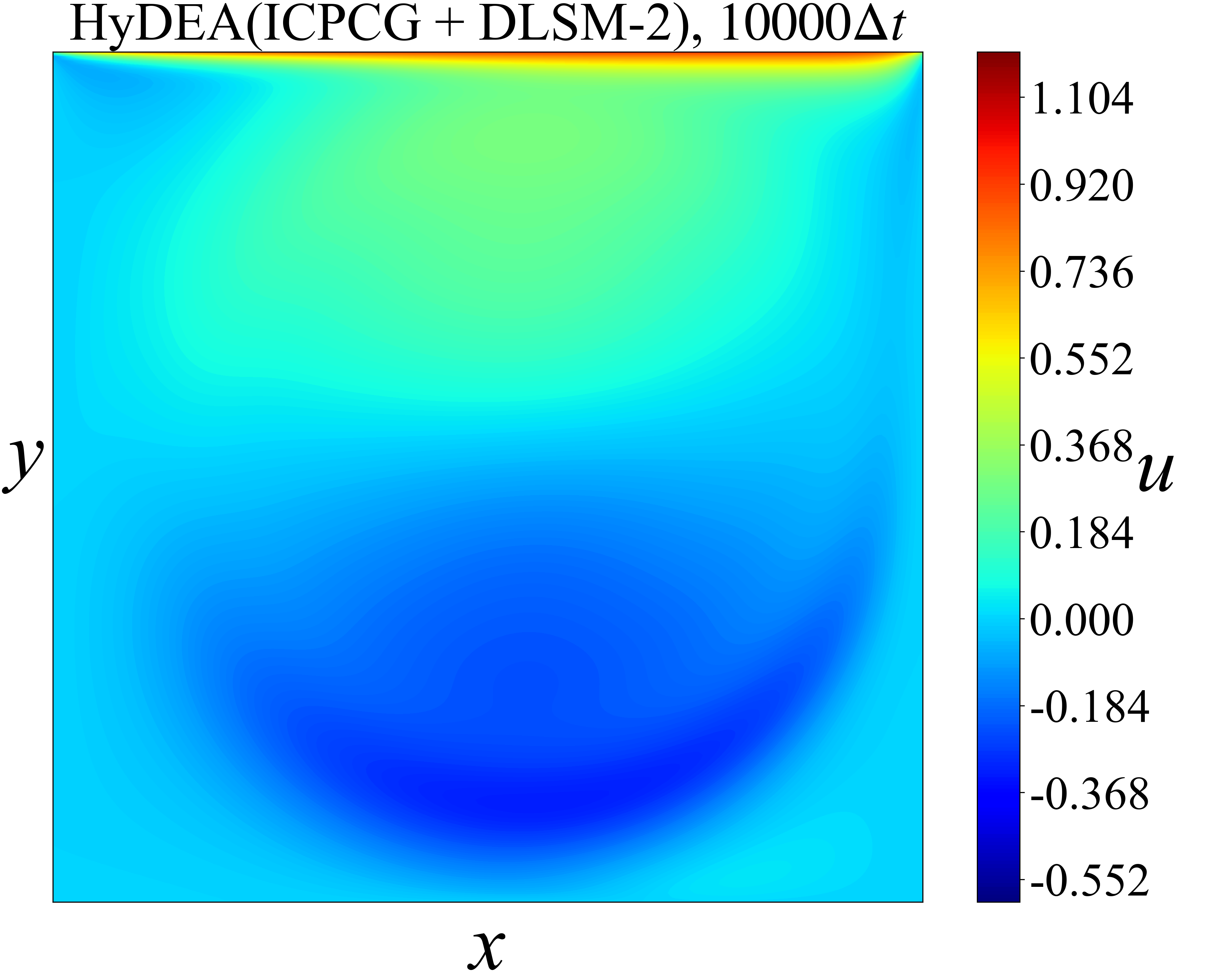}}
  \subfigure[]{
  \label{V_IC_10000step_Re3200}
  \includegraphics[scale=0.129]{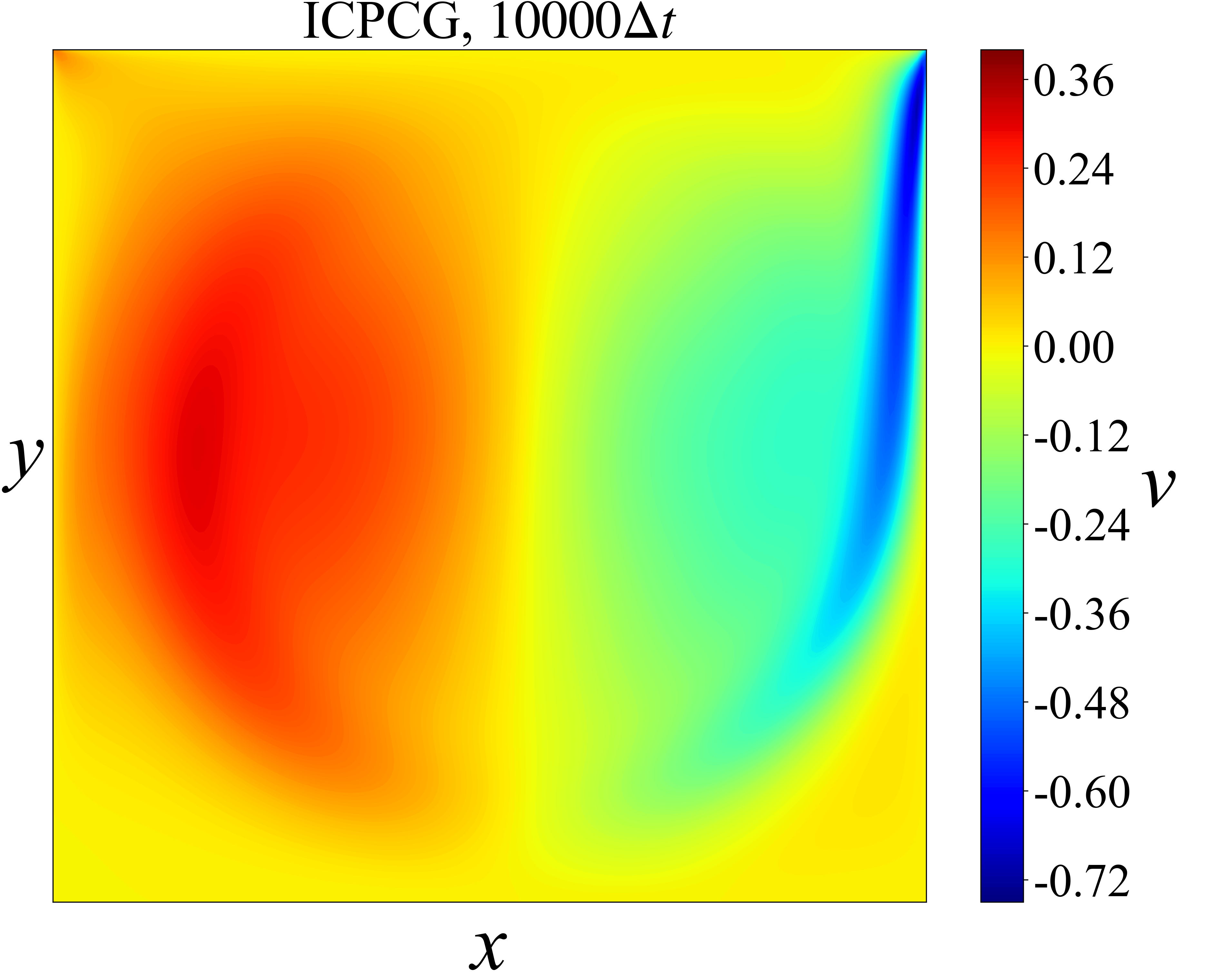}}
  \subfigure[]{
  \label{V_HIC_10000step_Re3200}
  \includegraphics[scale=0.129]{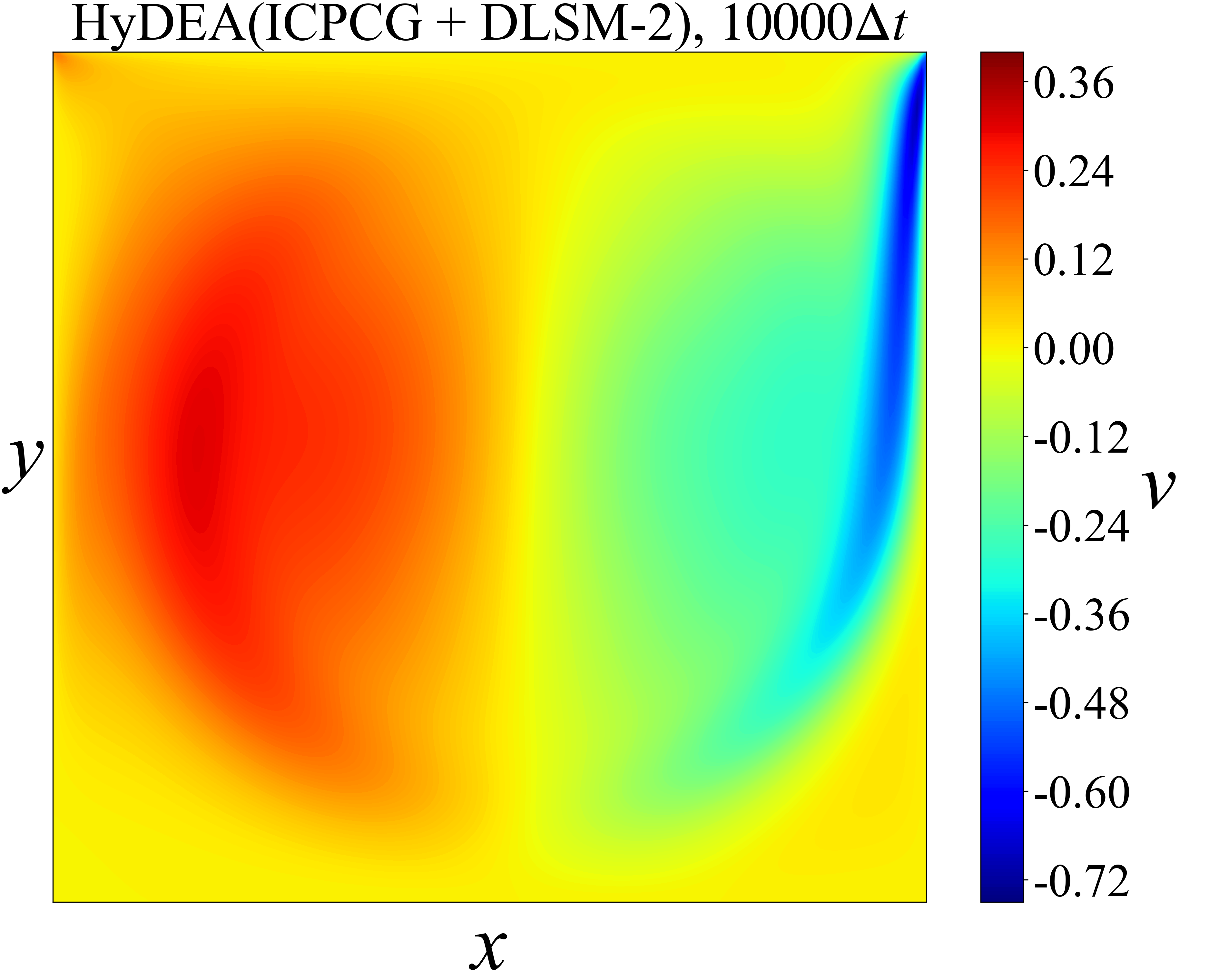}}
 \caption{Velocity fields for 2D lid-driven cavity flow at $Re=3200$ by ICPCG and HyDEA~(ICPCG+DLSM-2). (a)-(d) $u$ and $v$ at the $1500th$ time step. (e)-(h) $u$ and $v$ at the $10,000th$ time step.}
 \label{Flowfield192}
\end{figure}

Furthermore, as a representative case, HyDEA~(ICPCG+DLSM-2) demonstrates excellent agreement with the benchmark results of Ghia et al.~\cite{ghia1982}, as evidenced by the steady-state velocity profiles in Fig.~\ref{Re3200_Ghia_compare}. 
The velocity contours at the $1500th$ and $10,000th$ time steps in Fig.~\ref{Flowfield192} further validate HyDEA's accuracy, capturing the flow fields with high fidelity.

\subsubsection{Performance in mismatched discretizations} 
\label{128Inter}

\begin{figure}[htbp]
\centering
  \includegraphics[scale=0.15]{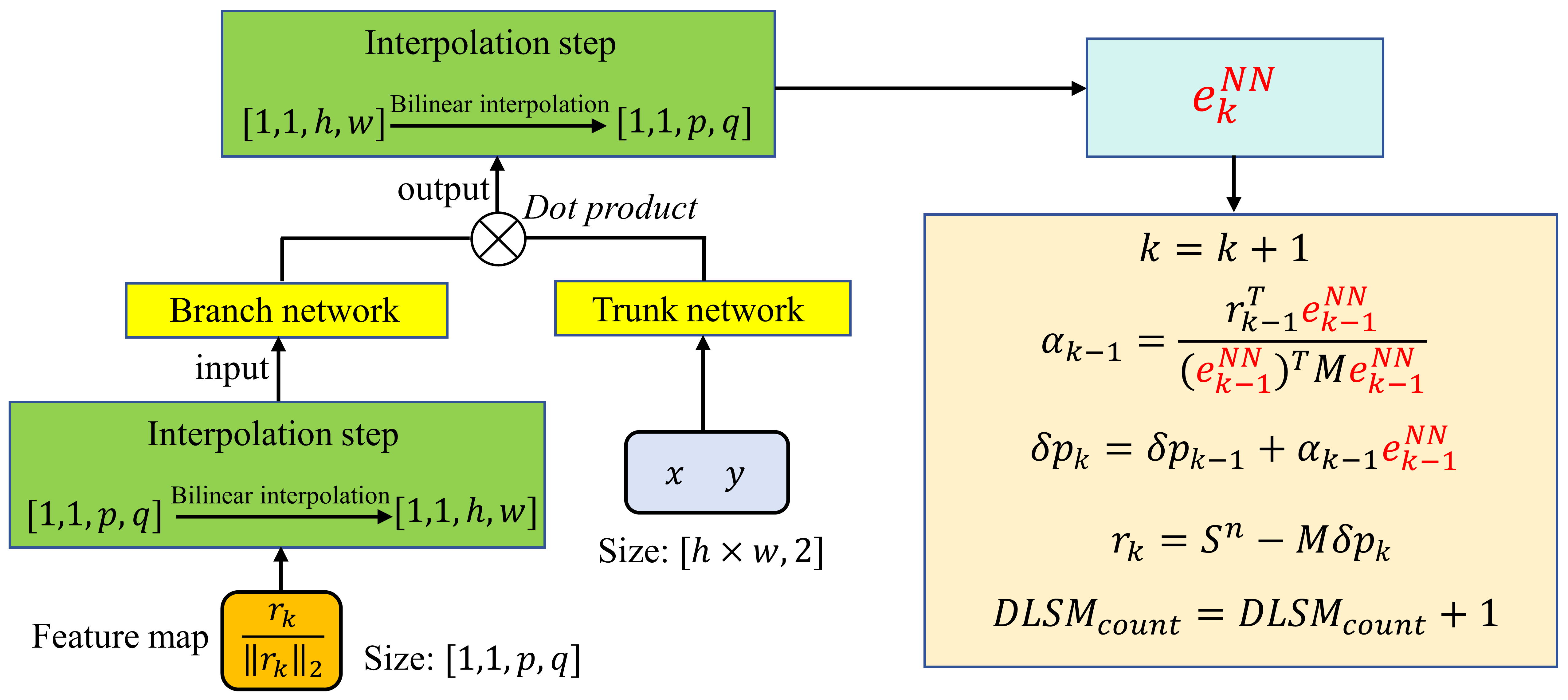}\\
  \caption{Downsmapling and upsampling for DeepONet within DLSM.}\label{adjust_HyDEA}
\end{figure}

In light of the significant computational overhead associated with dataset construction and neural network training for high-resolution simulations, this section examines HyDEA's applicability for super-resolutions.
Taking the version of Fig.~\ref{InitialPCG}) for example, where a CG-type method is taken as the initial solver, the workflow of HyDEA for super-resolution is sketched as follows:
\begin{enumerate}
    \item A DeepONet is trained {\it offline} based on a low-resolution dataset;
    \item A CG-type method conducts $Num_{CG-type}$ iterations on a high-resolution;
    \item The DLSM module receives the residual $r_k$ from the CG-type method and performs a downsampling to low-resolution $r_k^{low}$;
    \item The DeepONet predicts $e^{NN-low}_k$ for $r_k^{low}$ and performs an upsampling to high-resolution $e^{NN}_k$;
    \item The DLSM utizlied the $e^{NN}_k$ to finish the iteration.
\end{enumerate}
This HyDEA procedure repeats until convergence during each time step of the CFD.
We name this modified framework as super-resolution~(SR)-HyDEA for later reference.
To enable this capability, a pre-processing step and a post-processing step are introduced for DeepONet within the DLSM, as illustrated in Fig.~\ref{adjust_HyDEA}.
The input feature map is first downsampled via bilinear interpolation from high-resolution~($[1,1,p,q]$) to low-resolution~($[1,1,h,w]$) format before being fed into the DeepONet. Subsequently, the DeepONet's output is upsampled back to the original high resolution~($[1,1,p,q]$) via bilinear interpolation, after which the computation proceeds. 

As demonstration, we utilize DLSM-1 in Section~\ref{cavityRe1000} for SR-HyDEA,
where the DeepONet receives input feature maps with low resolution $[1,1,128,128]$,
corresponding to a CFD grid resolution of $128\times 128$.
$Num_{CG-type}=3$, and $Num_{DLSM}$ varies between $1$ and $2$. 
We evaluate the performace of this SR-HyDEA for four higher resolutions: $256 \times 256$, $384 \times 384$, $512 \times 512$ and $640 \times 640$. 
The CFD time step $\Delta t$ adjusts accordingly to be $0.002$, $0.002$, $0.0015$ and $0.0012$.
The geometric configuration and boundary conditions remain consistent with Section~\ref{cavityRe1000}, but we set $\nu=0.0001$ to achieve $Re=10,000$.

\begin{figure}[htbp] 
 \centering  
  \subfigure[]{
  \label{256size_128model_10step}
  \includegraphics[scale=0.21]{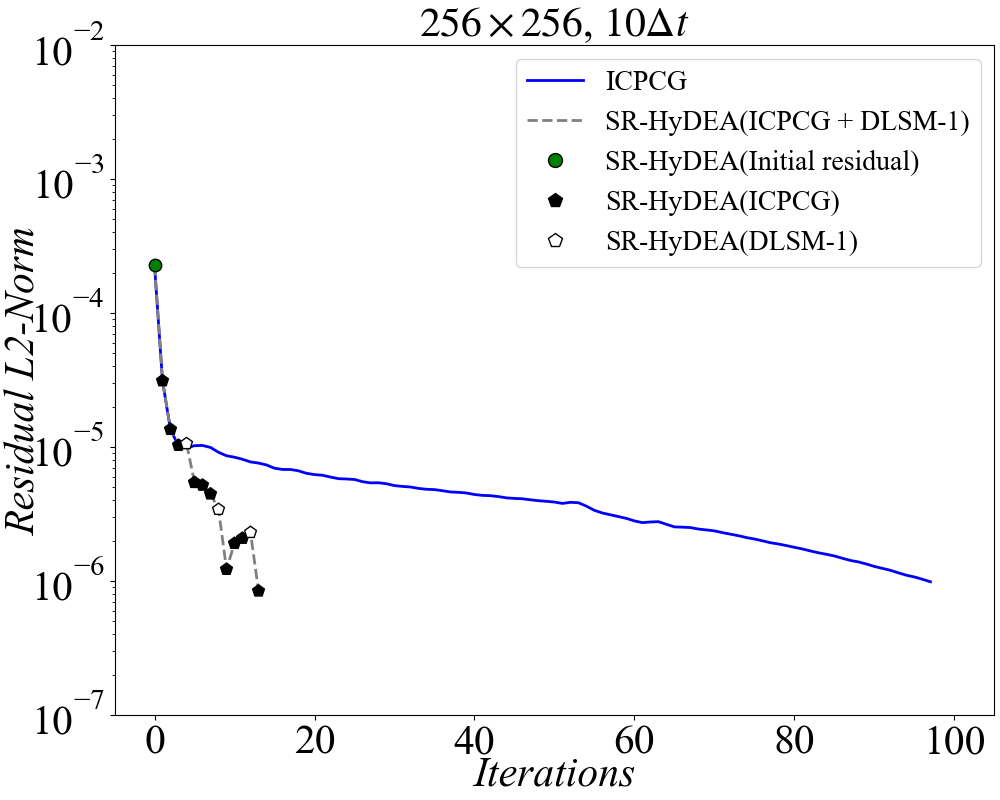}}
  \subfigure[]{
  \label{256size_128model_100step}
  \includegraphics[scale=0.21]{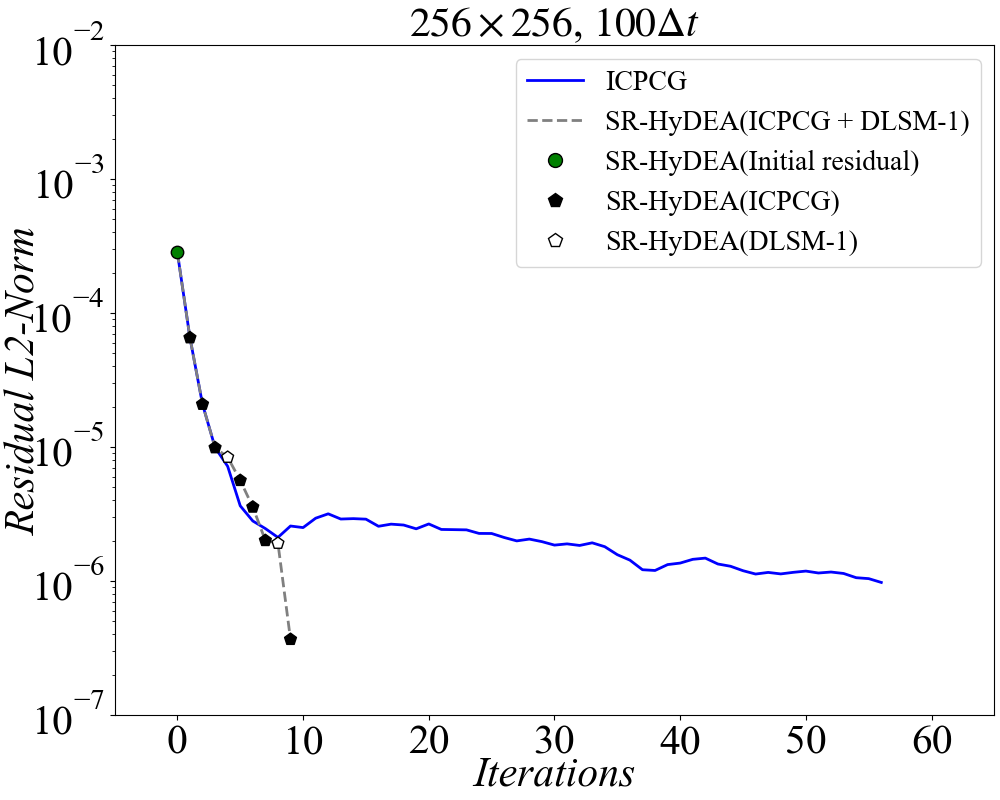}}
  \subfigure[]{
  \label{256size_128model_1000step}
  \includegraphics[scale=0.21]{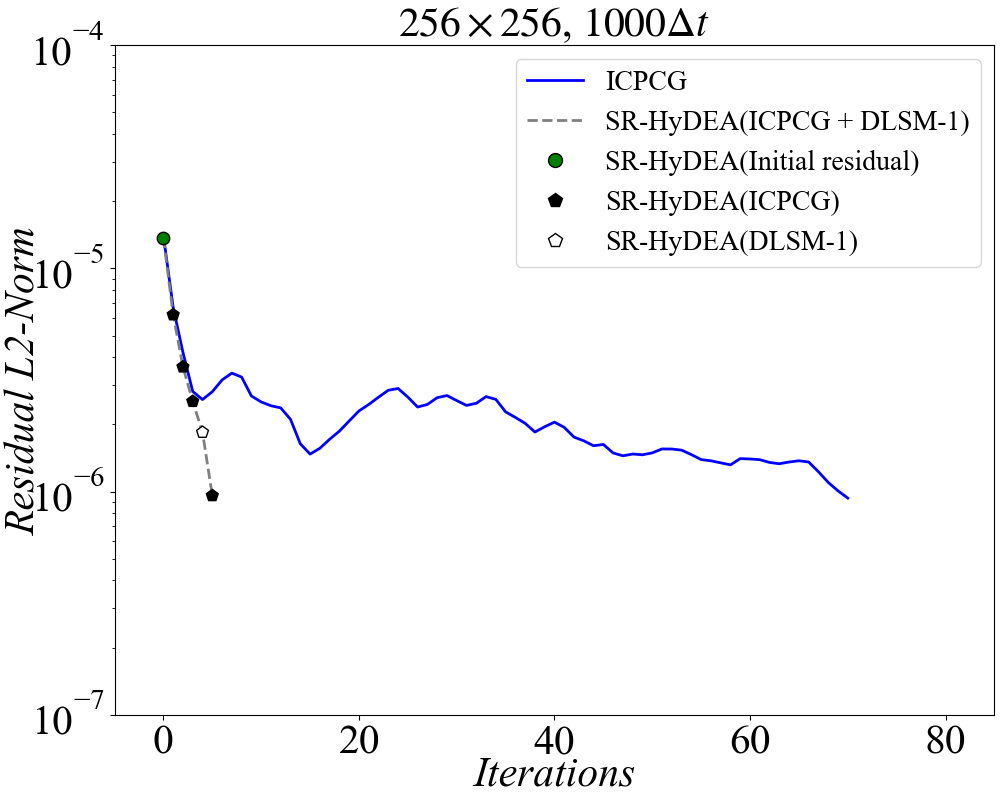}}
  \subfigure[]{
  \label{384size_128model_10step}
  \includegraphics[scale=0.21]{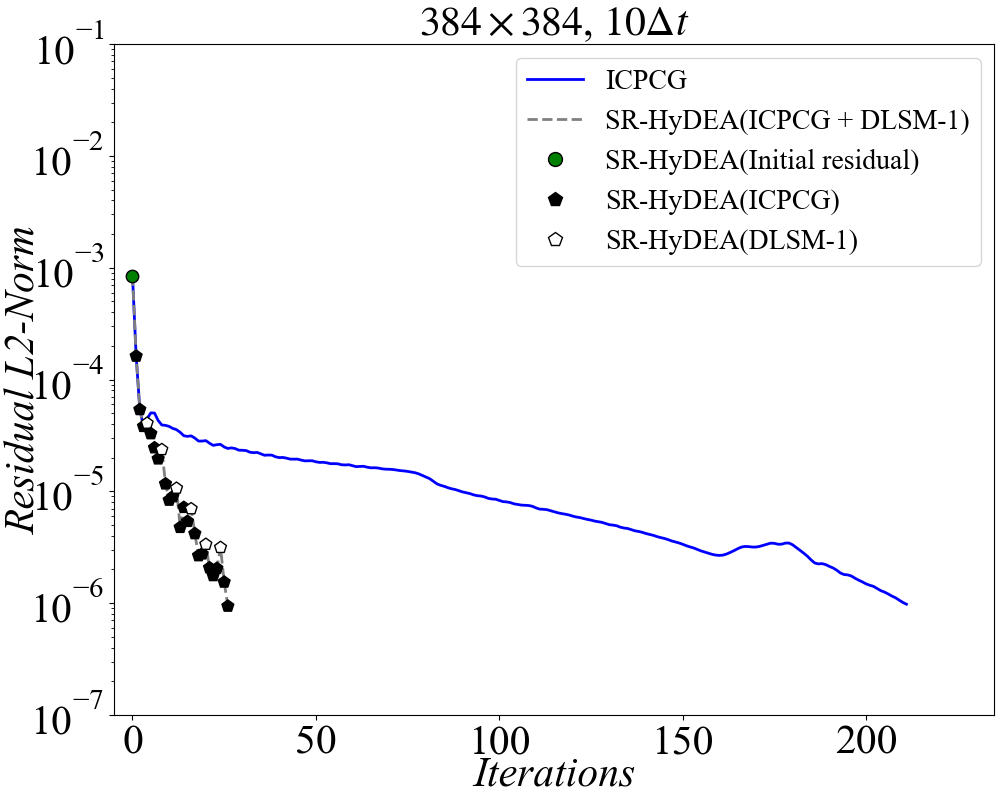}}
  \subfigure[]{
  \label{384size_128model_100step}
  \includegraphics[scale=0.21]{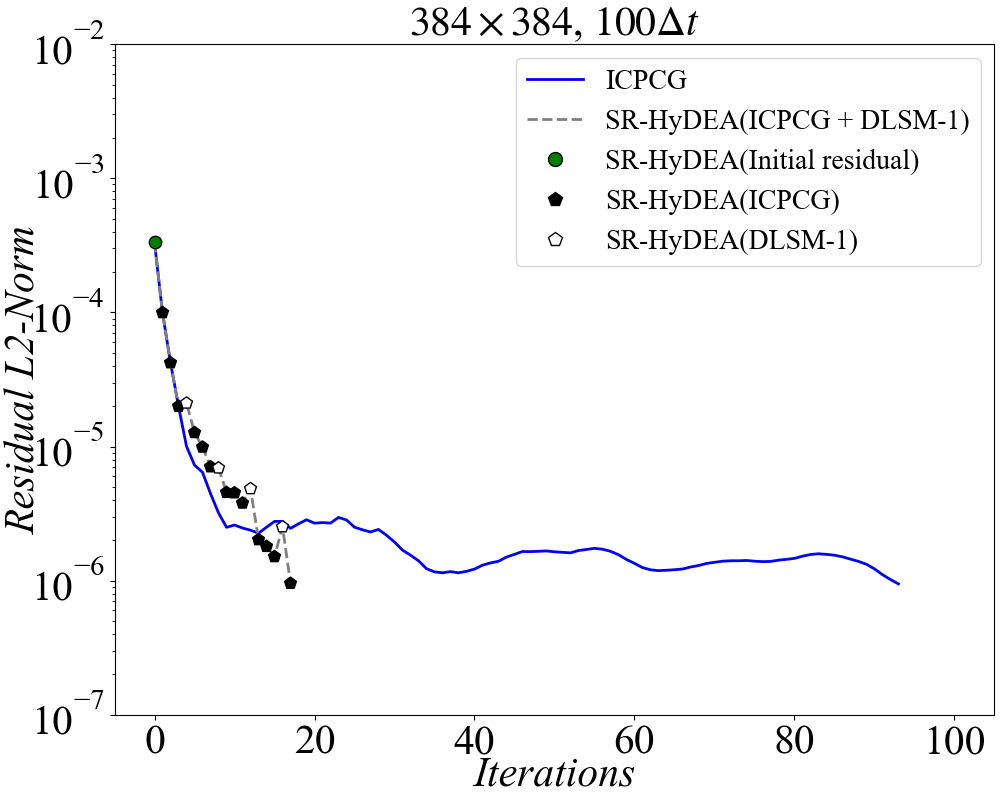}}
  \subfigure[]{
  \label{384size_128model_1000step}
  \includegraphics[scale=0.21]{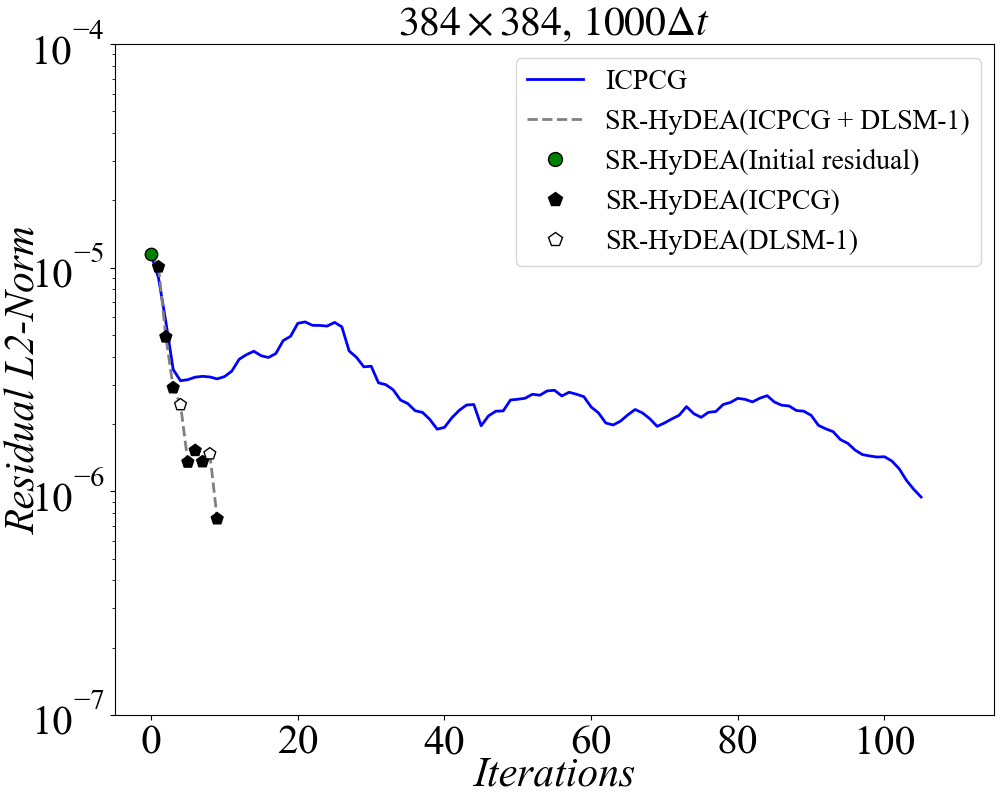}}
  \subfigure[]{
  \label{512size_128model_10step}
  \includegraphics[scale=0.21]{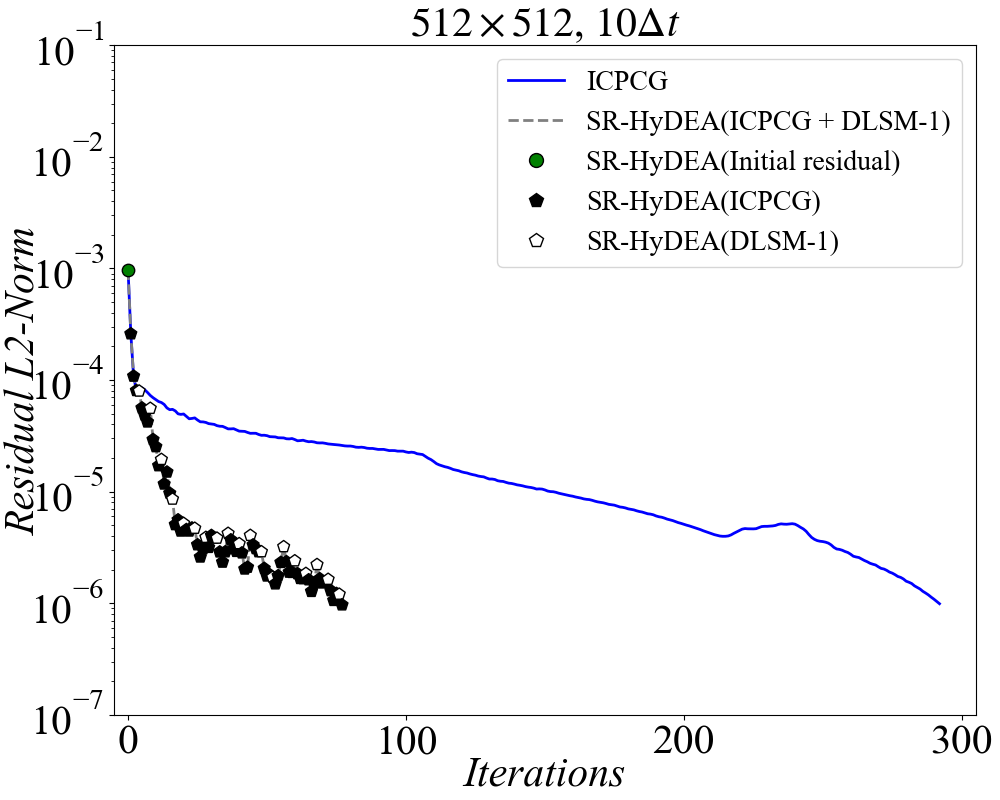}}
  \subfigure[]{
  \label{512size_128model_100step}
  \includegraphics[scale=0.21]{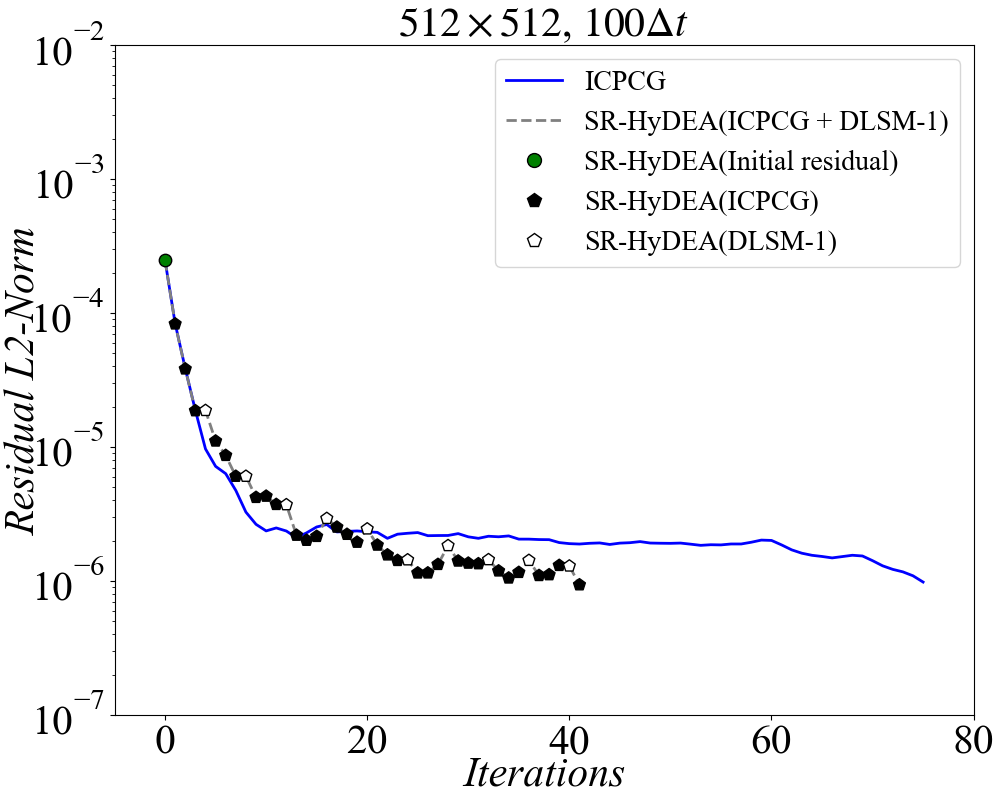}}
  \subfigure[]{
  \label{512size_128model_1000step}
  \includegraphics[scale=0.21]{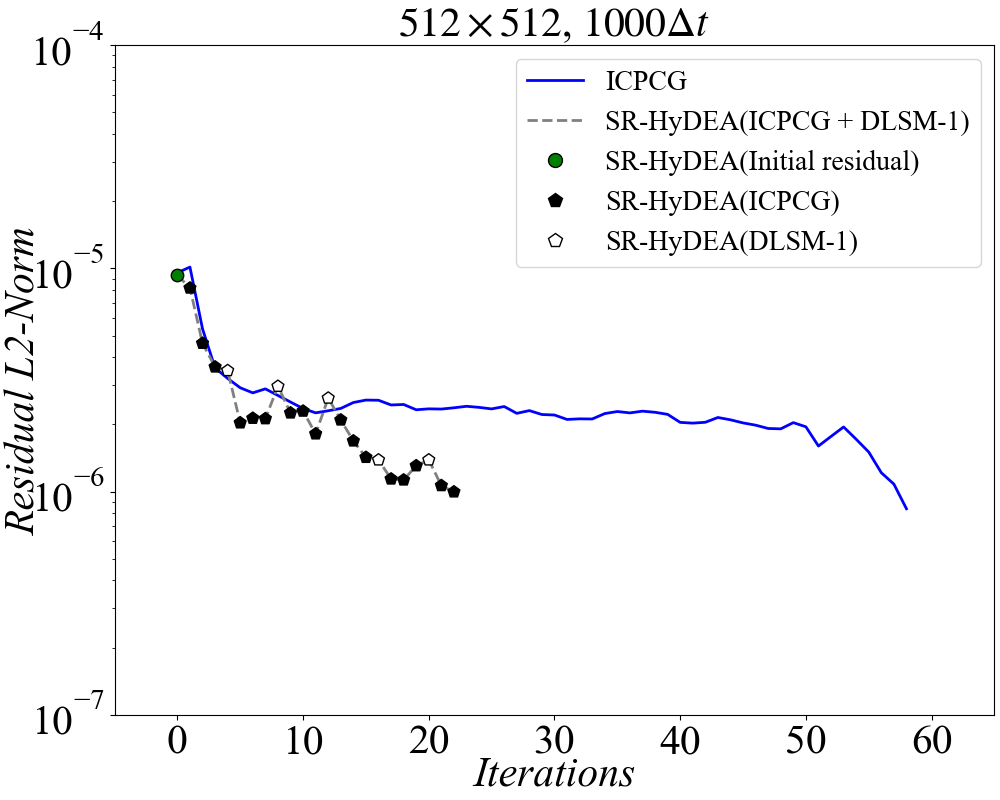}}
  \subfigure[]{
  \label{640size_128model_10step}
  \includegraphics[scale=0.21]{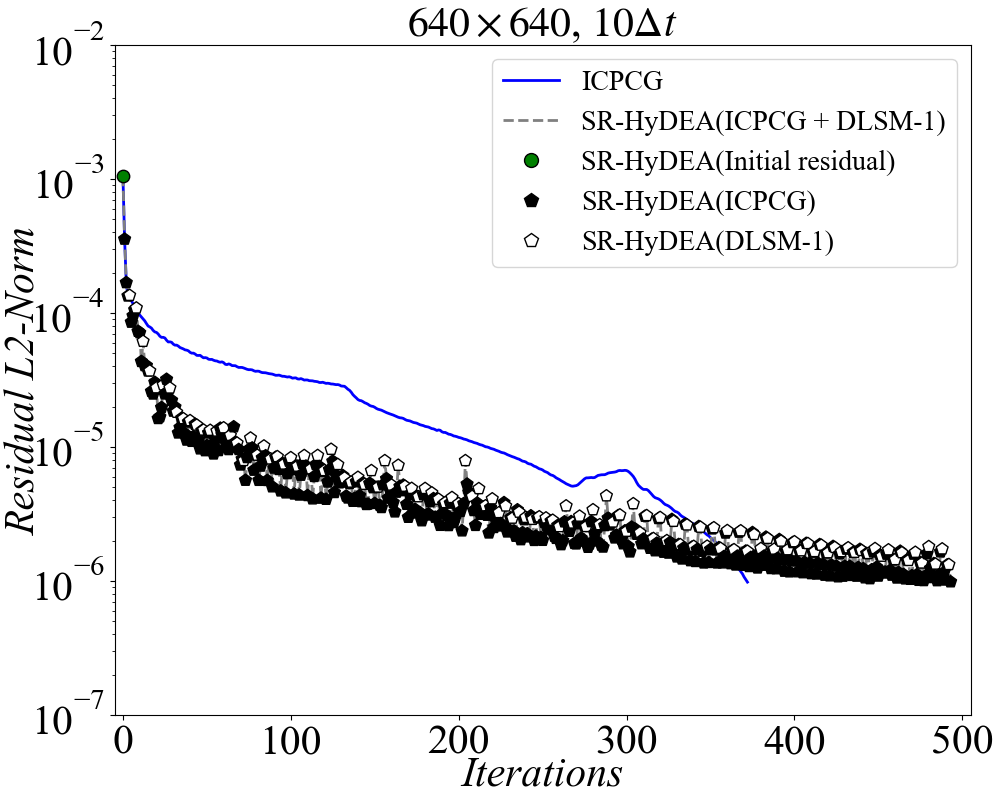}}
  \subfigure[]{
  \label{640size_128model_100step}
  \includegraphics[scale=0.21]{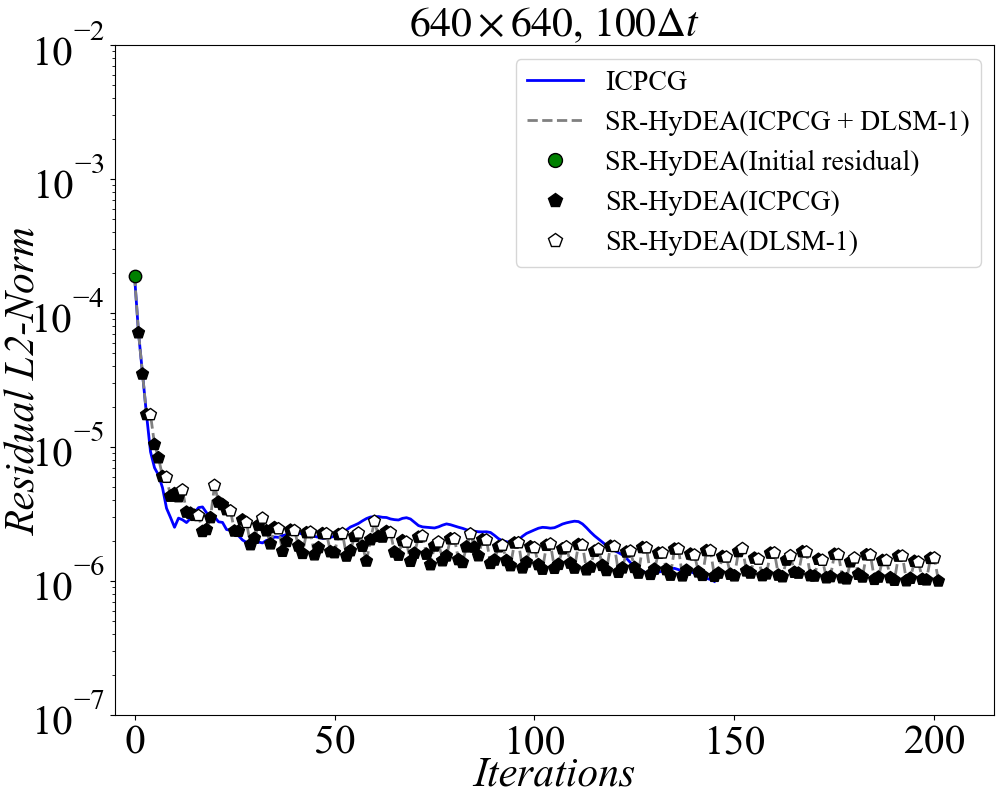}}
  \subfigure[]{
  \label{640size_128model_1000step}
  \includegraphics[scale=0.21]{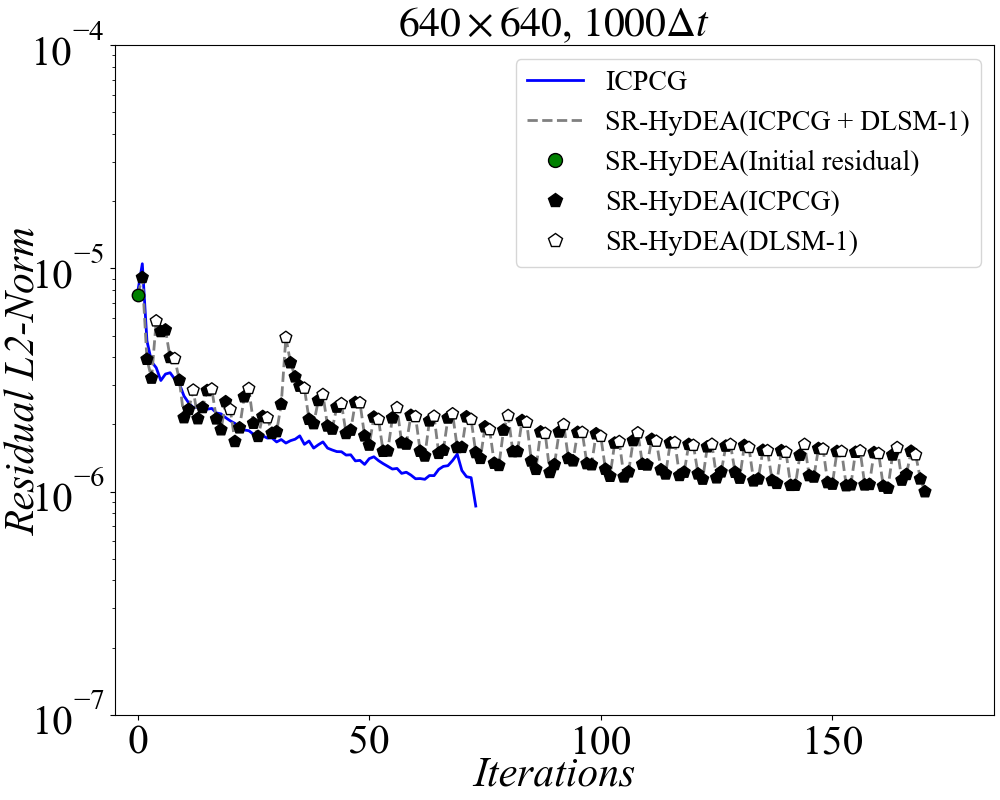}}
  \caption{Iterative residuals of solving the PPE for super-resolutions at the $10th$, $100th$ and $1000th$ time steps for 2D lid-driven cavity flow at $Re=10,000$. DeepONet is trained {\it offline} for $128\times 128$ and utilized {\it online} for (a)-(c) $256 \times 256$ (d)-(f) $384 \times 384$ (g)-(i) $512 \times 512$ (j)-(l) $640 \times 640$.}\label{128inter_highresolution}
\end{figure}

Here, ICPCG is taken as the baseline performance and compared against SR-HyDEA~(ICPCG+DLSM-1): the iterative residuals for solving the PPE at the $10th$, $100th$ and $1000th$ time steps with $Num_{CG-type}=3$ and $Num_{DLSM}=1$ are illustrated in Fig.~\ref{128inter_highresolution}. The SR-HyDEA demonstrates a significant reduction in the number of iterations compared to the ICPCG method alone, even when the resolution is increased by $4\times4=16$ times in two dimensions. 
However, the superior performance diminishes when the resolution is further increased to $5\times5=25$ times, revealing the inherent limitations of low-resolution neural network models in generalizing to high-resolution flow simulations.

\begin{figure}[htbp] 
 \centering  
  \subfigure[]{
  \label{128inter256}
  \includegraphics[scale=0.24]{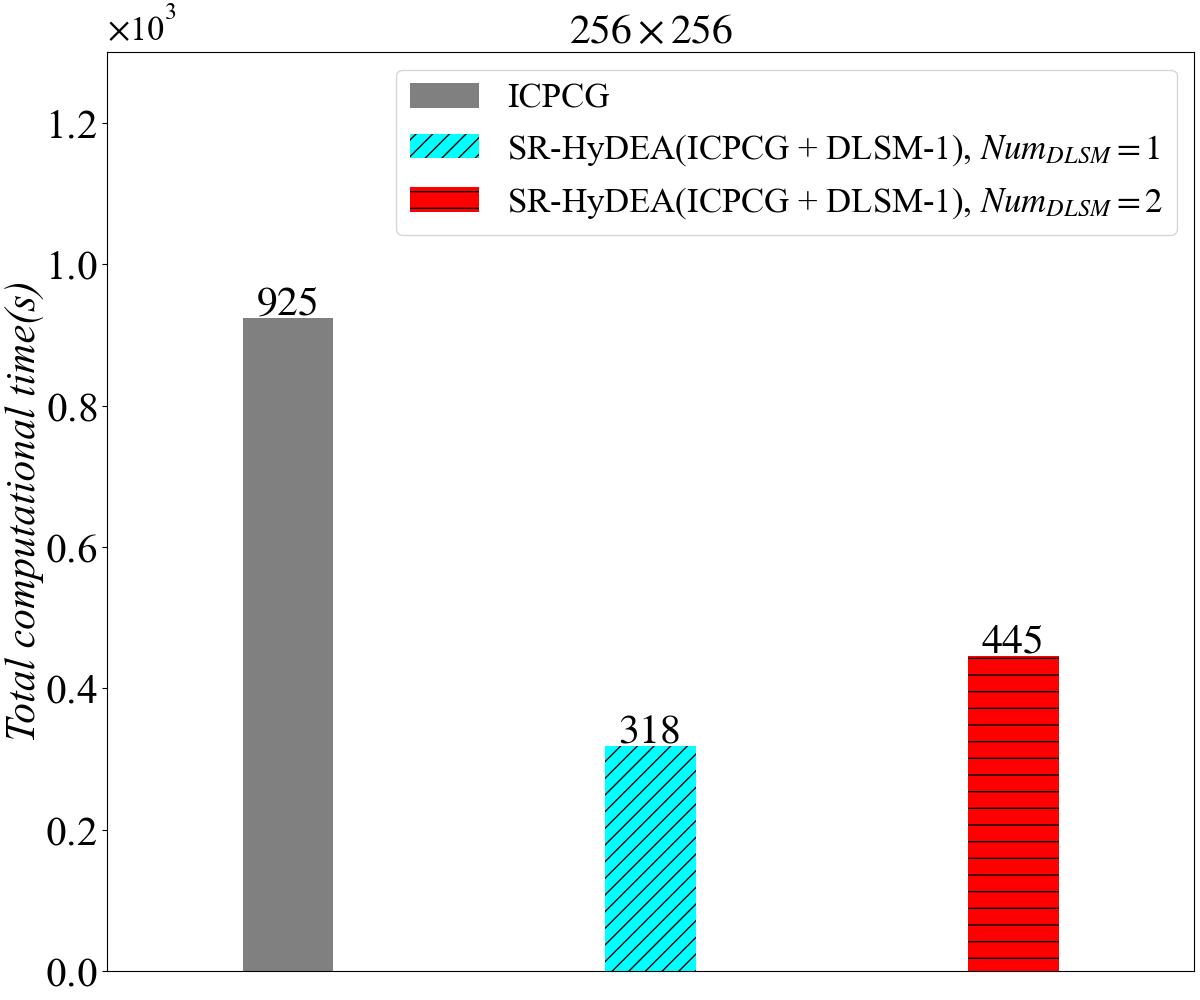}}
  \subfigure[]{
  \label{128inter384}
  \includegraphics[scale=0.24]{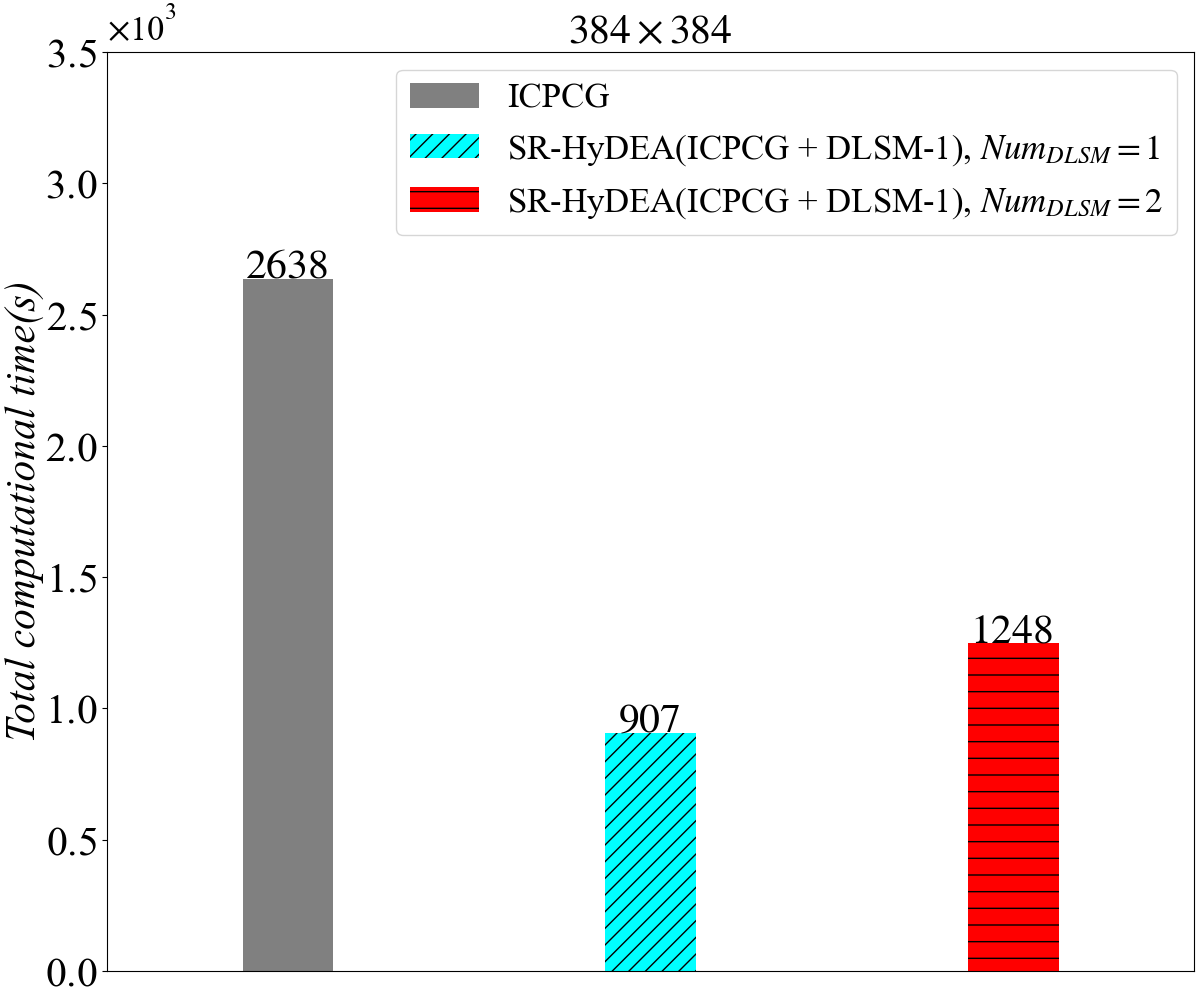}}
  \subfigure[]{
  \label{128inter512}
  \includegraphics[scale=0.24]{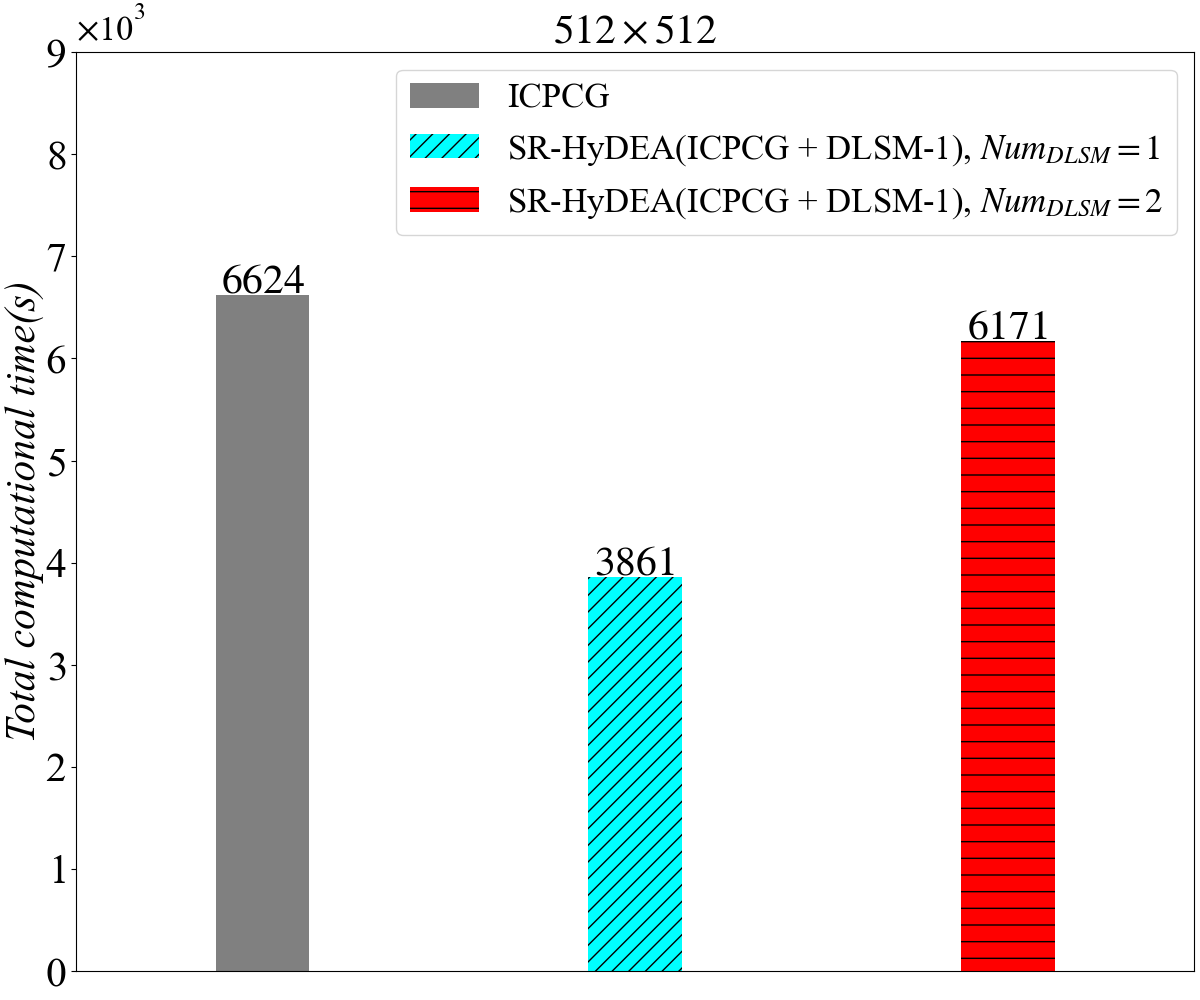}}
  \subfigure[]{
  \label{128inter640}
  \includegraphics[scale=0.24]{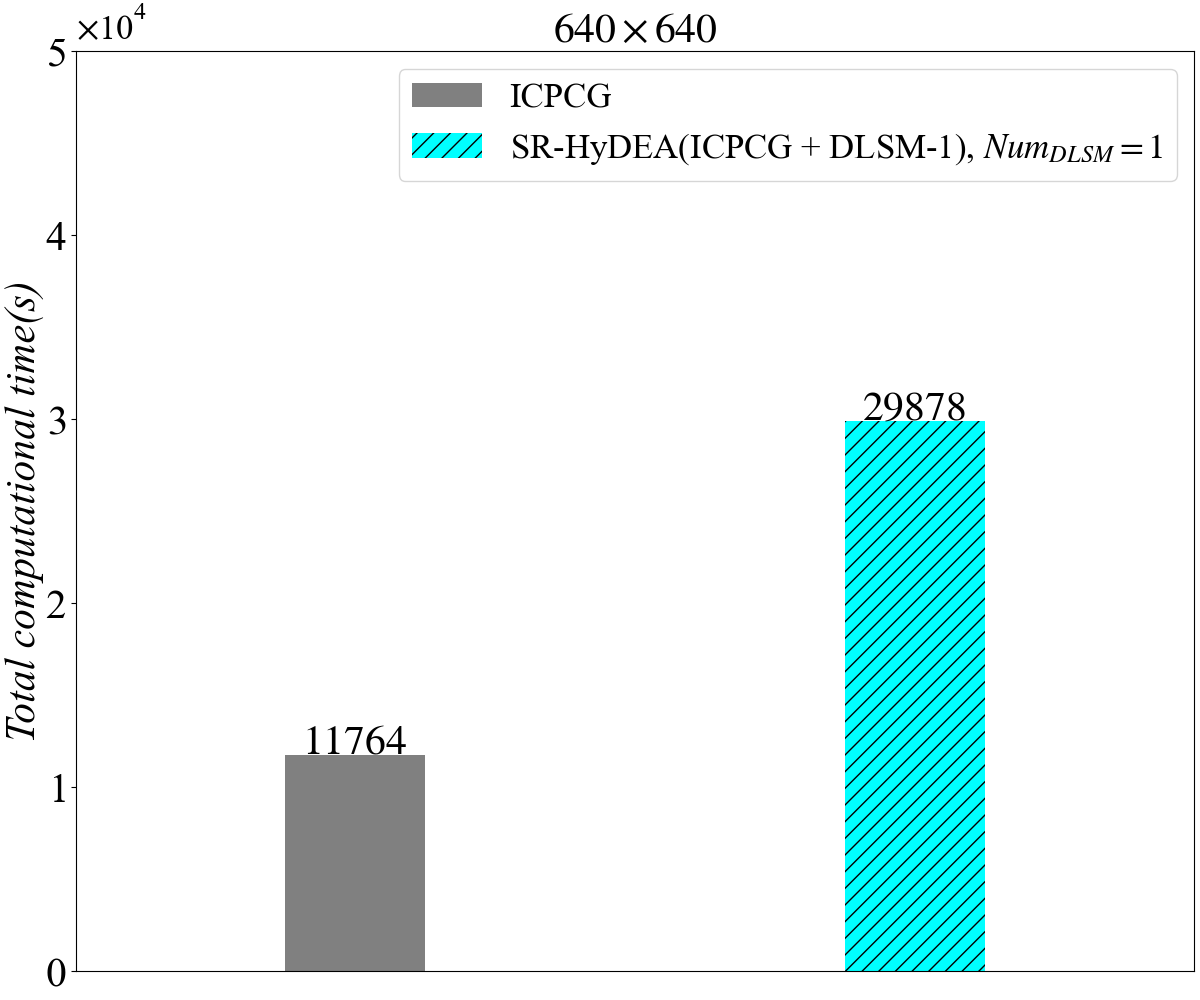}}
 \caption{Comparison of computational time required to solve the PPE over a duration of $10,000\Delta t$ for super-resolutions at $Re=10,000$. (a) $256 \times 256$ (b) $384 \times 384$ (c) $512 \times 512$ (c) $640 \times 640$.}
 \label{128Inter_time_compare}
\end{figure}

Fig.~\ref{128Inter_time_compare} further summarizes the comparison of computational time to solve the PPE over $10,000$ consecutive time steps.
In general, SR-HyDEA demonstrates gains in computational efficiency when $Num_{DLSM}=1$ across the first three high resolutions. However, when $Num_{DLSM}=2$, its performance benefit almost vanishes at resolution of $512 \times 512$. Consequently, we evaluate only for $Num_{DLSM}=1$ at the resolution of $640 \times 640$.
The results of computational time are consistent with those of iteration numbers shown
in Fig.~\ref{128inter_highresolution}: 
super-resolution works well if the upscaling factor is $\le 16$,
but a more aggressive operation deteriorates the performance.

\begin{figure}[htbp] 
 \centering  
  \subfigure[]{
  \label{U_IC_5000step_Re10000}
  \includegraphics[scale=0.129]{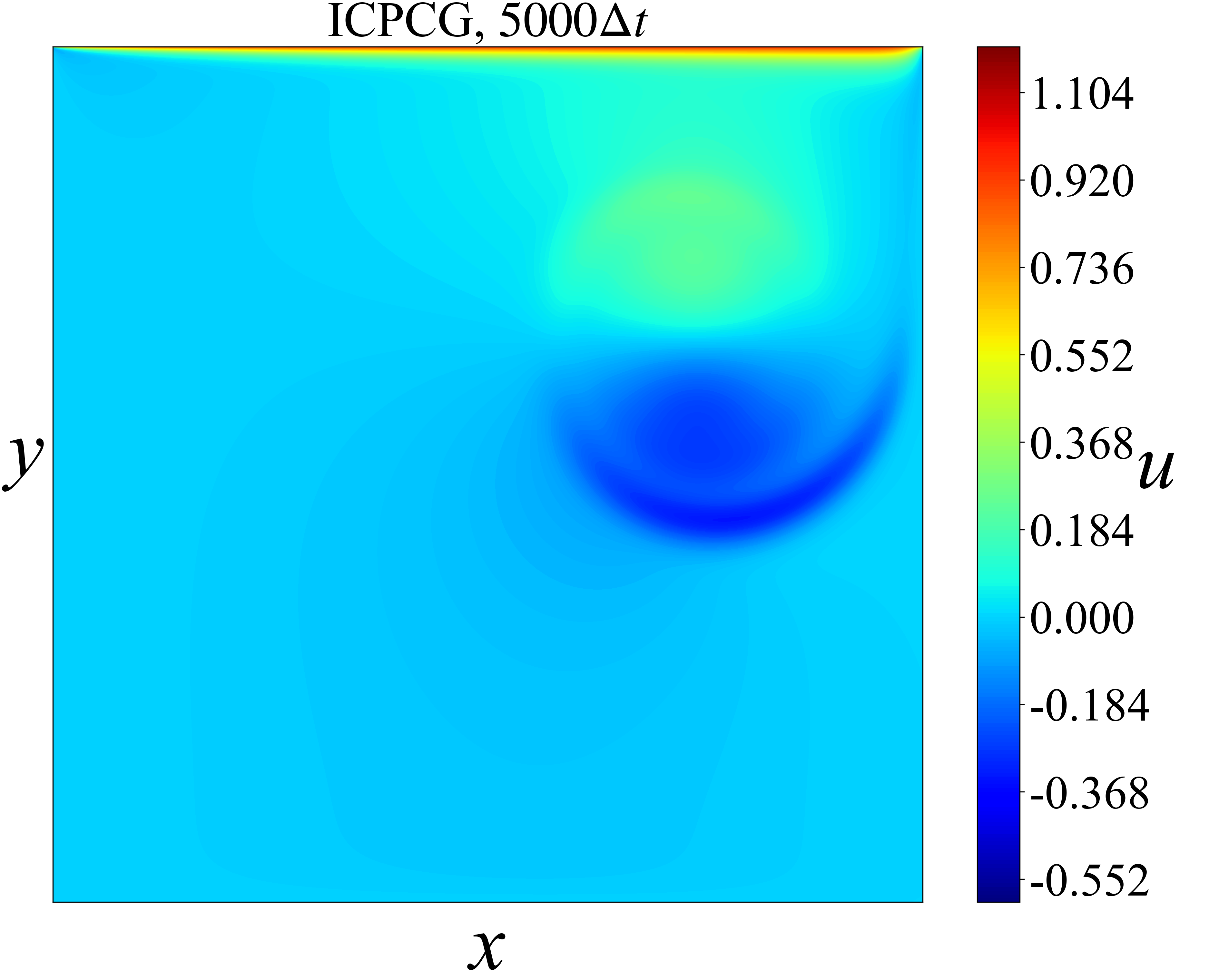}}
  \subfigure[]{
  \label{U_HIC_5000step_Re10000}
  \includegraphics[scale=0.129]{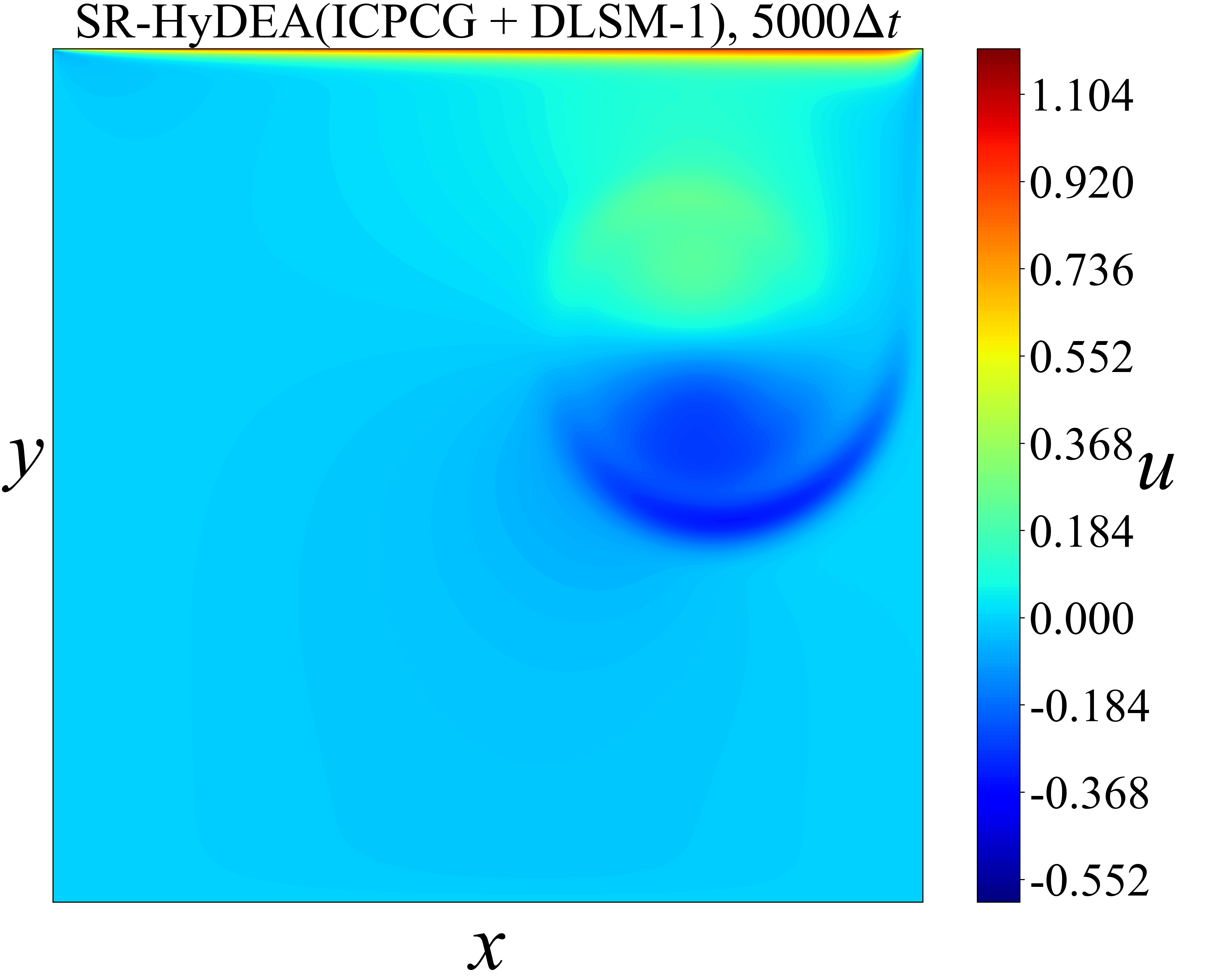}}
  \subfigure[]{
  \label{V_IC_5000step_Re10000}
  \includegraphics[scale=0.129]{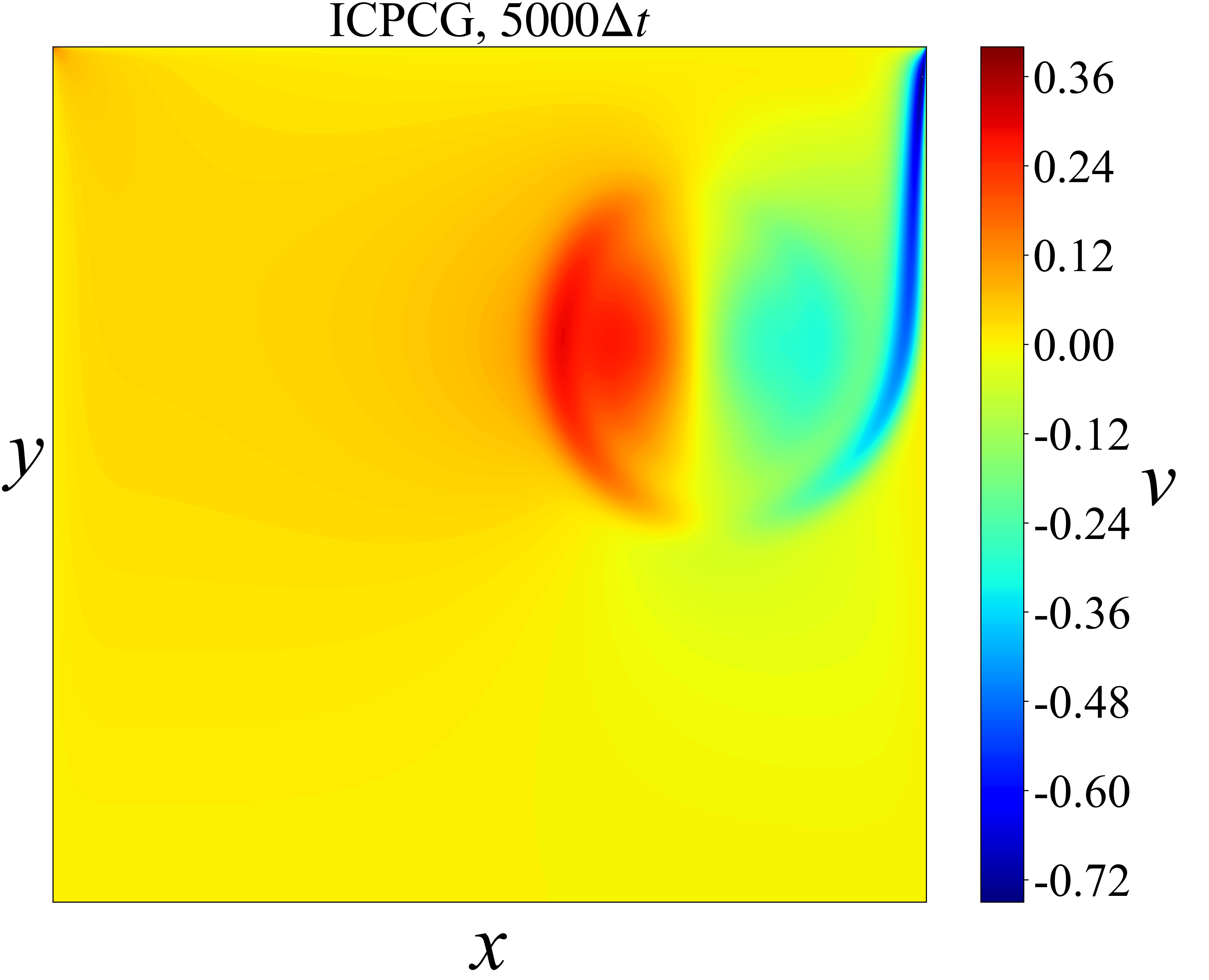}}
  \subfigure[]{
  \label{V_HIC_5000step_Re10000}
  \includegraphics[scale=0.129]{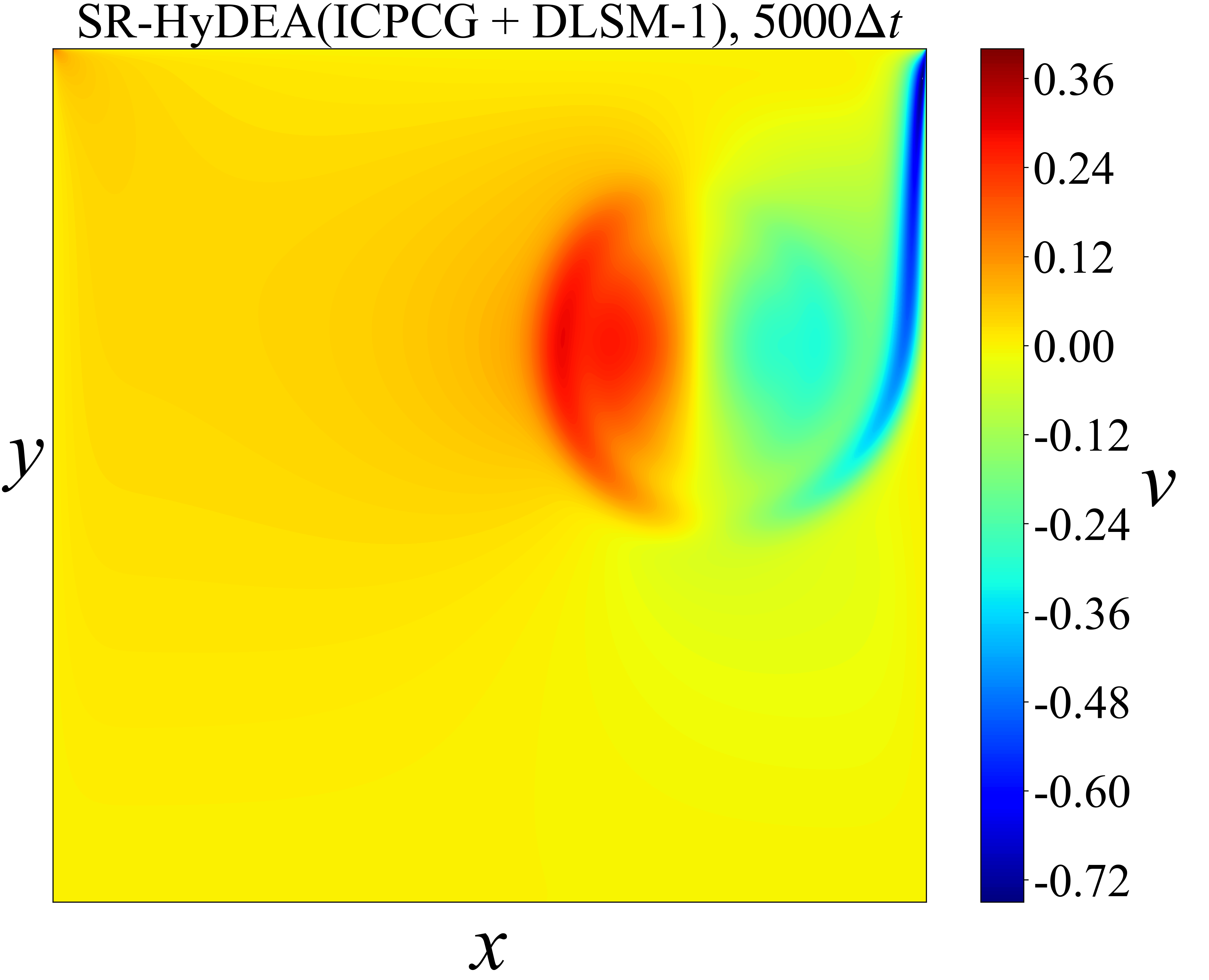}}
  \subfigure[]{
  \label{U_IC_10000step_Re10000}
  \includegraphics[scale=0.129]{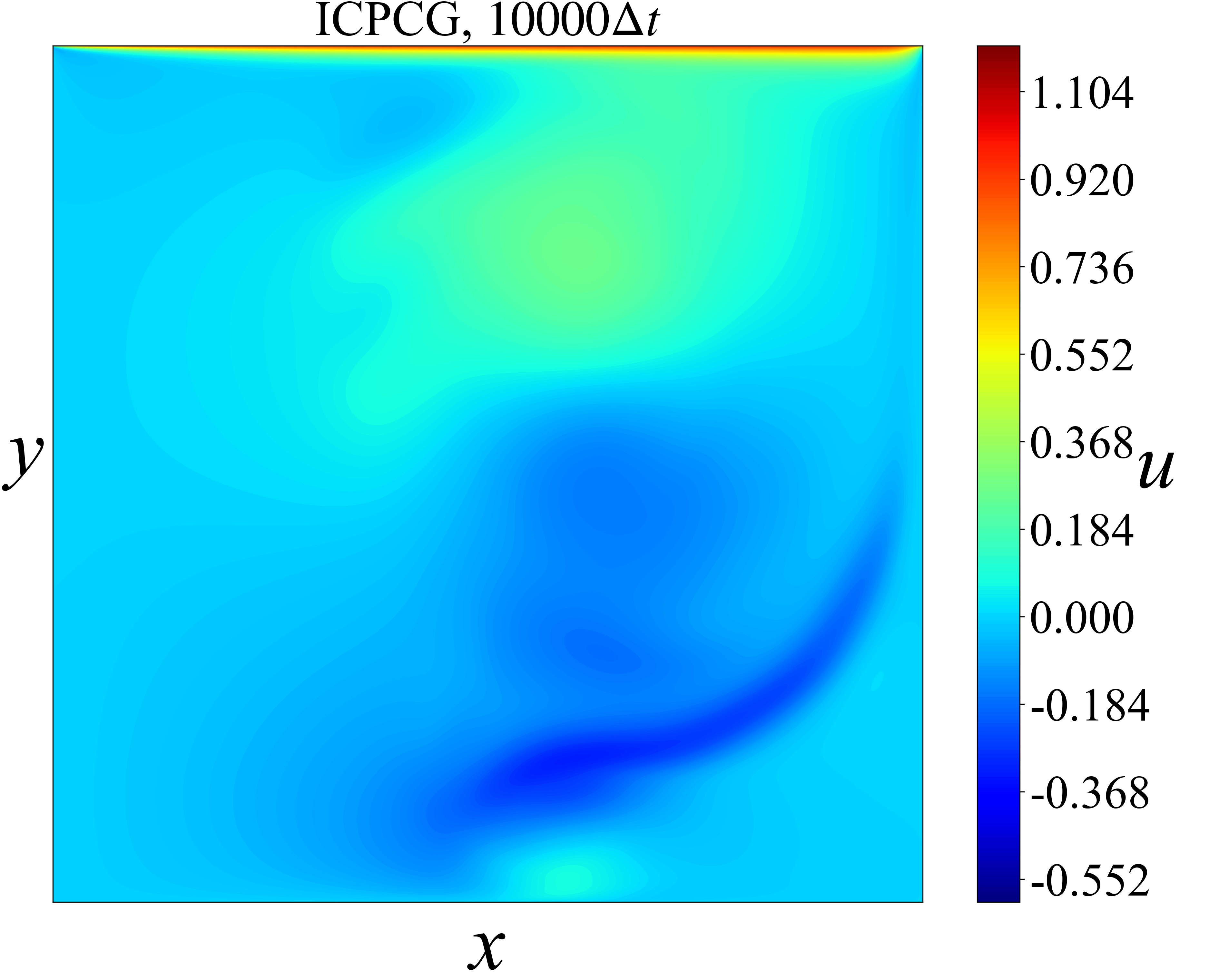}}
  \subfigure[]{
  \label{U_HIC_10000step_Re10000}
  \includegraphics[scale=0.129]{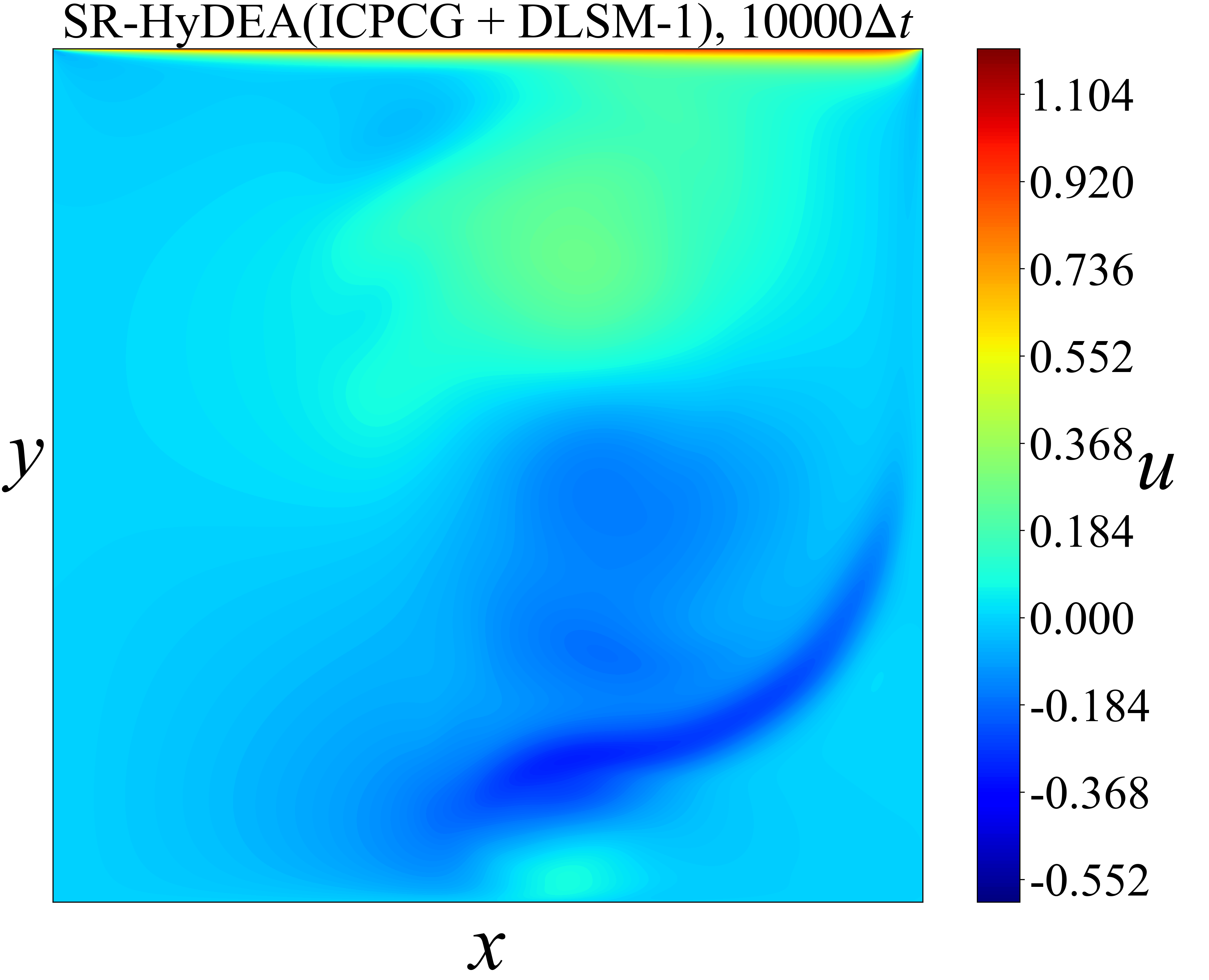}}
  \subfigure[]{
  \label{V_IC_10000step_Re10000}
  \includegraphics[scale=0.129]{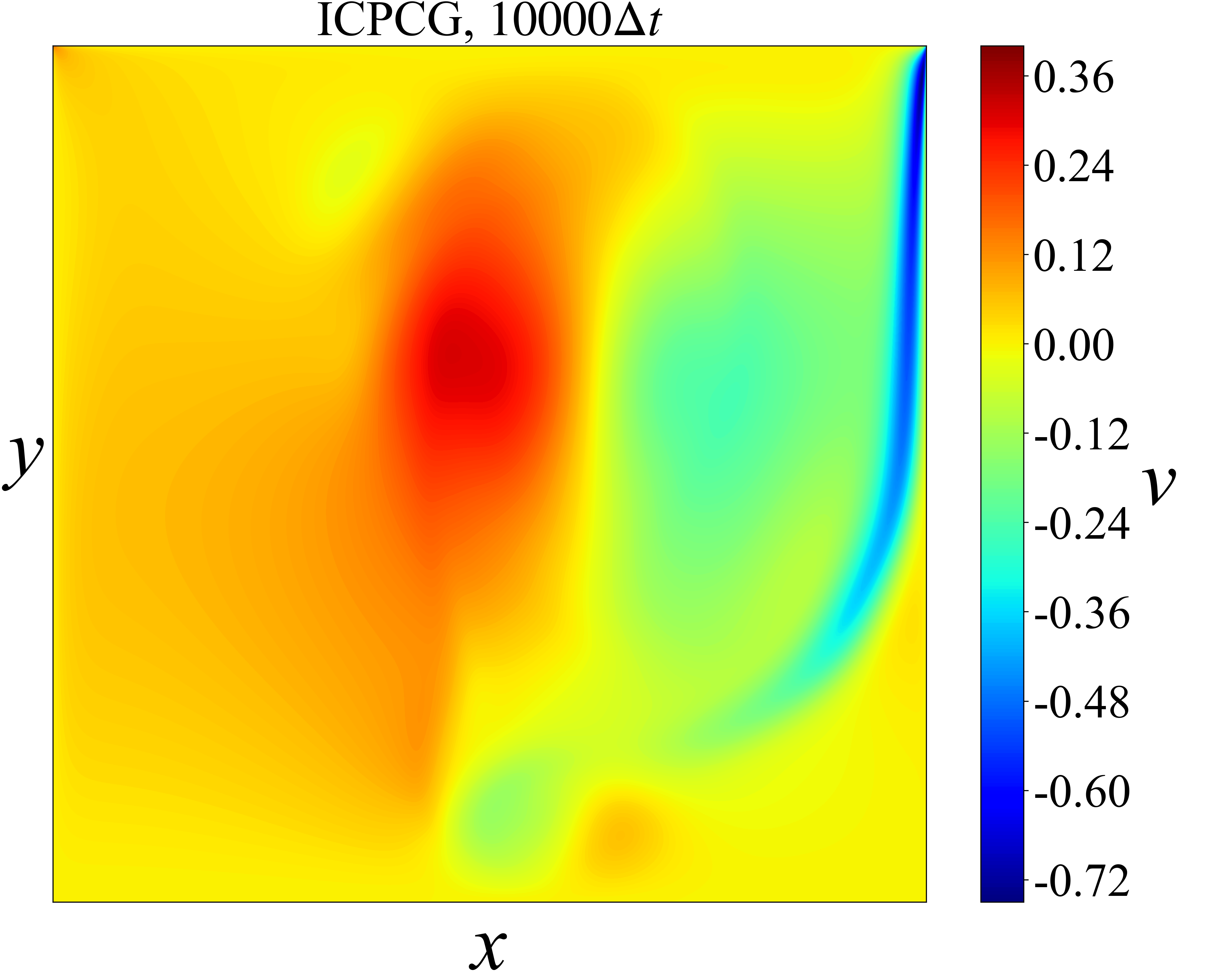}}
  \subfigure[]{
  \label{V_HIC_10000step_Re10000}
  \includegraphics[scale=0.129]{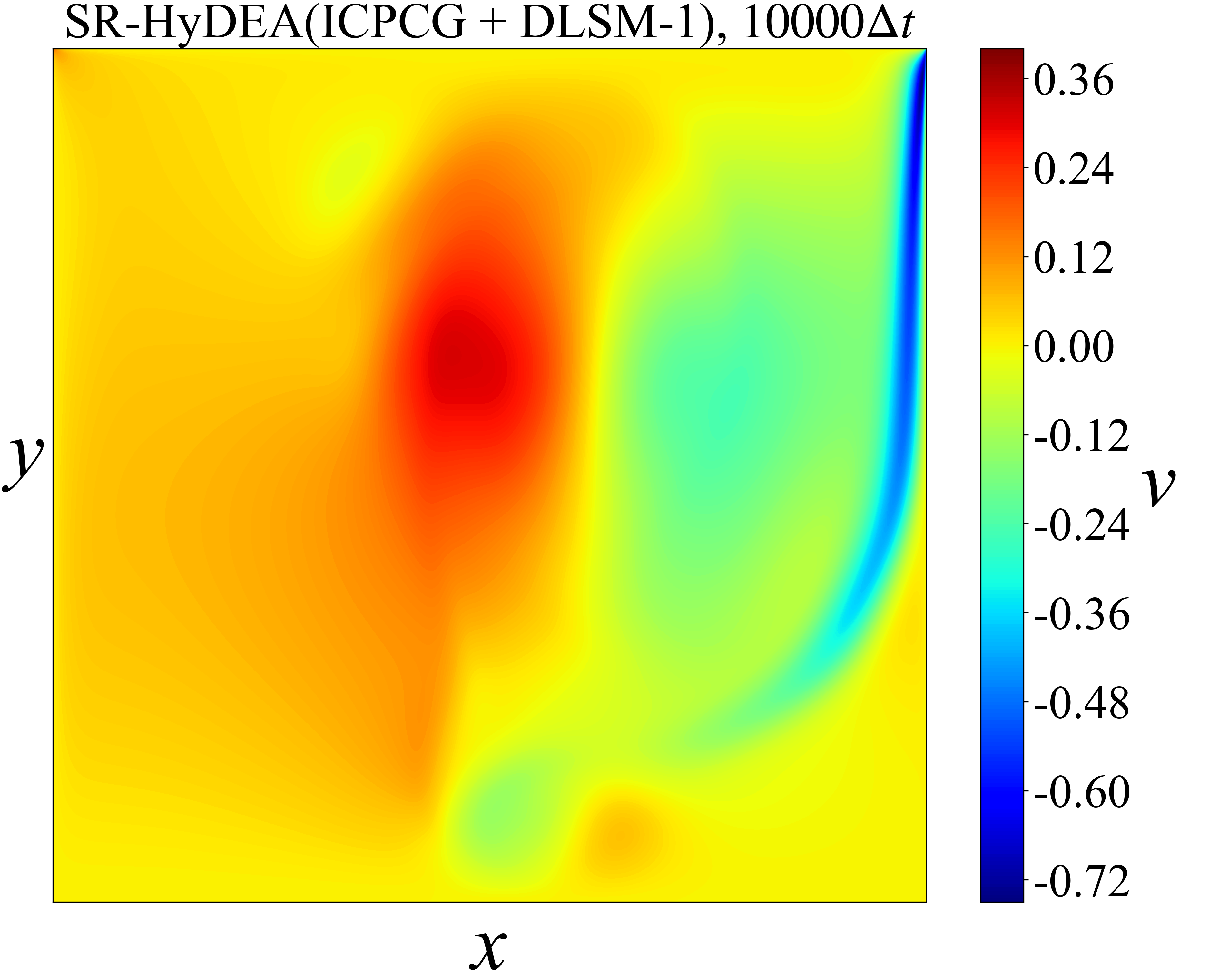}}
 \caption{Velocity fields for 2D lid-driven cavity flow at $Re=10,000$ by ICPCG and SR-HyDEA~(ICPCG+DLSM-1) with super-resolution of $512 \times 512$. (a)-(d) $u$ and $v$ at the $5000th$ time step. (e)-(h) $u$ and $v$ at the $10,000th$ time step.}
 \label{Flowfield512}
\end{figure}

Fig.~\ref{Flowfield512} presents velocity contours at the $5000th$ and $10,000th$ time steps for the resolution of $512 \times 512$. The result demonstrates that SR-HyDEA, despite employing a low-resolution~($128 \times 128$) DeepONet, achieves solution accuracy comparable to the high-resolution~($512 \times 512$) ICPCG method.

\subsubsection*{Comparison of DCDM, DLSM and HyDEA}
\begin{figure}[htbp] 
 \centering  
  \subfigure[]{
  \label{128inter_256_DCDM}
  \includegraphics[scale=0.21]{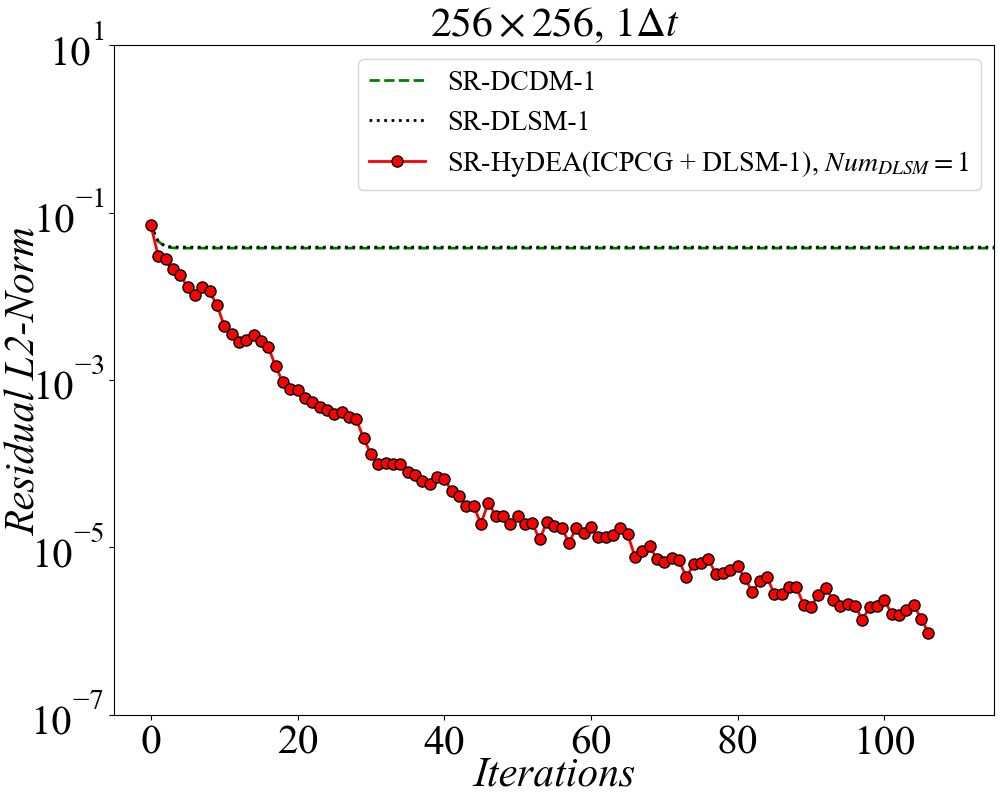}}
  \subfigure[]{
  \label{128inter_384_DCDM}
  \includegraphics[scale=0.21]{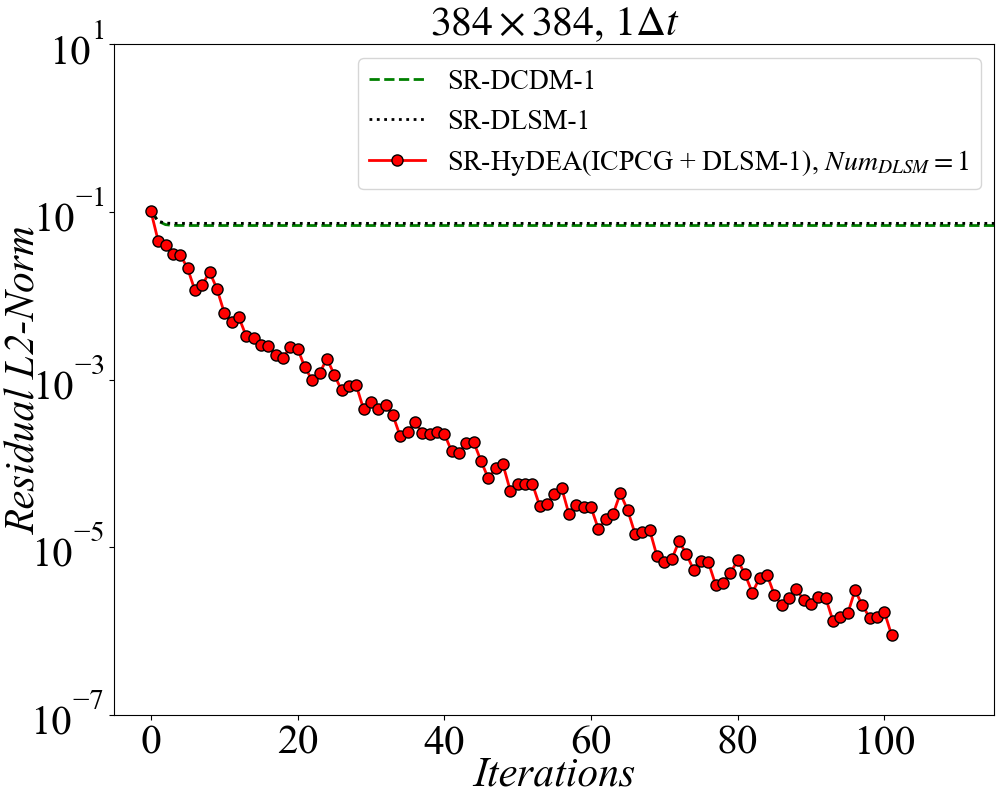}}
  \subfigure[]{
  \label{128inter_512_DCDM}
  \includegraphics[scale=0.21]{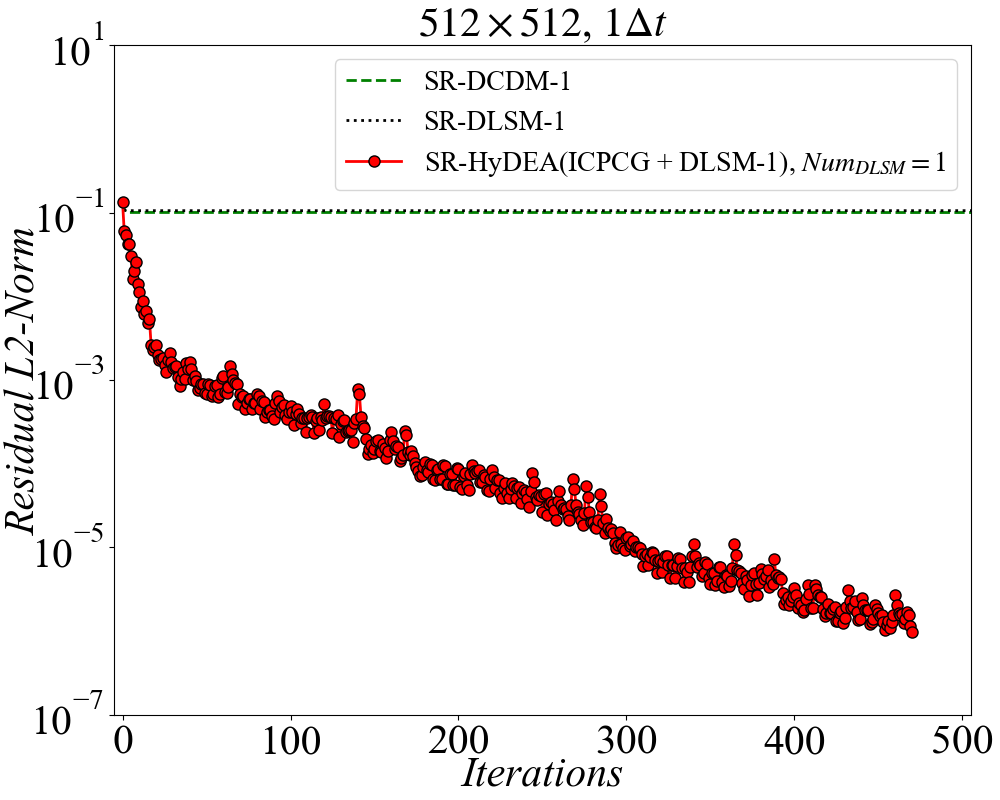}}
 \caption{Iterative residuals of solving the PPE for super-resolutions: comparisons among the SR-DCDM-1, SR-DLSM-1 and SR-HyDEA at the first time step for 2D lid-driven cavity flow at $Re=10,000$. DeepONet is trained {\it offline} for $128\times 128$ and utilized {\it online} for (a) $256 \times 256$. (b) $384 \times 384$. (c) $512 \times 512$.}
 \label{Rline_mis_d_DCDM_128inter}
\end{figure}

HyDEA combines neural networks with iterative methods to solve the PPE, while DCDM~\cite{kaneda2023DCDM} and DLSM rely exclusively on neural networks. 
As neural networks are incompetent to accurately capture high-frequency features,
the latter two methods may exhibit suboptimal performance in complex flow scenarios. 
Therefore, an investigation is conducted to evaluate whether the DCDM and DLSM implemented using DeepONet with low-resolution feature maps can effectively simulate high-resolution flow fields through bilinear interpolation, with the corresponding methods designated as SR-DCDM and SR-DLSM, respectively. Fig.~\ref{Rline_mis_d_DCDM_128inter} illustrates the iterative residuals of solving the PPE at the $1st$ time step for three high resolutions.
The iterative residuals of SR-DCDM and SR-DLSM show a modest initial decrease but quickly become plateau during further iterations.  This behavior suggests that completely replacing the traditional iterative solver with neural networks is not a suitable choice. In contrast, SR-HyDEA demonstrates a significantly superior performance compared to the SR-DCDM and SR-DLSM.

\subsection{Case 2: One stationary circular cylinder immersed in 2D lid-driven cavity flow}
\label{1Cylinder}

\begin{figure}[htbp]
\centering
  \includegraphics[scale=0.13]{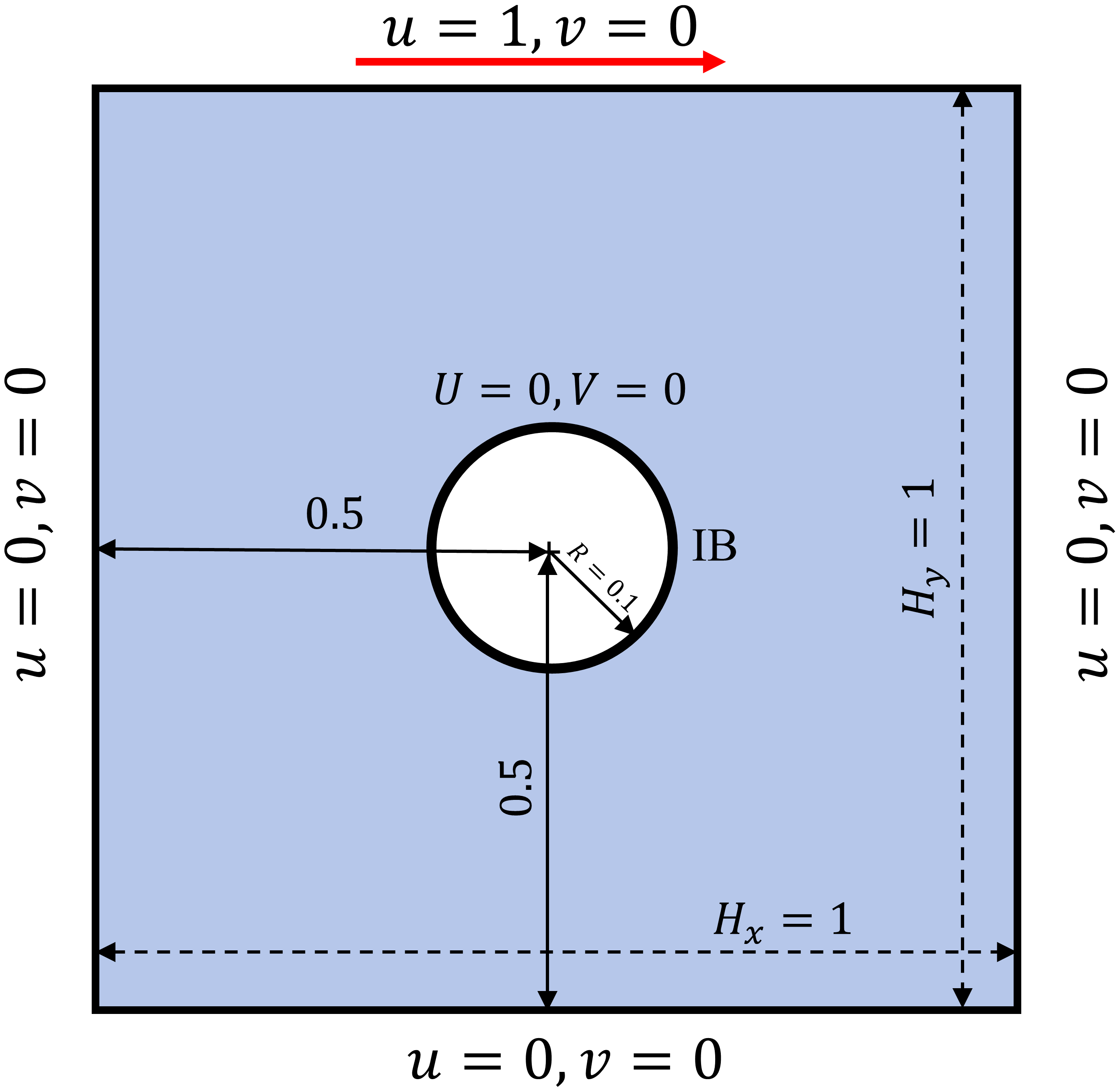}\\
  \caption{Schematic diagram of 2D lid-driven cavity flow containing an immersed stationary circular cylinder.}\label{Cavity_domain_1cylinder}
\end{figure}

This section examines the generalizability of HyDEA in simulating flows involving one stationary obstacle. 
A numerical experiment employs the same setup as in Section~\ref{cavityRe3200} 
with $Re=3200$ and grid resolution $192\times 192$, except that a stationary cylinder is immersed in the center of the computational domain as illustrated in Fig.~\ref{Cavity_domain_1cylinder}.

Furthermore, the DIBPM is activated for the cylinder in CFD computations and the exactly same DLSM-2 as in Section~\ref{cavityRe3200} is executed in HyDEA with $Num_{CG-type}=3$ and $Num_{DLSM}=2$, respectively.
We note that {\it the DeepONet within DLSM-2 remains the same network weights as in Section~\ref{cavityRe3200} and is not retrained. }

\begin{figure}[htbp] 
 \centering  
  \subfigure[]{
  \label{192_Rline_1Cylinder_IC_10steps}
  \includegraphics[scale=0.21]{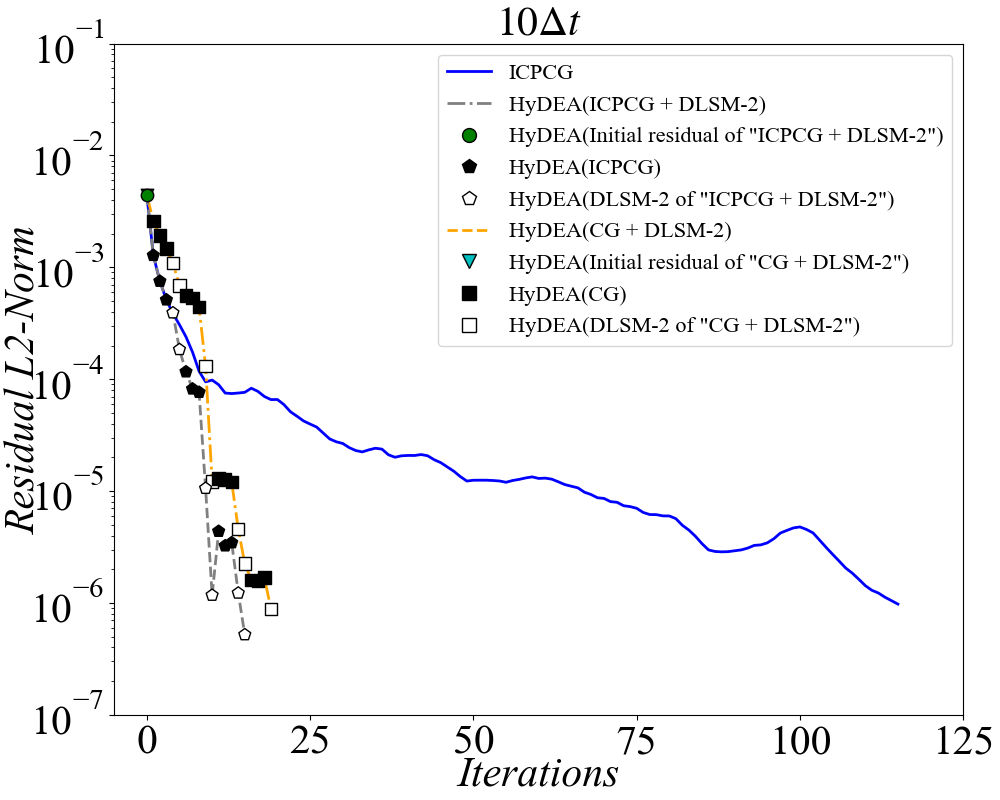}}
  \subfigure[]{
  \label{192_Rline_1Cylinder_IC_100steps}
  \includegraphics[scale=0.21]{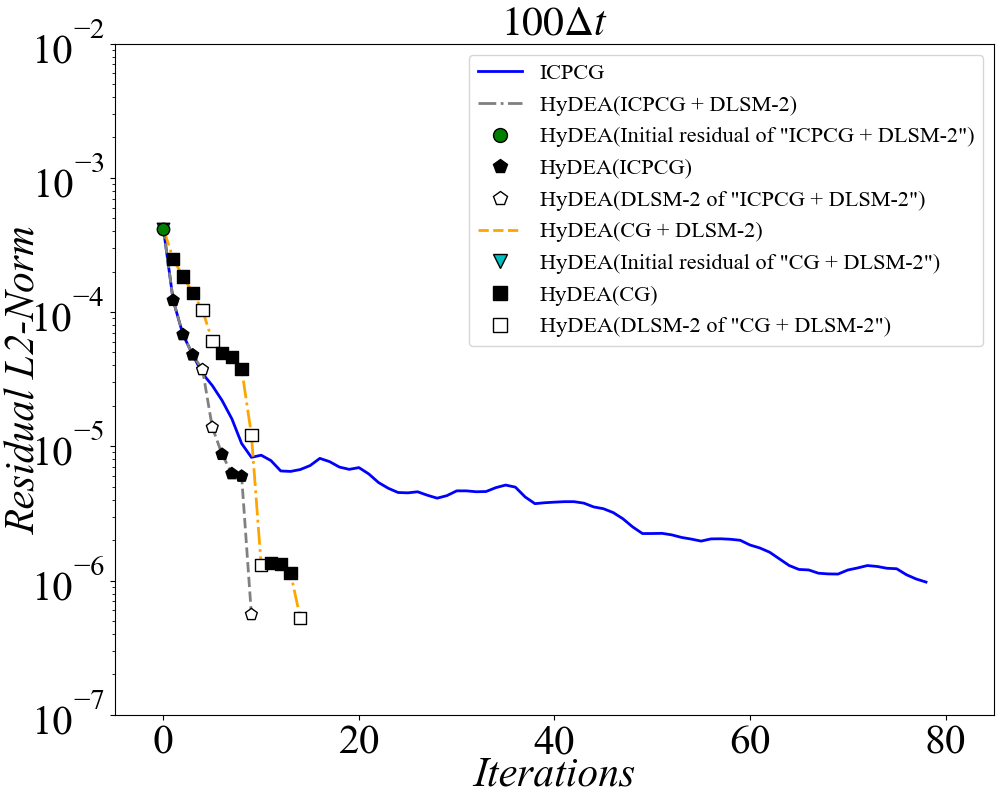}}
  \subfigure[]{
  \label{192_Rline_1Cylinder_IC_1000steps}
  \includegraphics[scale=0.21]{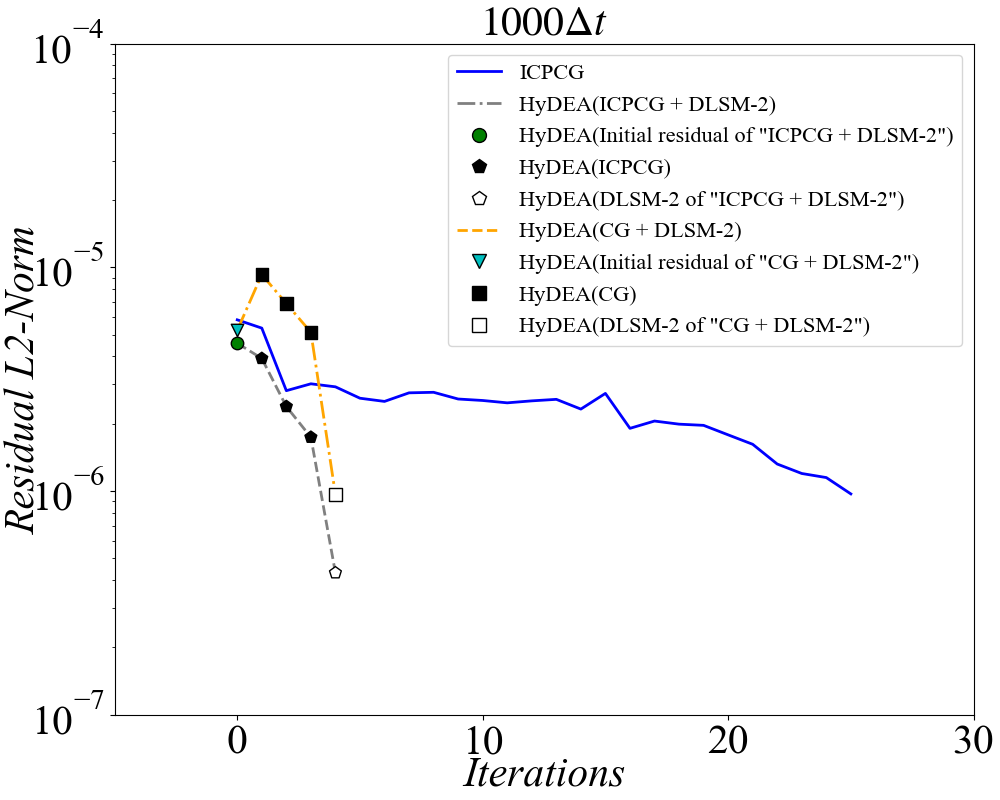}}
  \caption{Iterative residuals of solving the PPE for 2D lid-driven cavity flow with an immersed stationary circular cylinder at $Re=3200$ by ICPCG, HyDEA~(ICPCG+DLSM-2) and HyDEA~(CG+DLSM-2) . (a) $10th$ time step. (b) $100th$ time step. (c) $1000th$ time step.}\label{192_Rline_1cylinder_H}
\end{figure}

Taking HyDEA~(ICPCG+DLSM-2) and HyDEA~(CG+DLSM-2) as examples, the iterative residuals of solving the PPE at the $10th$, $100th$ and $1000th$ time steps are depicted in Fig.~\ref{192_Rline_1cylinder_H}.
For HyDEA~(ICPCG+DLSM-2), it takes $3$ rounds of the hybrid algorithm with $15$ iterations at $10\Delta t$,
$2$ rounds with $9$ iterations at $100\Delta t$,
and $1$ round with $4$ iterations at $1000\Delta t$, respectively.
For HyDEA~(CG+DLSM-2), it takes $4$ rounds of the hybrid algorithm with $19$ iterations at $10\Delta t$,
$3$ rounds with $14$ iterations at $100\Delta t$,
and $1$ round with $4$ iterations at $1000\Delta t$, respectively.
In general, HyDEA requires significantly fewer iterations than the ICPCG method alone, being consistent with the single-component flow simulation in Section~\ref{cavityRe3200}. 

\begin{figure}[htbp]
\centering
  \includegraphics[scale=0.24]{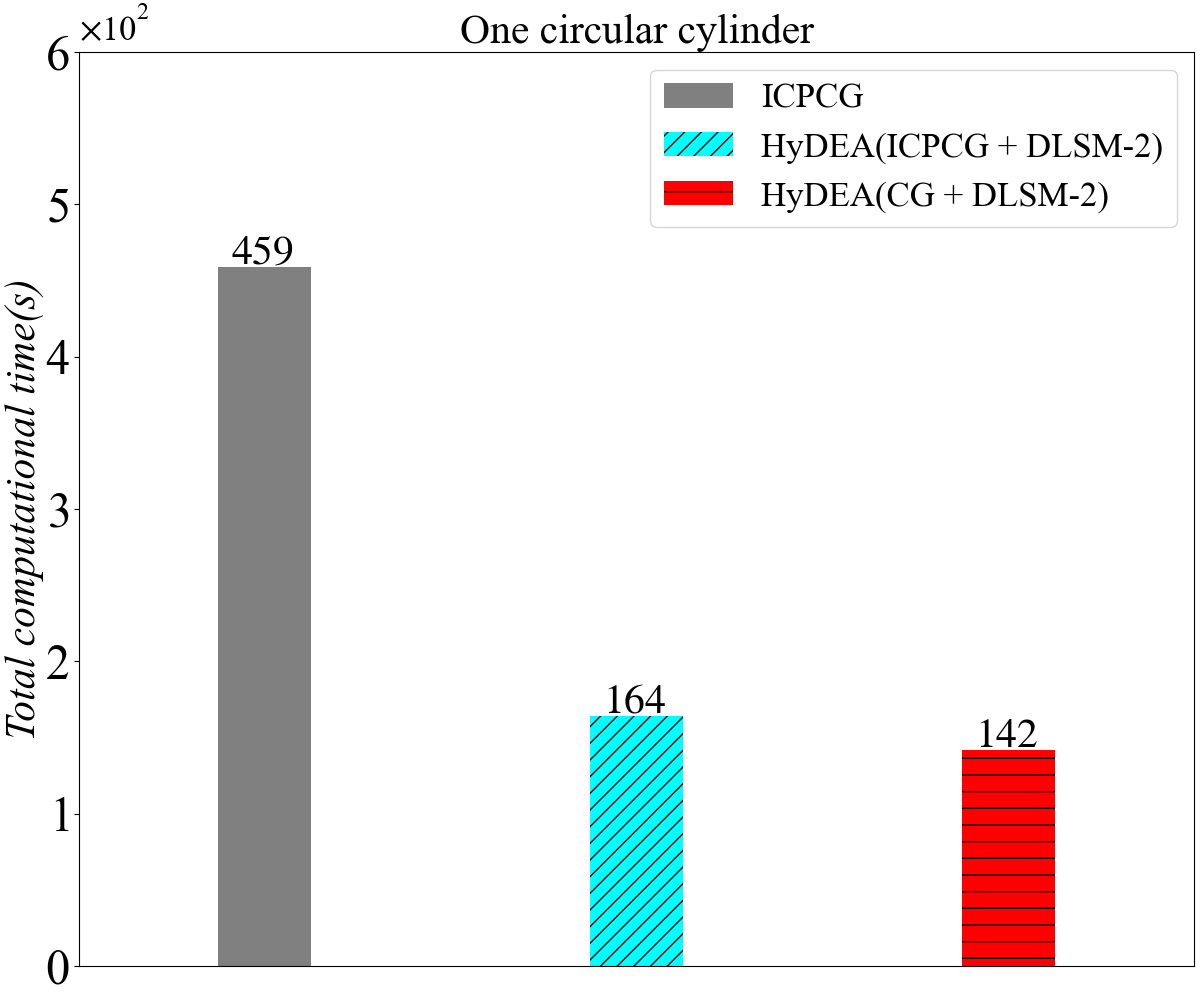}\\
  \caption{Computational time required to solve the PPE over a duration of $10,000\Delta t$ for 2D lid-driven cavity flow with an immersed stationary circular cylinder at $Re=3200$.}\label{Time_compare_1cylinder}
\end{figure}

Fig.~\ref{Time_compare_1cylinder} presents the computational time required to solve the PPE over $10,000$ consecutive time steps using both HyDEA and ICPCG methods. 
The results clearly indicates that HyDEA needs much less computational time compared to ICPCG method to achieve the same accuracy.

\begin{figure}[htbp] 
 \centering  
  \subfigure[]{
  \label{U_IC_1500step_1Cylinder}
  \includegraphics[scale=0.129]{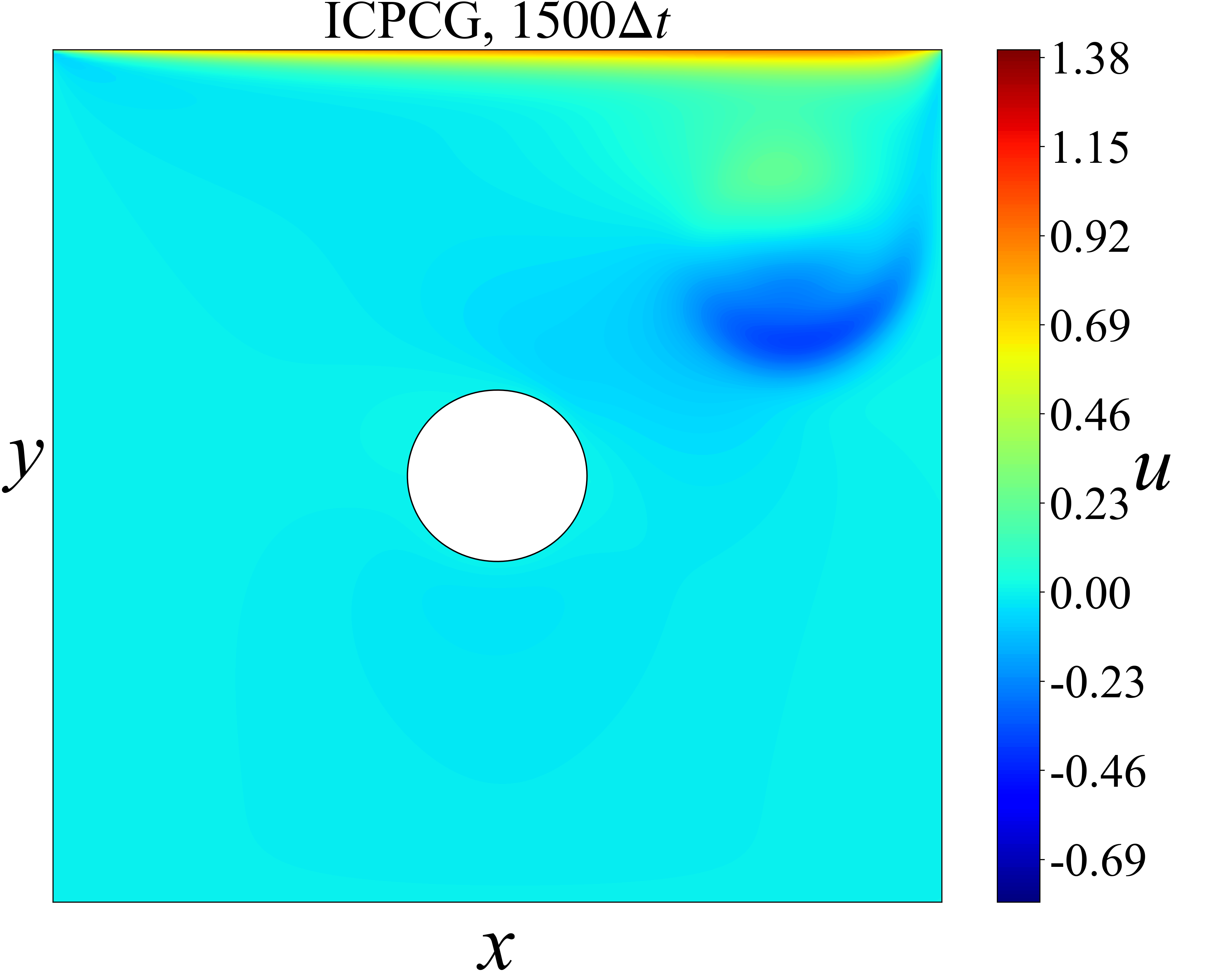}}
  \subfigure[]{
  \label{U_HIC_1500step_1Cylinder}
  \includegraphics[scale=0.129]{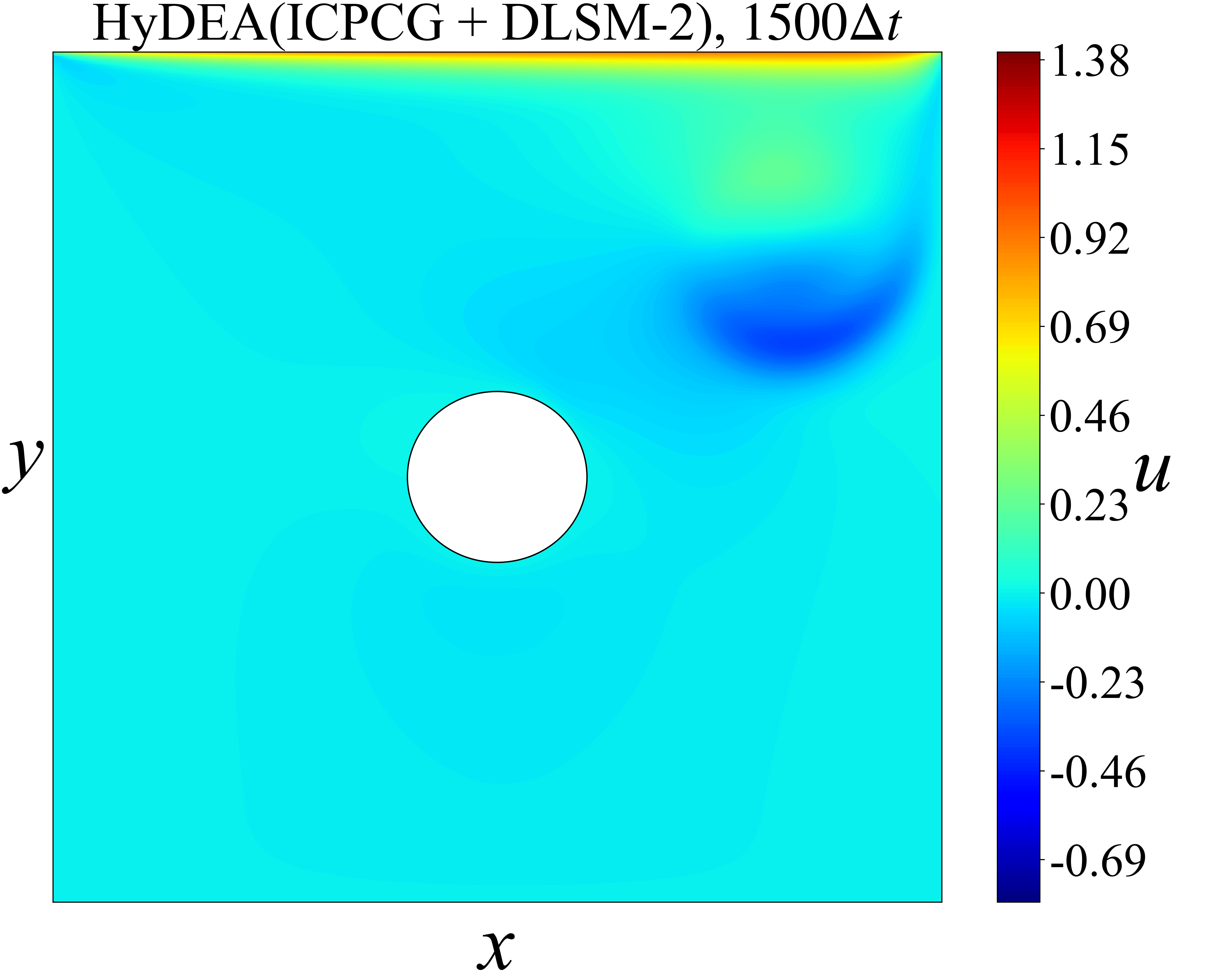}}
  \subfigure[]{
  \label{V_IC_1500step_1Cylinder}
  \includegraphics[scale=0.129]{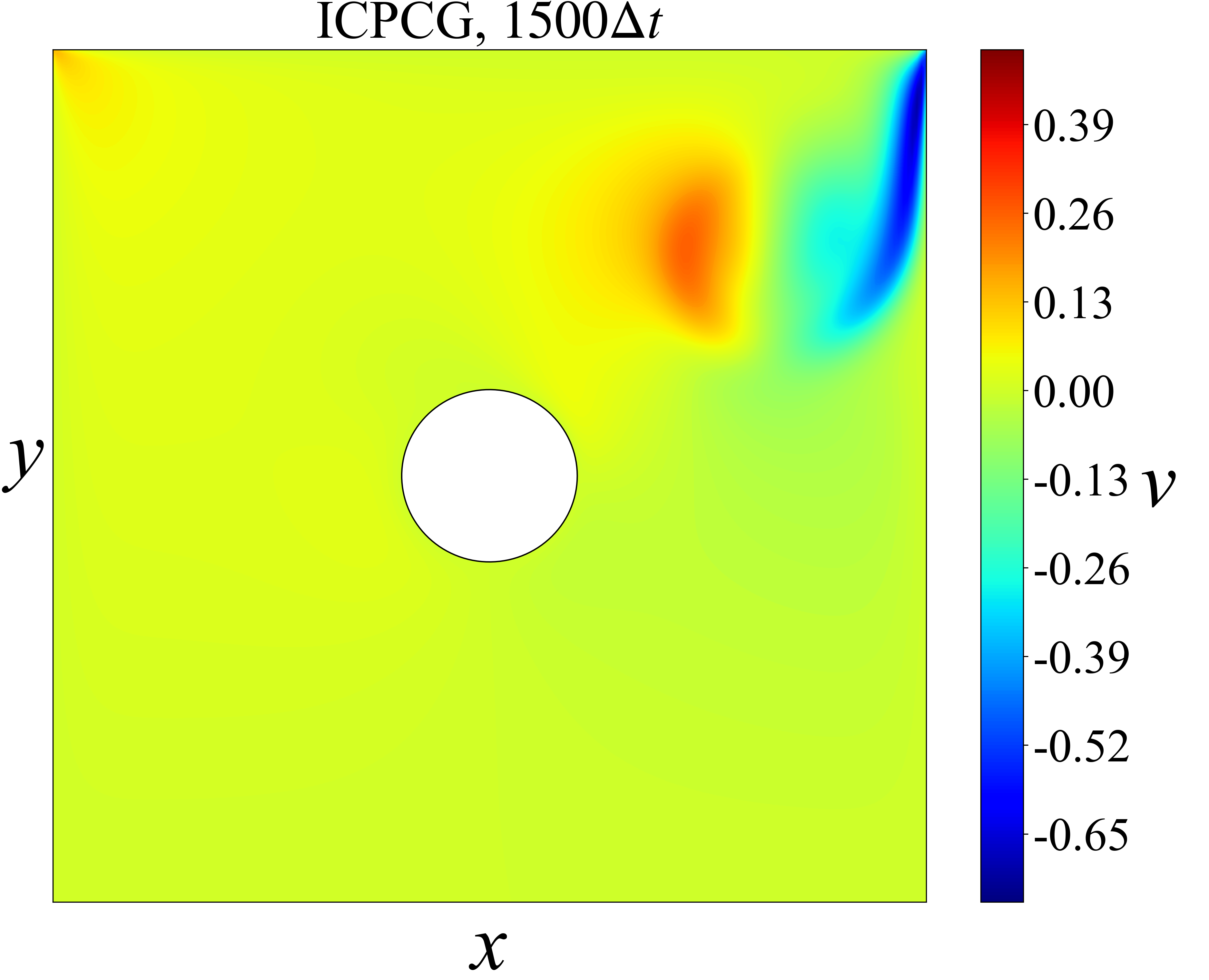}}
  \subfigure[]{
  \label{V_HIC_1500step_1Cylinder}
  \includegraphics[scale=0.129]{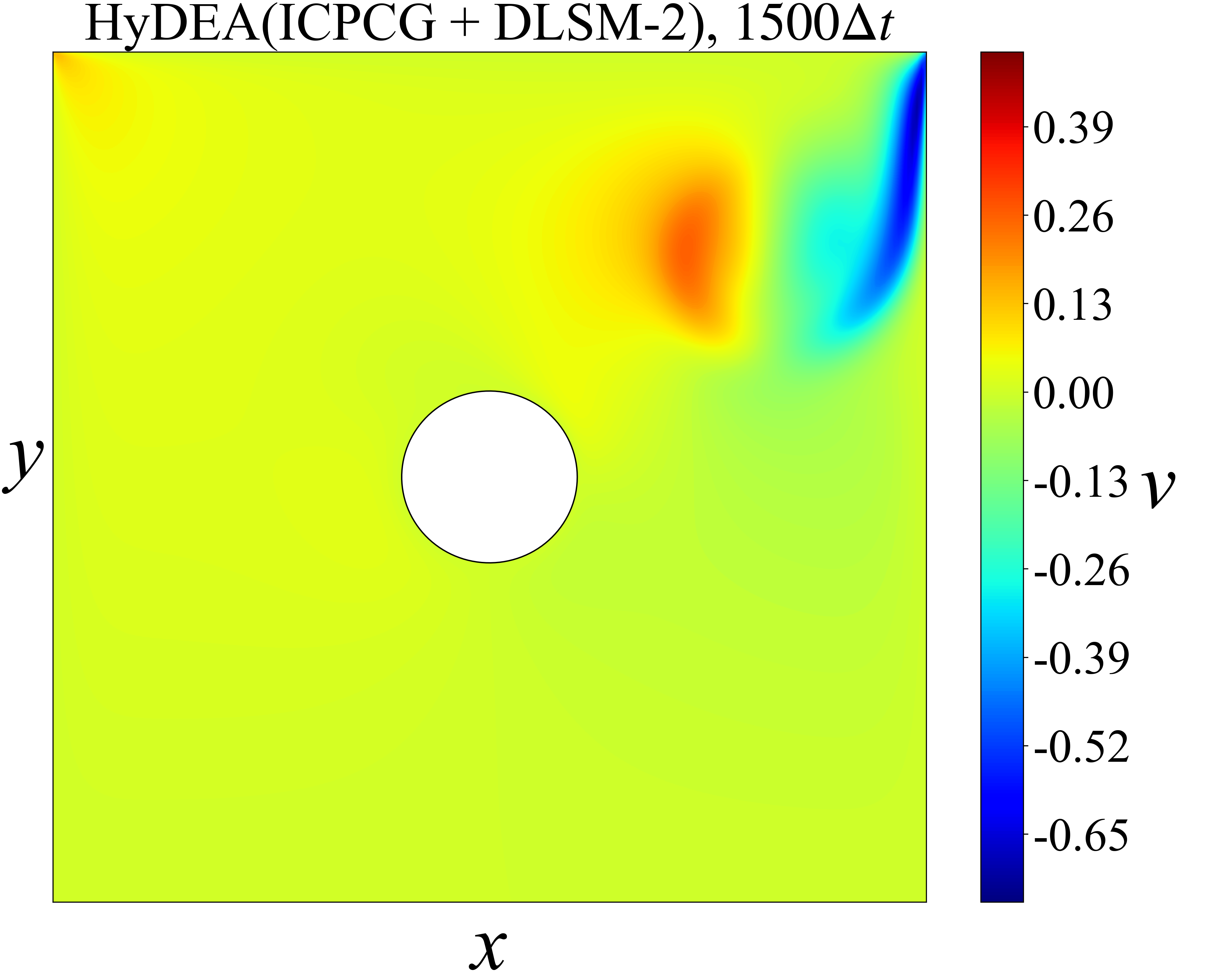}}
  \subfigure[]{
  \label{U_IC_6000step_1Cylinder}
  \includegraphics[scale=0.129]{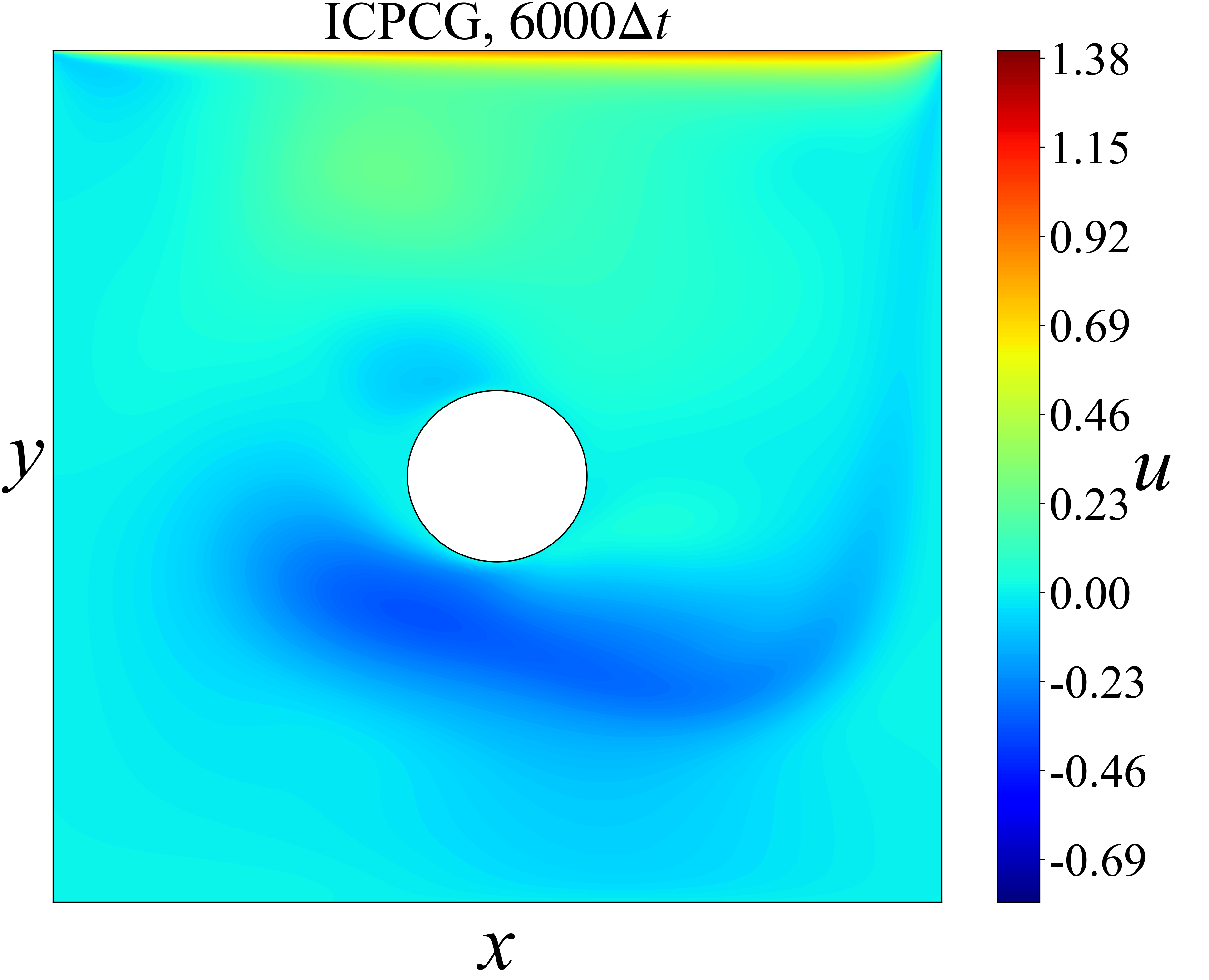}}
  \subfigure[]{
  \label{U_HIC_6000step_1Cylinder}
  \includegraphics[scale=0.129]{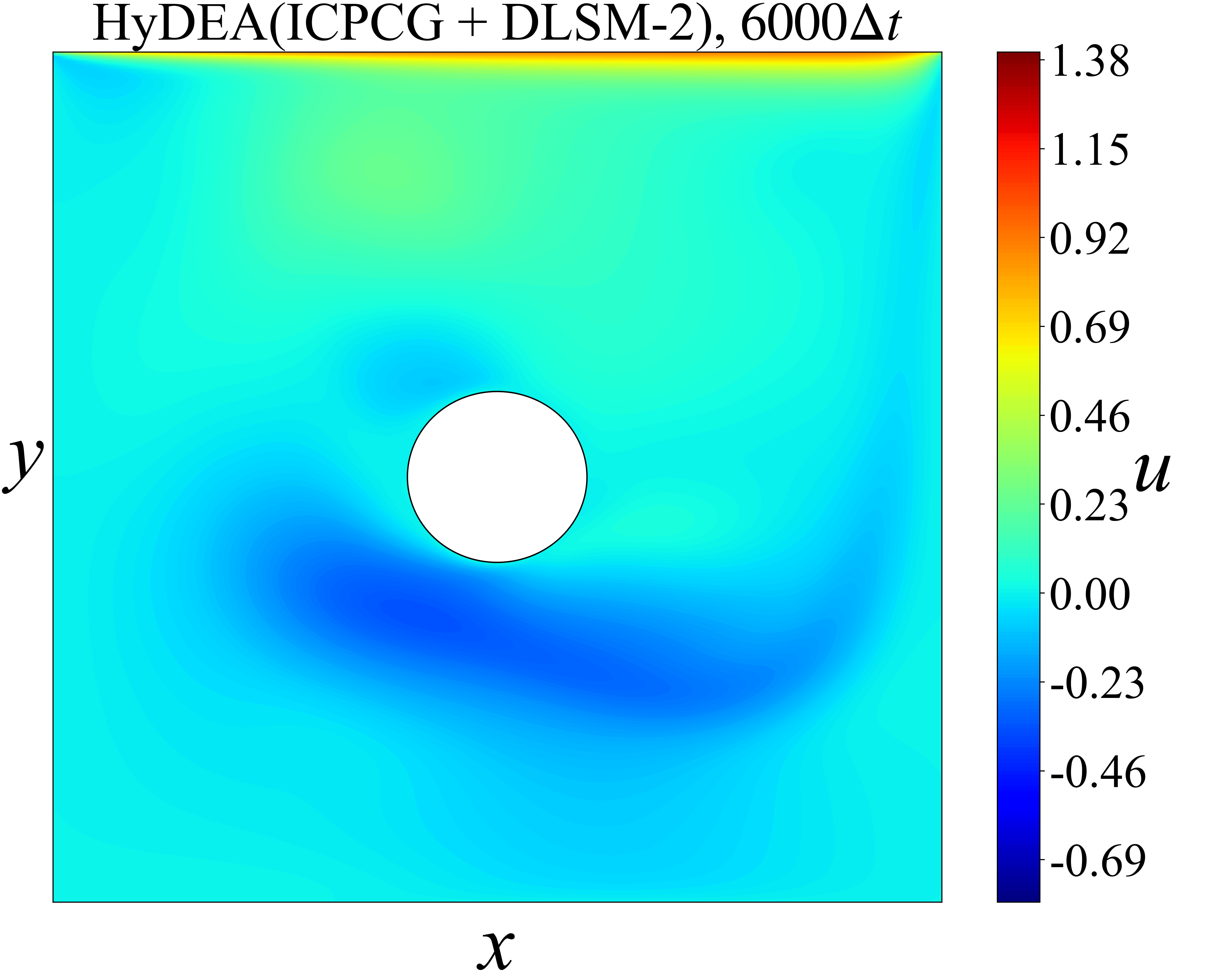}}
  \subfigure[]{
  \label{V_IC_6000step_1Cylinder}
  \includegraphics[scale=0.129]{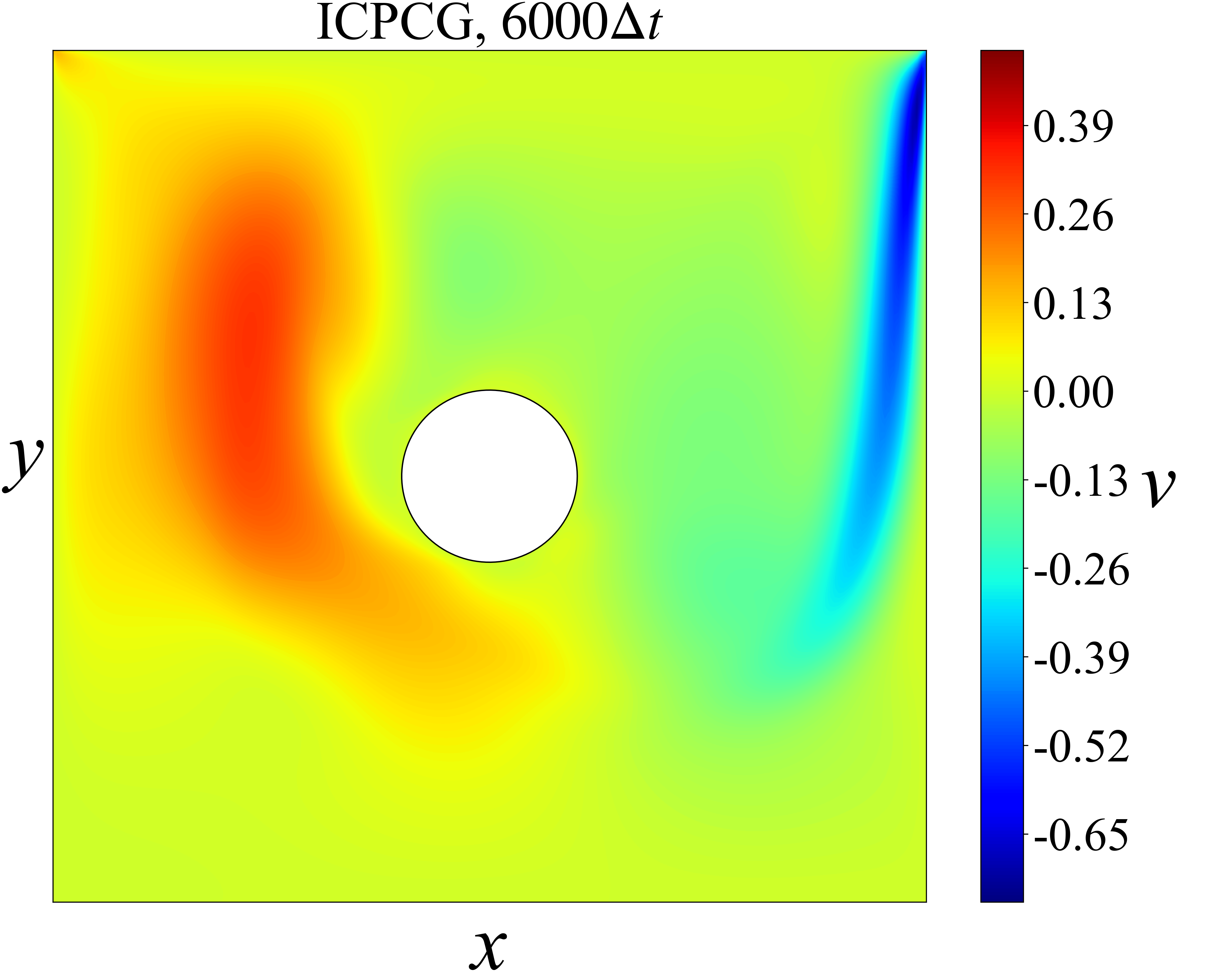}}
  \subfigure[]{
  \label{V_HIC_6000step_1Cylinder}
  \includegraphics[scale=0.129]{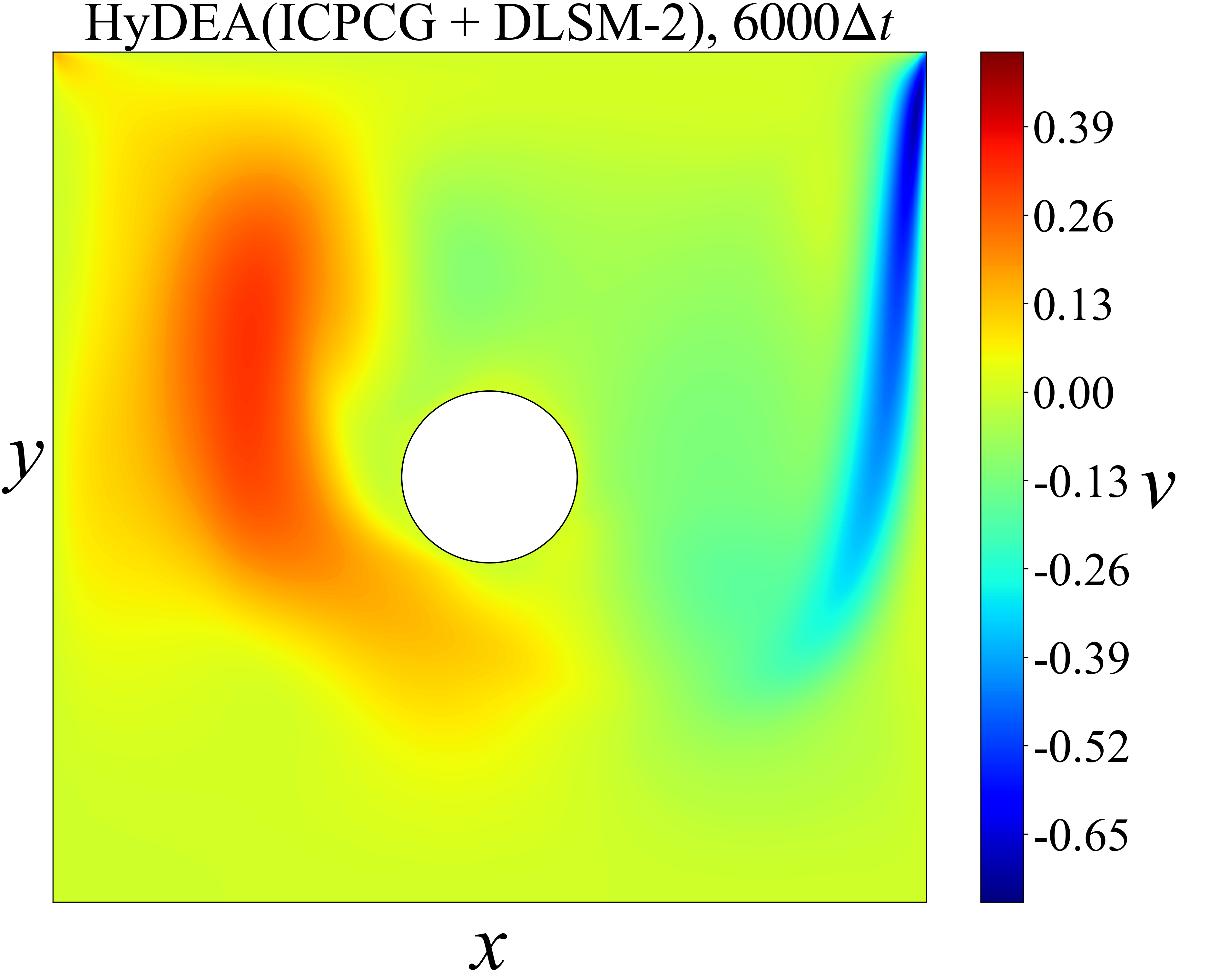}}
 \caption{Velocity fields for 2D lid-driven cavity flow with an immersed stationary circular cylinder at $Re=3200$ by ICPCG and HyDEA~(ICPCG+DLSM-2). (a)-(d) $u$ and $v$ at the $1500th$ time step. (e)-(h) $u$ and $v$ at the $6000th$ time step.}
 \label{Flowfield_1Cylinder}
\end{figure}

Furthermore, the velocity contours of ICPCG and HyDEA~(ICPCG+DLSM-2) at the $1500th$ and $6000th$ time steps are depicted in Fig.~\ref{Flowfield_1Cylinder}, demonstrating that the temporal evolution of the flow field is accurately captured.

\subsection{Case 3: Two stationary elliptical cylinders immersed in 2D lid-driven cavity flow}
\label{2Cylinder}

\begin{figure}[htbp]
\centering
  \includegraphics[scale=0.13]{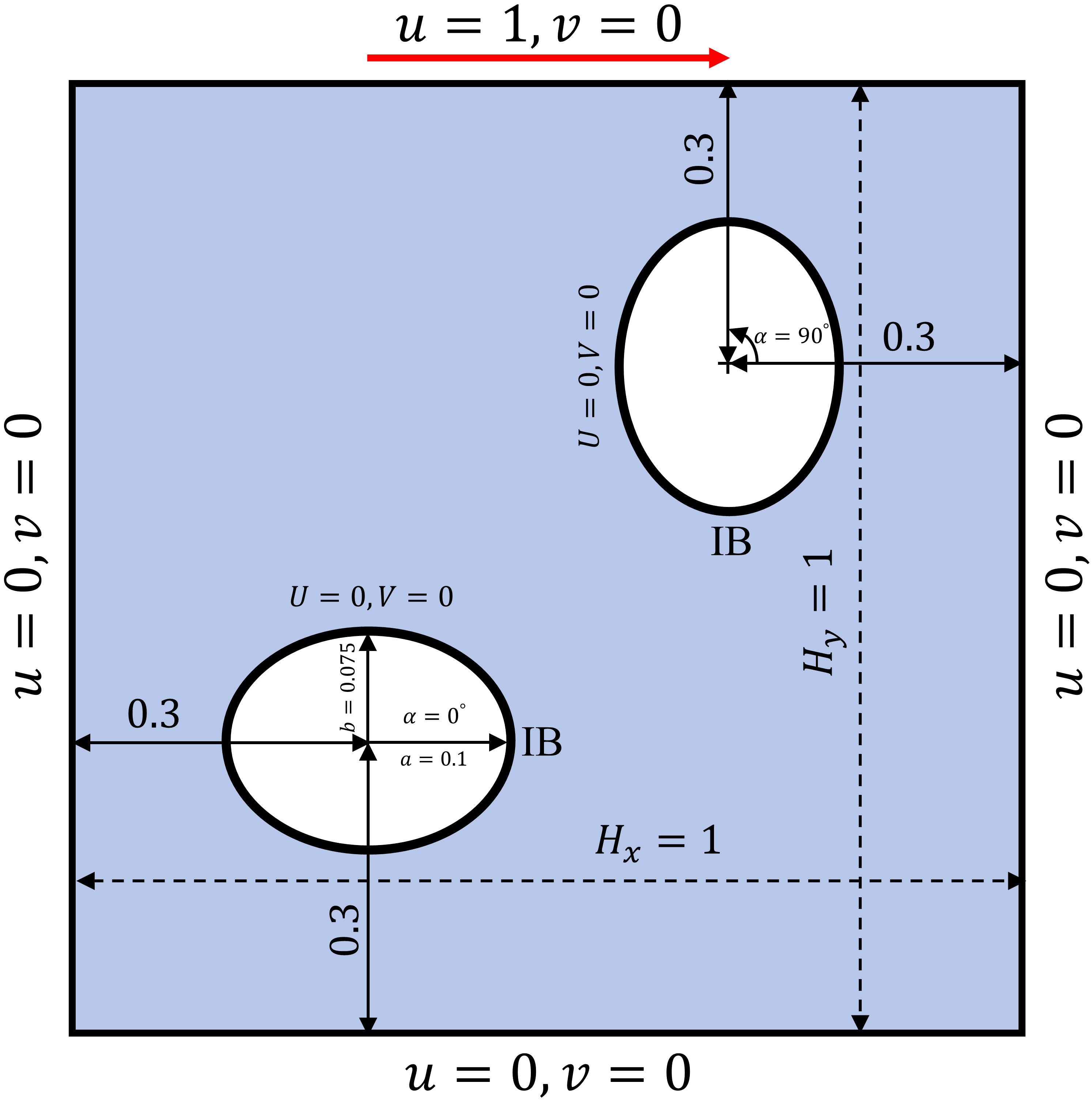}\\
  \caption{Schematic diagram of 2D lid-driven cavity flow containing two immersed stationary elliptical cylinders. $\alpha$ denotes the angle formed by the major axis of the elliptical cylinder and the x-direction.}\label{Cavity_domain_2ellipse}
\end{figure}

This section examines the generalizability of HyDEA in simulating flows involving two stationary obstacles. A numerical experiment employs the same setup as in Section~\ref{cavityRe3200} with $Re=3200$ and grid resolution $192\times 192$,
except that two stationary elliptical cylinders of aspect ratio $4:3$ are immersed in the computational domain as illustrated in Fig.~\ref{Cavity_domain_2ellipse}. 

Furthermore, the DIBPM is activated for the two elliptical cylinders in CFD computations and the exactly same DLSM-2 as in Section~\ref{cavityRe3200} is executed in HyDEA with $Num_{CG-type}=3$ and $Num_{DLSM}=2$, respectively.
We note that {\it the DeepONet within DLSM-2 remains the same network weights as in Sections~\ref{cavityRe3200} and \ref{1Cylinder}  and is not retrained. }

\begin{figure}[htbp] 
 \centering  
  \subfigure[]{
  \label{192_Rline_2Cylinder__10steps}
  \includegraphics[scale=0.21]{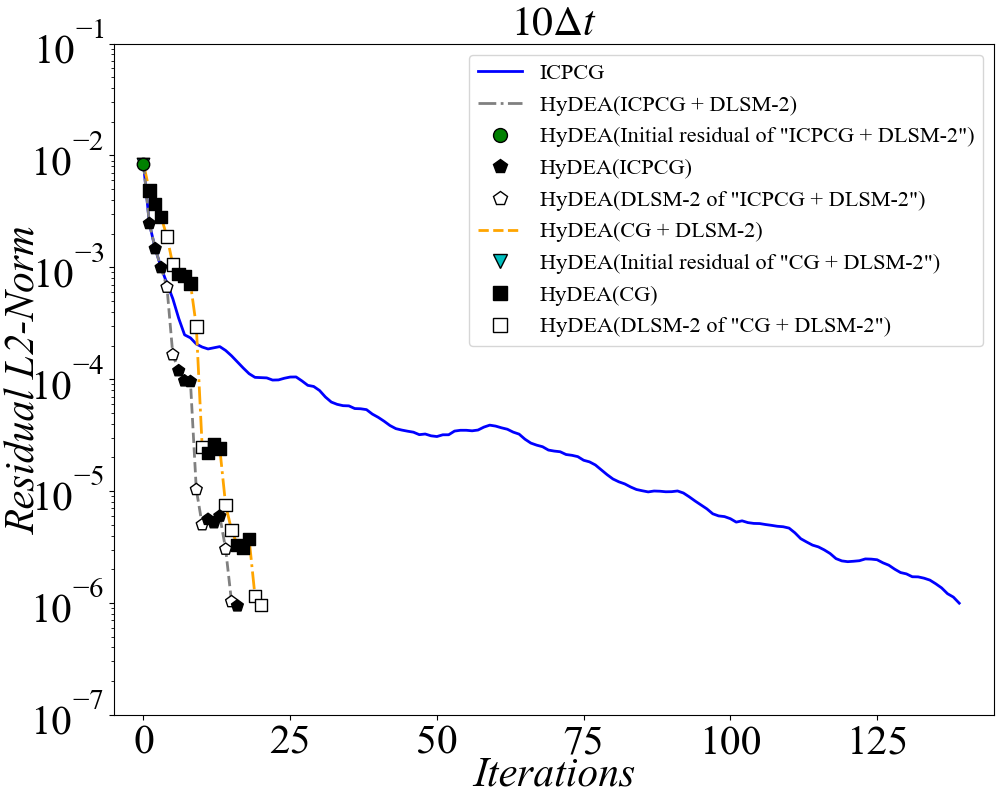}}
  \subfigure[]{
  \label{192_Rline_2Cylinder_100steps}
  \includegraphics[scale=0.21]{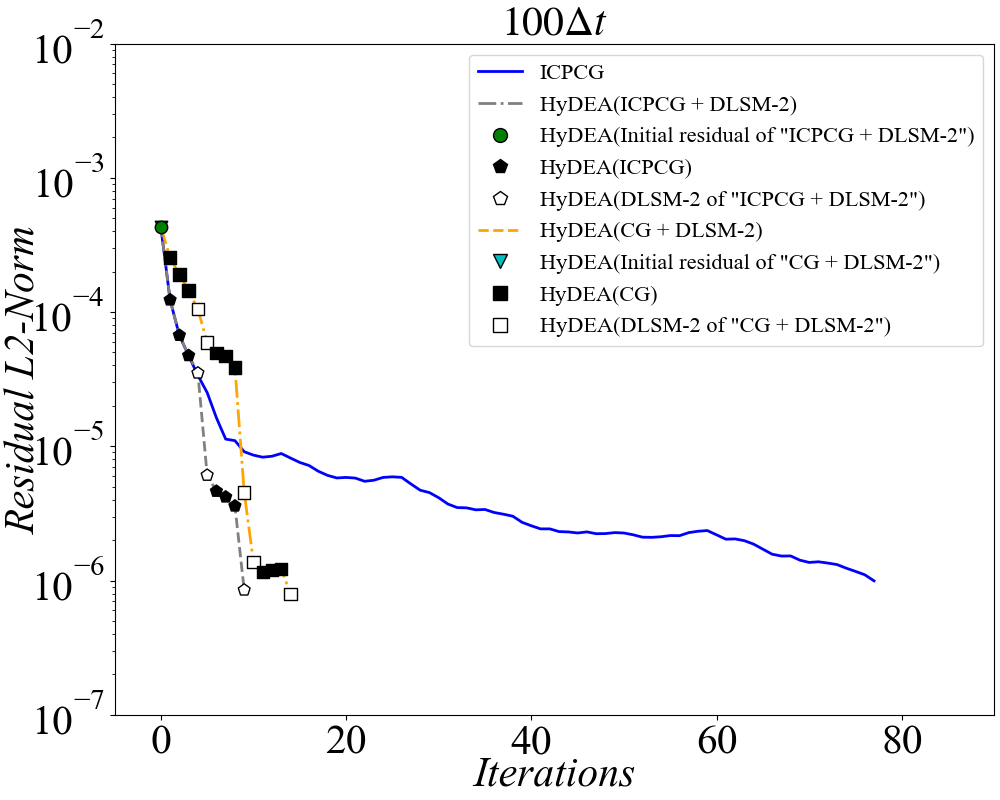}}
  \subfigure[]{
  \label{192_Rline_2Cylinder_1000steps}
  \includegraphics[scale=0.21]{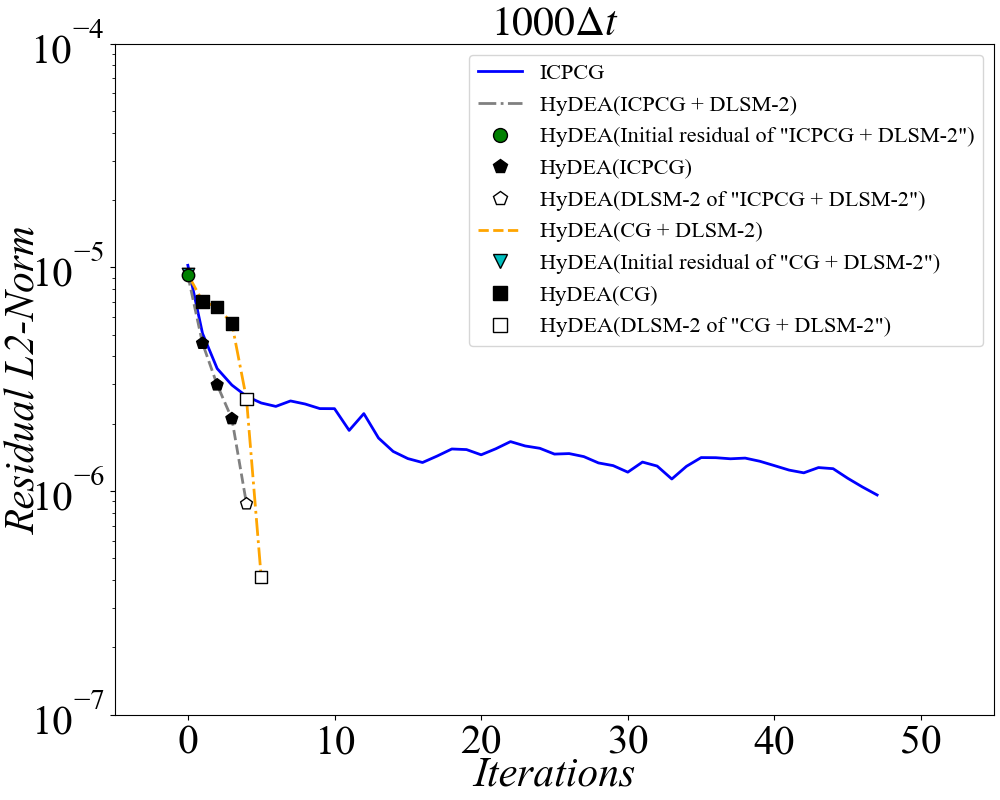}}
  \caption{Iterative residuals of solving the PPE for 2D lid-driven cavity flow with two immersed stationary elliptical cylinders at $Re=3200$ by ICPCG, HyDEA~(ICPCG+DLSM-2), and HyDEA~(CG+DLSM-2). (a) $10th$ time step. (b) $100th$ time step. (c) $1000th$ time step.}\label{192_Rline_2ellipse_H}
\end{figure}

Taking HyDEA~(ICPCG+DLSM-2) and HyDEA~(CG+DLSM-2) as examples, the iterative residuals of solving the PPE at the $10th$, $100th$ and $1000th$ time steps are depicted in Fig.~\ref{192_Rline_2ellipse_H}. 
For HyDEA~(ICPCG+DLSM-2), it takes $4$ rounds of the hybrid algorithm with $16$ iterations at $10\Delta t$,
$2$ rounds with $9$ iterations at $100\Delta t$,
and $1$ round with $4$ iterations at $1000\Delta t$, respectively.
For HyDEA~(CG+DLSM-2), it takes $4$ rounds of the hybrid algorithm with $20$ iterations at $10\Delta t$,
$3$ rounds with $14$ iterations at $100\Delta t$,
and $1$ round with $5$ iterations at $1000\Delta t$, respectively.
In general, HyDEA requires significantly fewer iterations than the ICPCG method alone, being consistent with the single-component flow simulation in Section~\ref{cavityRe3200}.
The results also align closely with those for one cylinder in Section~\ref{1Cylinder}, demonstrating HyDEA's robust generalization capability across flows with multiple obstacles.

\begin{figure}[htbp]
\centering
  \includegraphics[scale=0.24]{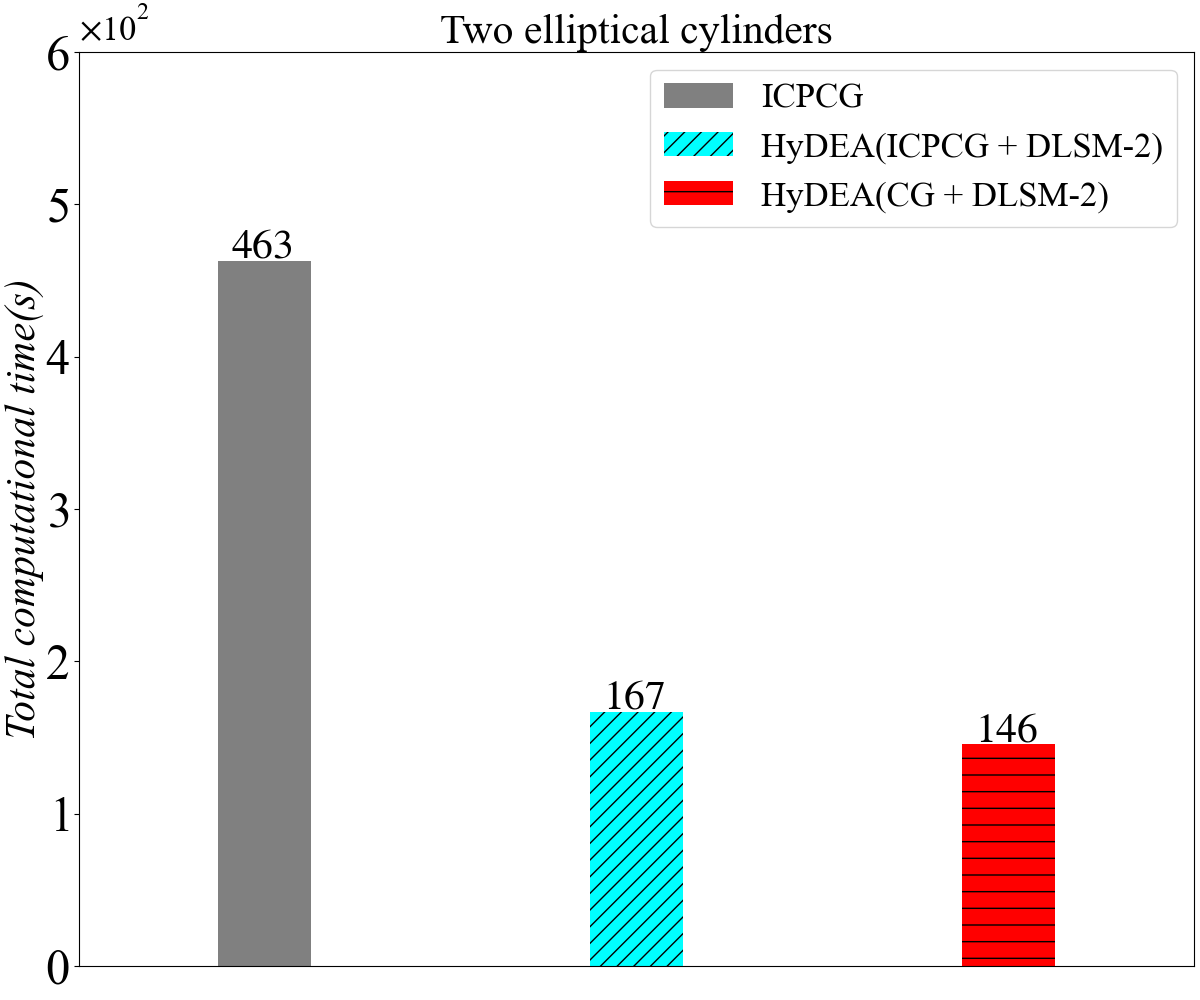}\\
  \caption{Computational time required to solve the PPE  over a duration of $10,000\Delta t$ for 2D lid-driven cavity flow with two immersed stationary elliptical cylinders at $Re=3200$.}\label{Time_compare_2cylinder}
\end{figure}

Fig.~\ref{Time_compare_2cylinder} depicts the computational time required to solve the PPE over $10,000$ consecutive time steps using both HyDEA and ICPCG method.
The results clearly indicates that HyDEA needs much less computational time compared to ICPCG method to achieve the same accuracy.

\begin{figure}[htbp] 
 \centering  
  \subfigure[]{
  \label{U_IC_1500step_2Cylinder}
  \includegraphics[scale=0.129]{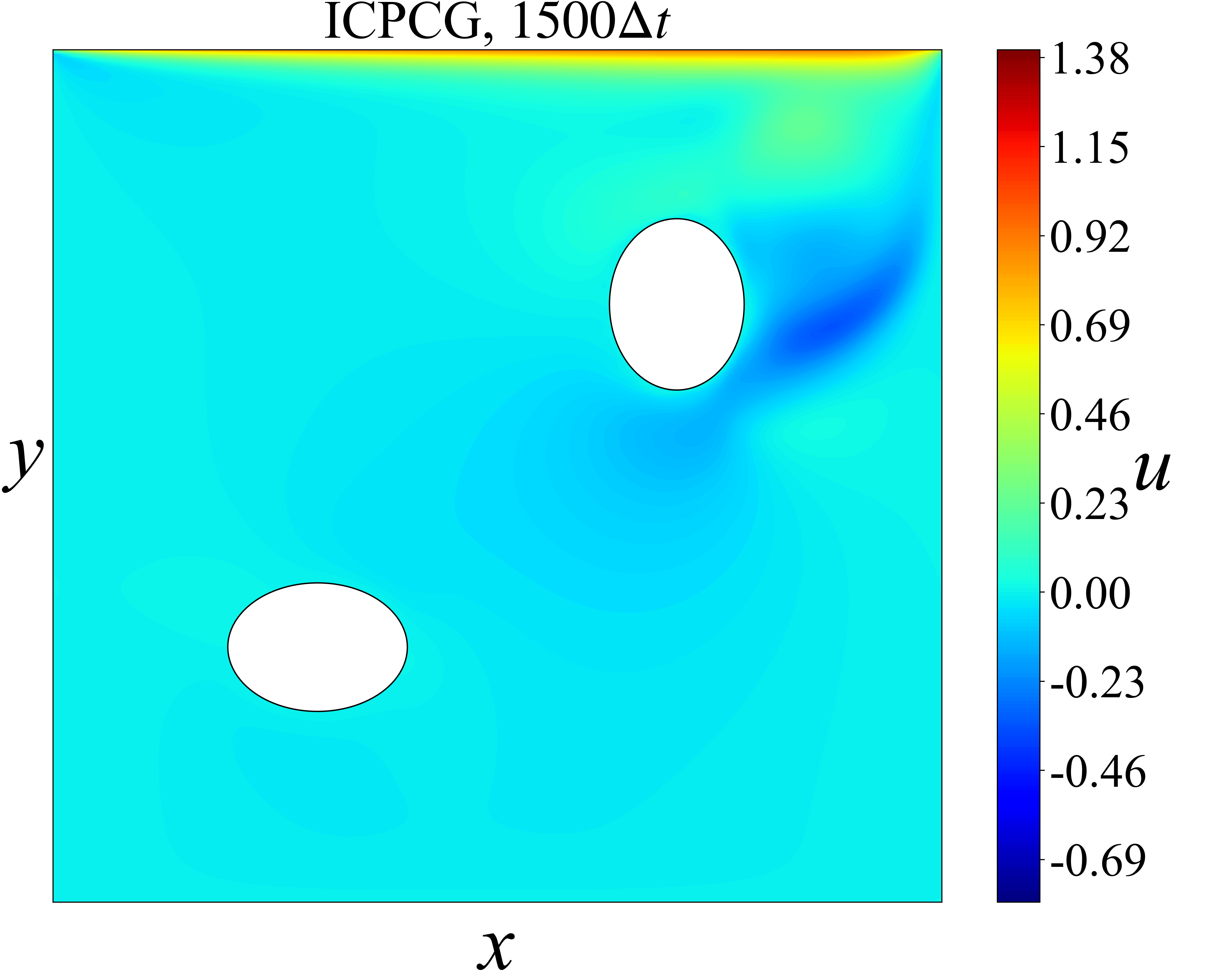}}
  \subfigure[]{
  \label{U_HIC_1500step_2Cylinder}
  \includegraphics[scale=0.129]{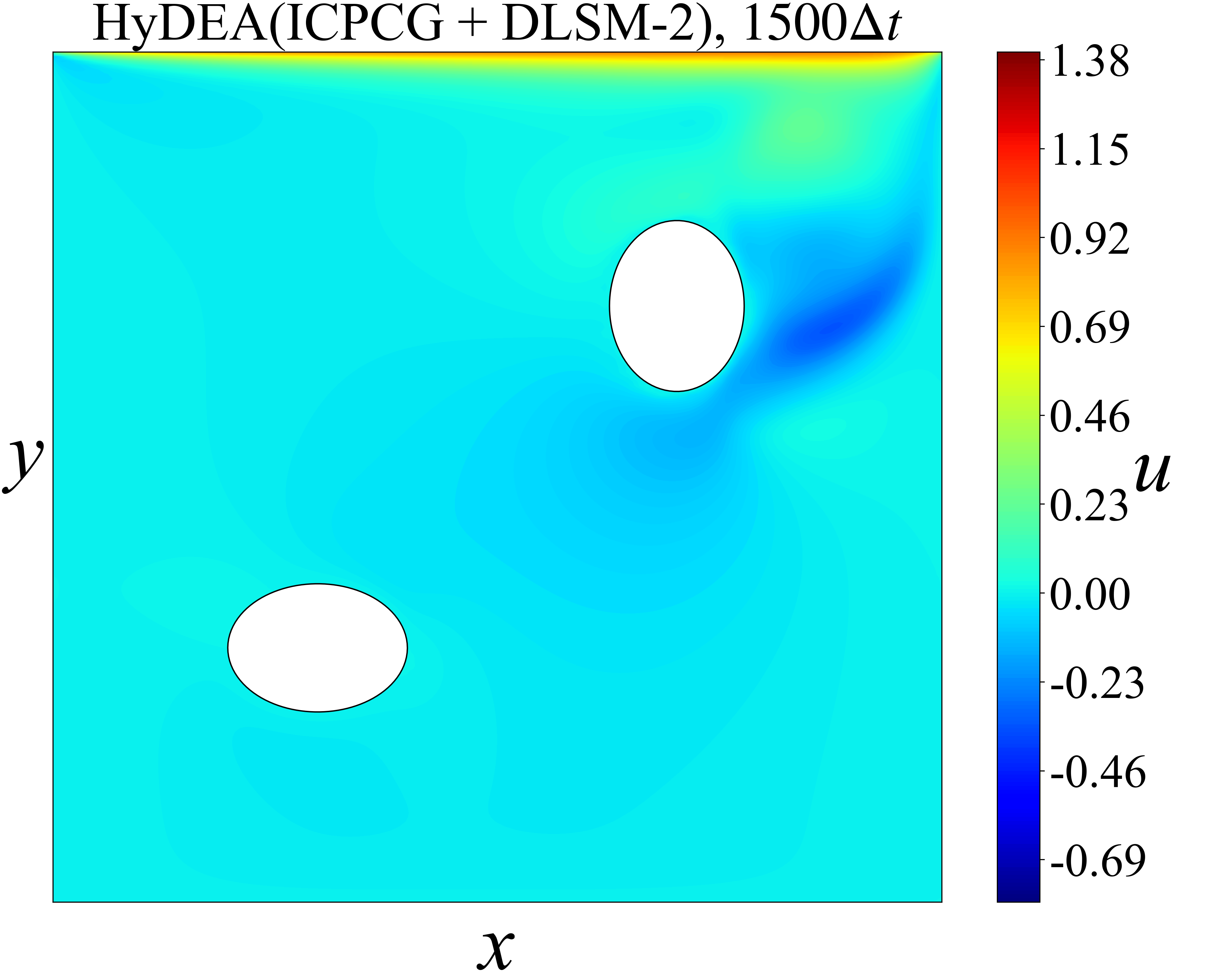}}
  \subfigure[]{
  \label{V_IC_1500step_2Cylinder}
  \includegraphics[scale=0.129]{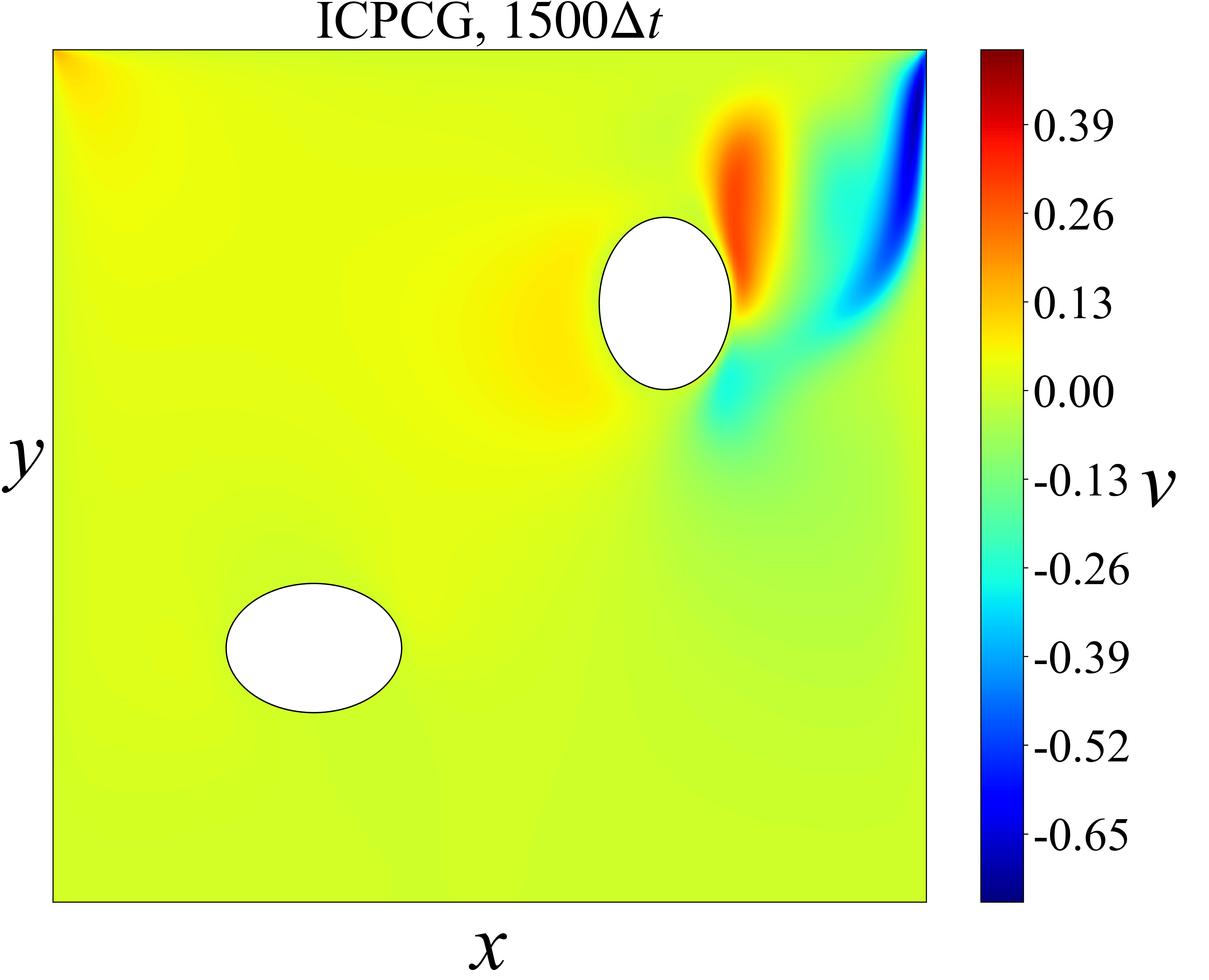}}
  \subfigure[]{
  \label{V_HIC_1500step_2Cylinder}
  \includegraphics[scale=0.129]{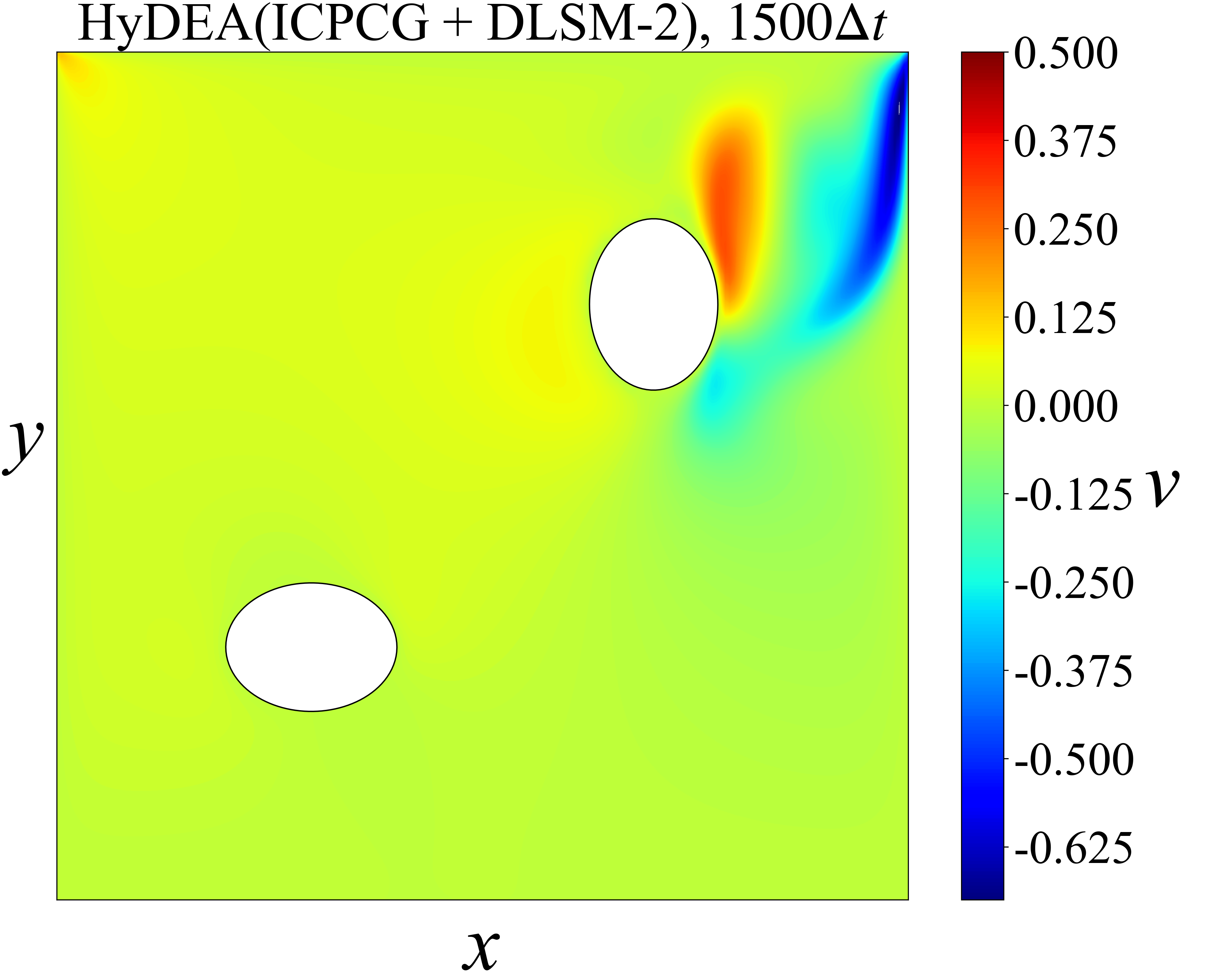}}
  \subfigure[]{
  \label{U_IC_6000step_2Cylinder}
  \includegraphics[scale=0.129]{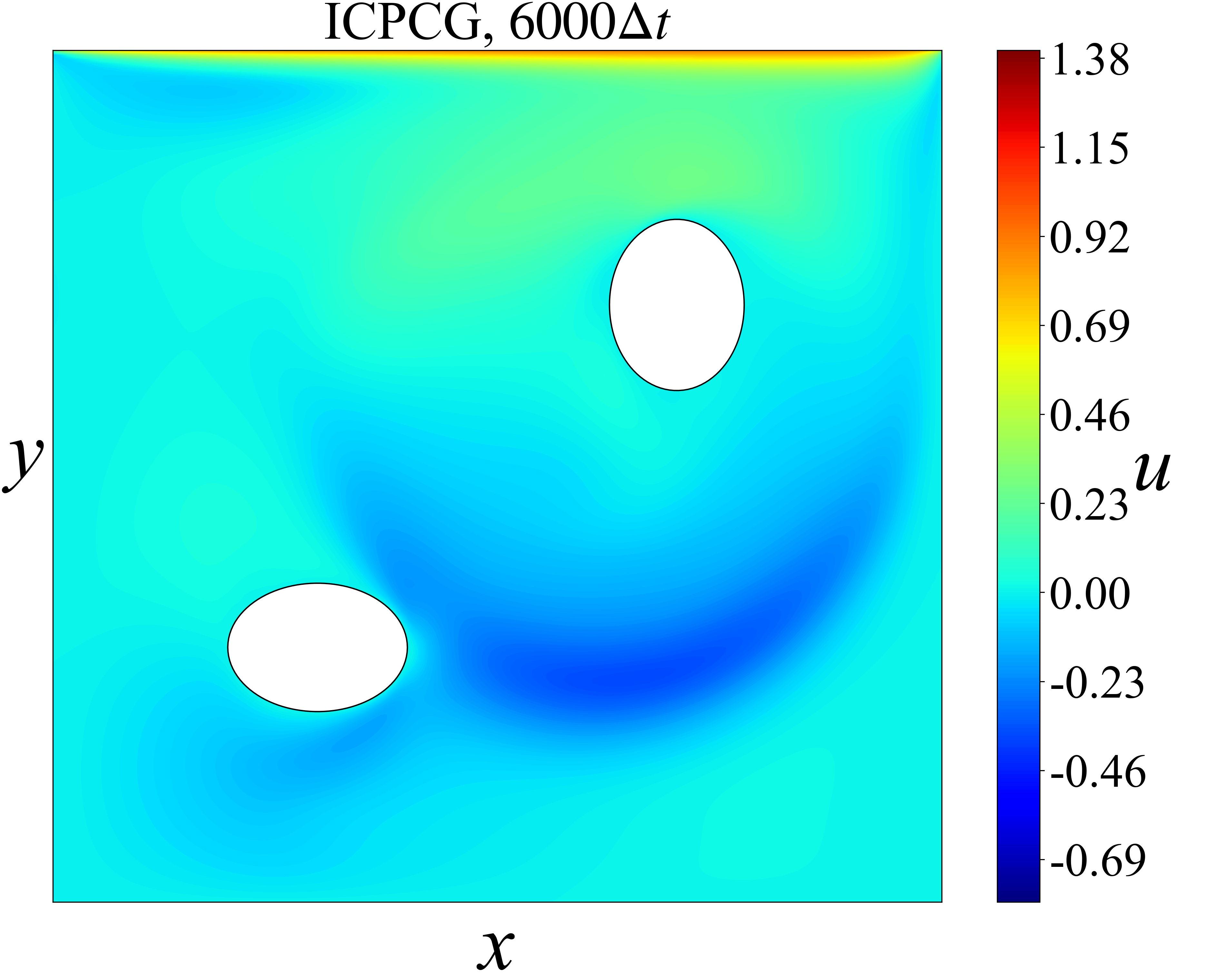}}
  \subfigure[]{
  \label{U_HIC_6000step_2Cylinder}
  \includegraphics[scale=0.129]{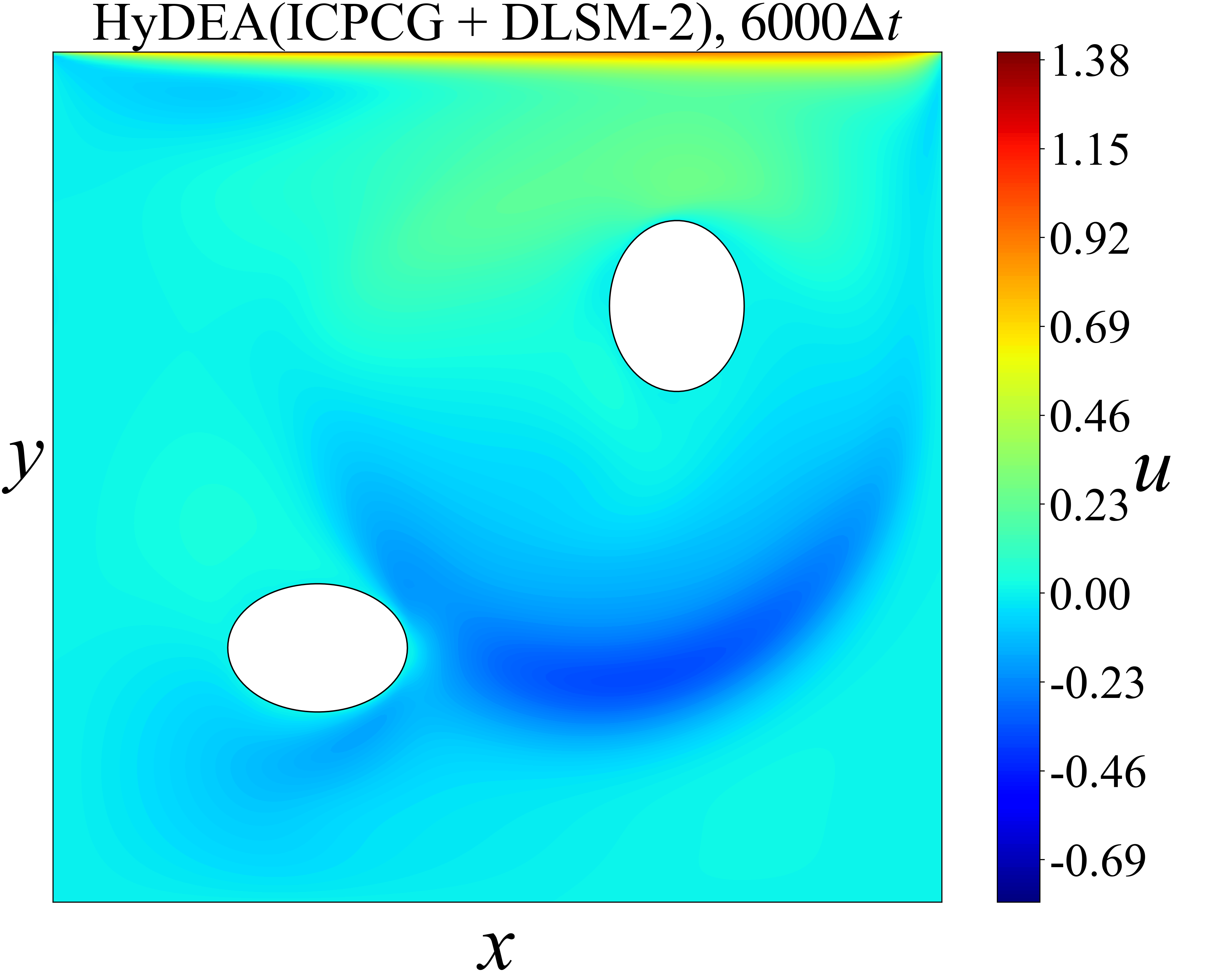}}
  \subfigure[]{
  \label{V_IC_6000step_2Cylinder}
  \includegraphics[scale=0.129]{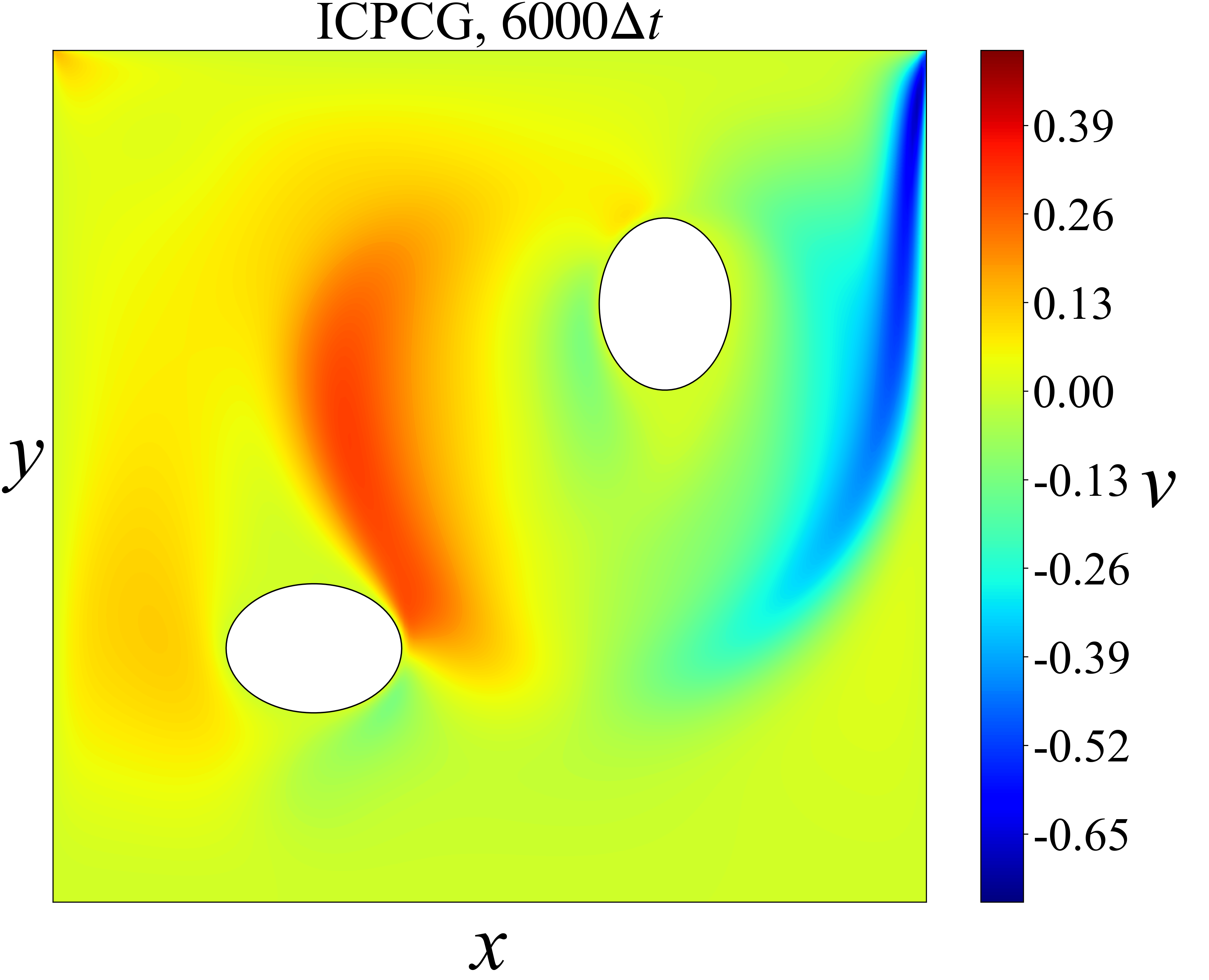}}
  \subfigure[]{
  \label{V_HIC_10000step_2Cylinder}
  \includegraphics[scale=0.129]{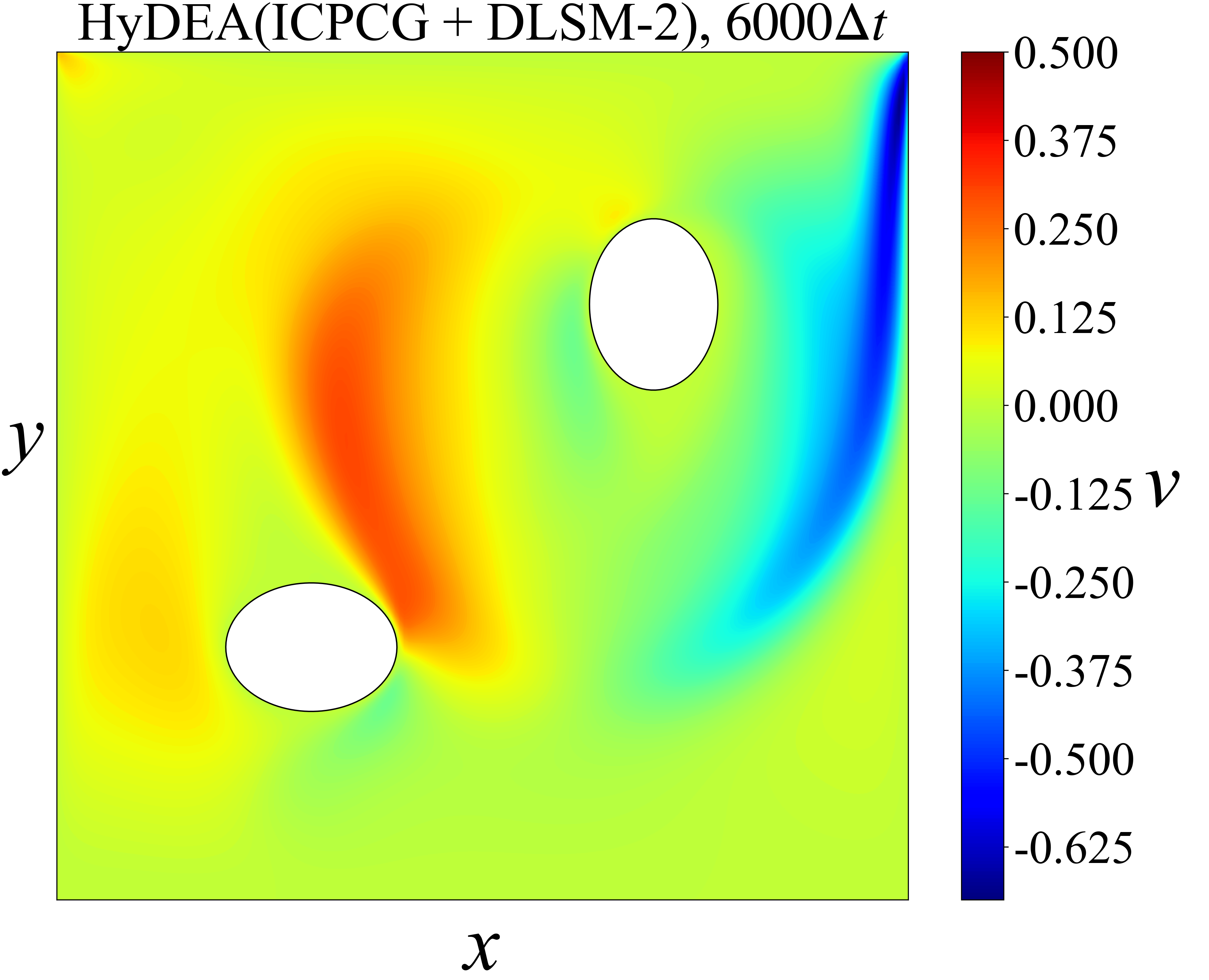}}
 \caption{Velocity contours for 2D lid-driven cavity flow with two immersed stationary elliptical cylinders at $Re=3200$ by ICPCG and HyDEA~(ICPCG+DLSM-2). (a)-(d) $u$ and $v$ at the $1500th$ time step. (e)-(h) $u$ and $v$ at the $6000th$ time step.}
 \label{Flowfield_2ellipse}
\end{figure}

Furthermore, velocity contours of ICPCG and HyDEA~(ICPCG+DLSM-2) at the $1500th$ and $6000th$ time steps are depicted in Fig.~\ref{Flowfield_2ellipse}, demonstrating that the temporal evolution of the flow field is accurately captured.

\subsection{Case 4: 2D flow around an inline oscillating cylinder at $Re=100$}
\label{OsCylinder}

\begin{figure}[htbp]
\centering
  \includegraphics[scale=0.13]{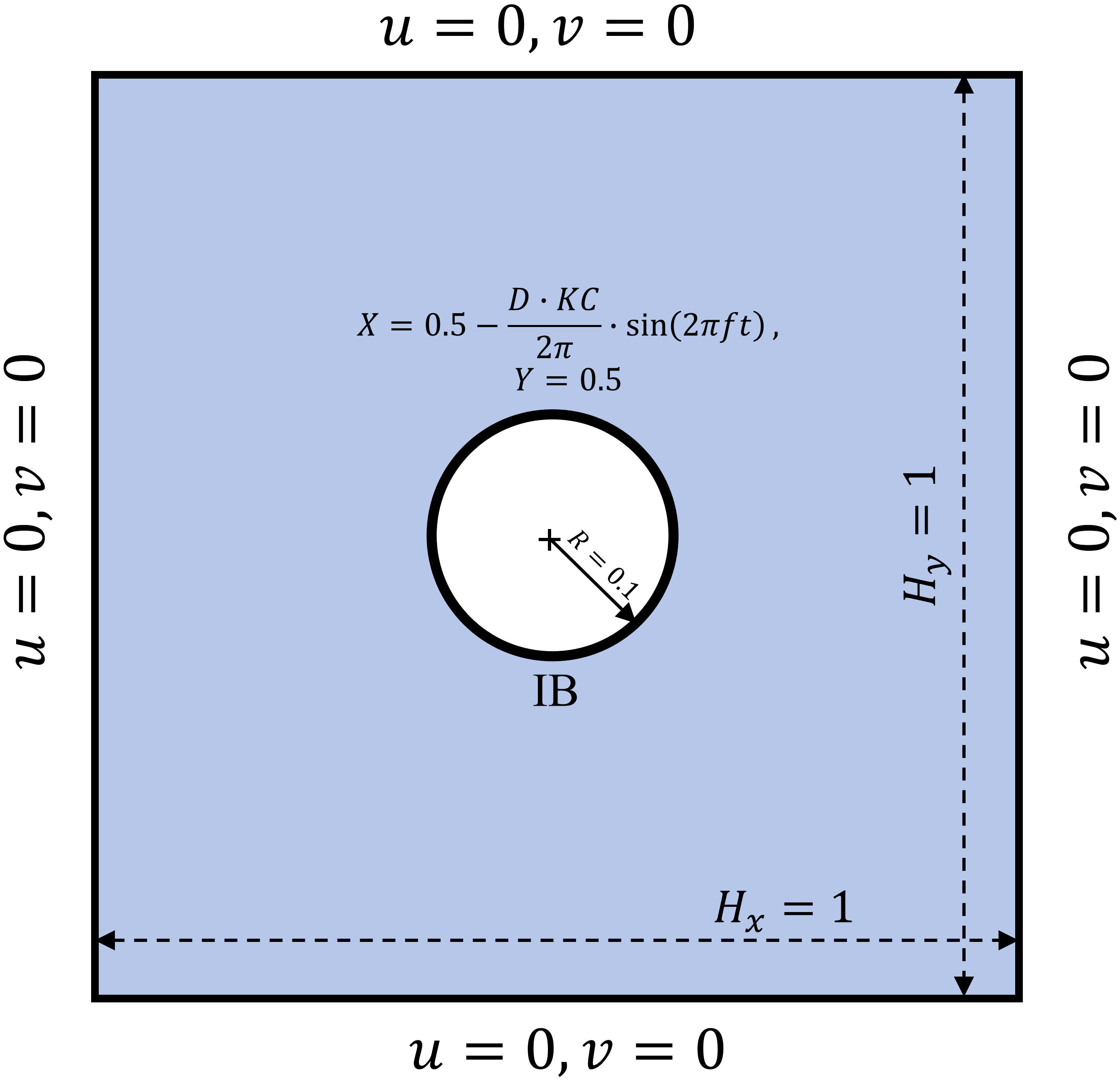}\\
  \caption{Schematic diagram of 2D flow around an inline oscillating cylinder.}\label{Oscylinder_domain}
\end{figure}

This section examines the generalizability of HyDEA in simulating flow around an inline oscillating cylinder. 
The geometric configuration and boundary conditions are illustrated in Fig.~\ref{Oscylinder_domain} with four walls being stationary.
Simulation setup resembles Section~\ref{cavityRe3200} with
$\Delta t=0.002$ and resolution of $192 \times 192$.
The the trajectory of the cylinder is governed by
\begin{eqnarray}
\label{Move trajectory of cylinder x}
 X &=& 0.5 - \frac{D \cdot KC}{2\pi} \cdot sin(2 \pi ft),
\\
\label{Move trajectory of cylinder y}
  Y &=& 0.5,
\end{eqnarray}
where ($X, Y$) is its center position and $t$ is time.
With frequency $f=0.5$, cylinder diameter $D=0.2$ and velocity amplitude $U_m=0.5$ in x-direction,
the Keulegan–Carpenter number $KC=U_m/fD=5$.
Moreover, with kinematic viscosity $\nu=0.001$, 
Reynolds number $Re=U_m D/ \nu=100$.

Furthermore, the DIBPM is activated for the oscillating cylinder in CFD computations and
the exactly same DLSM-2 as in Section~\ref{cavityRe3200} is executed with $Num_{CG-type}=3$ and $Num_{DLSM}=2$, respectively. 
A detailed analysis for other values of $Num_{CG-type}$ and $Num_{DLSM}$ is also provided in~\ref{appendixD}.

\begin{figure}[htbp] 
 \centering  
  \subfigure[]{
  \label{192_Rline_OsCylinder_IC_10steps}
  \includegraphics[scale=0.21]{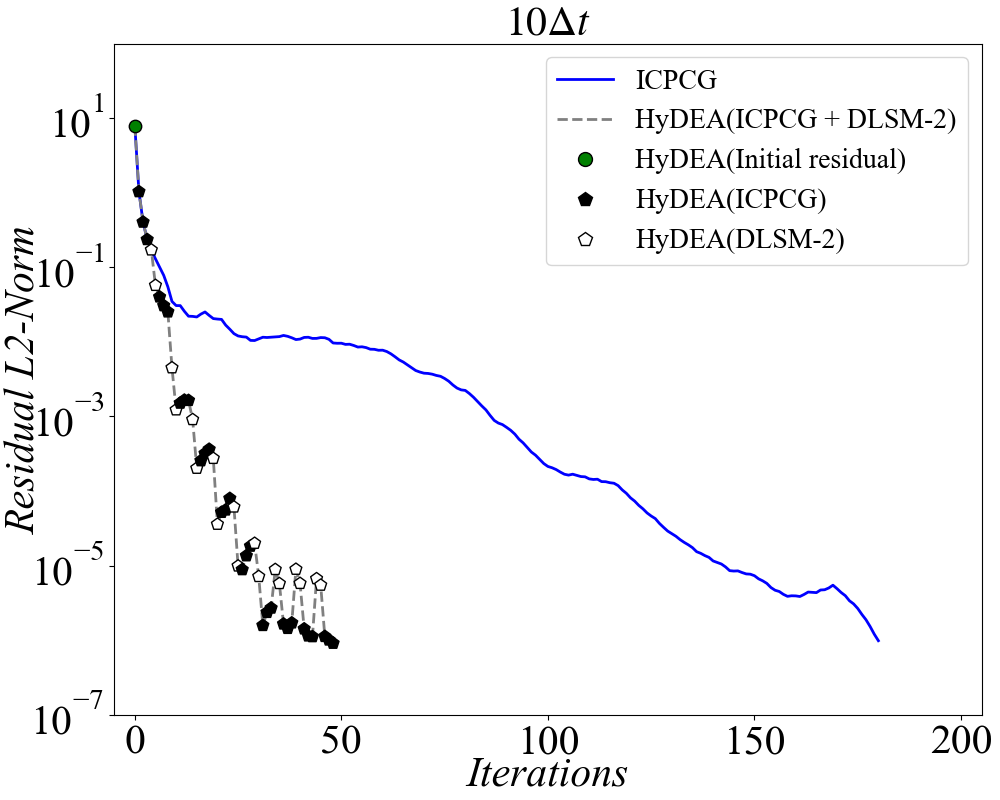}}
  \subfigure[]{
  \label{192_Rline_OsCylinder_IC_1000steps}
  \includegraphics[scale=0.21]{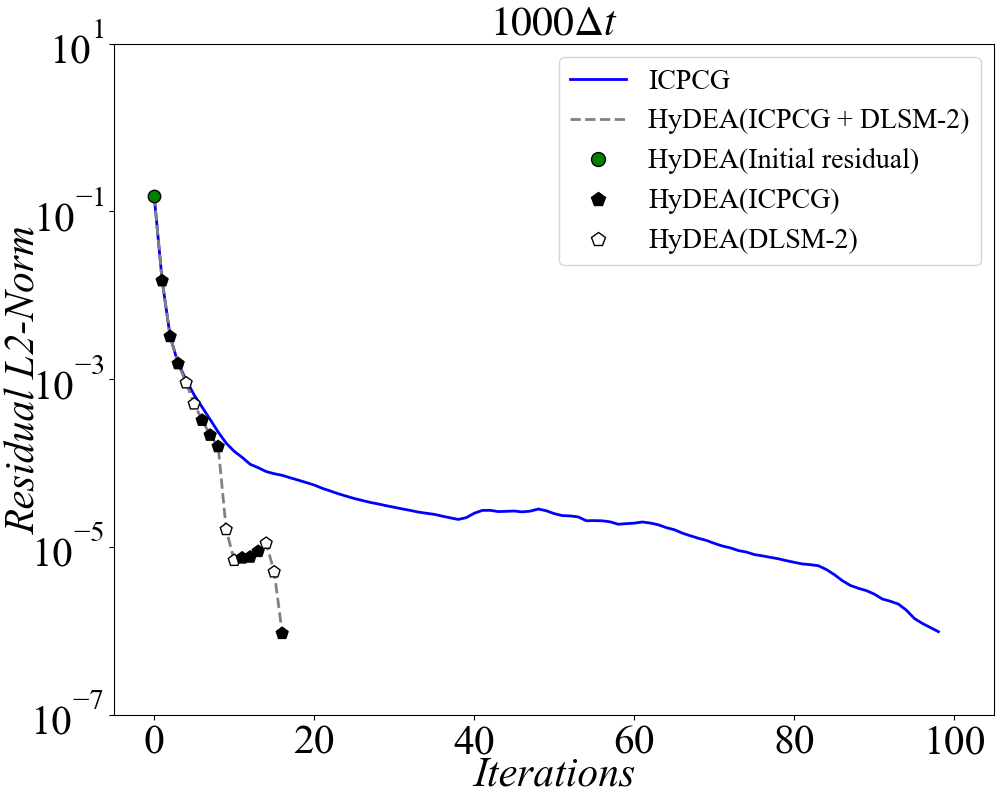}}
  \subfigure[]{
  \label{192_Rline_OsCylinder_IC_3400steps}
  \includegraphics[scale=0.21]{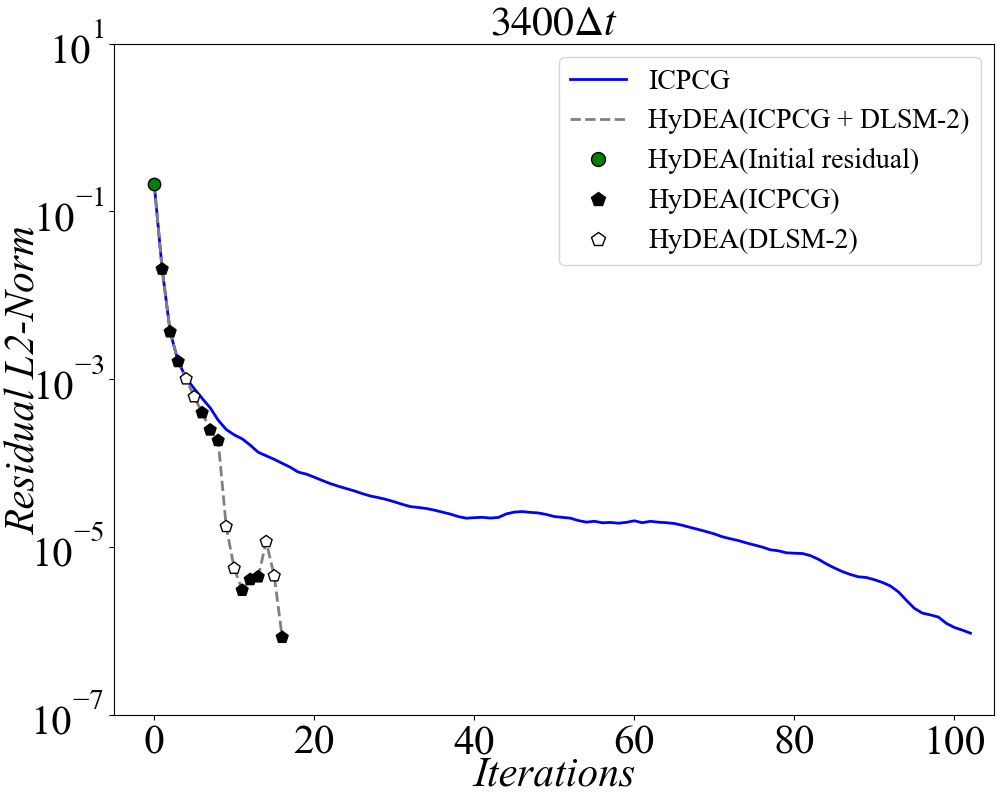}}
  \caption{Iterative residuals of solving the PPE for 2D flow around an inline oscillating cylinder at $Re=100$ and $KC=5$ by ICPCG and HyDEA~(ICPCG+DLSM-2). (a) $10th$ time step. (b) $1000th$ time step. (c) $3400th$ time step.}\label{192_Rline_Oscylinder_HIC}
\end{figure}

Taking ICPCG and HyDEA~(ICPCG+DLSM-2) as an example, Fig.~\ref{192_Rline_Oscylinder_HIC} presents the iterative residuals of solving the PPE at the $10th$, $1000th$ and $3400th$ time steps. 
For HyDEA~(ICPCG+DLSM-2), it takes
$10$ rounds of the hybrid algorithm with $48$ iterations at $10\Delta t$;
$4$ rounds with $16$ iterations at $1000\Delta t$;
$4$ rounds with $16$ iterations at $3400\Delta t$.
The results demonstrate that HyDEA achieves a substantial reduction in the number of iterations compared to the ICPCG method alone. 
Notably, the DeepONet weights employed in this case are identical to those used in Sections~\ref{cavityRe3200},~\ref{1Cylinder} and~\ref{2Cylinder}, 
which confirms HyDEA's robust generalization capability across diverse flow conditions.

\begin{figure}[htbp] 
 \centering  
  \subfigure[]{
  \label{U_IC_1000step_OsCylinder}
  \includegraphics[scale=0.129]{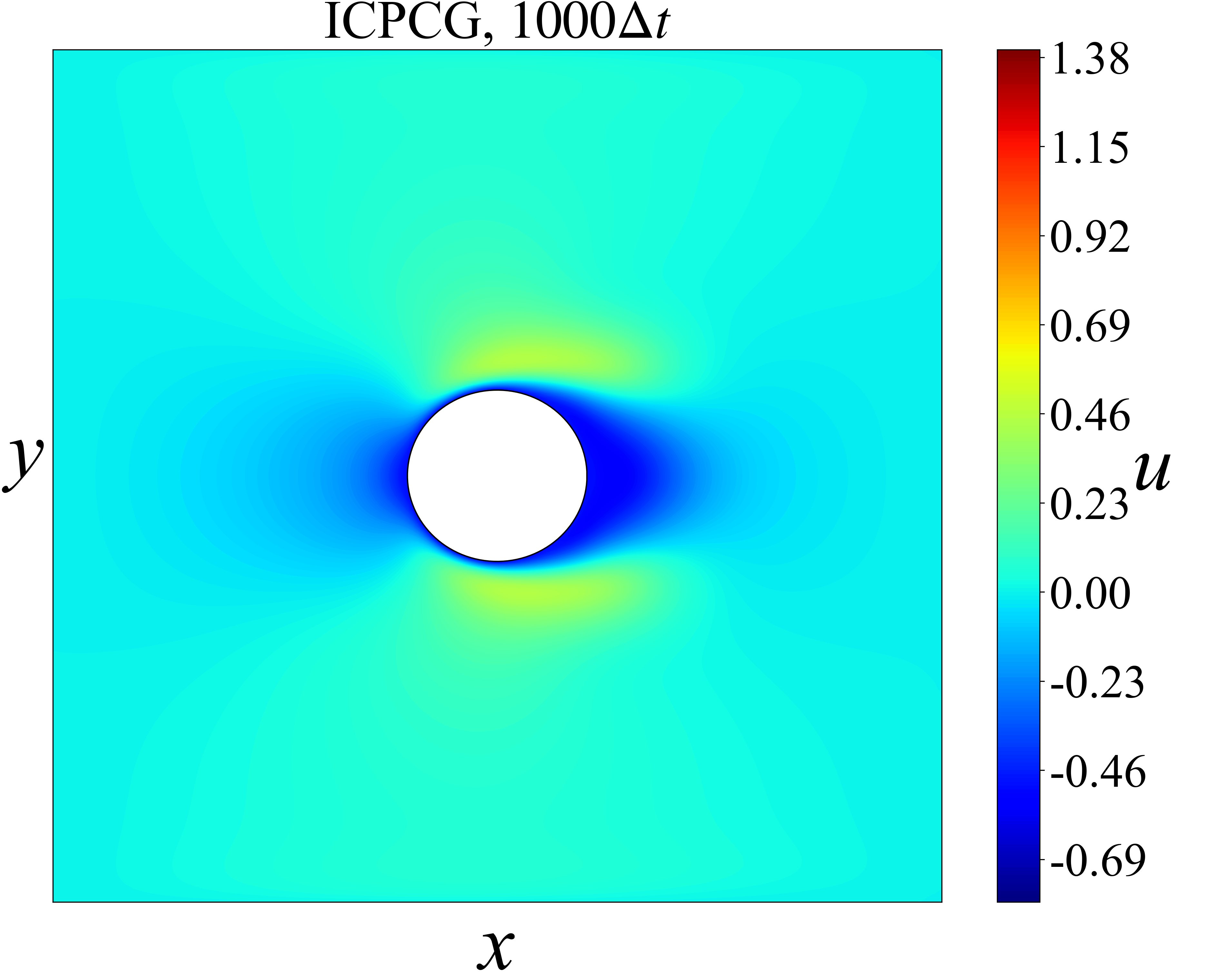}}
  \subfigure[]{
  \label{U_HIC_1000step_OsCylinder}
  \includegraphics[scale=0.129]{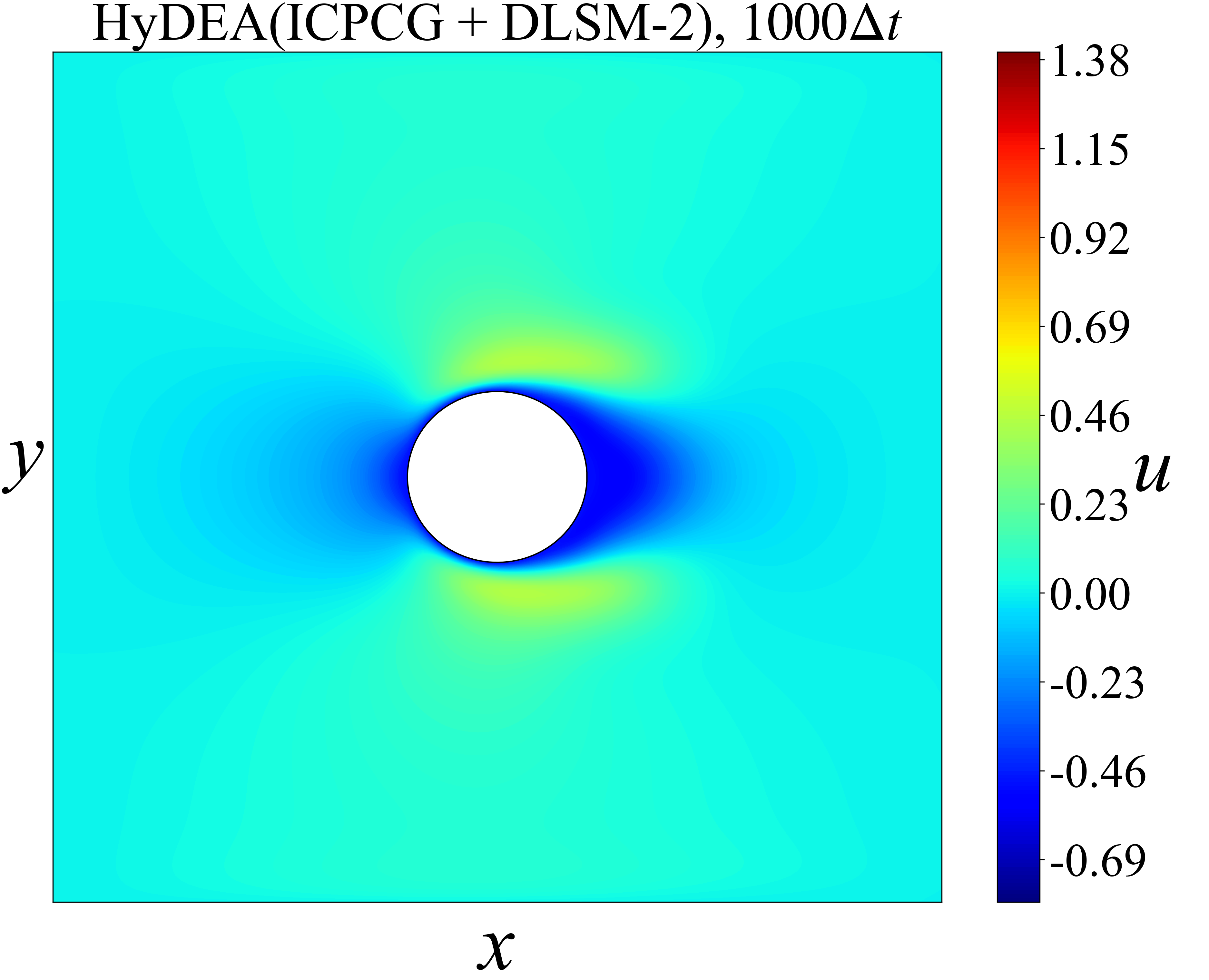}}
  \subfigure[]{
  \label{V_IC_1000step_OsCylinder}
  \includegraphics[scale=0.129]{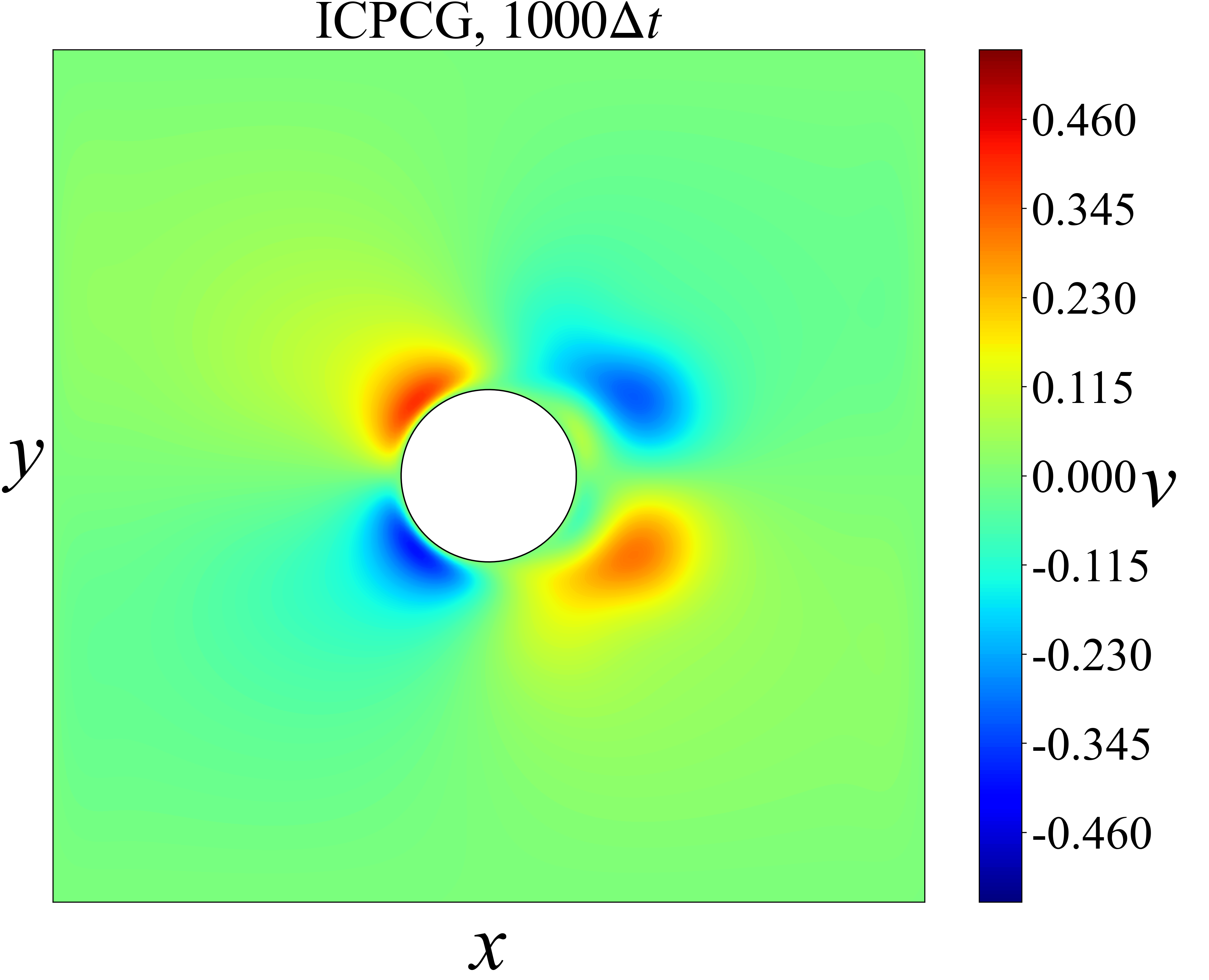}}
  \subfigure[]{
  \label{V_HIC_1000step_OsCylinder}
  \includegraphics[scale=0.129]{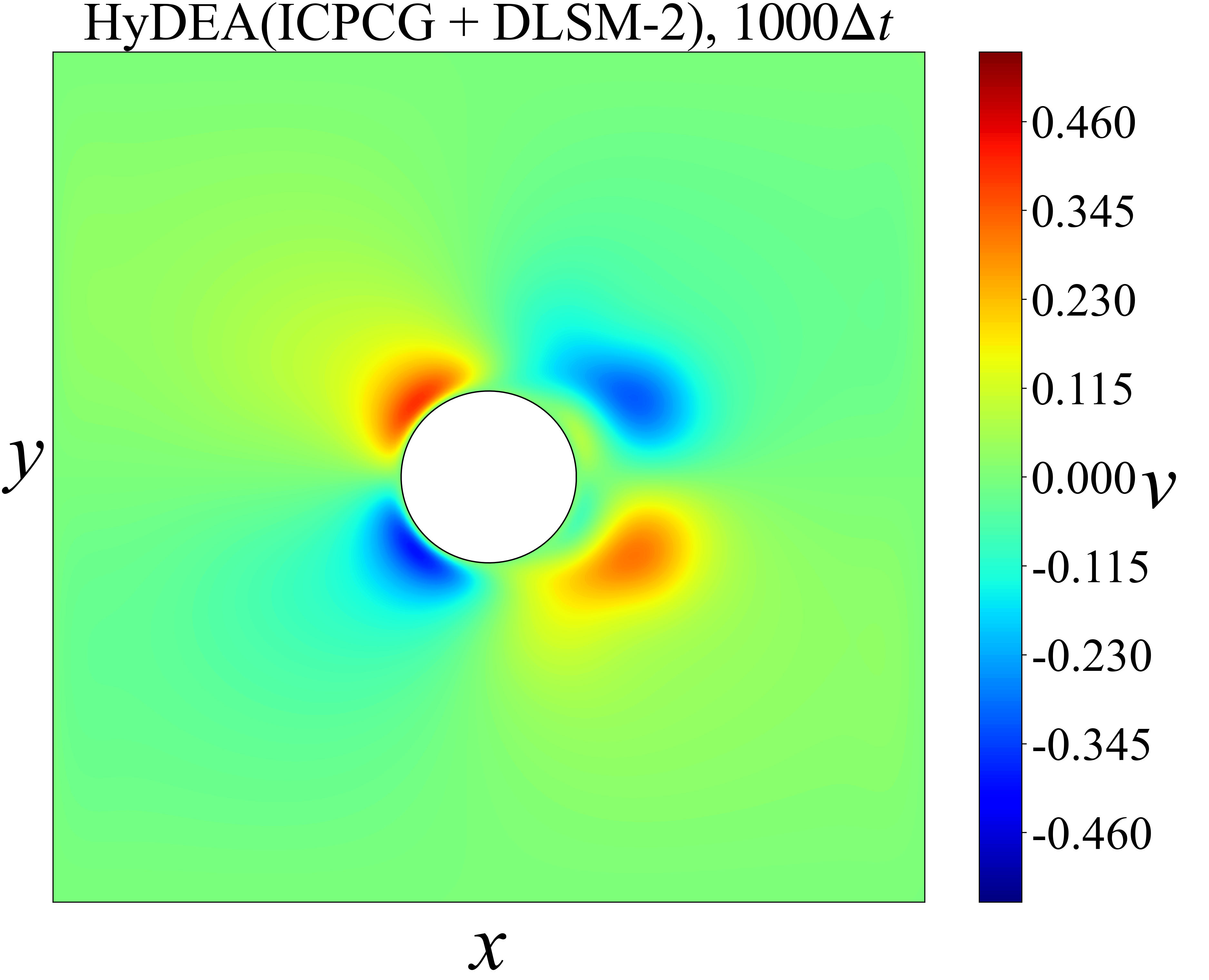}}
  \subfigure[]{
  \label{U_IC_3400step_OsCylinder}
  \includegraphics[scale=0.129]{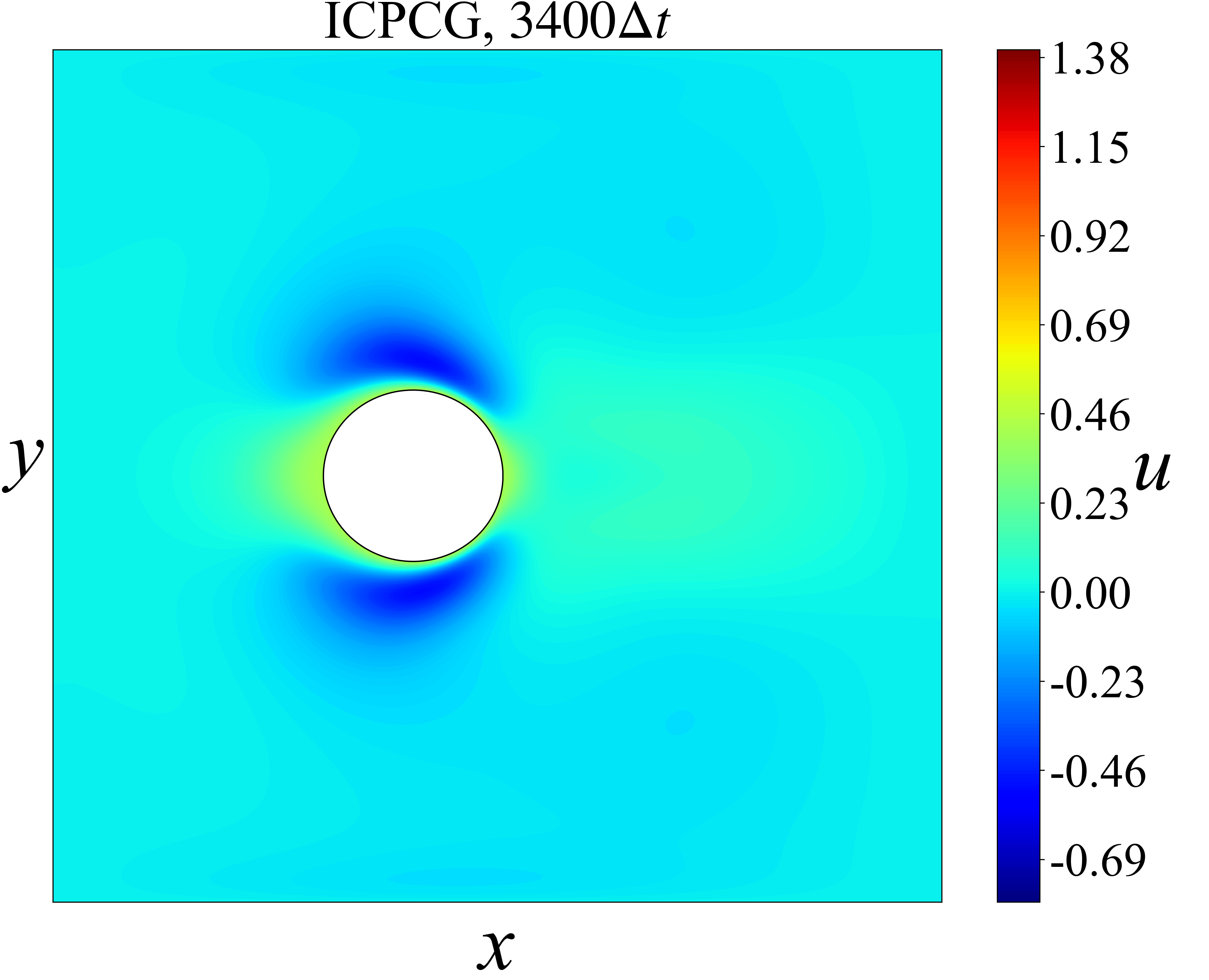}}
  \subfigure[]{
  \label{U_HIC_3400step_OsCylinder}
  \includegraphics[scale=0.129]{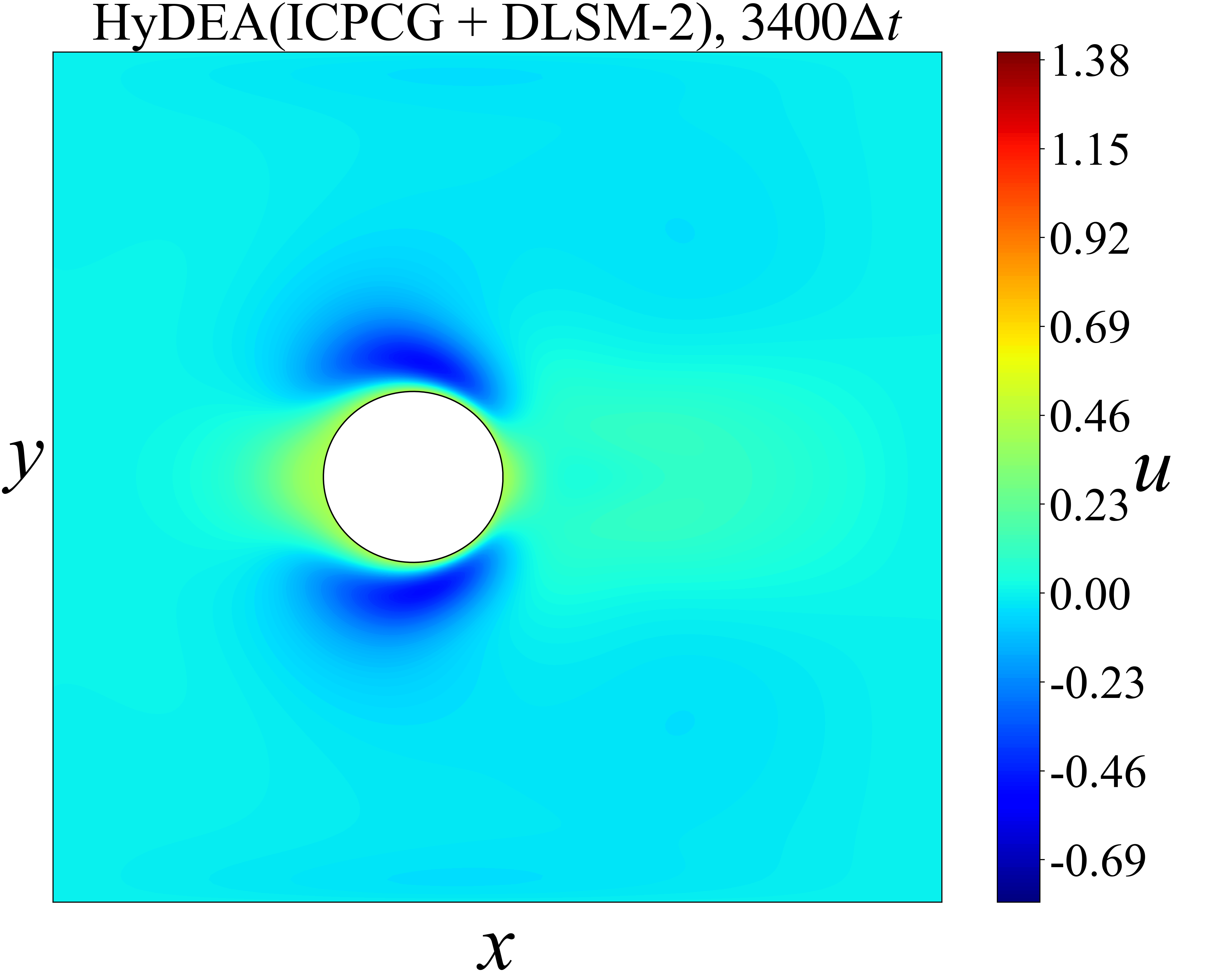}}
  \subfigure[]{
  \label{V_IC_3400step_OsCylinder}
  \includegraphics[scale=0.129]{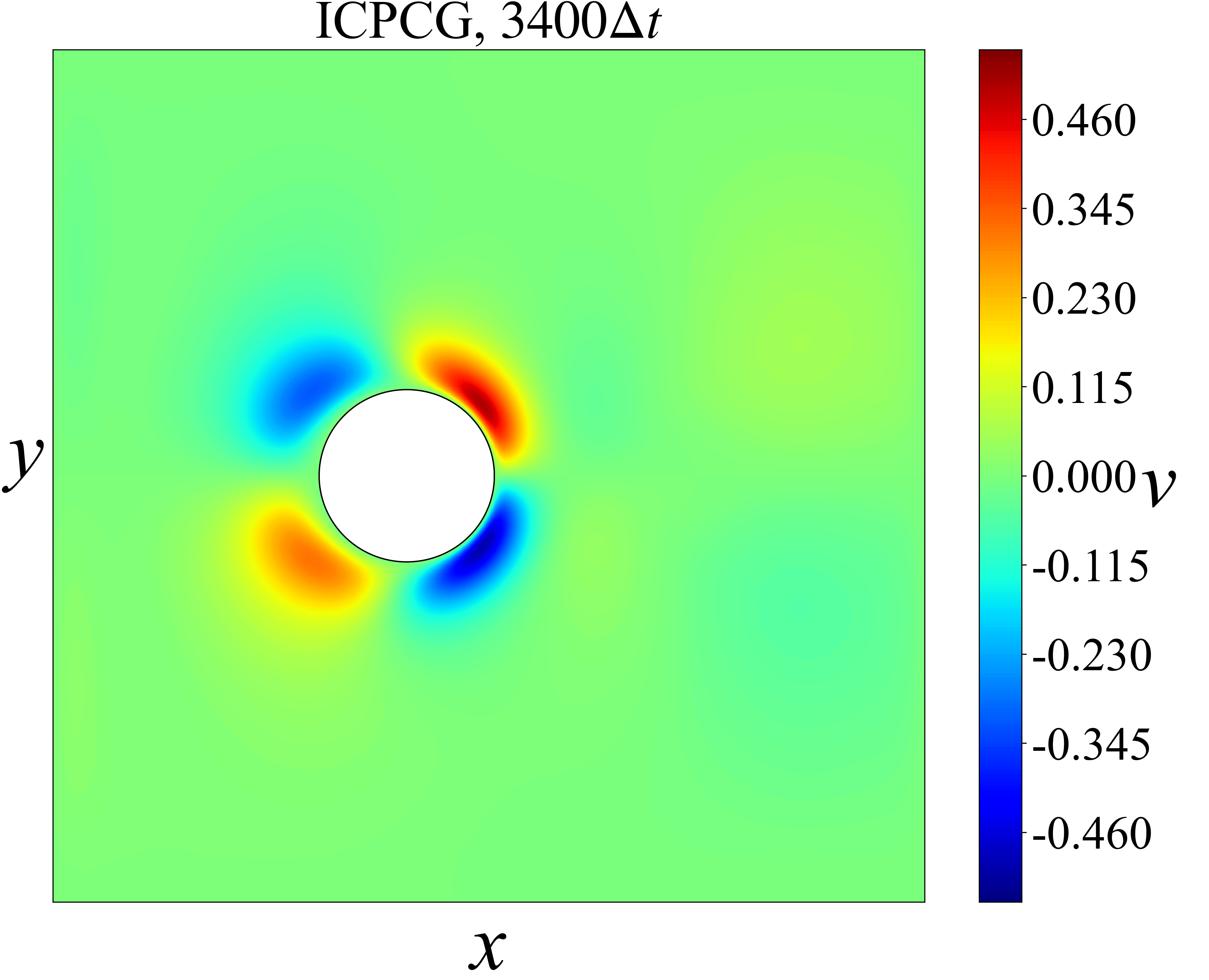}}
  \subfigure[]{
  \label{V_HIC_3400step_OsCylinder}
  \includegraphics[scale=0.129]{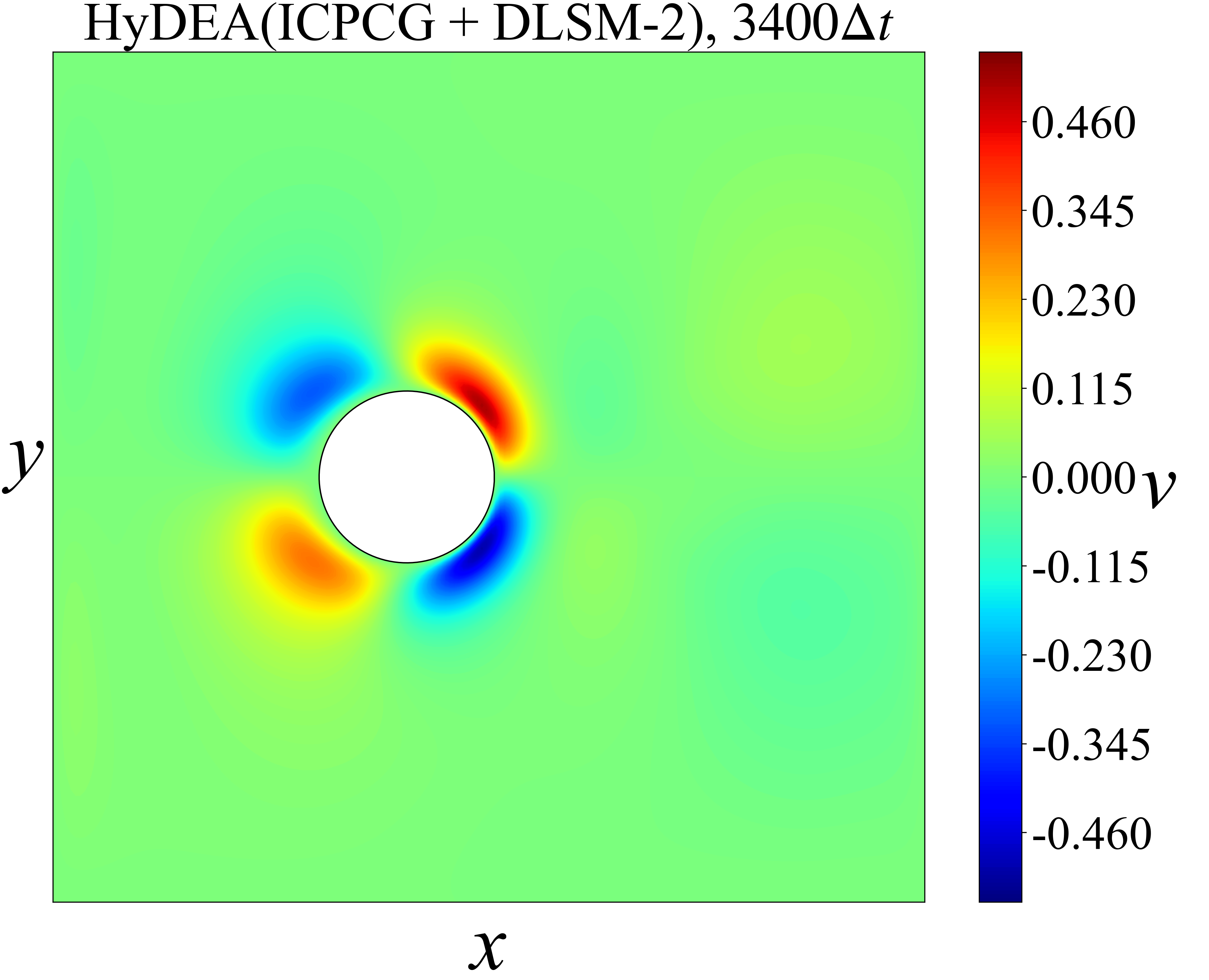}}
 \caption{Velocity fields for 2D flow around an inline oscillating cylinder at $Re=100$ and $KC=5$ by ICPCG and HyDEA~(ICPCG+DLSM-2). (a)-(d) $u$ and $v$ at the $1000th$ time step. (e)-(h) $u$ and $v$ at the $3400th$ time step.}
 \label{Flowfield_OsCylinder}
\end{figure}

Fig.~\ref{Flowfield_OsCylinder} presents the velocity contours of ICPCG and HyDEA~(ICPCG+DLSM-2) at the $1000th$ and $3400th$ time steps, clearly demonstrating the accurate temporal evolution of the flow fields.


\subsubsection*{Comparison of DCDM, DLSM and HyDEA}
\begin{figure}[htbp] 
 \centering  
 \subfigure[]{
  \label{192_Rline_OsCylinder_DCDMcompare_IC_3steps}
  \includegraphics[scale=0.26]{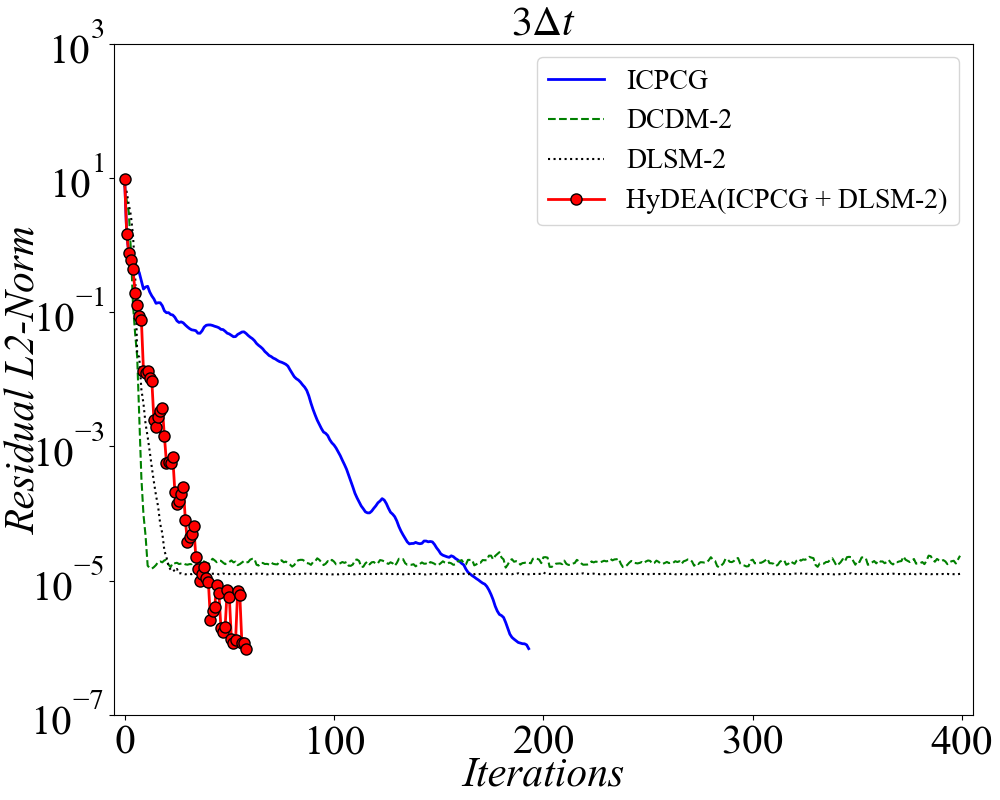}}
  \subfigure[]{
  \label{192_Rline_OsCylinder_DCDMcompare_IC_4steps}
  \includegraphics[scale=0.26]{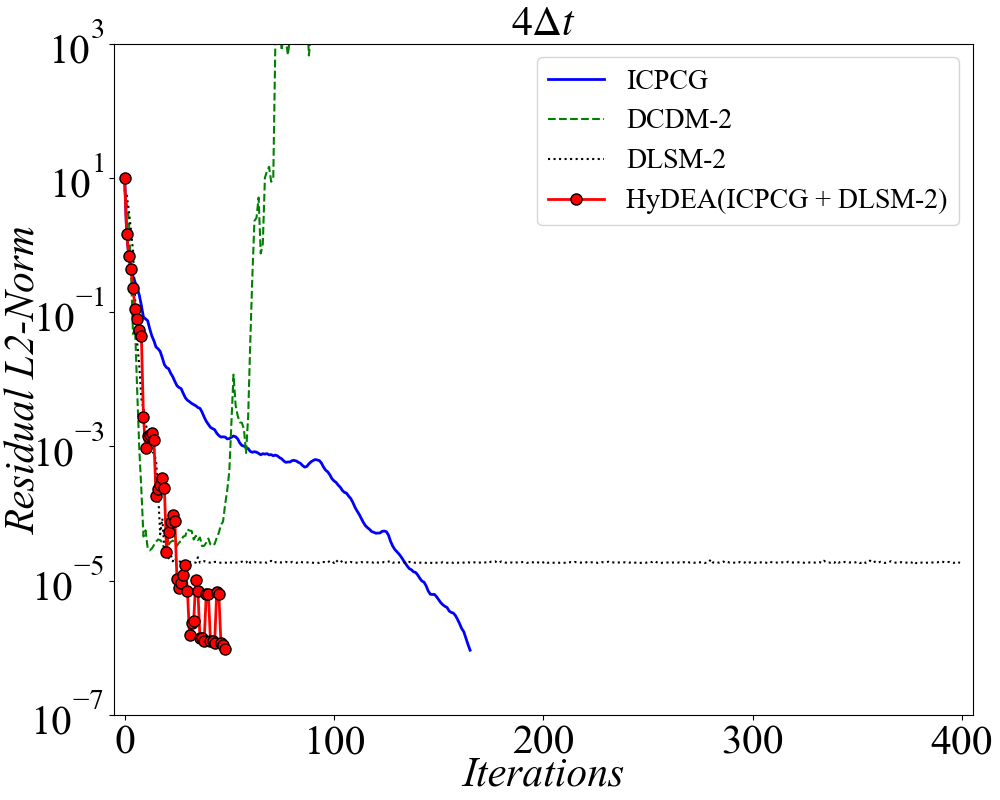}}
  \caption{Iterative residuals of solving the PPE for 2D flow around an inline oscillating cylinder at $Re=100$ and $KC=5$ with resolution $192\times 192$: comparisons among ICPCG, data-driven approaches of DCDM and DLSM, and HyDEA. (a) $3rd$ time step. (b) $4th$ time step.}\label{Os_DCDMcompare}
\end{figure}

We compare the performance of purely data-driven approaches  of DCDM and DLSM with HyDEA and presents the iterative residuals at the $3rd$ and $4th$ time steps, with maximum $400$ iterations in Fig.~\ref{Os_DCDMcompare}.
Both DCDM and DLSM show initially effective residual reduction, 
but their performance degrades significantly for further iterations
and cannot reach predefined tolerance of $10^{-6}$.
Notably, DLSM can sustain the residual at the plateau in the order $10^{-5}$,
while DCDM even diverges at $t=4\Delta t$.
In contrast, both ICPCG and HyDEA~(ICPCG+DLSM-2) can achieve desired convergence,
albeit with much less iterations for the latter.

\subsubsection*{Comparison of two versions of HyDEA}
\begin{figure}[htbp] 
 \centering  
  \subfigure[]{
  \label{192_Rline_3add2_diffHyDEA_Os_IC_10steps}
  \includegraphics[scale=0.26]{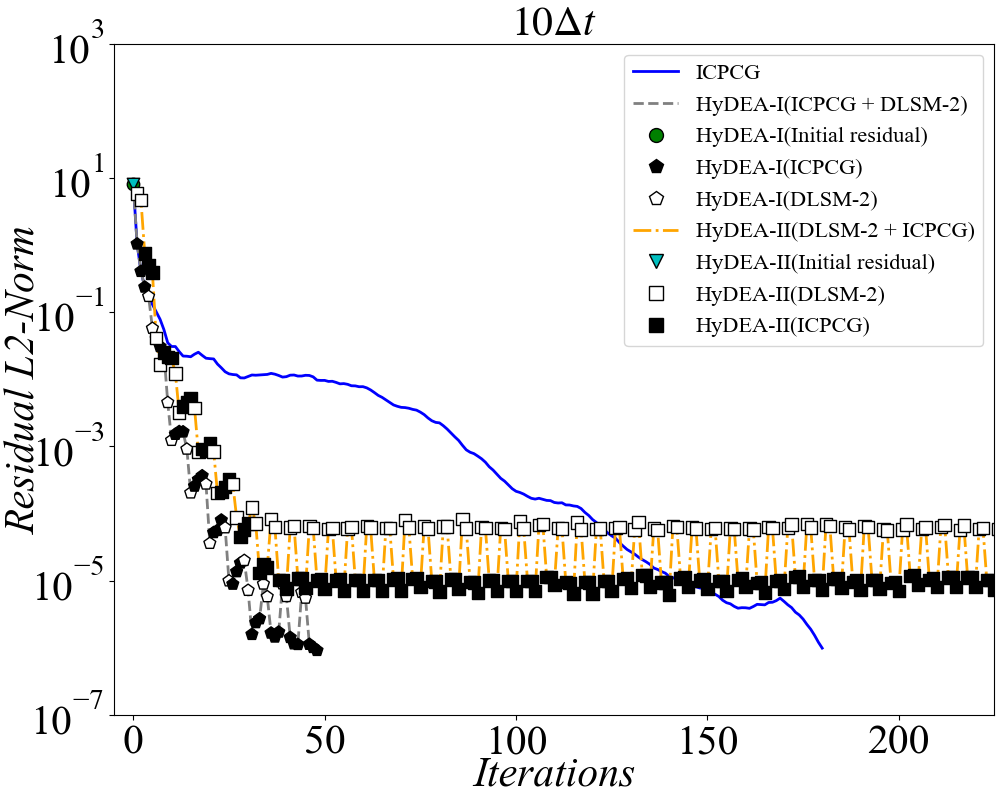}}
  \subfigure[]{
  \label{192_Rline_3add2_diffHyDEA_Os_IC_100steps}
  \includegraphics[scale=0.26]{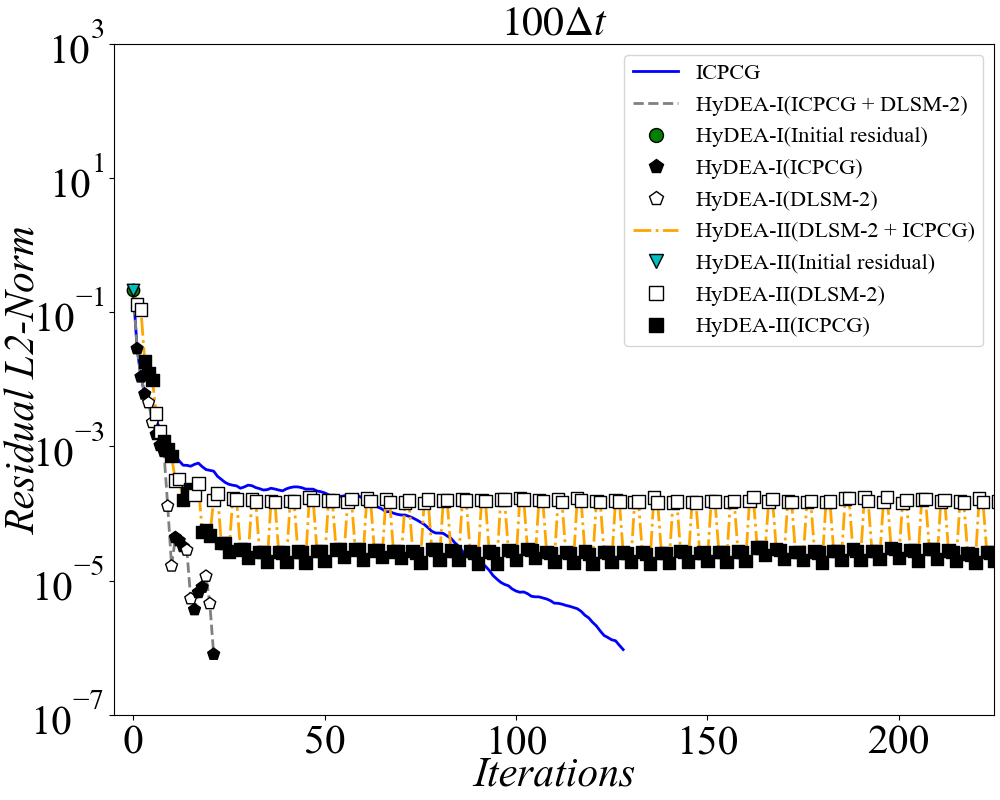}}
  \caption{Iterative residuals of solving the PPE for 2D flow around an inline oscillating cylinder at $Re=100$ and $KC=5$ by two different implementations of HyDEA . (a) $10th$ time step. (b) $100th$ time step.}\label{192_Rline_diffHyDEA_Os}
\end{figure}

As indicated in Fig.~\ref{HyDEA_process},  
HyDEA offers two distinct implementations shown in Fig.~\ref{InitialPCG} and~\ref{InitialNN} for each round of the hybrid algorithm.
We examplify the comparison of the two versions for the same scenario,
namely HyDEA-I~(ICPCG+DLSM-2) versus HyDEA-II~(DLSM-2+ICPCG) for:
\begin{itemize}
 \item CFD resolution of $192\times 192$;
 \item $Num_{CG-type}=3$ and $Num_{DLSM}=2$;
 \item maximum $400$ iteration at one time step.
\end{itemize}
Fig.~\ref{192_Rline_diffHyDEA_Os} presents the iterative residuals of solving the PPE using ICPCG, HyDEA-I and HyDEA-II at the $10th$ and $100th$ time steps. 
The results reveal a clear performance contrast, with HyDEA-I exhibiting superior convergence characteristics compared to that of HyDEA-II.
The observed performance superiority stems from the fact that in HyDEA-I, a CG-type method initially eliminates high-frequency errors to a certain extent before passing onto the DLSM module. 
Subsequently, DLSM/DeepONet receives an better input, which aligns with the training dataset,
and is able to predict more sensible line-search directions, ultimately leading to a faster convergence. 
Based on these discussions and demonstrated gains in performance, 
HyDEA-I emerges as the recommended version,
which was also extensively employed in previous benchmark cases
in Sections~\ref{cavity}, \ref{1Cylinder}, \ref{2Cylinder}, and \ref{OsCylinder}.

\subsubsection*{Comparison of DLSM and DCDM as module in HyDEA}

\begin{figure}[htbp] 
 \centering  
  \subfigure[]{
  \label{192_OS_VS_10steps}
  \includegraphics[scale=0.21]{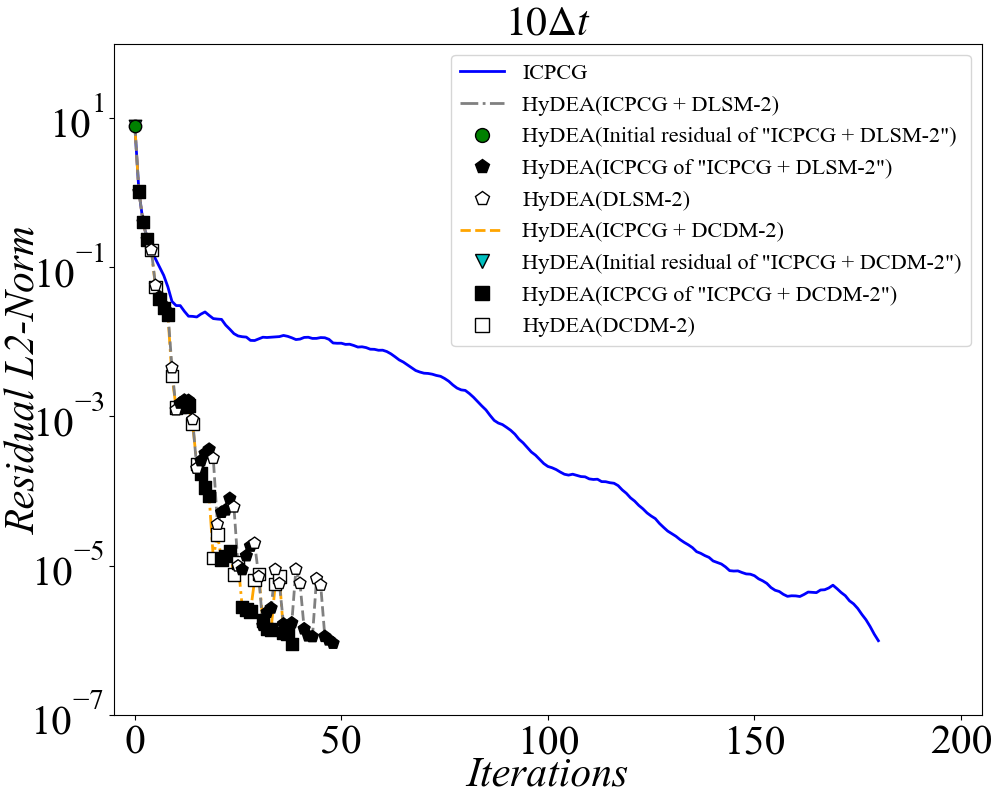}}
  \subfigure[]{
  \label{192_OS_VS_1000steps}
  \includegraphics[scale=0.21]{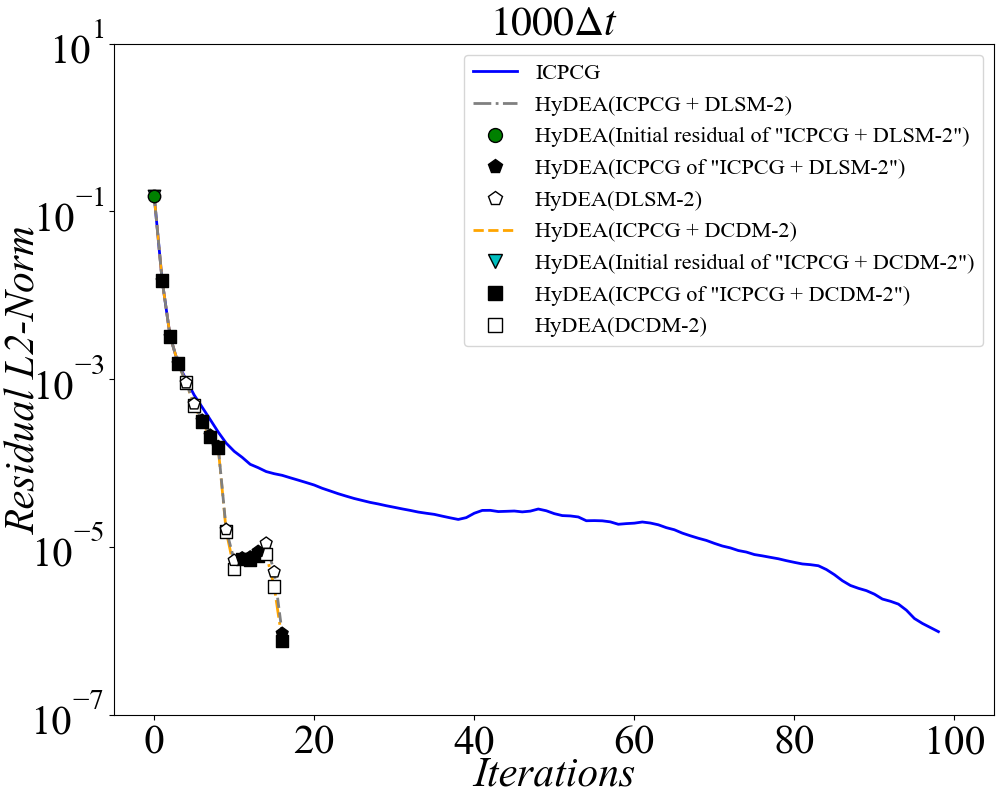}}
  \subfigure[]{
  \label{192_OS_VS_3400steps}
  \includegraphics[scale=0.21]{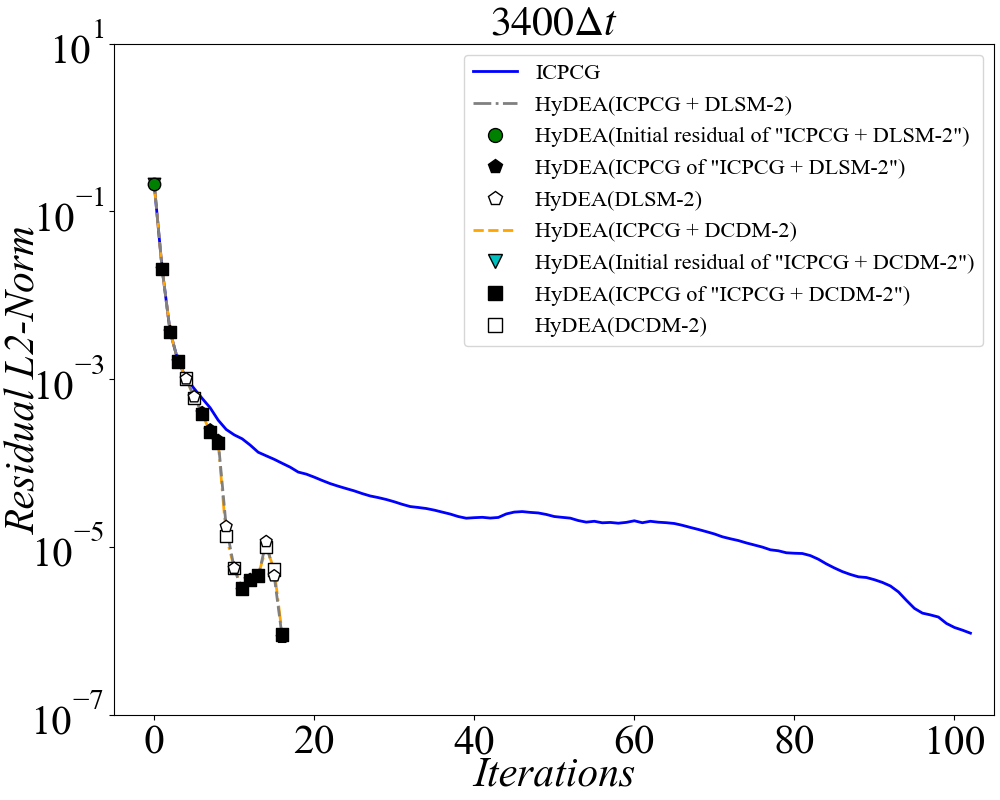}}
  \caption{Iterative residuals of solving the PPE for 2D flow around an inline oscillating cylinder at $Re=100$ and $KC=5$ by ICPCG, HyDEA~(ICPCG+DLSM-2), and HyDEA~(ICPCG+DCDM-2). (a) $10th$ time step. (b) $1000th$ time step. (c) $3400th$ time step.}\label{192_Os_VS_3+2}
\end{figure}

The DCDM of Kaneda et al.~\cite{kaneda2023DCDM} represents a special case of DLSM, where the line-search directions are made additionally $M$-orthogonal by a GS orthogonalization process. We evaluate respective performance of DLSM module and DCDM module within the HyDEA framework.
We compare the iterative residuals of HyDEA~(ICPCG+DLSM-2) and HyDEA~(ICPCG+DCDM-2)
for solving the PPE at the $10th$, $1000th$ and $3400th$ time steps in Fig.~\ref{192_Os_VS_3+2}. 
The results show that HyDEA achieves similar convergence rates when equipped with either DLSM-2 or DCDM-2, suggesting that GS orthogonalization of DeepONet-predicted line-search directions provides negligible benefits.

\section{Conclusion}
\label{conclusion}

We present HyDEA, a novel hybrid framework that synergizes deep learning with classical iterative solvers to accelerate incompressible flow simulations. By targeting the pressure Poisson equation (PPE) - the computational bottleneck in the fractional step method - HyDEA combines the strengths of a Deep Learning line-search method (DLSM) and a conjugate gradient (CG)-type solver. The DLSM module, powered by DeepONet, predicts optimal line-search directions to accelerate convergence for the large, sparse, symmetric positive-definite linear systems arising from finite difference discretizations. For flows with solid structures, the decoupled immersed boundary projection method seamlessly integrates into HyDEA while preserving the linear system’s structure.
HyDEA addresses the limitations of purely data-driven approaches by leveraging complementary strengths:
\begin{itemize}
\item
DeepONet captures efficiently large-scale solution features.
\item CG-type methods (including preconditioned variants with Incomplete Cholesky, Jacobi, and Multigrid) robustly resolve high-frequency errors.
\end{itemize}

Benchmark results demonstrate HyDEA’s superior performance and generalization:
\begin{itemize}
    \item Reduces iteration counts significantly compared to standalone CG-type methods across diverse flows (single-component, stationary/moving obstacles) using {\it fixed network weights}.
    \item Exhibits super-resolution capability, achieving accurate solutions on a $512 \times 512$ grid despite being trained on $128 \times 128$ data ($16 \times$ resolution increase).
 \item Outperforms standalone DLSM and purely data-driven methods (e.g., DCDM), proving neural networks augment but cannot replace classical solvers without sacrificing stability or accuracy.
\end{itemize}

The prototype implementation of HyDEA combines heterogeneous frameworks: $\mathtt{PyTorch}$~($\mathtt{Python}$) for DLSM module and $\mathtt{PETSc}$~(C) for CG-type solvers, which incurs computational overhead. Moreover, the loss function that incorporates the coefficient matrix leads to prohibitively high computational cost for large-scale simulations. Additionally, the CNN-based DeepONet restricts its applications to uniform Cartesian grids. Future efforts will focus on several key directions: code optimization to enhance the computational efficiency, extending HyDEA to non-uniform grids and parameterized linear systems for broader applicability, refining the dataset construction method for better compatibility with various traditional iterative solvers, and incorporating domain decomposition concepts to enable large-scale simulations.
HyDEA’s hybrid paradigm bridges the gap between data-driven efficiency and numerical robustness, offering a scalable pathway for high-fidelity flow simulations.

\section*{Acknowledgments}

The authors appreciate support from National Key R\&D Program of China (2022YFA1203200).

\appendix

\section{Analysis of convergence behavior of the CG method}
\label{appendixA}

This section analyzes the convergence behavior of the CG method, aiming to provide a rationale for the dataset construction method in Section~\ref{Dataset}. The CG method is an efficient algorithm for solving the quadratic vector optimization problem formulated in Eq.~(\ref{min_Vector_quadratic_F}). By substituting $\delta p=\delta p^{true}-e$, where $e$ denotes the iterative error vector, we can reformulate Eq.~(\ref{min_Vector_quadratic_F}) as:
\begin{eqnarray}
\label{min_Vector_quadratic_F_energyNorm}
 \min_{e \in \mathbf{R^a}} \{\frac{1}{2}e^{T}Me-\frac{1}{2}(\delta p^{true})^{T}M\delta p^{true}+c\}.
\end{eqnarray}
Since the second and third terms in Eq.~(\ref{min_Vector_quadratic_F_energyNorm}) are constant, the optimization problem reduces to:
\begin{eqnarray}
\label{min_energyNorm}
 \min_{e \in \mathbf{R^a}} \{e^{T}Me=\Vert e \Vert_{M}^{2}\},
\end{eqnarray}
where $\Vert e \Vert_{M}=\sqrt{e^{T}Me}$ denotes the energy norm of the error. The CG method progressively minimizes $\Vert e \Vert_{M}$ during its iterative solution process~\cite{Golub2013}.

The normalized eigenvectors $\{v_{0},v_{1},\cdots,v_{a-1}\}$ of the symmetric-positive-definite matrix $M$ form an orthonormal basis, enabling the initial error vector $e_{0}$ to be expressed as a linear combination of these normalized eigenvectors:
\begin{eqnarray}
\label{error_linear_combin_eigenvector}
 e_{0}=\sum_{j=0}^{a-1} \alpha_{j}v_{j},
\end{eqnarray}
where coefficient $\alpha_{j}$ represents the projection length of $e_{0}$ along the direction of $v_{j}$.

The error vector at the $k$th iteration can be expressed in the following form~\cite{shewchuk1994introductionCG}:
\begin{eqnarray}
\label{error_Pi_e}
 e_{k}=P_{k}(M)e_{0},
\end{eqnarray}
where $P_{k}$ is a $k$-degree polynomial satisfying $P_{k}(0)=1$. By substituting Eqs.~(\ref{error_linear_combin_eigenvector}) and~(\ref{error_Pi_e}) into the $\Vert e \Vert_{M}=\sqrt{e^{T}Me}$, the discrete formulation for $\Vert e \Vert_{M}^{2}$ at the $k$th iteration is obtained as follows:
\begin{eqnarray}
\label{errorNorm_dis}
 \Vert e_{k} \Vert_{M}^{2} = \sum_{j=0}^{a-1} \alpha_{j}^2 [P_{k}(\lambda_{j})]^{2}\lambda_{j},
\end{eqnarray}
where $\lambda_{j}$ is the eigenvalue corresponding to $v_j$. The CG method seeks an optimal polynomial $P_{k}$ that minimizes Eq.~(\ref{errorNorm_dis}). Furthermore, to analyze the convergence behavior, the upper bound for $\Vert e_{k} \Vert_{M}^{2}$ can be derived as follows:
\begin{eqnarray}
\label{errorNorm_maxminProblem}
 \Vert e_{k} \Vert_{M}^{2} \leq  \min_{P_{k}} \max_{\lambda \in \Lambda(M)}[P_{k}(\lambda)]^{2} \sum_{j=0}^{a-1} \alpha_{j}^2\lambda_{j} = \min_{P_{k}} \max_{\lambda \in \Lambda(M)}[P_{k}(\lambda)]^{2} \Vert e_{0} \Vert_{M}^{2}.
\end{eqnarray}
where $\Lambda(M)$ denotes the set of eigenvalues of $M$.

\begin{figure}[htbp] 
 \centering  
  \subfigure[]{
  \label{Chebyshev2}
  \includegraphics[scale=0.21]{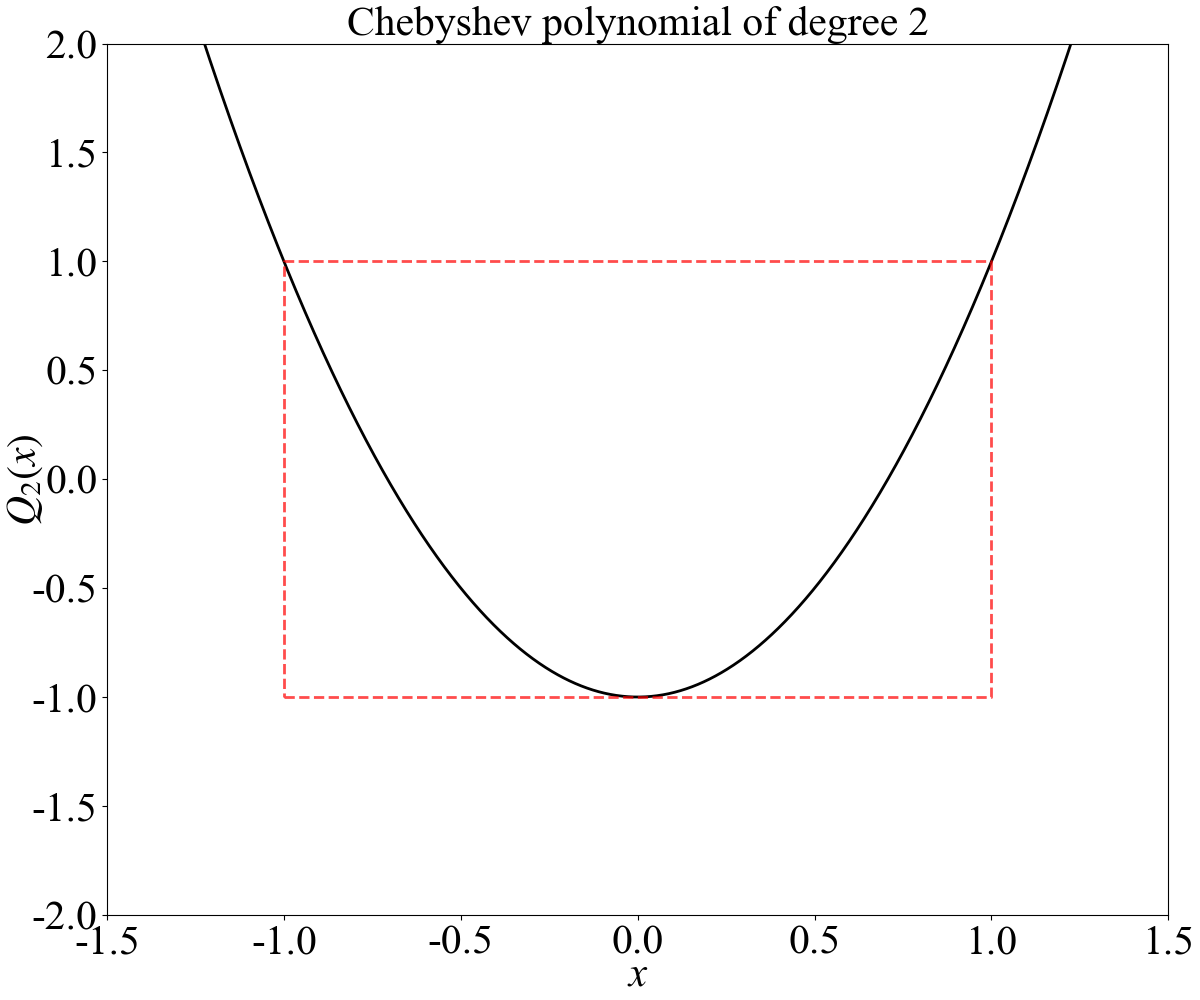}}
  \subfigure[]{
  \label{Chebyshev5}
  \includegraphics[scale=0.21]{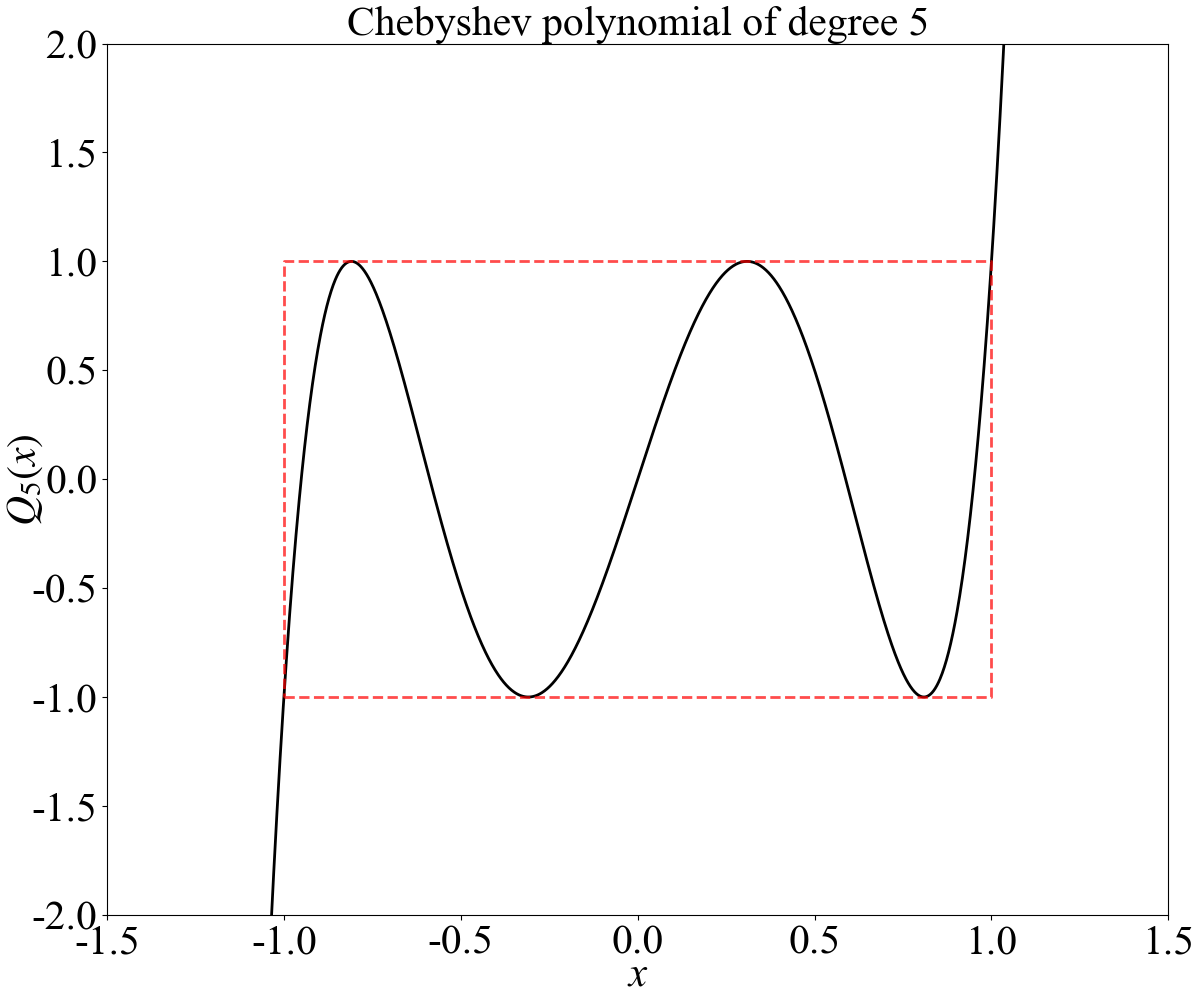}}
  \subfigure[]{
  \label{Chebyshev10}
  \includegraphics[scale=0.21]{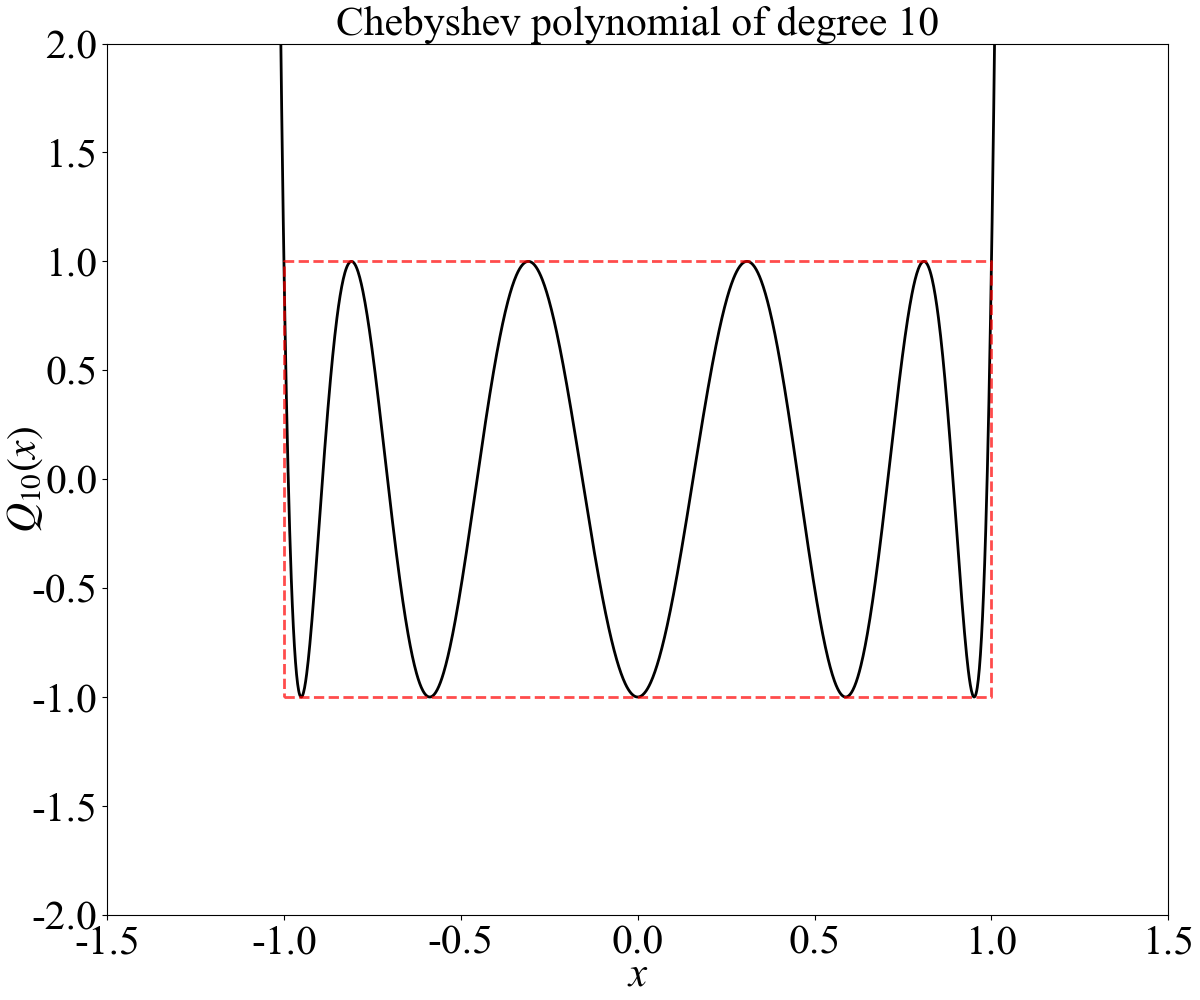}}
  \subfigure[]{
  \label{Chebyshev49}
  \includegraphics[scale=0.21]{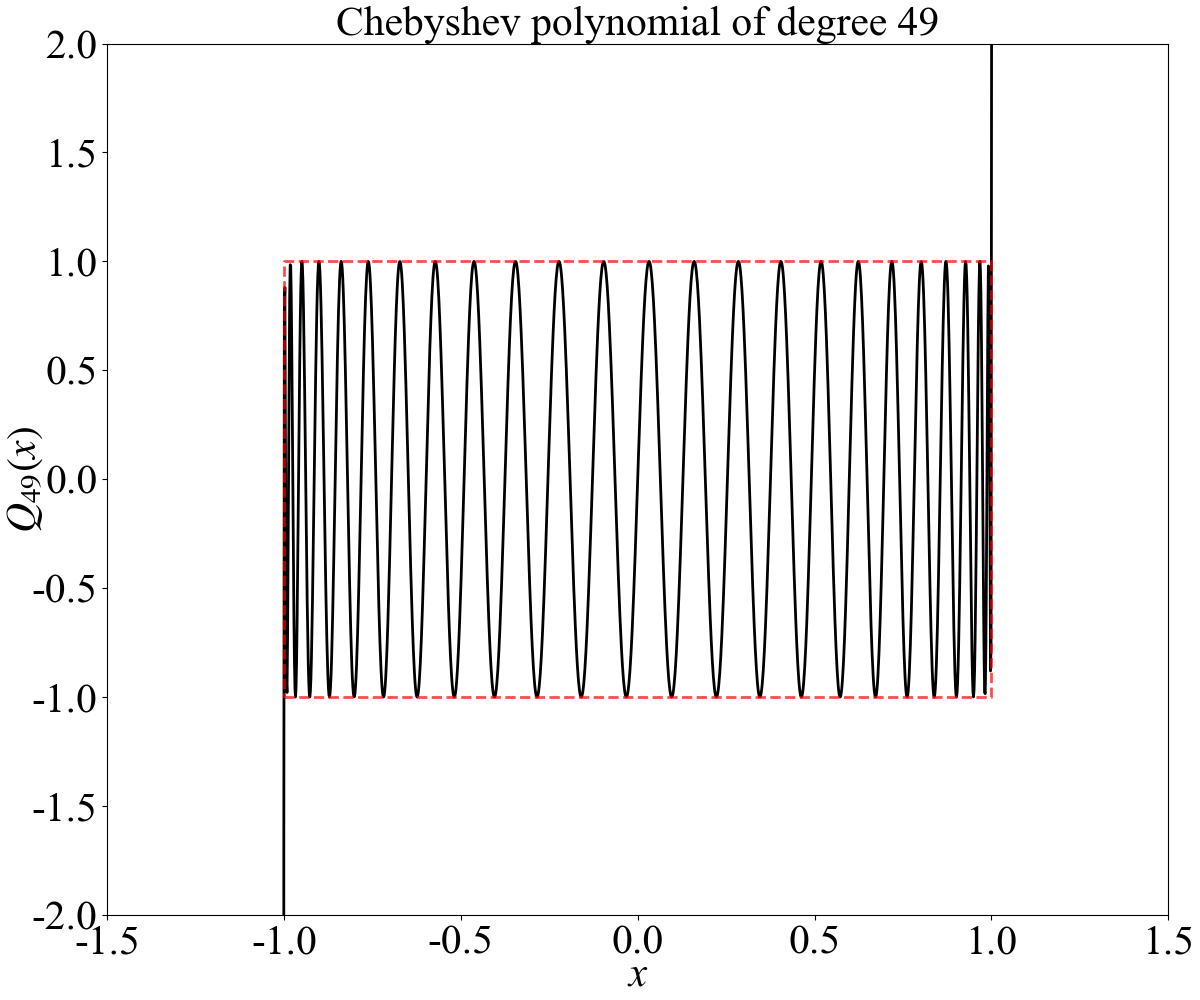}}
  \caption{Four representative Chebyshev polynomials of different degrees. (a) Degree 2. (b) Degree 5. (c) Degree 10. (d) Degree 49.}\label{Chebyshev}
\end{figure}

Remarkably, $\max_{\lambda \in \Lambda(M)}[P_{k}(\lambda)]^{2}$ is minimized when $P_{k}$ adopts the following form~\cite{shewchuk1994introductionCG, saad2003iterative}:
\begin{eqnarray}
\label{Pichoose}
 P_{k}(\lambda)=\frac{Q_{k}(\frac{\lambda_{max}+\lambda_{min}-2\lambda}{\lambda_{max}-\lambda_{min}})}{Q_{k}(\frac{\lambda_{max}+\lambda_{min}}{\lambda_{max}-\lambda_{min}})},
\end{eqnarray}
where $\lambda_{max}$ and $\lambda_{min}$ correspond to the largest and smallest eigenvalues of $M$, respectively. The polynomial $Q_{k}$ denotes the Chebyshev polynomial of degree $k$. Essentially, $P_{k}$ is constructed through a scaling transformation of $Q_{k}$. This enables analysis of the properties of $P_{k}$ through the well-established characteristics of $Q_{k}$. Four representative Chebyshev polynomials of different degrees are illustrated in Fig.~\ref{Chebyshev}. 

Analysis of Eq.~(\ref{Pichoose}) reveals that $P_{k}$ exhibits increasingly rapid oscillations with a higher zero point density as $\lambda$ approaches the spectral boundary regions near $\lambda_{max}$ and $\lambda_{min}$. Consequently, the eigenvalue distribution of $M$ exerts a fundamental influence on the convergence behavior of the CG method. 

\begin{figure}[htbp] 
 \centering  
  \subfigure[]{
  \label{Chebyshev2}
  \includegraphics[scale=0.24]{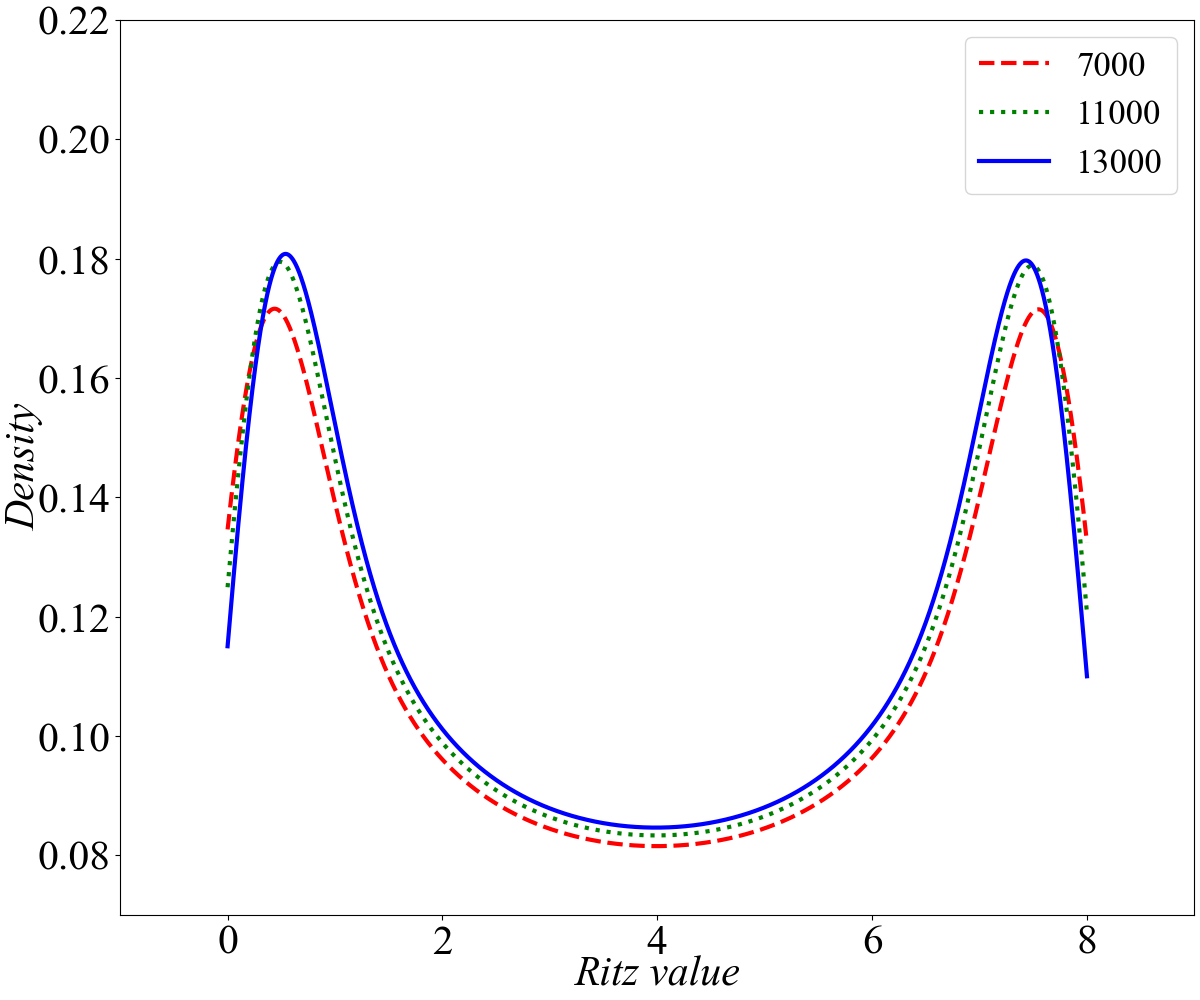}}
  \caption{Probability density functions of Ritz values at distinct Lanczos iteration counts.}\label{PDF_Ritz}
\end{figure}

The Ritz values derived from the Lanczos iteration approximate the eigenvalue distribution of $M$, with extreme eigenvalues converging rapidly during the early iterations~\cite{lanczos1950iteration}. Based on the uniform grid configuration~(detailed in Section~\ref{cavityRe3200} with grid resolution of $192\times192$), we calculate the Ritz values at three distinct Lanczos iteration numbers~($7,000$, $11,000$, and $13,000$), with the corresponding probability density function~(PDF) curves presented in Fig.~\ref{PDF_Ritz}. 
The results show that the eigenvalues of $M$ are predominantly concentrated in the spectral boundary regions.

Furthermore, we assume that the initial error $e_{0}$ is spectrally unbiased~(i.e. coefficients $\alpha_{j}$ in Eq.~(\ref{error_linear_combin_eigenvector}) are approximately equal across all indexes $j$). By analyzing:
\begin{itemize}
 \item the properties of $P_{k}$~(Eq.~(\ref{Pichoose})), 
 \item the PDF curves of Ritz values~(Fig.~\ref{PDF_Ritz}), and
 \item the discrete formulation of $\Vert e_{k} \Vert_{M}^{2}$~(Eq.~(\ref{errorNorm_dis})), 
\end{itemize}
we can conclude that the components of $\Vert e_{k} \Vert_{M}^{2}$ corresponding to the larger eigenvalues of $M$ decay more rapidly during the CG iterations, because larger eigenvalues enhance the damping effect of $P_{k}$. In contrast, the components of $\Vert e_{k} \Vert_{M}^{2}$ associated with smaller eigenvalues require more CG iterations to achieve significant reduction.

Following the formulation of Eq.~(\ref{errorNorm_dis}), the iterative residual vector at $k$th iteration can be expressed in the following discrete form:
\begin{eqnarray}
\label{rk_dis}
 r_{k} = Me_{k} = \sum_{j=0}^{a-1} \alpha_{j}P_{k}(\lambda_{j})\lambda_{j}v_{j}.
\end{eqnarray}
Being consistent with the established analysis, the components of $r_{k}$ associated with the larger eigenvalues of $M$ decay more rapidly during the CG iterations.

In light of the above analysis, the dataset construction method presented in Section~\ref{Dataset} provides a theoretically justified framework under the spectrally unbiased assumption of $e_0$. However, in practical applications to solve the PPE central to the fractional step method in incompressible flow simulations, the spectral properties of $e_0$ are intrinsically problem dependent and generally unknown a priori. This inherent uncertainty necessitates a parametric analysis of the key hyper-parameters in dataset construction.

\section{Sensitivity analysis of neural network parameters $A$, $B$, $C$ and $T$ on HyDEA's performance}
\label{appendixB}

This section analyzes the impact of neuron numbers $A$, $B$, and $C$ and channel number $T$ on HyDEA's performance. The analysis is conducted based on the flow configuration described in Section~\ref{cavityRe1000}, and more specifically:
\begin{itemize}
 \item 2D lid-driven cavity flow at $Re=1000$ using grid resolution of $128\times128$, with $b=0.6$, $m=3000$, $Num_{CG-type}=3$ and $Num_{DLSM}=2$.
\end{itemize}

To avoid over-extensive comparisons, the HyDEA~(CG+DLSM-1) and HyDEA~(ICPCG+DLSM-1) are employed in the forecasting stage. 

\subsubsection*{Sensitivity analysis of neuron numbers $A$, $B$, $C$}
We systematically analyze the impact of neuron numbers $A$, $B$, and $C$ on HyDEA’s performance, with the channel number $T=40$. The tested configurations summarized in Table~\ref{neuron FNN value}.

\begin{table}[htbp]
\renewcommand{\arraystretch}{1.5}
\normalsize
\centering
\caption{The values of neuron number in the FNN component of the DeepONet}
\begin{tabular}{cccccc}
\hline
   Neuron number  & Case I & Case II & Case III & Case IV & Case V \\
\hline
   $A$  & 50 & 75 & 100 & 150 & 250  \\

   $B$  & 100 & 150 & 200 & 300 & 500 \\

   $C$  & 50  & 75  & 100 & 150 & 250 \\
\hline 
\end{tabular}
\label{neuron FNN value}
\end{table}

\begin{figure}[htbp] 
 \centering  
  \subfigure[]{
  \label{128_diffSK_CG_time}
  \includegraphics[scale=0.21]{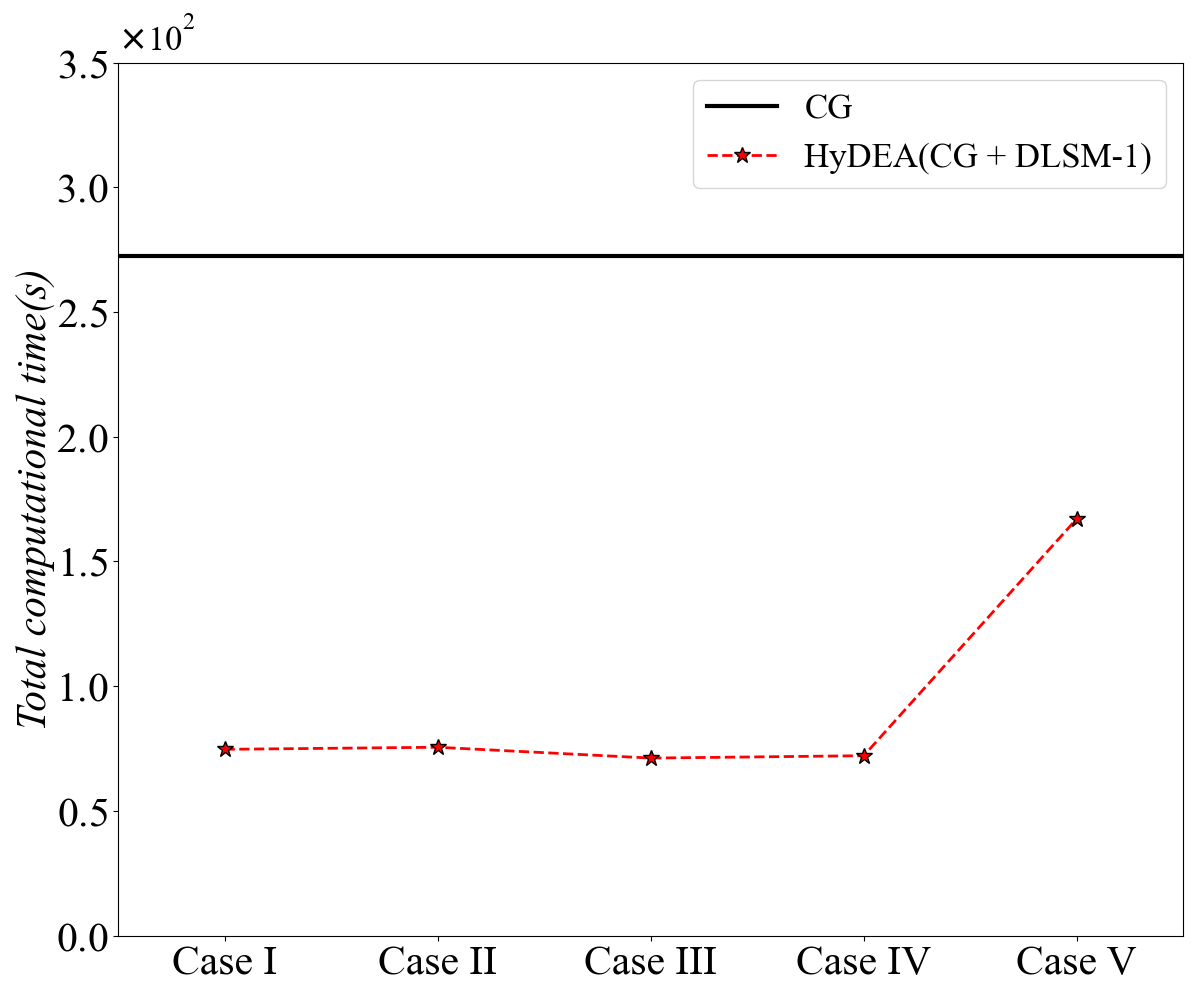}}
  \subfigure[]{
  \label{128_diffSK_IC_time}
  \includegraphics[scale=0.21]{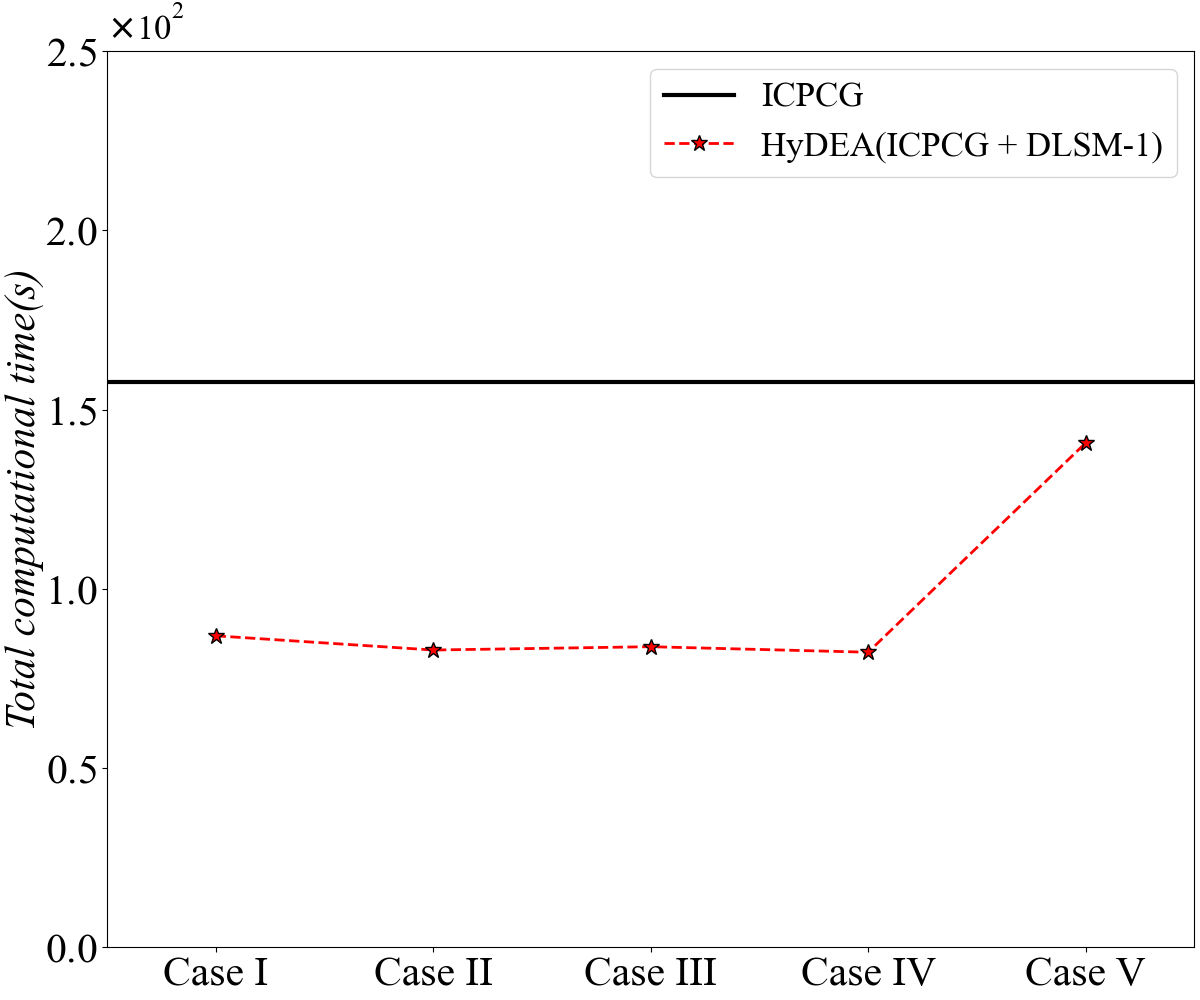}}
 \caption{Comparison of total computational time required to solve the PPE across Cases I to V. (a) HyDEA~(CG+DLSM-1). (b) HyDEA~(ICPCG+DLSM-1).}
 \label{Time compare 128 diffSK}
\end{figure}

\begin{figure}[htbp] 
 \centering  
  \subfigure[]{
  \label{128_Rline_diffSK_IC_10steps}
  \includegraphics[scale=0.21]{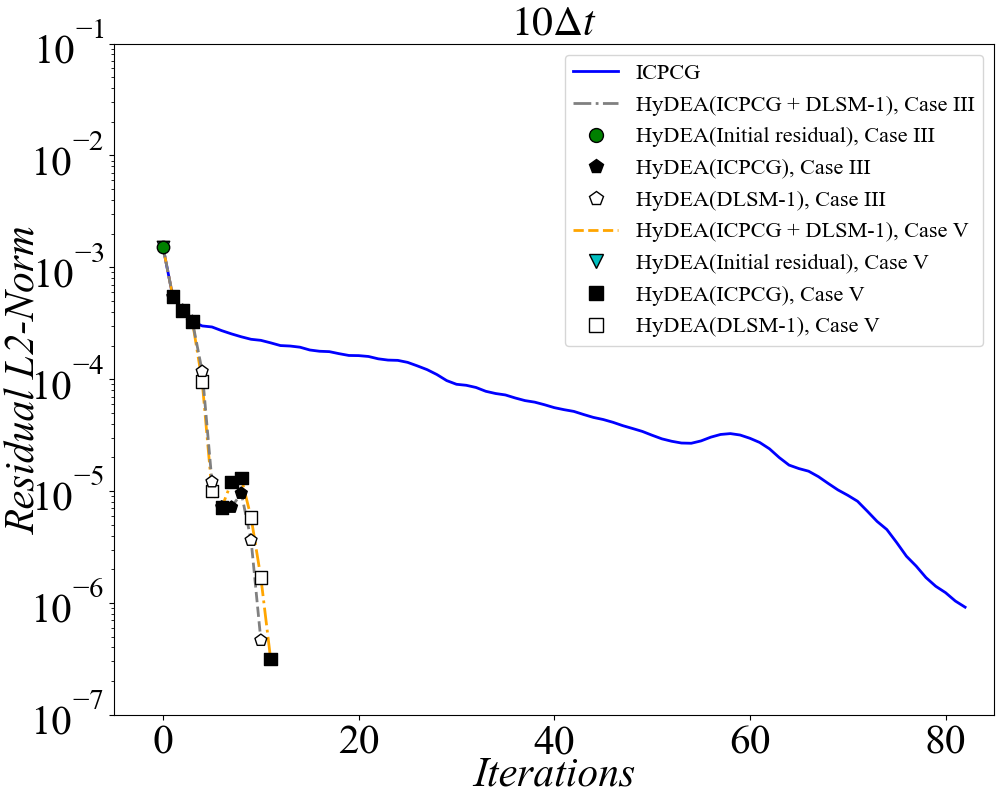}}
  \subfigure[]{
  \label{128_Rline_diffSK_IC_1000steps}
  \includegraphics[scale=0.21]{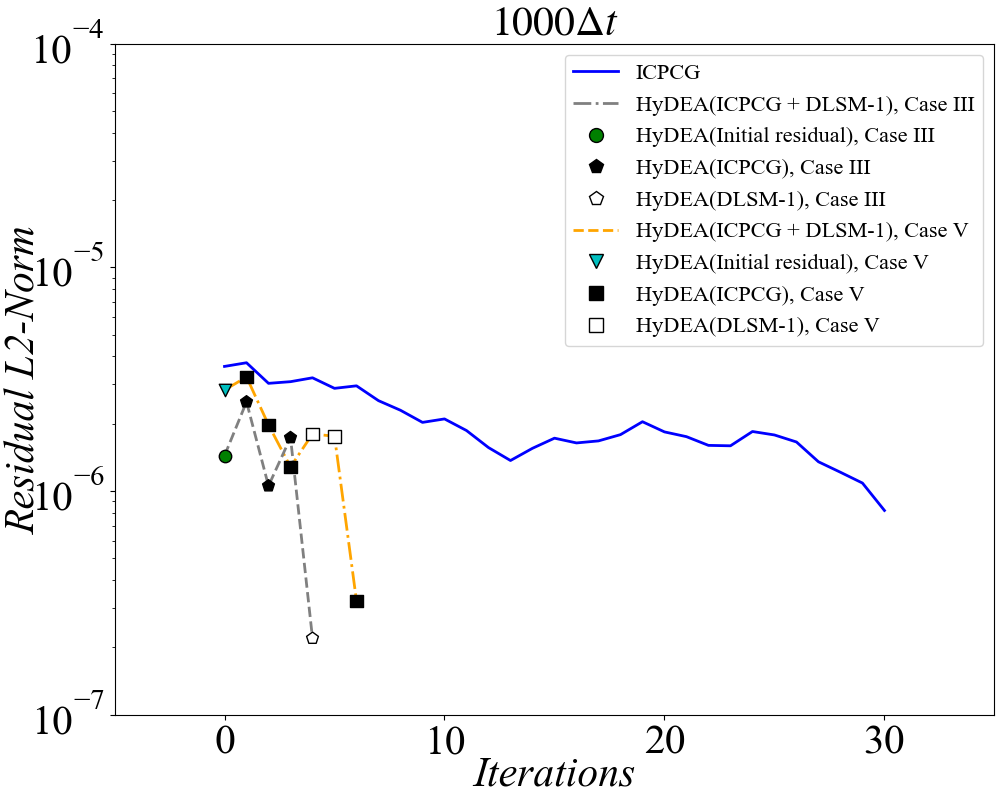}}
  \subfigure[]{
  \label{128_Rline_diffSK_IC_8000steps}
  \includegraphics[scale=0.21]{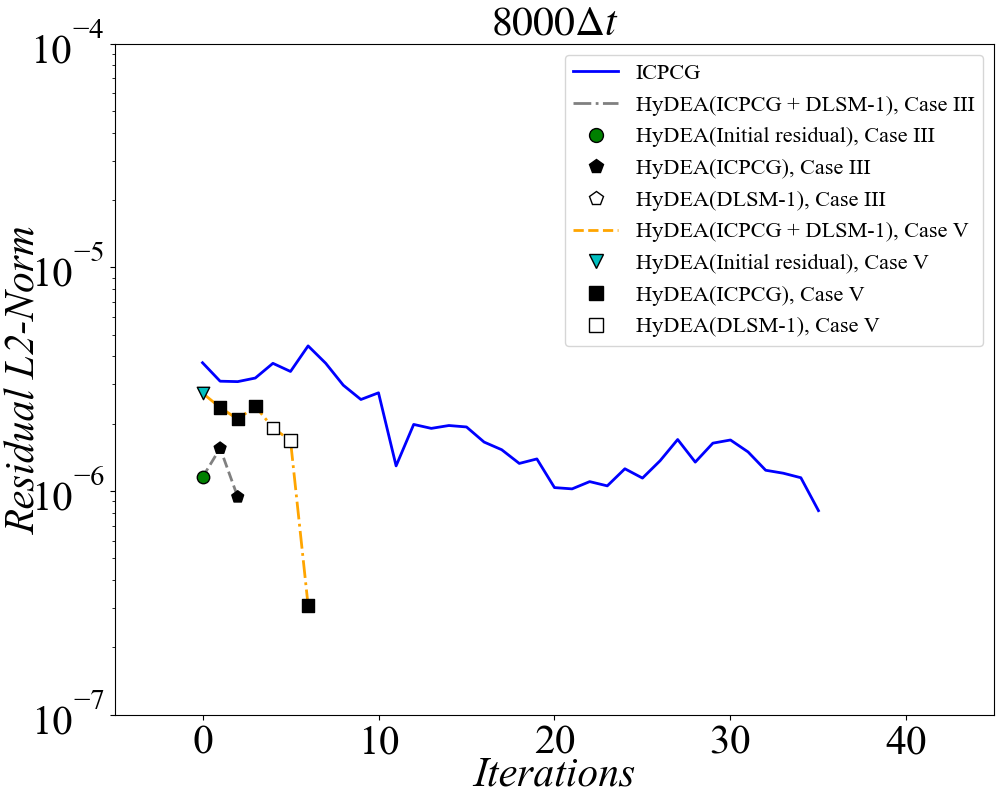}}
  \caption{Iterative residuals of solving the PPE for Cases III and V. (a) $10th$ time step. (b) $1000th$ time step. (c) $8000th$ time step.}\label{128_Rline_diffSK_3+2}
\end{figure}

Figs.~\ref{Time compare 128 diffSK} presents the total computational time over $10,000$ consecutive time steps required to solve the PPE for Cases I to V, and the iterative residuals of solving the PPE using HyDEA~(ICPCG+DLSM-1) for Cases III and V at three representative time steps are depicted in Fig.~\ref{128_Rline_diffSK_3+2}. 

The computational times for Cases I to IV are comparable, while Case V demonstrates notably inferior performance. This discrepancy likely stems from the increased number of network parameters, which hinders convergence to optimal network parameters within the limited number of training epochs. This limitation forces HyDEA to perform additional iterations to achieve the prescribed residual threshold, as demonstrated by the convergence curves in Fig.~\ref{128_Rline_diffSK_3+2}. Specifically, for case III, it takes $2$ rounds with $10$ iterations at $10\Delta t$, $1$ round with $4$ iterations at $1000\Delta t$, and $1$ round with $2$ iterations at $8000\Delta t$; for case V, it takes $3$ rounds with $11$ iterations at $10\Delta t$, $2$ rounds with $6$ iterations at $1000\Delta t$, and $2$ rounds with $6$ iterations at $8000\Delta t$.

Although additional training epochs may improve results for Case V, this would also induce higher computational cost. It is particularly noteworthy that HyDEA requires significantly fewer iterations than the ICPCG method to solve the PPE for both Case III and Case V, a result that warrants special emphasis.

\subsubsection*{Sensitivity analysis of channel number $T$}
\begin{figure}[htbp] 
 \centering  
  \subfigure[]{
  \label{128_diff_CG_timeCompare}
  \includegraphics[scale=0.21]{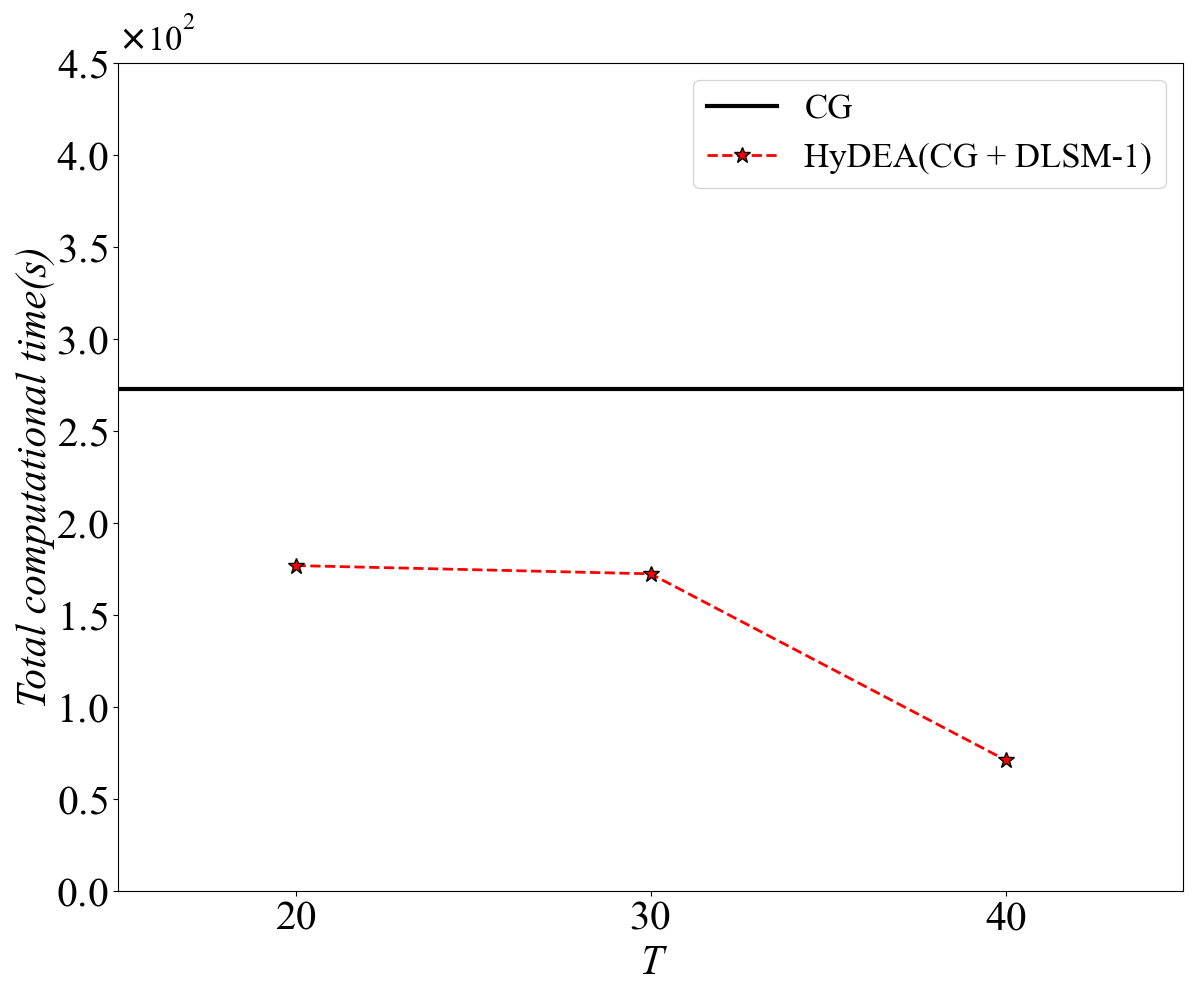}}
  \subfigure[]{
  \label{128_diffT_IC_timeCompare}
  \includegraphics[scale=0.21]{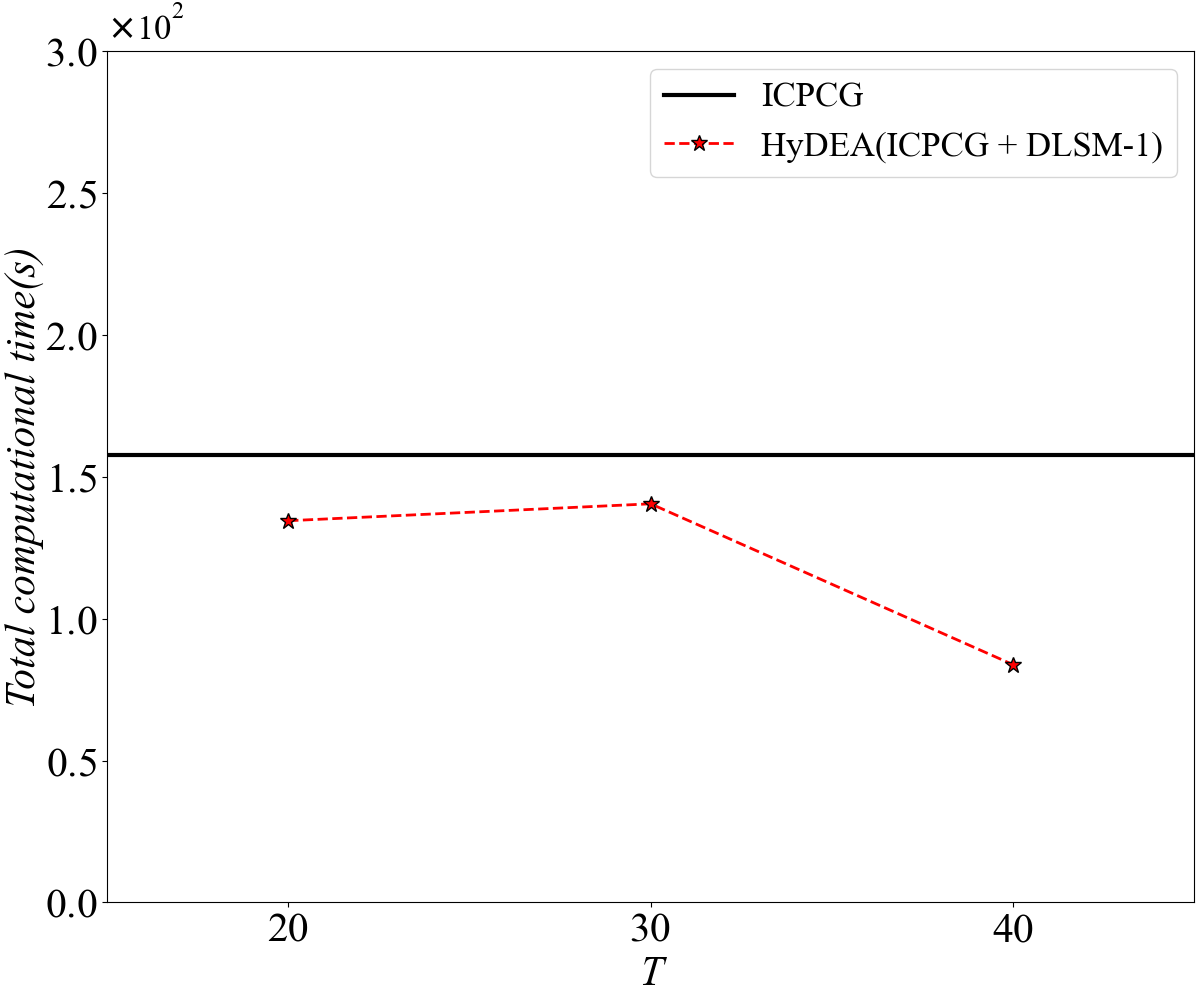}}
 \caption{Comparison of total computational time required to solve the PPE across varying channel number $T$. (a) HyDEA~(CG+DLSM-1). (b) HyDEA~(ICPCG+DLSM-1).}
 \label{Time compare 128 diffT}
\end{figure}

Subsequently, we analyze the impact of channel number $T$ by evaluating three configurations: $T=20$, $T=30$ and $T=40$. For all cases, we maintain identical neuron numbers fixed at $A=100$, $B=200$, and $C=100$. A comparison of the total computational time required to solve the PPE over $10,000$ consecutive time steps for these three cases is illustrated in Fig.~\ref{Time compare 128 diffT}.

\begin{figure}[htbp] 
 \centering  
  \subfigure[]{
  \label{128_Rline_diffT_IC_10steps}
  \includegraphics[scale=0.21]{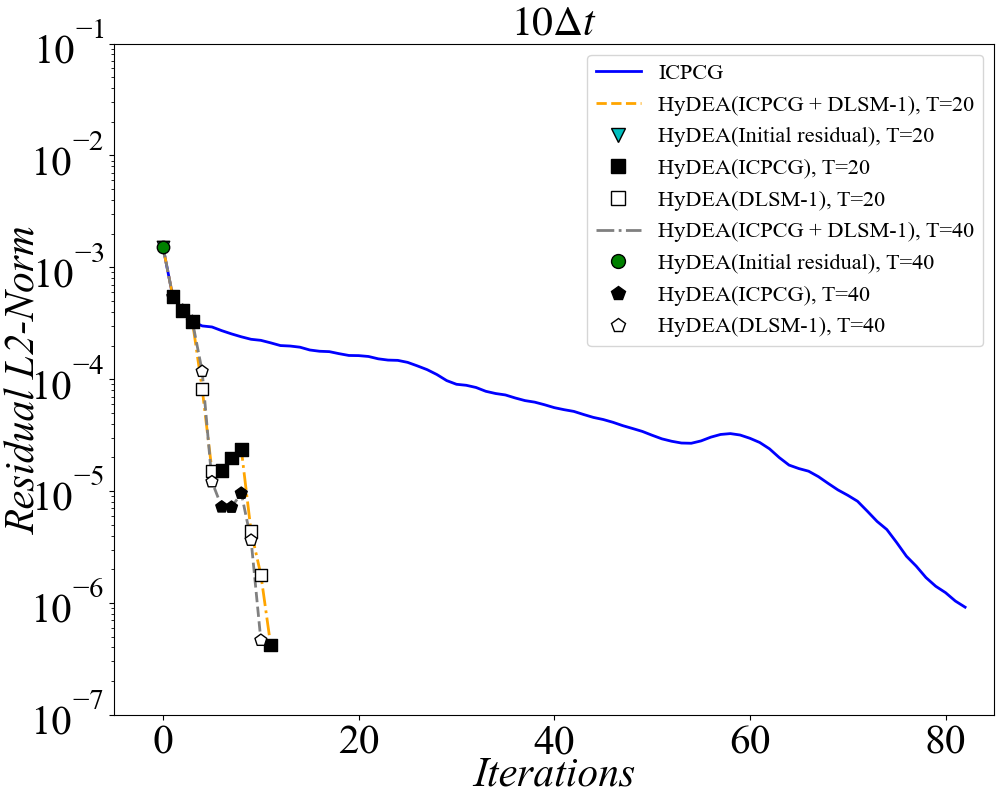}}
  \subfigure[]{
  \label{128_Rline_diffT_IC_1000steps}
  \includegraphics[scale=0.21]{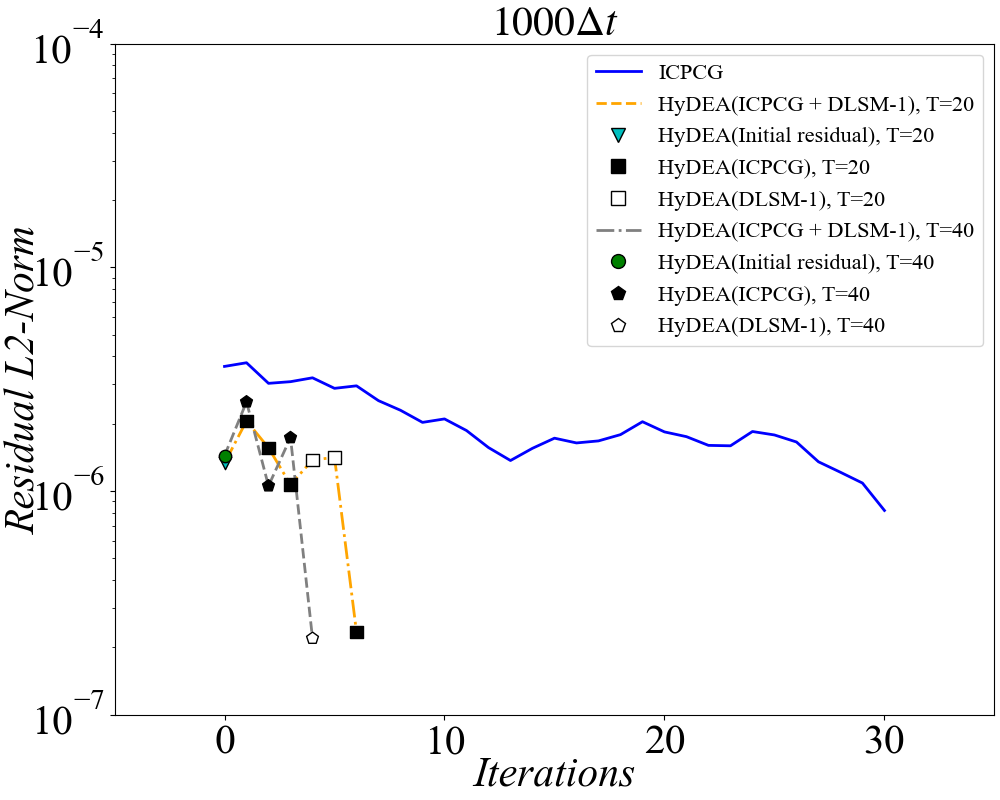}}
  \subfigure[]{
  \label{128_Rline_diffT_IC_8000steps}
  \includegraphics[scale=0.21]{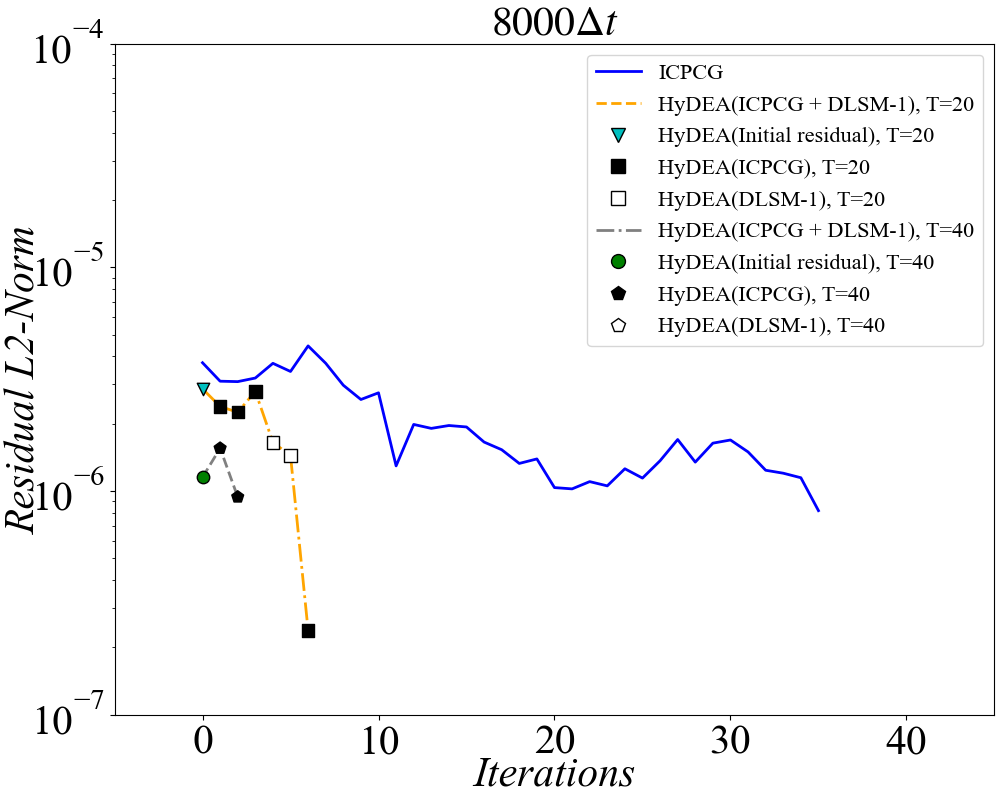}}
  \caption{Iterative residuals of solving the PPE for $T=20$ and $T=40$. (a) $10th$ time step. (b) $1000th$ time step. (c) $8000th$ time step.}\label{128_Rline_diffT_3+2}
\end{figure}

The results demonstrate that cases with $T=20$ and $30$ require higher computational time compared to $T=40$. The inferior performance of DeepONet at these lower channel numbers suggests that reduced feature space dimensionality compromises the model's ability to accurately learn the functional mapping from $r_{k}$ to $e_{k}$ spaces. This limitation forces HyDEA to perform additional iterations to achieve the prescribed residual threshold, as evidenced by the iterative residuals of $T=20$ and $40$ in Fig.~\ref{128_Rline_diffT_3+2}, ultimately increasing total computational time. Specifically, for $T=20$, it takes $3$ rounds with $11$ iterations at $10\Delta t$, $2$ rounds with $6$ iterations at $1000\Delta t$, and $2$ rounds with $6$ iterations at $8000\Delta t$; for $T=40$, it takes $2$ rounds with $10$ iterations at $10\Delta t$, $1$ round with $4$ iterations at $1000\Delta t$, and 1 round with 2 iterations at $8000\Delta t$.

\section{Sensitivity analysis of dataset construction parameters $m$ and $b$ on HyDEA's performance}
\label{appendixC}

This section conducts a systematic sensitivity analysis to quantify the influence of dataset construction parameters $m$ and $b$ on HyDEA's performance. The analysis is conducted based on the 2D lid-driven cavity flows described in Sections~\ref{cavityRe1000} and~\ref{cavityRe3200}, which employ two grid resolutions. More specifically:
\begin{itemize}
 \item $Re=1000$ using grid resolution of $128\times128$, with $Num_{CG-type}=3$ and $Num_{DLSM}=2$. 
 \item $Re=3200$ using grid resolution of $192\times192$, with $Num_{CG-type}=3$ and $Num_{DLSM}=2$. 
\end{itemize}

\subsubsection*{2D lid-driven cavity flow at $Re=1000$}
We first analyze the configuration of Section~\ref{cavityRe1000}. The specific values of $m$ and $b$ are summarized in Table~\ref{Data parameter 128}. The combination of the two parameters generates $15$ groups of the training dataset.

\begin{table}[htbp]
\renewcommand{\arraystretch}{1.5}
\normalsize
\centering
\caption{The values of dataset construction parameters $m$ and $b$ in the case of Section~\ref{cavityRe1000}}
\begin{tabular}{ccccccc}
\hline
   Parameter  &     Value      \\
\hline
   $m$  & 2000, 3000, 5000, 7000, 9000 \\

   $b$  & 0.4, 0.5, 0.6 \\
\hline
\end{tabular}
\label{Data parameter 128}
\end{table}

\begin{figure}[htbp] 
 \centering 
  \subfigure[]{
  \label{128_SK200_CG}
  \includegraphics[scale=0.22]{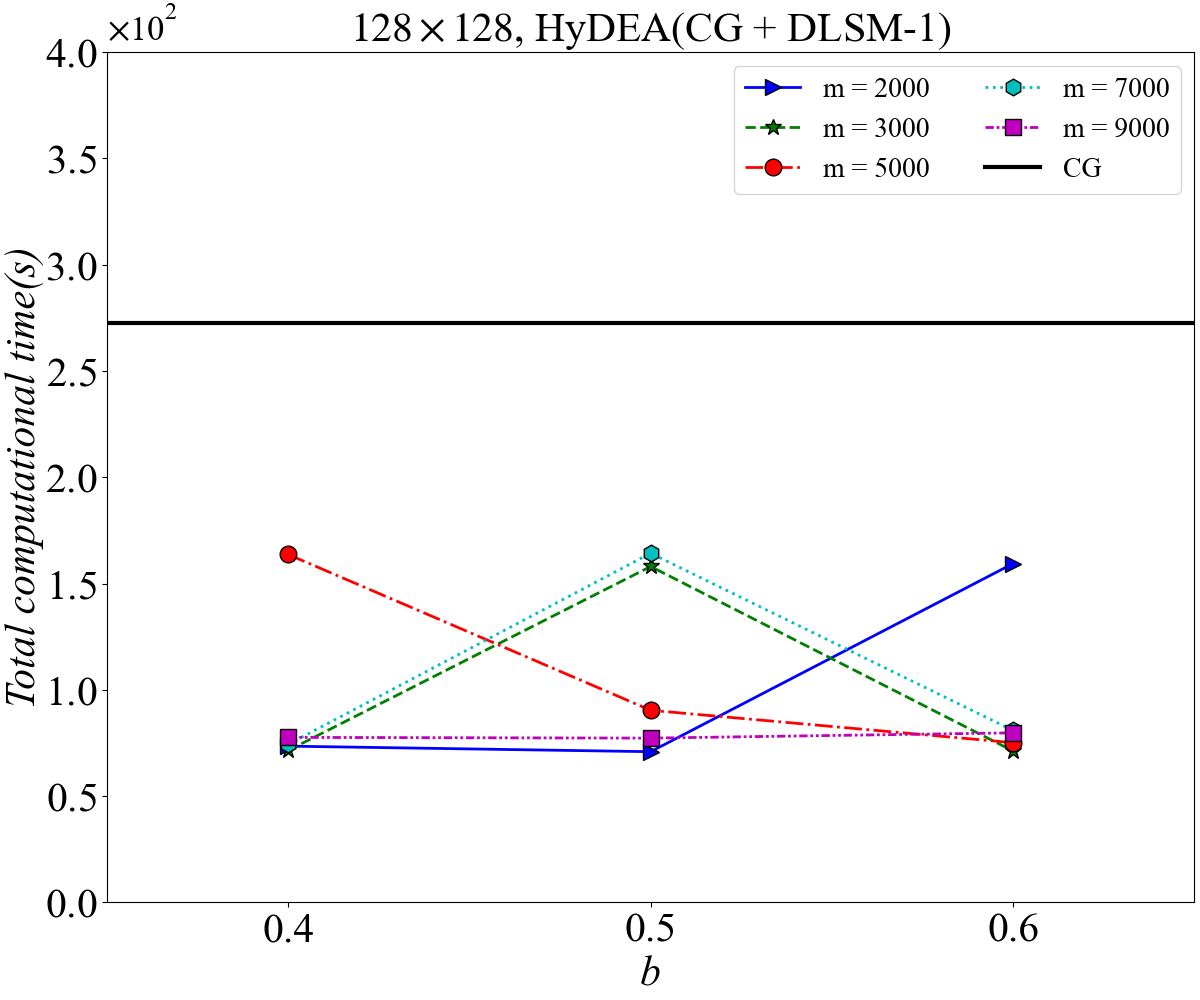}}
  \subfigure[]{
  \label{128_SK200_IC}
  \includegraphics[scale=0.22]{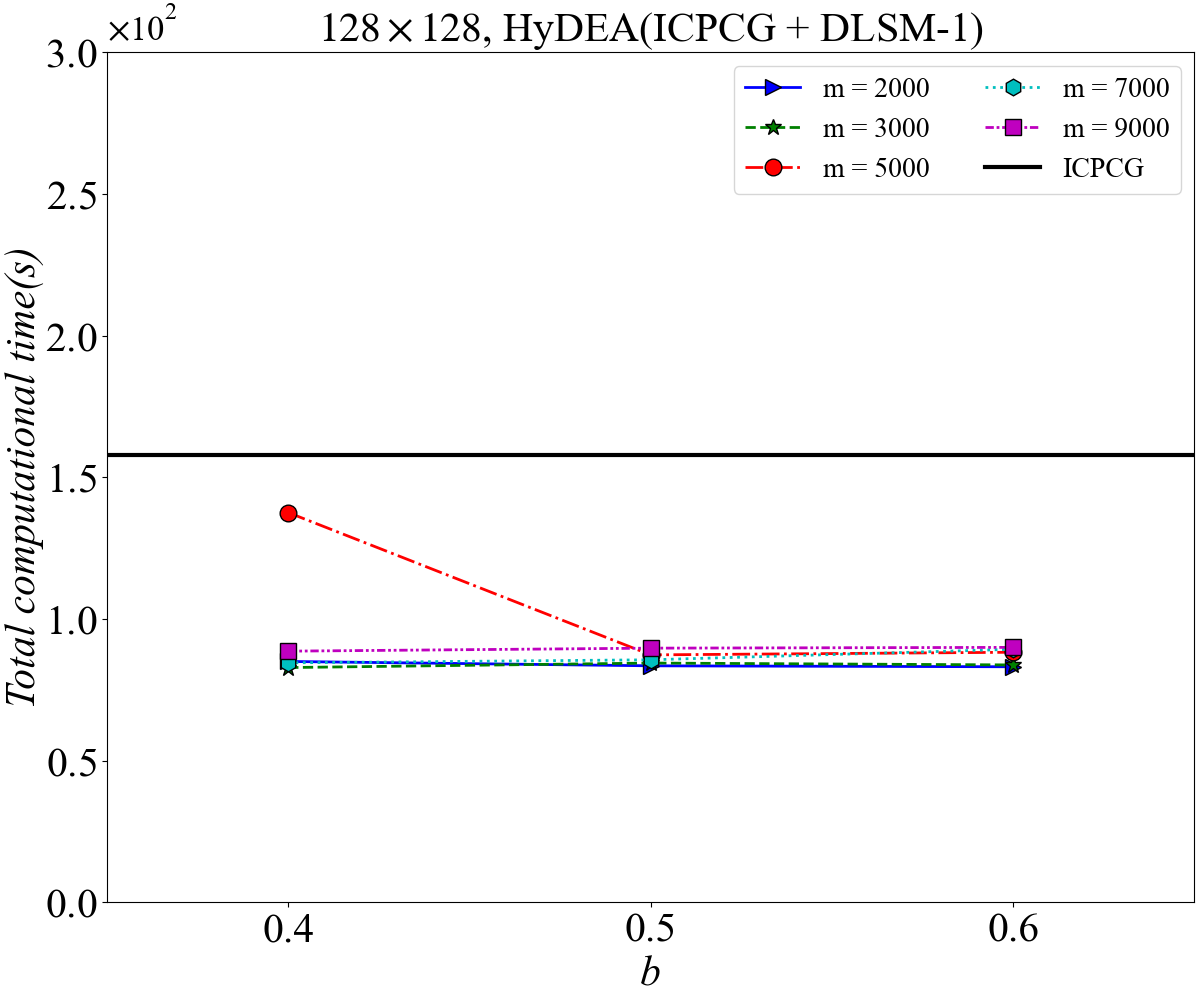}}
  \subfigure[]{
  \label{128_SK200_J}
  \includegraphics[scale=0.22]{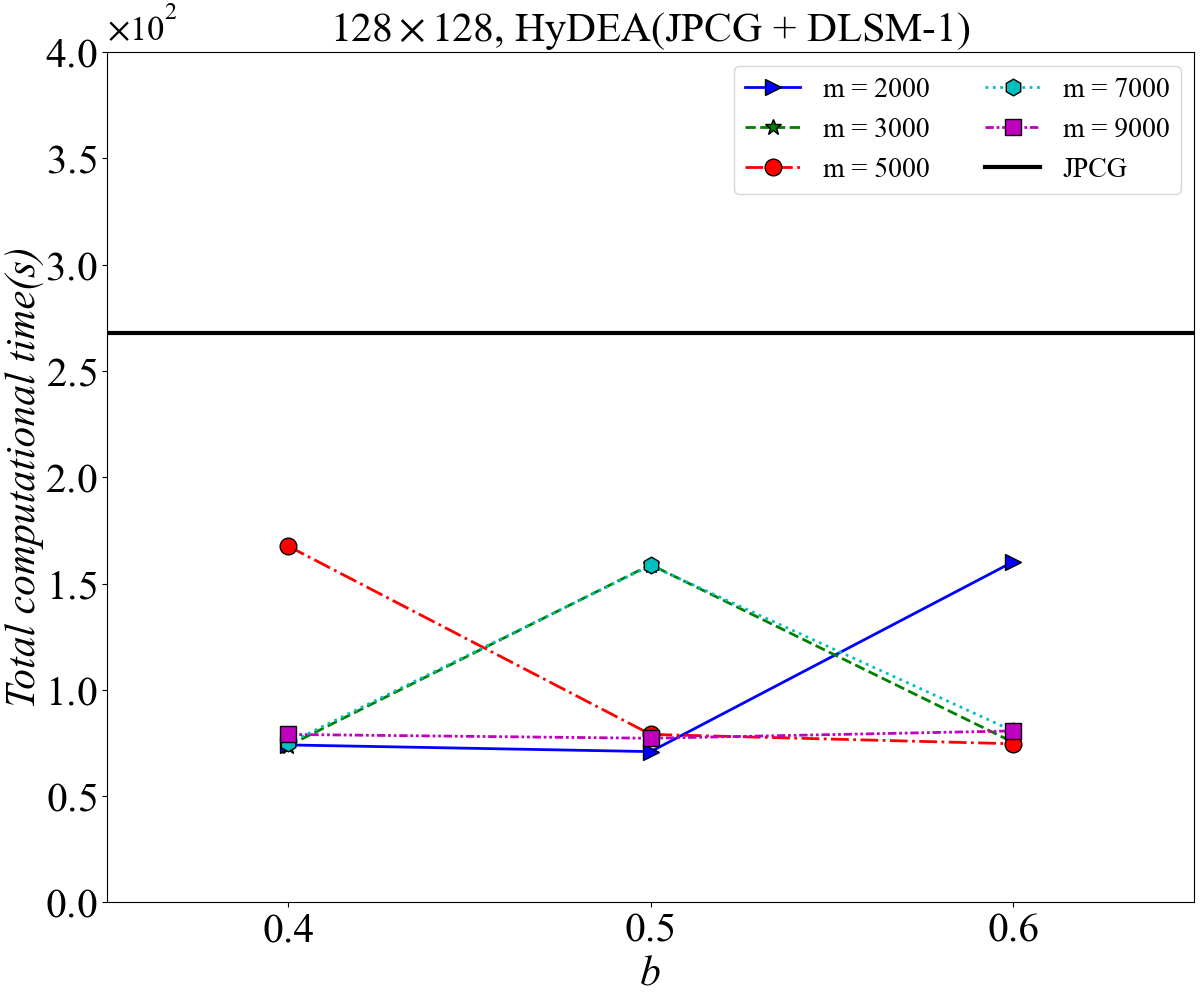}}
  \subfigure[]{
  \label{128_SK200_MG}
  \includegraphics[scale=0.22]{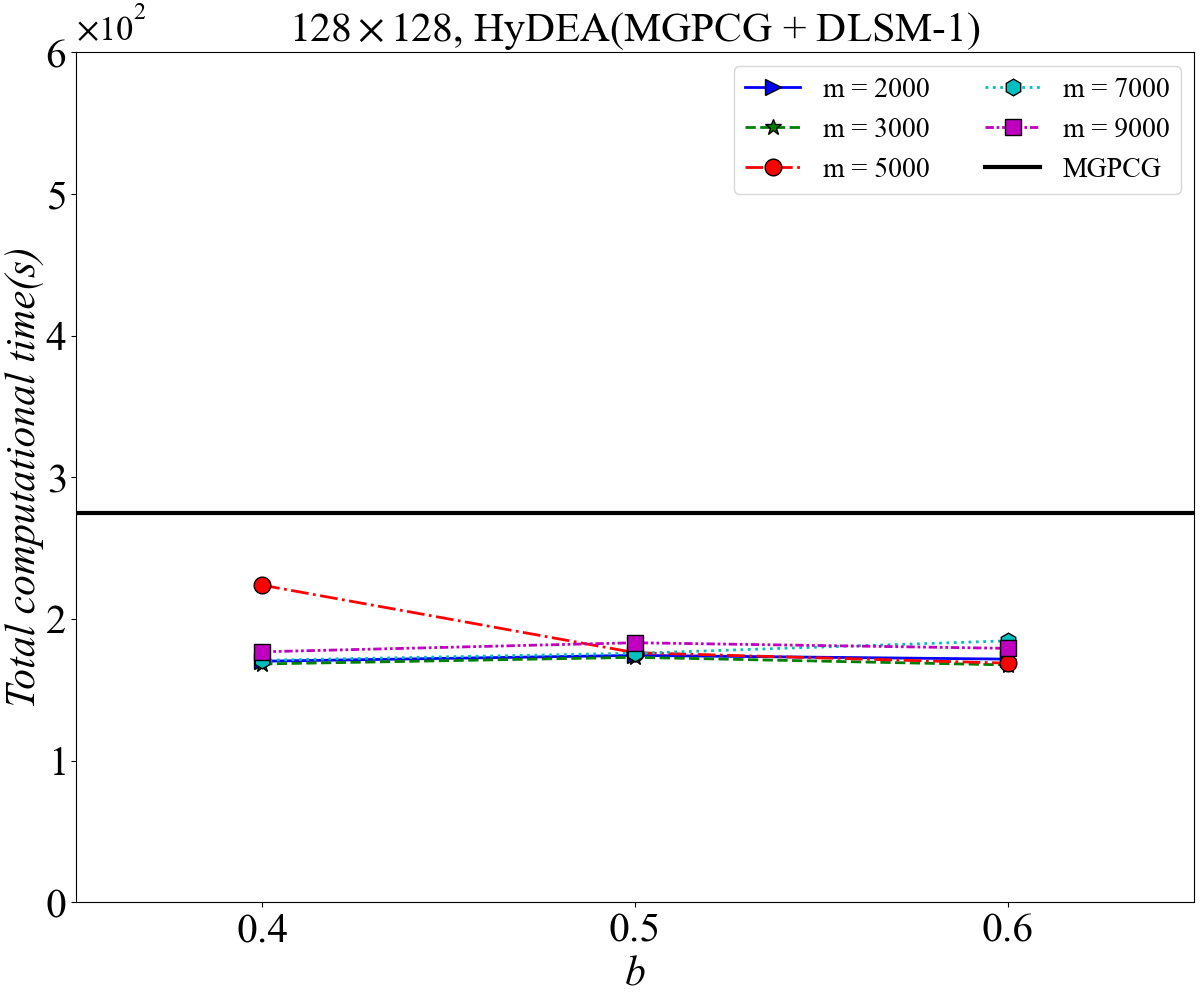}}
 \caption{Comparison of total computational time required to solve the PPE for 2D lid-driven cavity at $Re=1000$: influence of different values of $m$ and $b$. (a) HyDEA~(CG+DLSM-1). (b) HyDEA~(ICPCG+DLSM-1). (c) HyDEA~(JPCG+DLSM-1). (d) HyDEA~(MGPCG+DLSM-1).}
 \label{Time compare 128 SK200}
\end{figure}

Fig.~\ref{Time compare 128 SK200} presents a comprehensive comparison of total computational time between HyDEA~(CG-type+DLSM-1) and CG-type methods required to solve the PPE over $10,000$ consecutive time steps under varying dataset construction parameters.

With appropriate parameter selection, HyDEA exhibits lower computational time compared to CG-type methods. Notably, when $b=0.6$, the results of HyDEA demonstrate enhanced performance in most values of $m$.

\subsubsection*{2D lid-driven cavity flow at $Re=3200$}
\begin{figure}[htbp] 
 \centering 
  \subfigure[]{
  \label{192_SK200_CG}
  \includegraphics[scale=0.21]{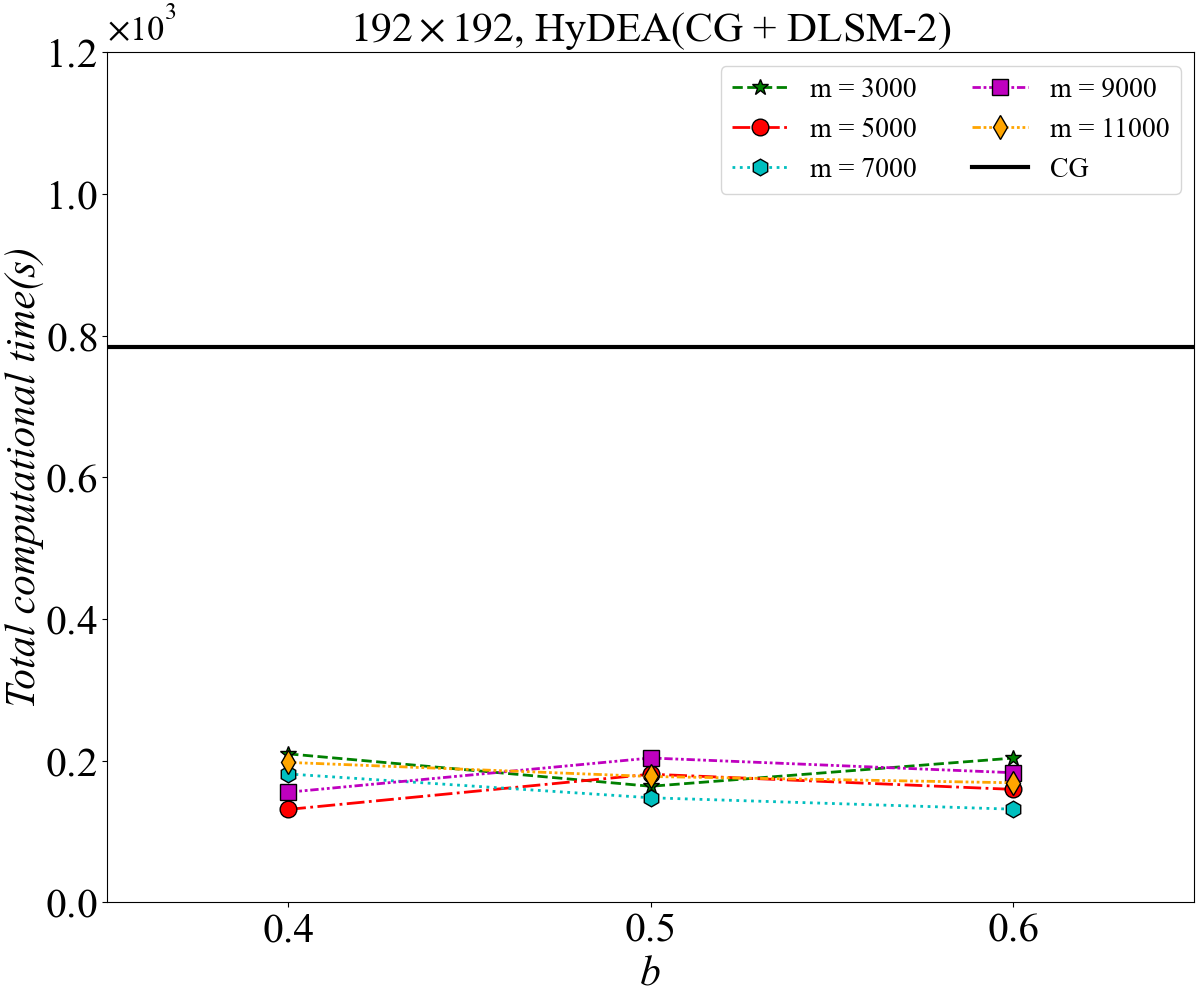}}
  \subfigure[]{
  \label{192_SK200_IC}
  \includegraphics[scale=0.21]{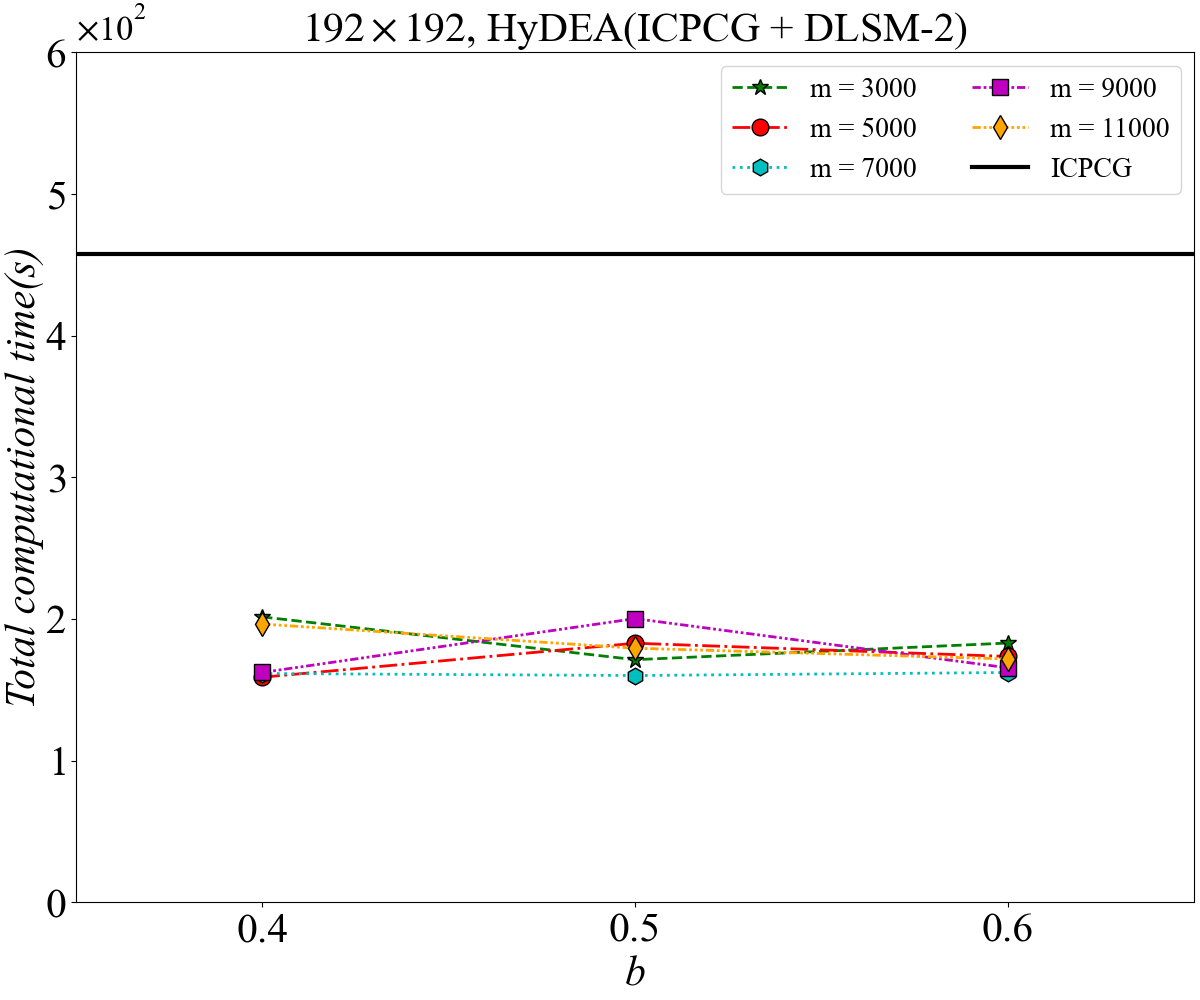}}
  \subfigure[]{
  \label{192_SK200_J}
  \includegraphics[scale=0.21]{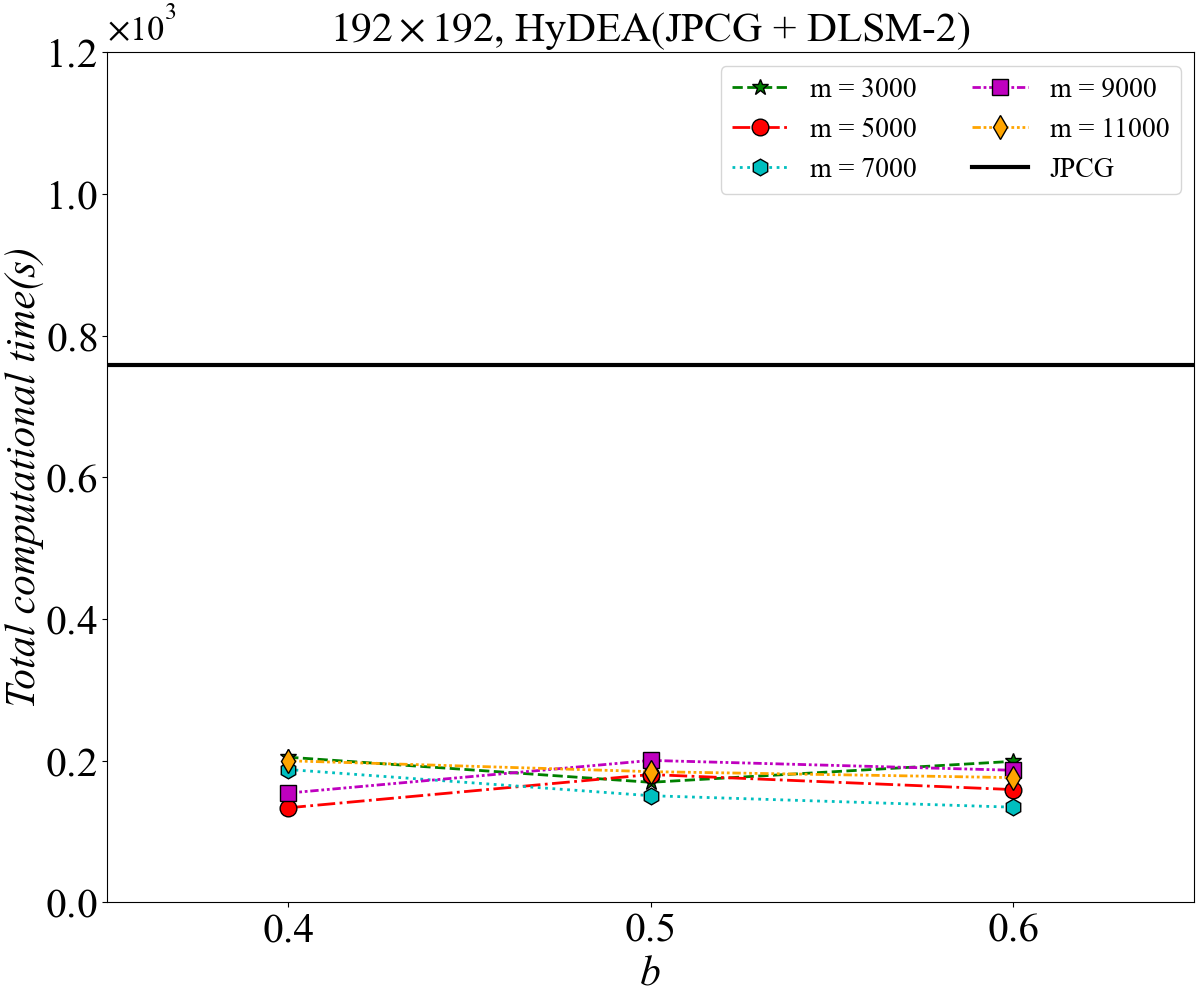}}
  \subfigure[]{
  \label{192_SK200_MG}
  \includegraphics[scale=0.21]{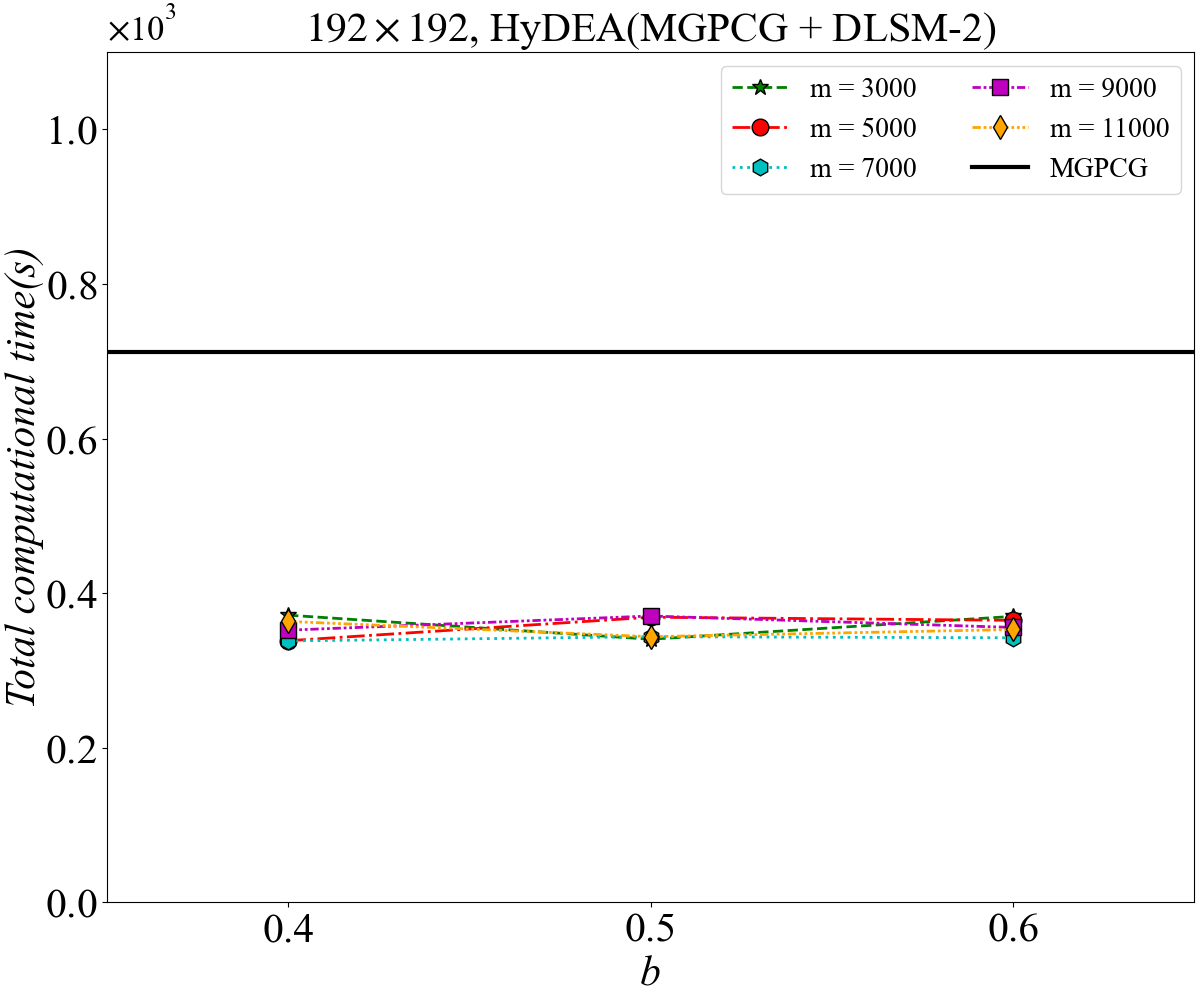}}
 \caption{Comparison of total computational time required to solve the PPE for 2D lid-driven cavity at $Re=3200$: influence of different values of $b$ and $m$. (a) HyDEA~(CG+DLSM-2). (b) HyDEA~(ICPCG+DLSM-2). (c) HyDEA~(JPCG+DLSM-2). (d) HyDEA~(MGPCG+DLSM-2).}
 \label{Time compare 192 SK200}
\end{figure}

Furthermore, for the configuration in Section~\ref{cavityRe3200}, the specific values of $m$ and $b$ are summarized in Table~\ref{Data parameter 192}, and a comparison of the total computational time over $10,000$ consecutive time steps required to solve the PPE is illustrated in Fig.~\ref{Time compare 192 SK200}.  

\begin{table}[htbp]
\renewcommand{\arraystretch}{1.5}
\normalsize
\centering
\caption{The values of dataset construction parameters $m$ and $b$ in the case of Section~\ref{cavityRe3200}}
\begin{tabular}{ccccccc}
\hline
   Parameter  &  Value \\
\hline
   $m$  &  3000, 5000, 7000, 9000, 11000 \\

   $b$  & 0.4, 0.5, 0.6 \\
\hline
\end{tabular}
\label{Data parameter 192}
\end{table}

The results demonstrate that HyDEA consistently outperforms CG-type methods in computational efficiency across all tested parameter combinations, and shows particularly more stable time costs compared to low-resolution scenarios. This robust performance advantage underscores HyDEA's enhanced adaptability to high-resolution simulations, as evidenced by its greater tolerance to variations in dataset construction parameters.

\section{Sensitivity analysis of $Num_{CG-type}$ and $Num_{DLSM}$ on HyDEA's performance}
\label{appendixD}

This section presents a systematic sensitivity analysis to quantify the influence of key computational parameters $Num_{CG-type}$ and $Num_{DLSM}$ on HyDEA's performance. The analysis is conducted based on two flow configurations described in Sections~\ref{cavityRe1000} and~\ref{OsCylinder}, and more specifically:
\begin{itemize}
 \item 2D lid-driven cavity flow at $Re=1000$ with grid resolution of $128\times128$, which evolves toward a steady state. The dataset is constructed using $m=3000$ and $b=0.6$.
 \item 2D flow around an inline oscillating cylinder at $Re=100$ with grid resolution of $192\times192$, which maintains inherently unsteady characteristics. The dataset is constructed using $m=7000$ and $b=0.6$.
\end{itemize}

\subsubsection*{2D lid-driven cavity flow at $Re=1000$}
We first analyze the configuration of Section~\ref{cavityRe1000}, holding $Num_{DLSM}$ constant at $2$ while varying $Num_{CG-type}$ from $1$ to $5$ to investigate its impact. 

\begin{figure}[htbp] 
 \centering 
  \subfigure[]{
  \label{128_Madd2_SK200_CG_time}
  \includegraphics[scale=0.20]{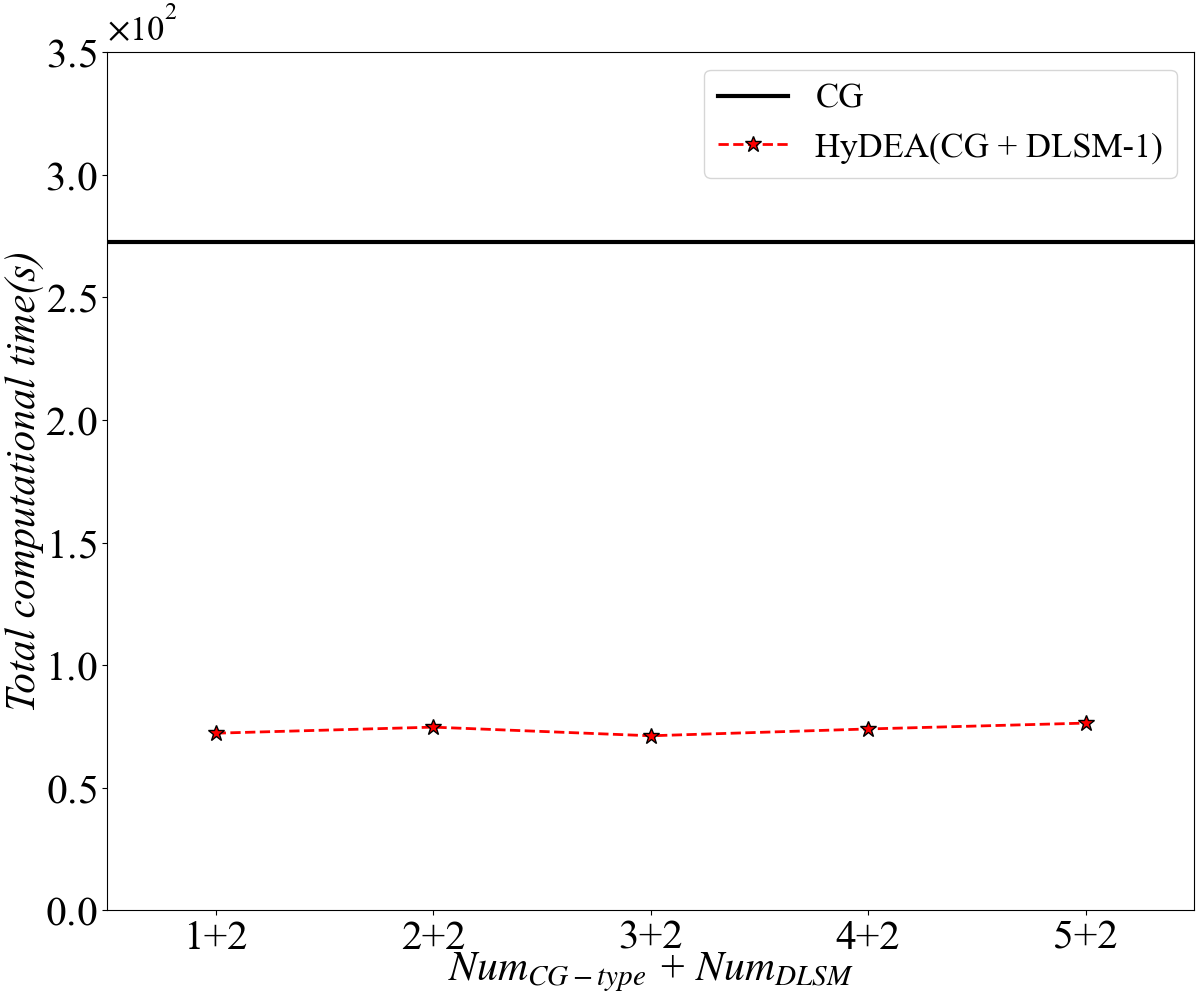}}
  \subfigure[]{
  \label{128_Madd2_SK200_IC_time}
  \includegraphics[scale=0.20]{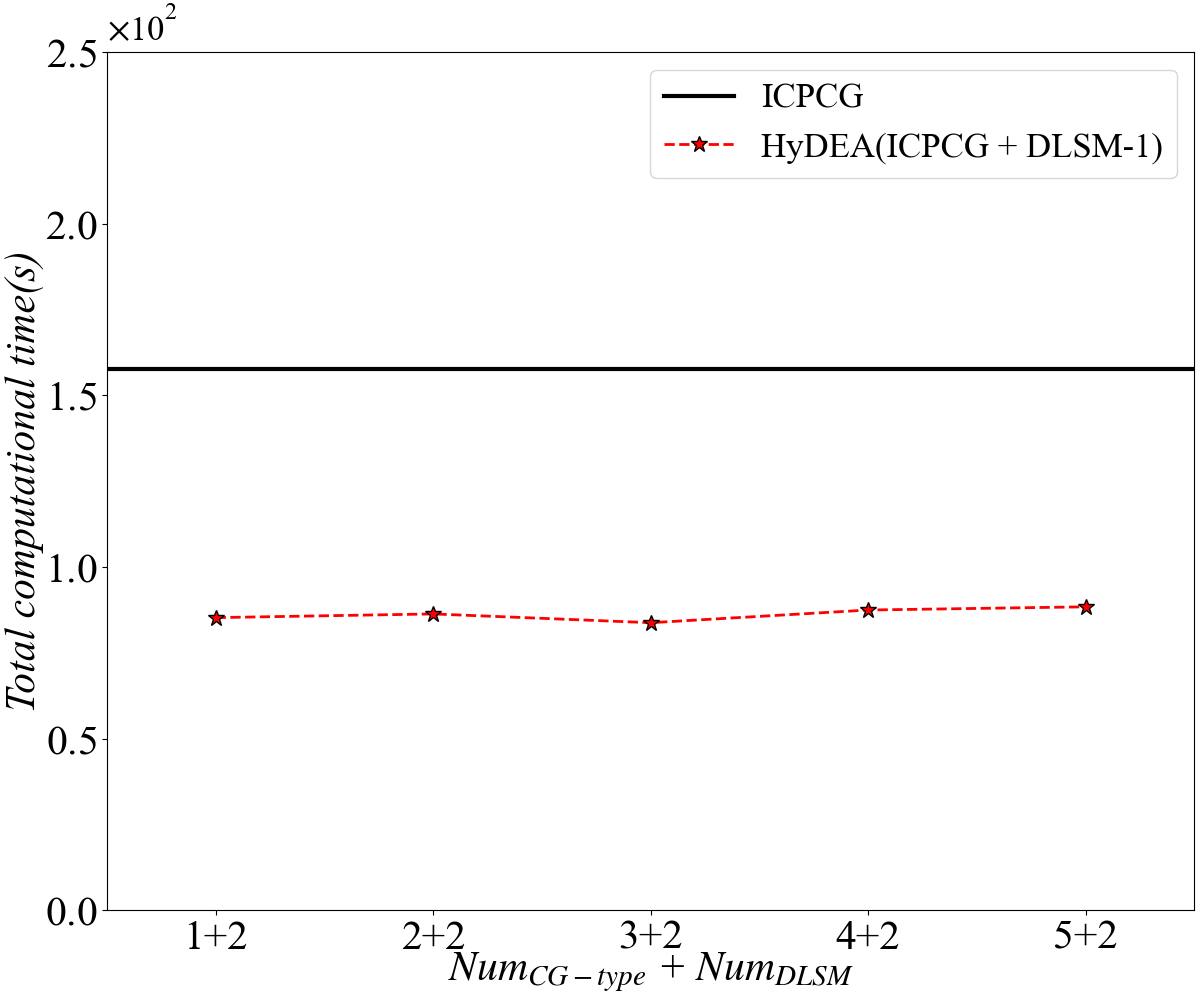}}
 \caption{Comparison of total computational time required to solve PPE across varying values of $Num_{CG-type}$ for 2D lid-driven cavity flow at $Re=1000$. (a) HyDEA~(CG+DLSM-1). (b) HyDEA~(ICPCG+DLSM-1).}
 \label{Time compare 128 Madd2}
\end{figure}

Taking HyDEA~(CG+DLSM-1) and HyDEA~(ICPCG+DLSM-1) as representative cases, Fig.~\ref{Time compare 128 Madd2} presents the total computational time required to the PPE over $10,000$ consecutive time steps. The results demonstrate that the computational time exhibits a consistent trend across varying values of $Num_{CG-type}$ in this case.

\begin{figure}[htbp] 
 \centering  
  \subfigure[]{
  \label{128_3addM_SK200_CG_time}
  \includegraphics[scale=0.20]{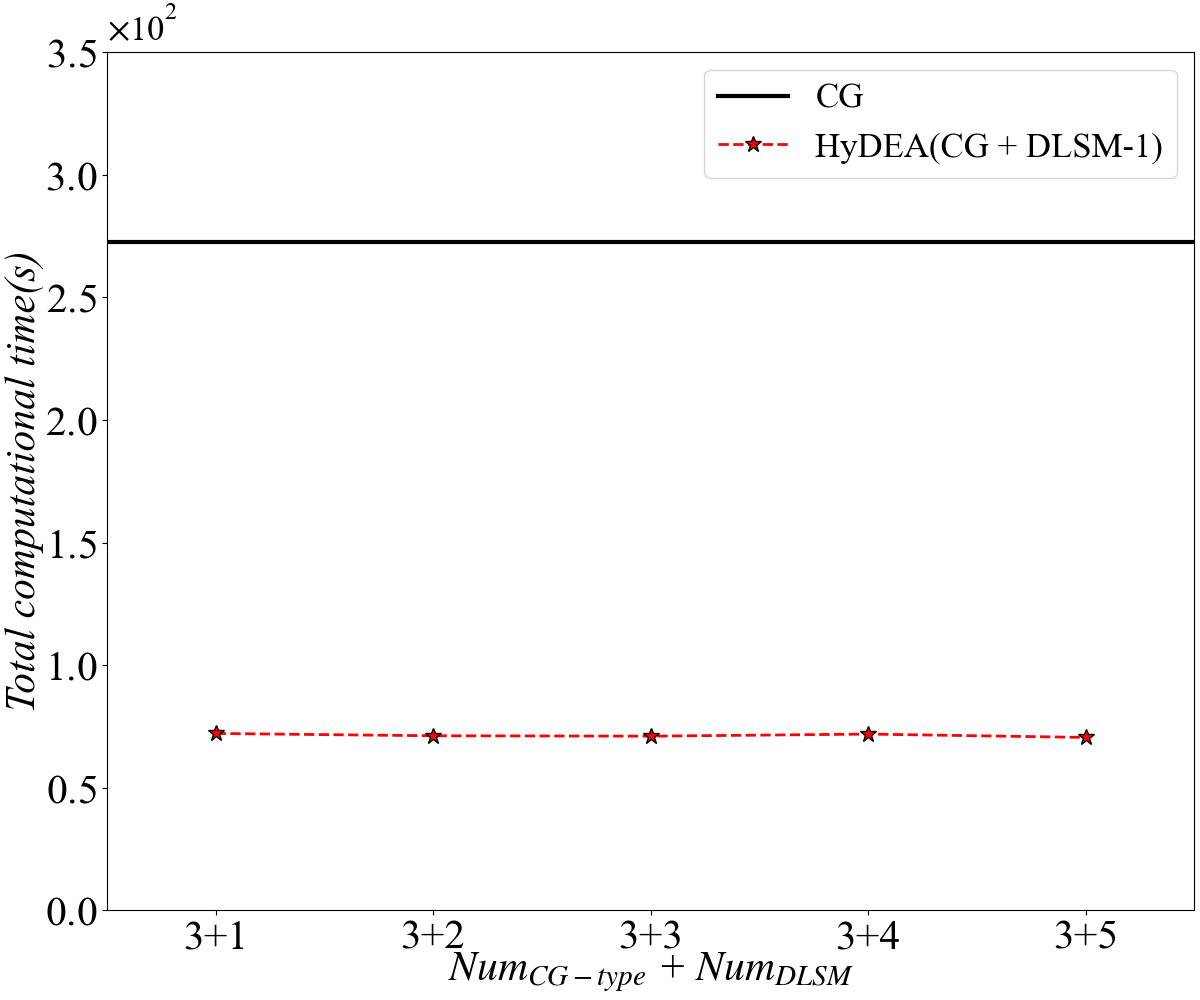}}
  \subfigure[]{
  \label{128_3addM_SK200_IC_time}
  \includegraphics[scale=0.20]{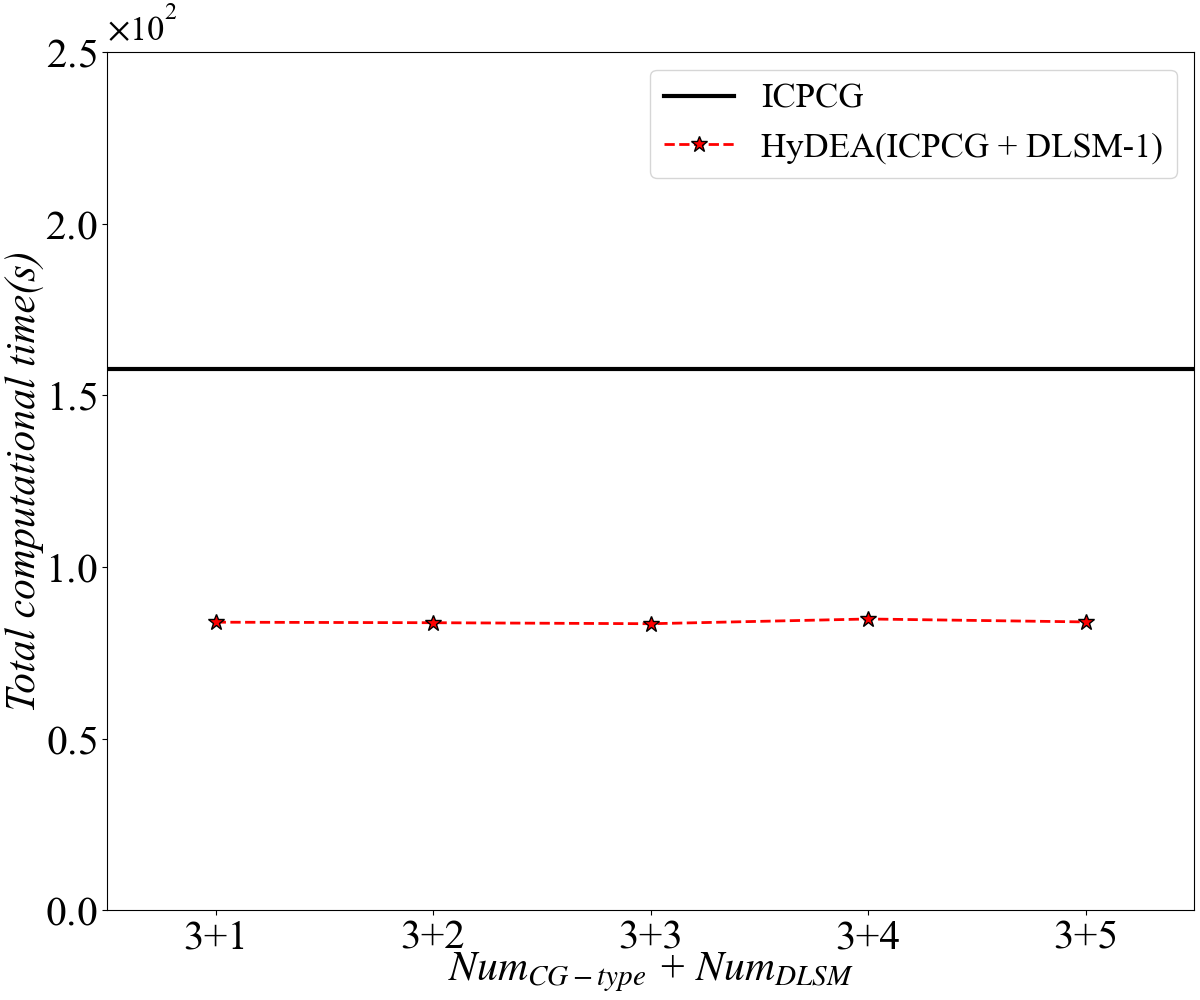}}
 \caption{Comparison of computational time for different values of $Num_{DLSM}$ for 2D lid-driven cavity flow at $Re=1000$. (a) HyDEA~(CG+DLSM-1). (b) HyDEA~(ICPCG+DLSM-1).}
 \label{Time compare 128 3addM}
\end{figure}

Furthermore, $Num_{CG-type}$ is held constant at $3$ while varying $Num_{DLSM}$ from $1$ to $5$ to investigate its impact. The total computational time required to solve the PPE over $10,000$ consecutive time steps is presented in Fig.~\ref{Time compare 128 3addM}. The results demonstrate that varying $Num_{DLSM}$ has a negligible impact on the total computational time in this case. This insensitivity occurs because the flow field approaches the steady state  over time, the DLSM module typically meets the residual threshold before reaching the specified maximum $Num_{DLSM}$ iteration limit for the PPE solution.

\subsubsection*{2D flow around an inline oscillating cylinder at $Re=100$ and $KC=5$}
Subsequently, we perform a parametric analysis to evaluate the influence of $Num_{CG-type}$ and $Num_{DLSM}$ in the case presented in Section~\ref{OsCylinder}. 

\begin{figure}[htbp]
\centering
  \subfigure[]{
  \label{192Os_Madd2_timecompare}
  \includegraphics[scale=0.21]{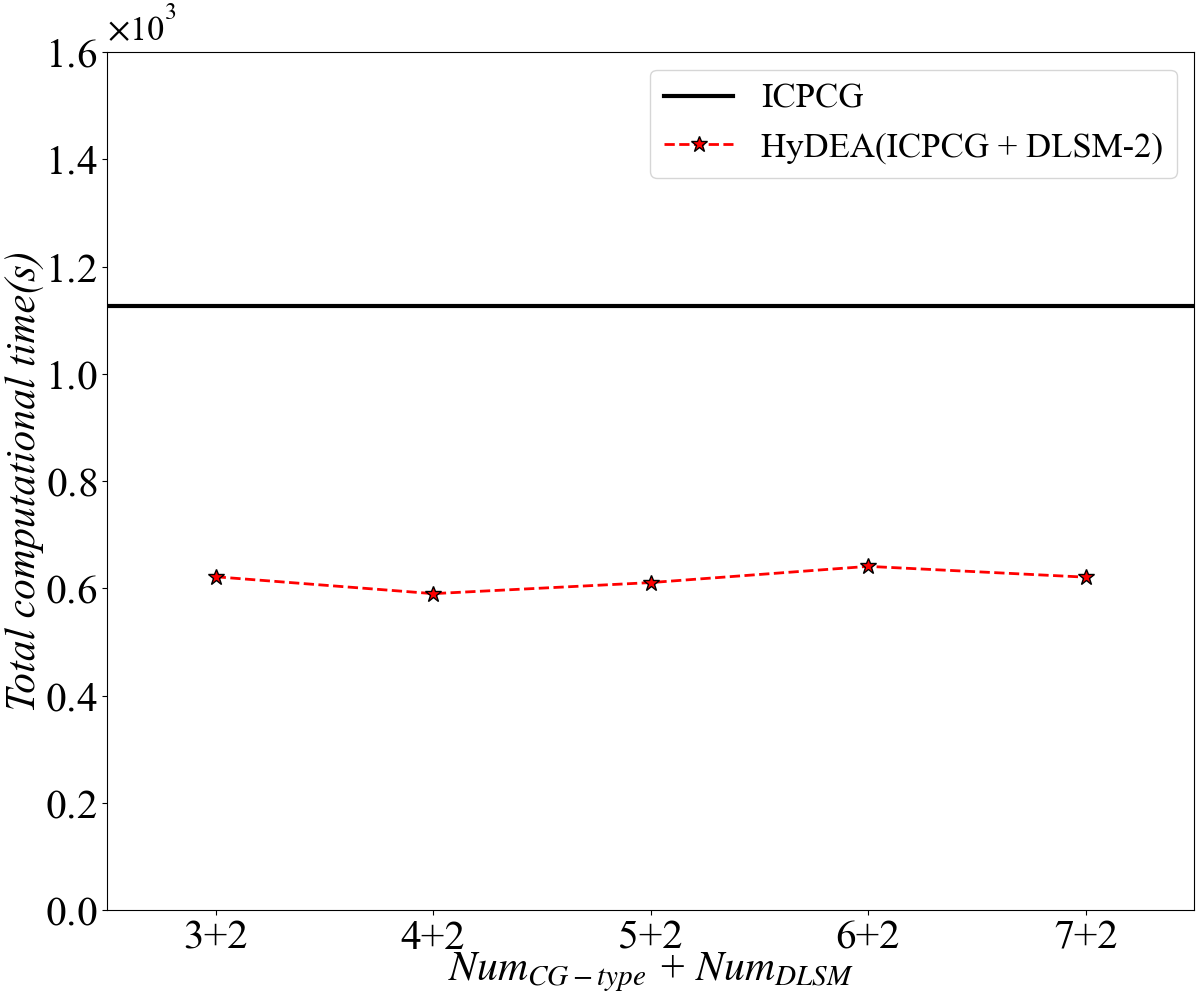}}
  \subfigure[]{
  \label{192Os_3addM_timecompare}
  \includegraphics[scale=0.21]{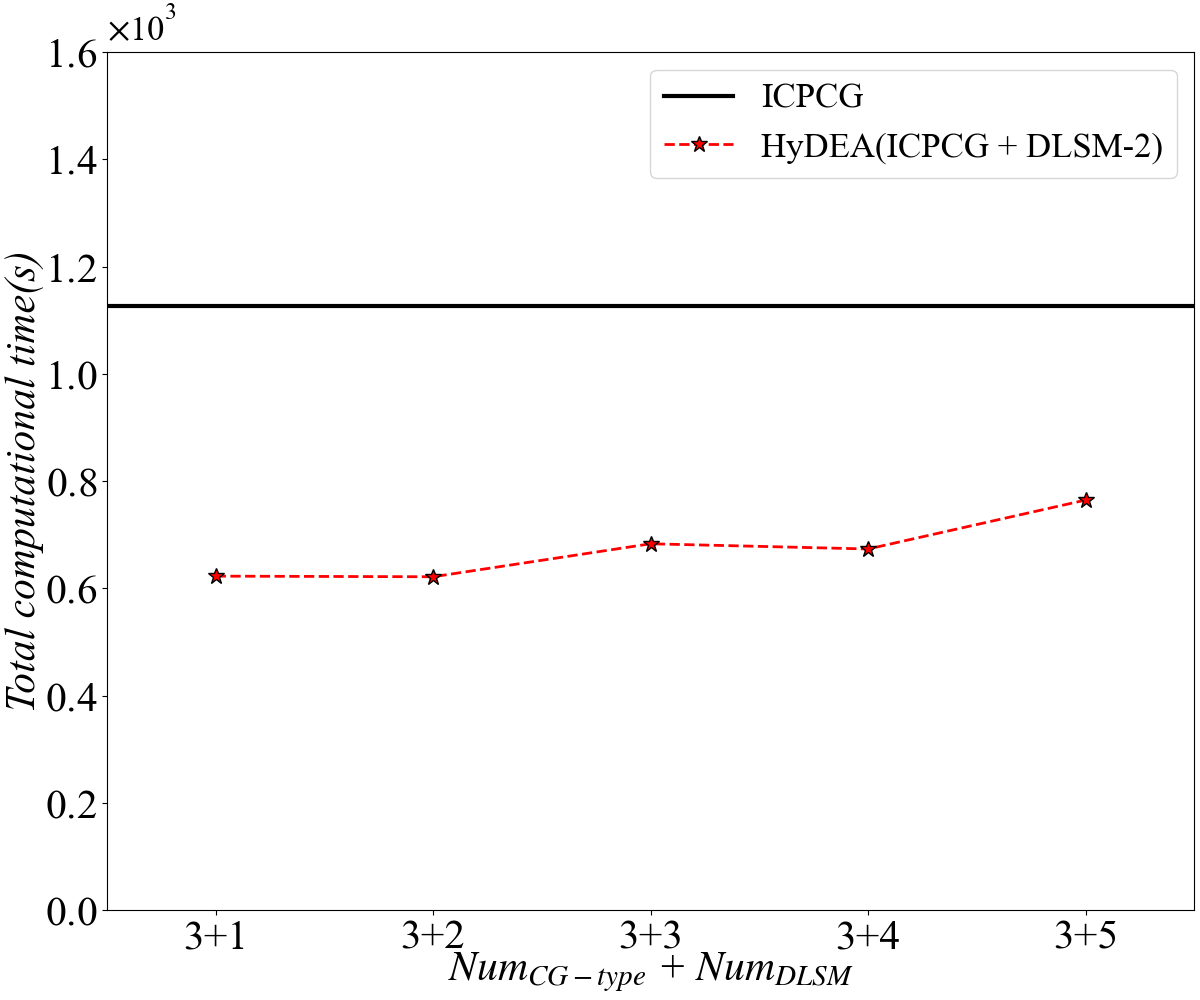}}
  \caption{Comparison of total computational time required to solve the PPE across varying values of $Num_{CG-type}$ and $Num_{DLSM}$ for 2D flow around an inline oscillating cylinder. (a) $Num_{DLSM}=2$, $Num_{CG-type}$ varies across the values $3, 4, 5, 6$ and $7$. (b) $Num_{CG_type}=2$, $Num_{DLSM}$ varies across the values $1, 2, 3, 4$ and $5$.}\label{Oscylinder_timecompare}
\end{figure}

The $Num_{DLSM}$ is fixed at $2$ while varying $Num_{CG-type}$ from $1$ to $7$ to investigate its impact, with the result presented in Fig.~\ref{192Os_Madd2_timecompare}. Notably, we exclude the results of $Num_{CG-type}=1$ and $2$ from the analysis due to their unsatisfactory performance. For values of $Num_{CG-type}$ above $2$, the computational efficiency are comparable.

Furthermore, the $Num_{CG-type}$ is initially fixed at $3$ while varying $Num_{DLSM}$ from $1$ to $5$ to investigate its impact. Fig.~\ref{192Os_3addM_timecompare} presents the total computational time required to solve the PPE over $10,000$ consecutive time steps. The results demonstrate that increasing $Num_{DLSM}$ leads to a slight reduction in overall computational efficiency. This degradation occurs because more frequent consecutive DLSM invocations during the computation. Interestingly, this trend contrasts with the observations in Fig.~\ref{Time compare 128 3addM}, because the lid-driven cavity flow will gradually reaches steady-state conditions. 

Based on the above results, we recommend set $Num_{CG-type}$ to be larger than $Num_{DLSM}$ in HyDEA for complex flow simulations to maintain optimal computational performance.

\section{Sensitivity analysis of iterative termination criterion on HyDEA’s performance}
\label{appendixE}
This section presents a sensitivity analysis to evaluate the impact of iterative termination criterion on the performance of HyDEA. The analysis is conducted based on the flow configuration described in Section~\ref{cavityRe1000}, and more specifically:
\begin{itemize}
 \item 2D lid-driven cavity flow at $Re=1000$ using grid resolution of $128\times128$, with $Num_{CG-type}=3$ and $Num_{DLSM}=2$.
\end{itemize}

To avoid over-extensive comparisons, the HyDEA (MGPCG+DLSM-1) are employed in the forecasting stage.

\begin{figure}[htbp] 
 \centering  
  \subfigure[]{
  \label{128_Rline_1e-5_HMGPCG_10step}
  \includegraphics[scale=0.21]{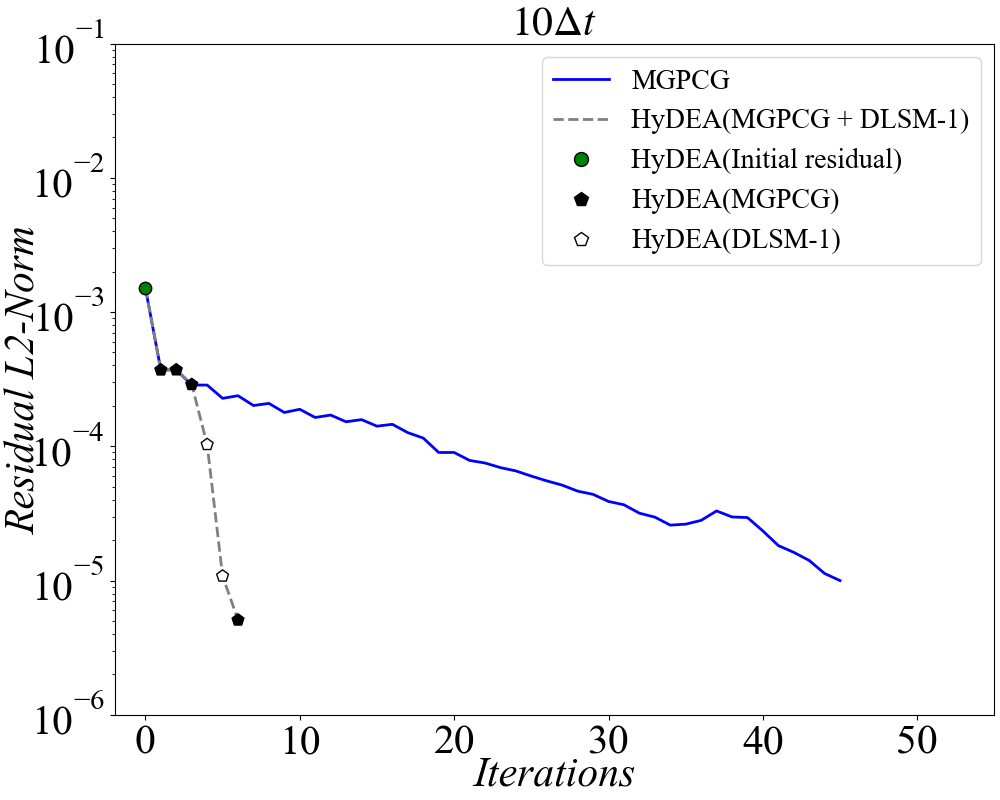}}
  \subfigure[]{
  \label{128_Rline_1e-5_HMGPCG_100step}
  \includegraphics[scale=0.21]{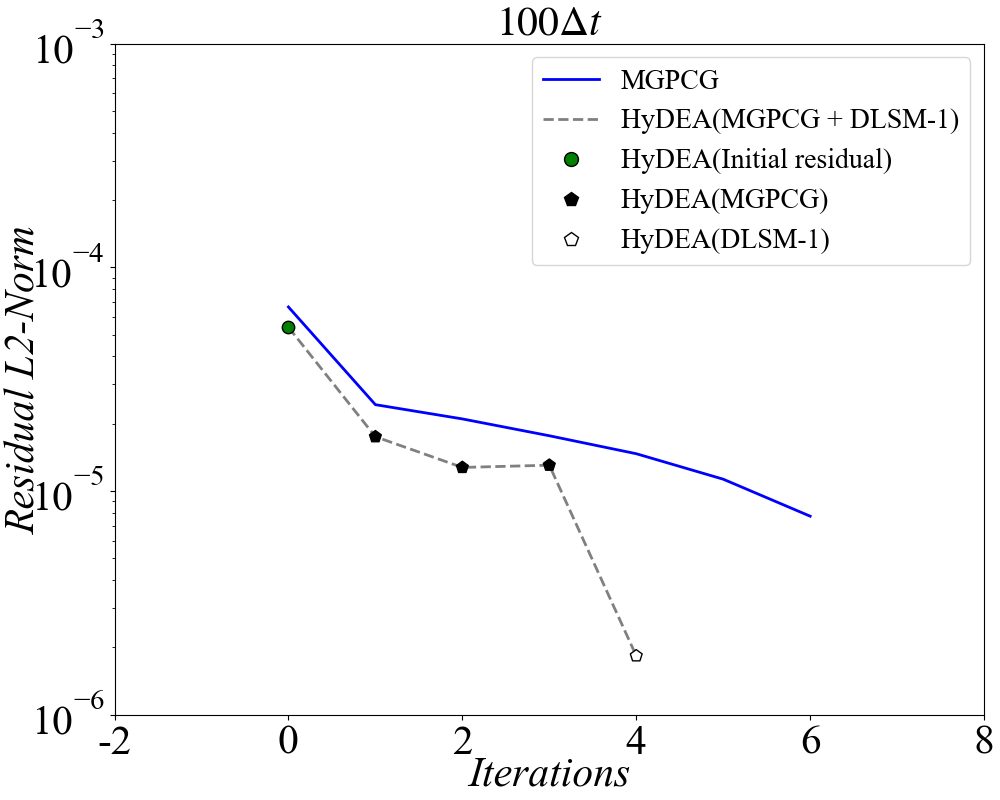}}
  \subfigure[]{
  \label{128_Rline_1e-5_HMGPCG_1000step}
  \includegraphics[scale=0.21]{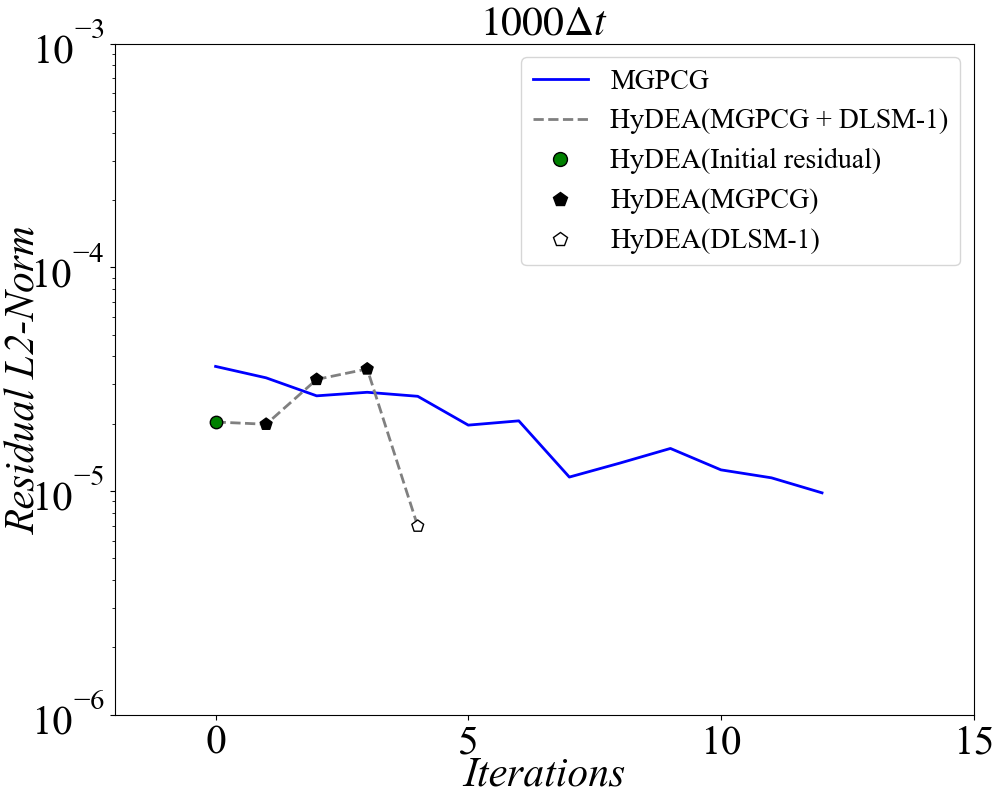}}
  \subfigure[]{
  \label{128_Rline_1e-6_HMGPCG_10step}
  \includegraphics[scale=0.21]{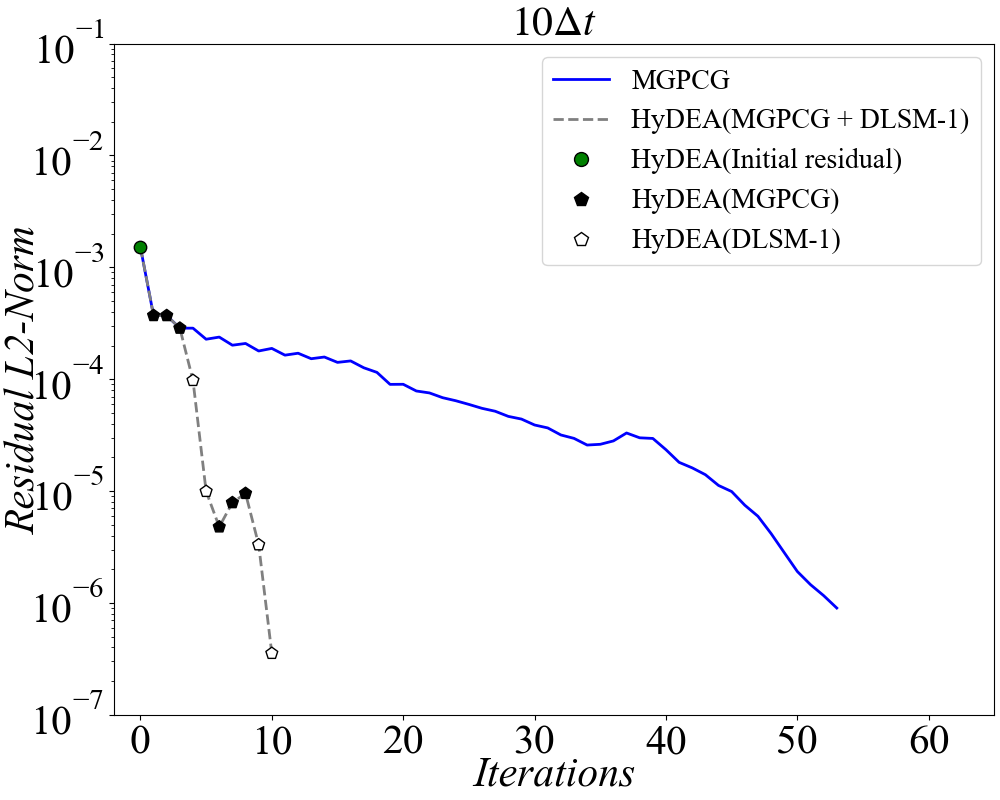}}
  \subfigure[]{
  \label{128_Rline_1e-6_HMGPCG_100step}
  \includegraphics[scale=0.21]{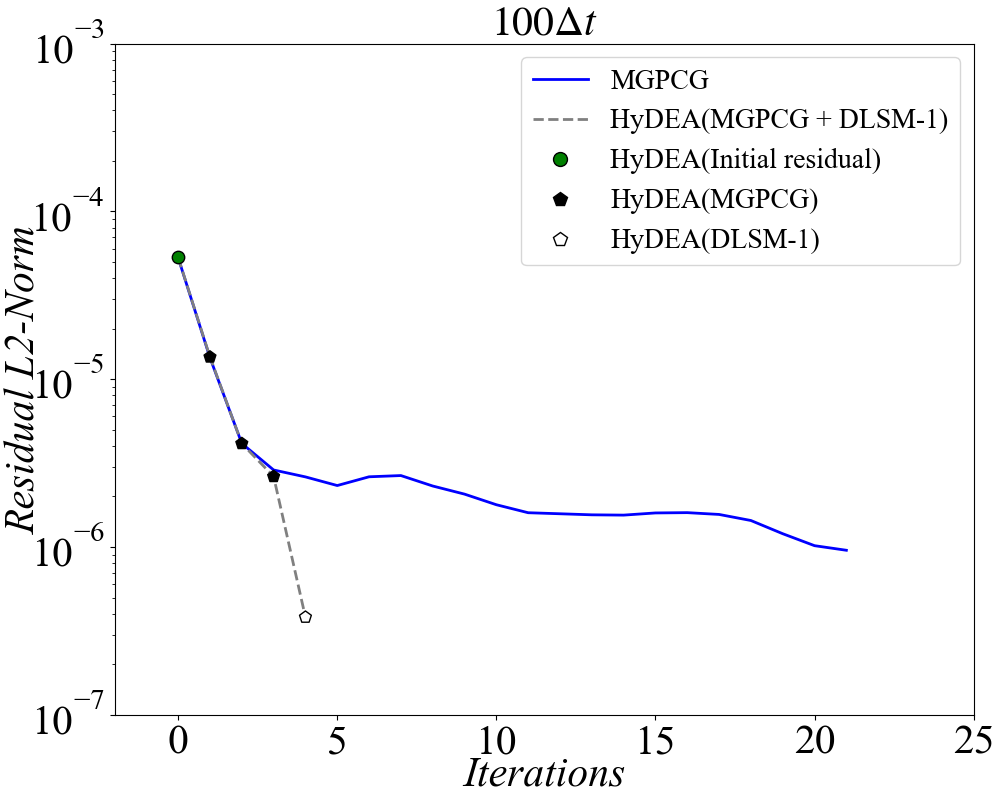}}
  \subfigure[]{
  \label{128_Rline_1e-6_HMGPCG_1000step}
  \includegraphics[scale=0.21]{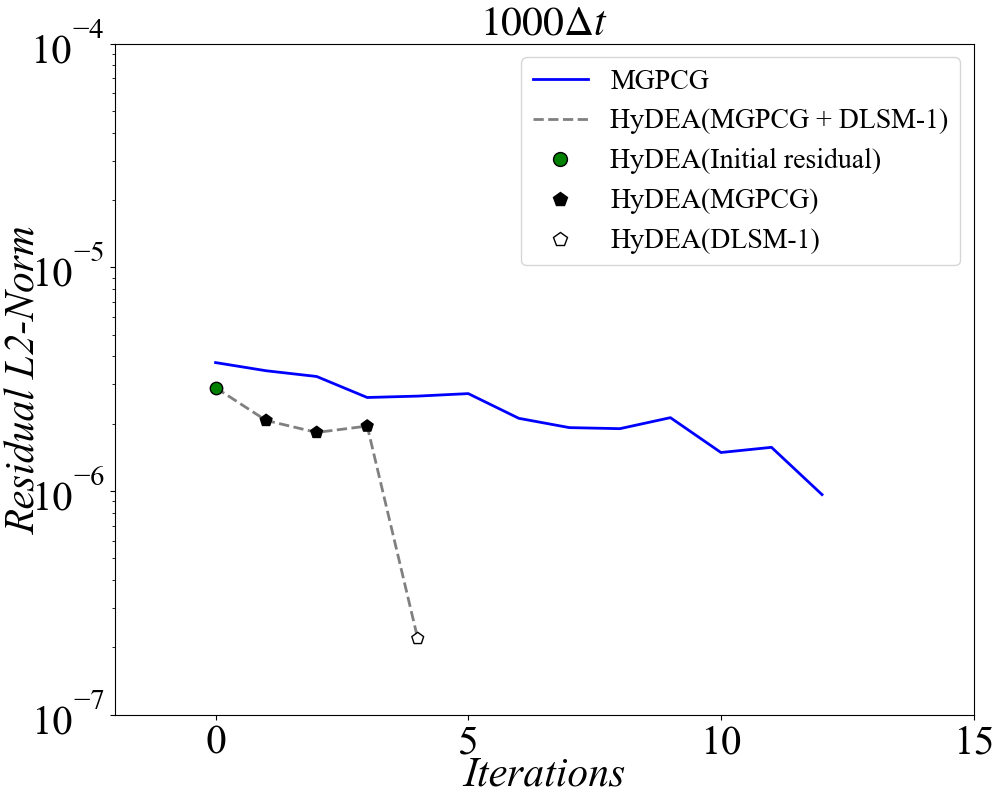}}
  \subfigure[]{
  \label{128_Rline_1e-7_HMGPCG_10step}
  \includegraphics[scale=0.21]{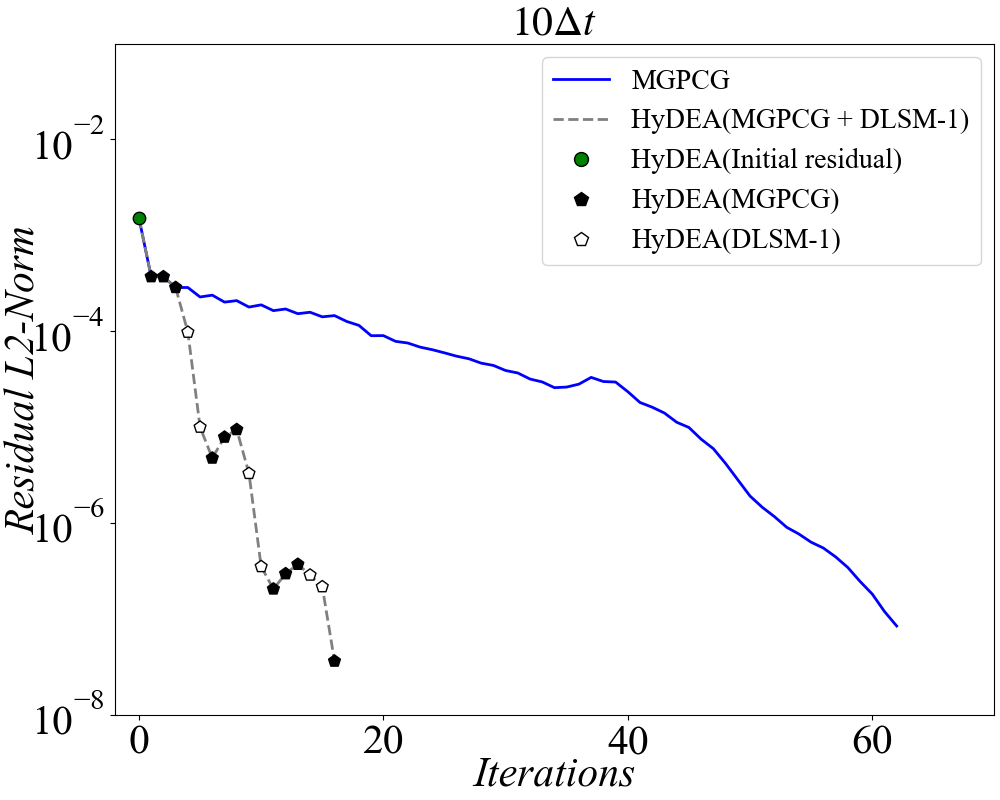}}
  \subfigure[]{
  \label{128_Rline_1e-7_HMGPCG_100step}
  \includegraphics[scale=0.21]{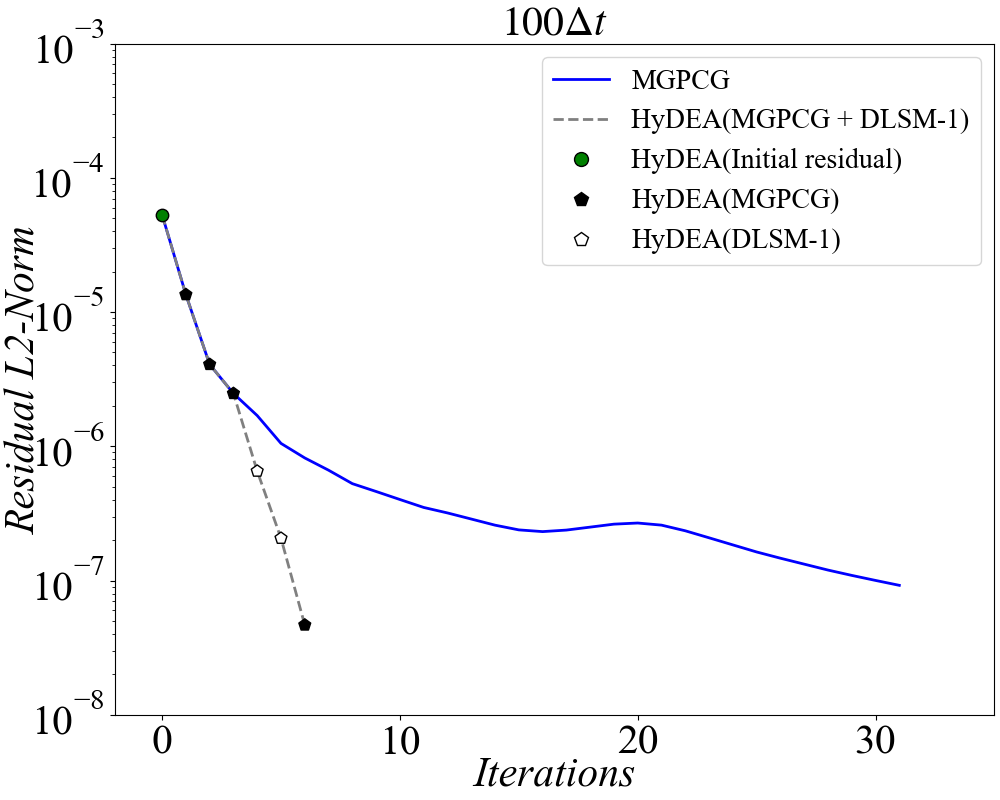}}
  \subfigure[]{
  \label{128_Rline_1e-7_HMGPCG_1000step}
  \includegraphics[scale=0.21]{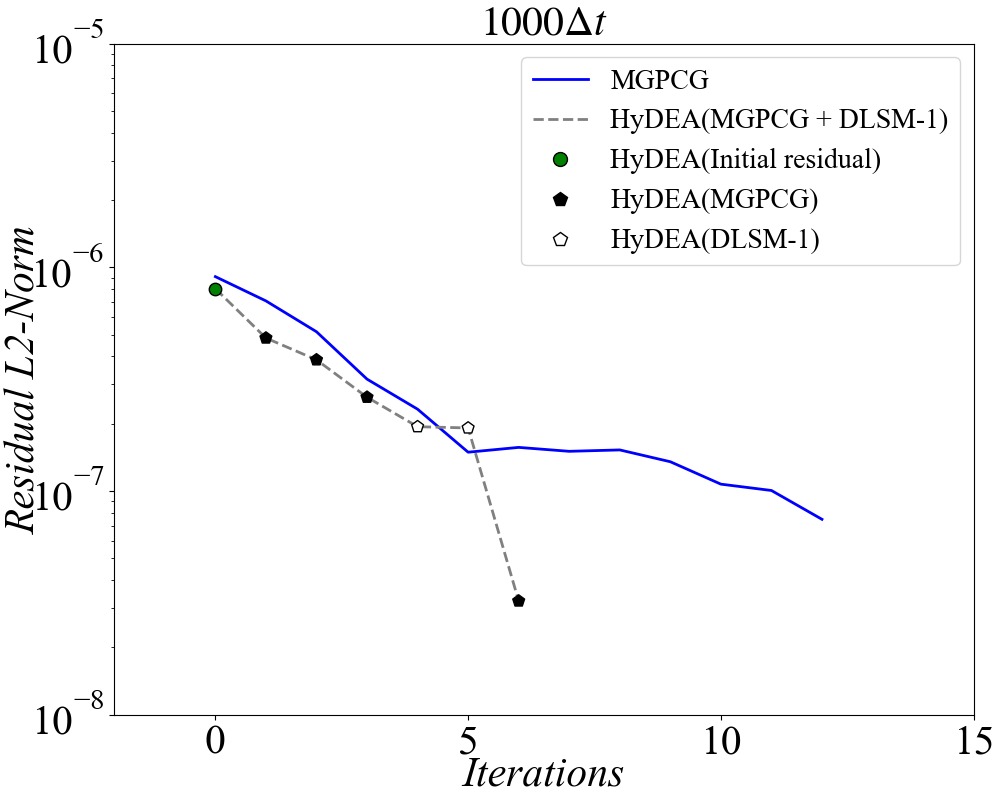}}
  \caption{Iterative residuals of solving the PPE under different iterative termination criteria at the $10th$, $100th$ and $1000th$ time steps.  (a)-(c) $\epsilon=10^{-5}$. (d)-(f) $\epsilon=10^{-6}$. (g)-(i) $\epsilon=10^{-7}$.}\label{128_Rline_HMG_diffcriteria}
\end{figure}

We analyze the impact of the iterative termination criterion expressed in $L2$-norm on the performance of HyDEA by evaluating three different thresholds: $\epsilon=10^{-5}$, $\epsilon=10^{-6}$ and $\epsilon=10^{-7}$. The iterative residuals of solving the PPE using MGPCG and HyDEA (MGPCG+DLSM-1) at three representative time steps are presented in Fig.~\ref{128_Rline_HMG_diffcriteria}.

The iterative residuals of solving the PPE exhibits noticeable discrepancies in later time steps under different iterative termination criteria. These differences stem from the fact that the initial guess for the PPE in each new time step is inherited from the previous step's solution. Notably, HyDEA requires fewer iterations than MGPCG in all cases, demonstrating its superior performance.


\bibliographystyle{elsarticle-num} 
\bibliography{main.bib}





\end{document}